\documentclass[12pt]{article}
\usepackage{amsmath}
\usepackage{amssymb}
\usepackage{amsthm}
\usepackage{graphicx}
\usepackage{graphics}
\usepackage{caption}
\usepackage{subcaption}
\numberwithin{equation}{section}
\usepackage[left=0.95in,right=0.95in,bottom=1.5in]{geometry}
\usepackage{setspace}
\usepackage [linktocpage]{hyperref}
\begin{document}
\title{\bf Comparing Some Nucleon-Nucleon Potentials \\
\ \  }
\author{{\bf M. Naghdi \footnote{E-Mail: m.naghdi@mail.ilam.ac.ir} } \\
\textit{Department of Physics, Faculty of Basic Sciences}, \\
\textit{University of Ilam, Ilam, West of Iran.}}
\date{\today}
 \setlength{\topmargin}{0.1in}
 \setlength{\textheight}{9.2in}
  \maketitle
  \vspace{0.3in}
    \thispagestyle{empty}
    \begin{center}
\textbf{Abstract}
\end{center}

The aim is to compare a few Nucleon-Nucleon (NN) potentials especially Reid68, Reid68-Day, Reid93, UrbanaV14, ArgonneV18, Nijmegen 93, Nijmegen I and Nijmegen II. Although these potentials have some likenesses and are almost phenomenological, they include in general different structures and their own characteristics. The potentials are constructed in a manner that fit the NN scattering data or phase shifts and are compared in this way. A high-quality scale of a potential is that it fits the data with $\chi^{2}/N_{data} \approx 1$, describes well the deuteron properties and gives satisfactory results in nuclear-structure calculations. However, these scales have some failures. Here, we first compare many potentials by confronting them with the data. Then, we try to compare the potential forms by considering the potential structures directly and therefore regarding their substantial bases somehow. To do so, we note that since the potentials are written in different schemas, it is necessary to write them in a unique schema. On the other hand, because three major terms in the NN interaction are central, tensor and spin-orbit terms; so, to perform a reduction plan and arrive at a common structure, we choose the Reid's potential form. Next, we compare the potentials for some states and address some other related issues as well.

  \newpage
  \setlength{\topmargin}{-0.7in}
  \pagenumbering{arabic} 
  \setcounter{page}{2} 

\section{Introduction}
In the past decades, near a century, various models to describe nucleon-nucleon interaction have been framed. In general, one can divide the main models into four categories---Look at \cite{M1N} for a brief typical review on the subject including references therein. \\
For the models based on Quantum Chromo Dynamics (QCD), the main examples are the "constituent quark models" (first in \cite{FredMyhrer}), "skyrme model" (first in \cite{nucl-th/0007051}), Nambu-Jona-Lasinio (NJL) model \cite{Rashdan}, "lattice QCD" models (first in \cite{Beane1}), Moscow-type potentials \cite{Kukulin}, the Oxford potential \cite{Oxford1} and many others. In these models/potentials, in general, the aim is to describe hadron-hadron processes in terms of Quark and Gluon degrees of freedom. \\
Effective Field Theory (EFT) is another outstanding approach to NN problem; Look at \cite{1210.0992, 1302.3241} for recent studies and references therein. By breaking chiral symmetry of QCD Lagrangian in low energies, the main degrees of freedom are not quarks and gluons but there are pions and nucleons. Then, one employees a Chiral Perturbation Theory (CHPT) expansion in terms of the fields up to some orders. Some important $\chi$EFT potentials are presented by \footnote{It is notable that we refer to the first presented version of the potentials by the various groups, while we quote some particular versions depended on the need.} Texas-group (first in \cite{Ordonez1}), Brazil-group (first in \cite{Robilotta1}), Munich-group (first in \cite{Munich1}), Idaho-Group (first in \cite{Idaho1}) and Bochum-Julich-Group (first in \cite{BochumJulich2}). Chiral EFT potentials are nowadays more interesting to stand as the standard NN potentials. \\
The Boson Exchange (BE) models, as the name implies, use various meson exchanges in three NN interaction parts \footnote{The NN interaction was divided into three parts first in \cite{Tokyo1}. The long-range (LR) part ($r\gtrsim 2 fm$) is always represented by a One-Pion-Exchange Potential (OPEP) tail attached to other parts. The intermediate/medium-range (MR) part ($1fm\lesssim r \lesssim 2fm$) is always owed to the scalar-meson exchanges (two pions and heavier mesons). The short-range (SR) part ($r\lesssim 1 fm$) is considered as the vector-bosons exchanges (heavier mesons and multi-pion exchanges as well as the QCD effects).}. Most potentials in this category always employ the field-theoretical and dispersion-relations techniques. Among the first versions are Partovi-Lomon model (first in \cite{PartoviLomon}), Stony Brook-group' potential (first in \cite{Jackson}), the super-soft-core potentials (first in \cite{deTourrei0l}), Funabashi potentials (first in \cite{Funabashi01}), Paris-group (first in \cite{Cottingham}), Bonn-group (first in \cite{Machleidt}), Padua-group (first in \cite{Minelli}), Nijmegen-group (first in \cite{Nijm78}) and Hamburg-group (first in \cite{Jaede}) potentials. The typical potentials such as the Virginia-Group potential \cite{Virgina1}, the Bochum-Group potential \cite{Bochum1} and Tubingen-Group \cite{Tubingen1} are also mentionable. \\
The almost pure phenomenological NN potentials have many free parameters to be fitted with the experimental data. Even with less physical meanings, they are still important and applicable to nuclear-structure calculations. Some famous examples are Hamada-Johnston potential (first in \cite{HamadaJohnston}), Yale-group potential \cite{Yale}, Reid potentials (Reid68 \cite{Reid68}, Reid68-Day \cite{Reid68-Day} and Reid93 \cite{Nijm93} by Nijmegen-group), Urbana-group potentials (e.g. UrbanaV14 \cite{UV14}), Argonne-group potentials (e.g. ArgonneV14 \cite{AV14} and ArgonneV18 \cite{AV18}) and some Nijmegen-group potentials \cite{nucl-th/9509024}. \\
In addition, there are some other typical and special NN potential models. Among them are the potentials based on "Mean Field Theory" (MFT) (first in \cite{Serra}) which are interesting particularly in nuclear many-body calculations. The "Renormalization Group" (RG) approach (first in \cite{nucl-th/0108041}) to NN interaction is also creditable. In RG potentials, by integrating out the high-momentum components of the various potentials, one could arrive in some model-independent low-momentum interactions with satisfactory results.

A definite fact about the models based on QCD (and even the old EFT models) is that they still need more quantitative improvements. These models describe the characteristic phenomena saw in the nucleon-nucleon, pion-nucleon and pion-pion scattering very well qualitatively but they almost fail quantitatively. Still, lattice QCD models give better quantitative results as the previous qualitative descriptions for the short-range part mostly. In general, common features of the "QCD-inspired" models that decrease the demand for them are cumbersome mathematics, large number of parameters and limitations in applications essentially to very low energies.  Nevertheless, nowadays the hybrid quark-meson models give satisfactory results; where for the LR and MR parts, they use the potentials from the other phenomenological and boson-exchange models while just the SR part is discussed with the QCD techniques; as some examples look at \cite{VinhMau1}, \cite{JapanQCDNN}, \cite{Shimizu001}, and \cite{nucl-th/0212044} for a review of the QCD-inspired models. \\
Chiral EFT models have had a successful growth as they now show themselves as the standard two-nucleon and few-nucleon potentials. Their new high-quality potentials, such as the Idaho-Group \cite{Idaho1} and the Bochum-Julich-Group \cite{BochumJulich2} potentials, give the results very well as the famous high-quality phenomenological potentials next to having more theoretical and physical grounds. Indeed, they are becoming the best candidates to describe the NN interaction both qualitatively and quantitatively. Still, the proper renormalization of the chiral nuclear potentials and few-nucleon forces especially in the higher-orders of the chiral expansion are remained to be well addressed with these models; look at \cite{1110.3761, 1210.0992, 1302.3241}.\\
The boson-exchange models have even a further old intimacy with the NN interaction facts. For example, in One-Boson-Exchange (OBE) potentials, for each set of the mesons, a role is given in one part of the interaction. In general, six non-strange bosons, which are the pseudo-scalar mesons $\pi$ and $\eta$, the vector mesons $\rho$ and $\omega$, and two scalar bosons $\delta$ and $\sigma$, where the first meson in each group is isovector while the second is isoscalar, with the masses below 1 GeV, are always considered. The $\pi$ meson provides the tensor force, which is reduced at SR to the $\rho$ meson. $\omega$ creates the spin-orbit force and SR repulsion, and $\sigma$ is responsible for the MR attraction and also provides a good parametrization of $2\pi$ system in S-state. Therefore, it is easy to understand why a model that includes theses four mesons can reproduce the major properties of the nuclear force \cite{Machleidt00}. In these models, besides the mentioned mesons, other different meson exchanges may be also included depending on the case. Then, the strength for any not considered meson exchanges (e.g., multi-meson exchanges) is left as a free parameter to be determined by fitting the NN scattering data. Among the best BE potentials are the parameterized Paris \cite{Paris2} potential, CD Bonn \cite{Bonn4} potential and Nijmegen 93, Nijmegen I, Nijmegen II \cite{Nijm93} potentials.

On the other hand, the most important feature of the phenomenological NN potentials is their simplicity. General form of a potential allowed by the symmetries like rotation, translation, isospin and so on is always considered. There, the SR and MR parts are always determined in a phenomenological manner while for the LR part, OPEP is often used. There are, however, some undesirable problems yet. Three-body forces and relativistic effects are included implicitly in these potentials. Further, the phenomenological models don't give much information about NN dynamics and physics. For example, in a phenomenological potential that uses the Yukawa-type functions as $Y(r)=\frac{g}{4\pi r} \exp{\big(\frac{-m c}{\hbar} r\big)}$, the masses in the exponent and other free parameters are to be earned by fitting the NN scattering data; and similar for more complicated or other type functions. The same is true in most boson-exchange potentials in which some parameters which have physical meaning are free to be determined by fitting the data. That is a weakness because the nuclear force, in principle, should not depend on the external and by hand controls so much. Nevertheless, some failures are indeed unavoidable though they could not decrease from the successes obtained by these potentials. Well reproducing the NN scattering data and neutron properties as well as giving satisfactory results in many nuclear-structure calculations are striking merits of the phenomenological potentials which make them still interesting.

Anyway, among many high-quality potentials, we try to a basic comparison of some NN potentials as another step toward understanding the potential differences and similarities \footnote{We have here considered some early almost phenomenological NN potential forms commonly used in nuclear calculations. Further studies, by including even more new realistic potentials, are to be done later.}. Here, we handle the coordinate-space potentials of Reid68 \cite{Reid68}, Reid68-Day \cite{Reid68-Day}, UrbanaV14 \cite{UV14}, Reid93, Nijmegen 93, Nijmegen I, Nijmegen II \cite{Nijm93} and ArgonneV18 \cite{AV18}. In fact, we recast the exact potentials in one common form, and then compare the potential shapes in some channels to find likenesses and differences. This way may be applicable to many other potentials but probably with other methods to turn them into a nearly common form. This task seems to be important in that is fairly expected to have more acceptable NN models and potentials as the various comparisons then guide us to find the better ones.

The reaming parts of this note are organized as follows. In section 2, we first study the common and important criteria to measure the potential's quality. Then, we compare many old and new potentials by confronting them with the experimental data, where giving a $\chi^{2}/N_{data}$ near the perfect value of 1 is the main criterion. There, we see that almost all famous potentials from 1990's on give satisfactory results with addressing some high-quality potentials. In section 3, we sketch our reduction schema to compare some almost high-quality phenomenological NN potentials. We discuss on the main involved potential structures and details and how to recast them into our wished form. In section 4, based on the plots got for many different channels of the potentials, we make some comparisons among them with addressing their likenesses and differences after reduction of course. There, we also see more features and weakness of the potentials when they are plotted together. In section 4, we make some closing remarks on nucleon-nucleon models, comparing the potentials, problems, challenges and future directions.

\section{Evaluating a Few NN Potentials}
\subsection{The Main Criteria} \label{sub2.1}
The quality measurement of the various NN potentials is possible through several methods. Giving satisfactory results in nuclear-structure calculations and deuteron properties (such as the ratio of D-wave to S-wave, quadratic magnetic-moment, electric quadruple-moment and binding energy) are two outstanding ways. It is of course necessary to mention that, in some potentials, these experimental parameters are used to fit the potential parameters and forms. Reproducing the phase shifts in different channels and comparing them with experimental values is another method for the potential-quality measurement (PQM). Measuring the cross sections and polarizations in \emph{np} scattering at high energies and analyzing power in low energies, next to many spin observations in \emph{pp} scattering, are also the tasks a potential should respond them. Especially, $\chi^{2}$ associated with fitting the experimental NN data by a potential is another desirable parameter for PQM as it is always considered in evaluating many potentials.

In the case of using $\chi^{2}$ for PQM, however, there are some problems \cite{nucl-th/9301019}. For instance, a notable point is that $\chi^{2}$ is not a magic number as its relevance to the "quality" of a potential is indeed limited. For example, for a potential with many parameters but a weak theory with little physics, one may even gain the best fit of the data resulting in the optimized value of $\chi^{2}/datum\approx 1$. On the other hand, there may be a model or potential based on a tight physical theory but includes few parameters with each parameter having a physical meaning. Then, a $\chi^{2}/datum$ of 2 or 3 may be reasonable. Therefore, $\chi^{2}$ is just one aspect among many others that one can consider simultaneously to judge about a NN potential's quality. Other equally important criteria are the theoretical bases of a potential and its off-shell behavior to be tested with the off-shell NN data. The latter aspect is important when one uses the NN potential in nuclear calculations. In fact, a $\chi^{2}/datum$ between 1 and 6 doesn't affect drastically the nuclear-structure results, while the off-shell differences are more important. Meanwhile, occasionally with the high-precision experimental data, $\chi^{2}$ may reflect more or less the inerrancy of the data than the quality of the base theory. Another discussion is that if one can consider the $\chi^{2}$ for PQM, one should include both the \emph{pp} and \emph{np} data and not one of them. An important point here is that a \emph{pp} potential must only be confronted with \emph{pp} data and a \emph{np} potential with \emph{np} data. It is also mentionable that although the NN scattering data are improved in the recent decades for the energy regions of 350 MeV $\leq T_{lab}\leq$ 2 GeV, many potentials are valid up to the pion-production threshold energy of $T_{lab}\approx$ 280 MeV and don't necessarily include inelastic channels.

\subsection{Confronting the Potentials with Data}
\subsubsection{Some Old Potentials} \label{2.2.1}
In a work performed in 1992 \cite{nucl-th/9211013} by Nijmegen-group, some potential forms i.e., Hamada-Johnston potential \cite{HamadaJohnston}, Reid soft-core (Reid68) potential \cite{Reid68}, a super-soft-core potential \cite{deTourreil}, Funabashi potential \cite{Funabashi01, Funabashi02}, Nijm78 potential \cite{Nijm78}, parameterized Paris (Paris80) potential \cite{Paris2}, UrbanaV14 (Urb81) potential \cite{UV14}, ArgonneV14 (Arg84) potential \cite{AV14}, coordinate-space Full-Bonn (or Bonn87) potential \cite{Machleidt} and Bonn89 potential \cite{Haidenbauer}, \cite{Machleidt00} were compared with some \emph{pp} scattering data below $T_{lab}=$350 MeV. Later, they faced the potentials with the \emph{np} (indeed all \emph{pp}+\emph{np}) data in \cite{nucl-th/9411002}. \\
Some potentials don't give good descriptions of the low-energy data mostly below 2 MeV. Although this is partly because of the inaccurate ${}^1S_0$ phase shift, there are other reasons such as fitting to the old data and mainly weak structures of the potentials. Still, one should note that because some potentials are originally fitted to each \emph{pp} or \emph{np} data, it is not so strange to give poor descriptions of the opposite data. \\
The Hamada-Johnston (HJ62) hard-core potential \cite{HamadaJohnston}, such as most similar phenomenological potentials, uses OPEP in the LR part and some potential terms composed of the operators based on the symmetries with radial functions, which in turn have some free parameters to be fitted to the scattering data, in terms of inter-nucleon distances for the MR and SR parts. The HJ62 potential is fitted to its time both \emph{pp} and \emph{np} data; and when meet the Nijmegen 1993 \emph{pp} database , it gives $\chi^{2}/N_{data}=6.1$ in the energy range of 2-350 MeV (we use for most \emph{pp} potentials here) and $\chi^{2}/N_{data}=3.7$ for the \emph{np} data in the energy range of 5-350 MeV (we use for most \emph{np} potentials here). \\
The Ried68 soft-core potential \cite{Reid68} is fitted to both \emph{pp} and \emph{np} data of that time and next to the LR OPEP part, it uses the Yukawa-type functions by the pion masses for each partial-wave up to the angular-momentum of $J\leq2$. The potentials are not regular at the origin because of a $r^{-1}$ singularity. Later, Day extended the Reid68 potential for the upper partial waves. Day81 potential \cite{Reid68-Day} describes the high-energy \emph{pp} data good with $\chi^{2}/N_{data}=1.9$ and \emph{np} data with $\chi^{2}/N_{data}=10.7$ so bad! \\
The super-soft-core (TRS75) potential in \cite{deTourreil} is fitted to both \emph{pp} and \emph{np} data and includes various meson exchanges and uses some step-like cut-off functions to regulate the potentials at the origin. It describes all \emph{pp} data with $\chi^{2}/N_{data}\approx 3.3$ and high-energy \emph{np} data with $\chi^{2}/N_{data}\approx 3.6$. The Funabashi potential \cite{Funabashi01, Funabashi02} is a similar meson-exchange field-theoretical potential as the former \cite{deTourreil} with the bad overall behavior of $\chi^{2}/N_{data}\approx 20$. \\
The charge-independent (CI) Nijm78 potential \cite{Nijm78} includes various meson-exchanges besides using Pomeron and other Regge-pole trajectories. It has both coordinate- and momentum-space versions and with 13 parameters gives a good description of \emph{pp} data with $\chi^{2}/N_{data}\approx 2$. \\
The parameterized Paris potential (Paris80) \cite{Paris2} is a meson-exchange potential that uses the dispersion theory to estimate the intermediate Tow-Pion-Exchange Potential (TPEP). The $\omega$-meson exchange in the SR part is included as a part of three-pion exchange with a special repulsive soft-core potential. The potential includes some static Yukawa functions with 13 originally needed parameters to fit the time \emph{pp}+\emph{np} data. It gives a reasonable description of the low-energy and especially high-energy data in the energy range of 5-350 MeV with $\chi^{2}/N_{data}=2.2$ for \emph{pp} data and with $\chi^{2}/N_{data}=3.8$ for \emph{np} data.\\
The UrbanaV14 (Urb81) potential \cite{UV14} is almost full phenomenological. It includes 14 different potential types which are central, spin-spin, tensor, spin-orbit, quadratic spin-orbit, centrifugal, centrifugal spin-spin as well as other seven ones with dependence on the isospin. For the LR part, as usual, OPEP is used while the MR part is parameterized with a TPEP with 14 parameter; and for the SR part, two Woods-Saxon potentials with 20 parameters are employed. The potentials are regulated with special cut-off functions. Describing \emph{pp} data in the energy range of 5-350 MeV are bad with $\chi^{2}/N_{data}=5.9$; whereas for \emph{np} data in the same range; it is fair with $\chi^{2}/N_{data}=2.7$. That is because the potential was originally fitted to \emph{np} data and not \emph{pp} data. \\
The ArgonneV14 (Arg84) potential \cite{AV14} has similar structure as the Urb81 potential but with fewer parameters and that it is fitted to the time \emph{np} data in the energy range of 25-400 MeV. It provides as improvement compared with Urb81 in the energy range of 5-350 MeV just for the \emph{np} data with $\chi^{2}/N_{data}=2.1$.\\
The Bonn-group comprehensive meson-exchange potentials use various field-theoretical techniques. The potentials are in terms of multiple OBEP and special TPEP (by an energy-independent $\sigma$-meson exchange) parts. The form factors truncate the potentials in the short distances and the SR repulsion comes from the $\omega$-meson exchange. The first version named as Full-Bonn (Bonn87) potential \cite{Machleidt} is in coordinate-space and uses various meson and two-pion exchanges, and is regularized at the origin by the dipole form factors. This potential doesn't describe good all data with giving $\chi^{2}/N_{data}>10$. Its updated version (Bonn89) \cite{Machleidt00}, \cite{Haidenbauer}, gives good description of \emph{pp} and \emph{np} data in the energy range of 5-350 MeV with $\chi^{2}/N_{data}=1.8$ and $\chi^{2}/N_{data}=3$ respectively; while the low-energy data descriptions are not so good. Other Bonn potentials (Bonn-A and Bonn-B) \cite{Machleidt00} with small differences from Bonn87 \cite{Machleidt} are also not satisfactory in describing data.

So far, we see that from the older potentials, only Nijm78 and Bonn89 give satisfactory descriptions of \emph{pp} scattering data in the energy range of 0-350 MeV. By excluding the data of 0-2 MeV, Reid68 and Paris80 give a fair description as well. These potentials reproduce $\chi^{2}/N_{data}\approx 2$ as they encounter the Nijmegen 1992 data \cite{nucl-th/9211013}. When confronting with the Nijmegen 1994 \emph{np} scattering data \cite{nucl-th/9411002}, just Arg84 and Nijm93 (we describe later) give $\chi^{2}/N_{data}\approx 2$ while Urb81 and Bonn89 give $\chi^{2}/N_{data}\approx 3$ for the energies of 5-350 MeV. The other almost old potentials give a large or very large contribution to $\chi^{2}$ especially in the low-energy region.\\
A reason to don't reproduce so much good results by some potentials, when facing either \emph{pp} or \emph{np} data, is that they are fitted only to \emph{np} or \emph{pp} data respectively, or to \emph{pp}+\emph{np} data. Second, some data, to which the original potentials are fitted, are old and incomplete nowadays; and third and maybe the most important one is that, some potentials have weak theoretical structures and bases. Fourth, about different results for very low energies, we first note that the \emph{pp} $^{1}S_{0}$ phase shift in the energies of KeV-2 MeV is very good known; therefore, a small deviation for $^{1}S_{0}$ predicted by a potential give rises to a large contribution to $\chi^{2}$. Nevertheless, the last contribution should not be too large because most potentials suppose to give good descriptions of the scattering-length and effective-range parameters. This means that the other phase shifts should often be improved to earn a better fit.\\
Indeed, one should note that, to give reasonable results, the potentials are necessary to fit both \emph{pp} and \emph{np} data because a good fit to \emph{pp} (\emph{np}) data dose not automatically guarantees a good fit to \emph{np} (\emph{pp}) data. An important conclusion is that only the potentials that are explicitly fitted to \emph{pp} (\emph{np}) data give reasonable descriptions of \emph{pp} (\emph{np}) data. Therefore, we assume that some potentials are not in fact NN potential but \emph{pp} or \emph{np} potential. For instance, Reid68, Nijm78, Paris80 and Bonn89 may be called \emph{pp} potentials while Urb81 and Arg84 may be called \emph{np} potentials. Meanwhile, we should again mention that the potentials such as HJ62, Bonn87 and Bonn A and Bonn B don't describe well the \emph{pp} and \emph{np} scattering data.

\subsubsection{New High-Quality Potentials} \label{2.2.2}
Nijmegen Partial-Wave-Analysis (PWA) \cite{NijmPWA} improved more the NN phase shift analysis. The analysis was indeed a potential analysis, where the final phase shifts were the ones predicted by some "optimized" partial-wave potentials. In PWA's, a SR and a LR part with a separation line in $r=1.4 fm$ are considered. The LR part in turn includes a detailed electromagnetic part and a detailed nuclear part---It is notable that in Nijmegen potentials, the mass differences between the charged and neutral pions and between proton and neutron are included; and because of their special SR parameterizations, the potentials are in contact with QCD. In the overall Nijmegen analysis in 1993 (PWA93) \cite{NijmPWA}, for 1787 \emph{pp} data and 2514 \emph{np} data below $T_{lab}=$350 MeV, the "perfect" result of $\chi^{2}/N_{data}=0.99$ is obtained. Later, they performed another PWA up to the energy of 500 MeV that is above the pion-production threshold \cite{NijmPWA3}, where the more updated data, inelasticity's and some other effects are included as well. For two newer PWA of \emph{pp}+\emph{np} data look also at \cite{0706.2195} and \cite{1304.0895}. \\
Other generation of the Nijmegen-group potentials are Nijm93, NijmI, NijmII \cite{Nijm93}. These potentials are based on the soft-core Nijm78 potential \cite{Nijm78}. Nijm93, as a nonrelativistic meson-exchange potential, is an updated version of Nijm78, where the low-energy NN interaction is based on Regge-Pole theory. This potential includes the charge-dependent (CD) terms, 13 parameters and exponential form factors. It gives a good description of both \emph{pp} and \emph{np} data from 0-350 MeV with $\chi^{2}/N_{data}\approx 1.9$. The NijmI potential includes momentum-dependent terms that lead to nonlocal structure of the potential in the configuration-space. In other words, the local representation of OPE part is preserved while the tracks of non-localities are included in the MR and SR parts by computing the second-order Feynman diagrams of the OBE parts. On the other hand, the NijmII nonrelativistic potential is fully local. In both the latter potentials, all 41 parameters are adjusted separately for each partial-wave, and at very short distances the exponential form factors are used for regularization. The potentials fit all data well with $\chi^{2}/N_{data}\approx 1.03$ and so have high-quality. For a more recent generation of the high-quality Nijmegen (extended-soft-core) potentials look at \cite{NijmESC2}.

On the other hand, the first disadvantage of the Reid68 \cite{Reid68} potential is the poor quality of the \emph{np} data at the time of its construction. Another point is its $r^{-1}$ singularity and then its Fourier transform into the momentum-space. To transform, the singularities are regularized by dipole form factors in Reid93 \cite{Nijm93}. Here, OPEP is included besides the mass difference between the neutral and charged pions. In Reid93, the potentials are parameterized for each partial-wave separately by combinations of the central, spin-orbit and tensor parts (with the local Yukawa functions) including the associated operators, while in Nijm93 the potential forms are same for all partial waves. With 5 phenomenological parameters, it gives a good description of all data with $\chi^{2}/N_{data}\approx 1.03$ and deuteron properties as the other high-quality Nijmegen potentials.

The ArgonneV18 (Arg94) potential \cite{AV18} is a local potential that includes an electromagnetic (EM) part, a proper OPEP for the LR part that is regularized at the short distances, and a phenomenological parameterizations for the MR and SR parts with aid of the local Woods-Saxon potentials. The EM part is similar to that used in the Nijmegen PWA93 next to including the short-range terms and finite-size effects. The core functions are effective in $r=0.5fm$. The operators in Arg94 are more (eighteen) compared with a typical nonrelativistic OBE potential and also compared with the similar older phenomenological potentials such as Urb81 \cite{UV14} and Arg84 \cite{AV14}. With 40 adjustable parameters, it gives $\chi^{2}/N_{data}\approx 1.03$ for 4301 \emph{pp} and \emph{np} data in the energy range of 0-350 MeV. In a later study \cite{AV18pq}, another extension of the Arg94 potential was made (called ArgonneV18pq potential), where various quadratic momentum-dependence in the NN potentials were included to fit the data in the high partial waves with their effects in some nuclear applications.

And the last and best version on the trail of the Bonn-group potentials is the CD-Bonn potential \cite{Bonn4}. It is again based on relativistic meson-exchange theory. The charge-dependence (CD) and charge-symmetry breaking are included in all partial waves with $J\leq4$. The charge-symmetry breaking is because of the OPE part of the potential and differences between the neutral- and charged-pion masses. This potential has a nonlocal structure arising from the covariant Feynman amplitudes. The potential may be called phenomenological because of fine-tuning the partial waves to earn a wished $\chi^{2}$ per datum. It fits 2932 \emph{pp} data below 350 MeV available in 2000 with $\chi^{2}/N_{data}=1.01$ and 3058 \emph{np} data with $\chi^{2}/N_{data}=1.02$, and so has high-quality as well.

Now, among these newer potentials, which are fitted to \emph{pp}+\emph{np} data, the Nijm93 potential has indeed the lowest quality with $\chi^{2}/N_{data}\approx 2$. Other potentials, which are NijmI, NijmII, Reid93, CD-Bonn and Arg94 potentials as well as the Nijmegen PWA93, all give $\chi^{2}/N_{data}\approx 1$ that marks their high-quality. Still, there are some other typical potentials that we address briefly in the next paragraph.\\
The Padua-group NN potential \cite{Minelli}, which is based on meson-exchange theory by employing the phenomenological terms, describes the phase shifts and deuteron properties similar to the Paris80 \cite{Paris2}, Arg84 \cite{AV14} and Bonn-A \cite{Machleidt00} potentials. The Virginia-group potentials \cite{Virgina1}, as some special relativistic OBE models, have almost the same quality to fit the data as the Bonn87 \cite{Machleidt} and Arg84 \cite{AV14} potentials. The Bonn-B potential \cite{Machleidt00} was starting point to build one-solitary-boson-exchange potential (OSBEP) by Hamburg-group (Ham95) \cite{Jaede}. It is shown \cite{Hamburg2a} that Ham95 potential describes the deuteron properties and the scattering data by Arndt et al. \cite{nucl-th/9706003} similar to the Bonn-B potential. In fact, with 8 parameters, it describes 1292 \emph{pp} data in the energy range of 1-300 MeV with $\chi^{2}/N_{data}= 6.8$ and 2719 \emph{np} data in the energy range of 0-300 MeV with $\chi^{2}/N_{data}= 4.1$ near the Nijm93 results. Other potentials such as the Bochum-group potential \cite{Bochum1} that uses meson exchange for the long distances and takes attention to the nuclear structure in the shorter distances, Moscow-group potentials \cite{Kukulin}, \cite{Moscow3} that use a hybrid of the quark-model of QCD and the meson-exchange picture and the Oxford potential \cite{Oxford1} as a QCD-inspired potential, claim to provide good descriptions of NN data. So, in general, there are many high-quality models and potentials based on meson exchanges, QCD, and especially more recent chiral EFT potentials that we mention some others bellow.

\subsubsection{A Summary}
In summary, for the approach in which the criteria in subsection \ref{sub2.1} and mainly $\chi^{2}$ is considered for the potential-quality measurement (PQM), we can make the following statements. Great progress on the NN data quality was achieved by Nijmegen-group in 1990's when more focus was started on the quantitative aspects of the NN potentials as well. Even the best NN potentials of 1980's, such as Paris80 \cite{Paris2}, Urb81 \cite{UV14}, Arg84 \cite{AV14} and Bonn89 \cite{Haidenbauer}, \cite{Machleidt00}, fit NN data typically with at least $\chi^{2}/N_{data}= 2$ that is above the perfect or wished value of $\chi^{2}/N_{data}= 1$. A more completed and updated NN database by Nijmegen-group \cite{NijmPWA}, \cite{nucl-th/9211013}, \cite{nucl-th/9411002} made more opportunities to build better potentials as discrepancies in the predictions could not be blamed on the bad fitting of the scattering data. Then, some new CD NN potentials were constructed in the early and mid 1990's. There are NijmI, NijmII and Reid93 potentials \cite{Nijm93} by Nijmegen-group, Arg94 potential \cite{AV18} and CD-Bonn potential \cite{Bonn4}. All these potentials have about 45 parameters and fit NN data with nearly $\chi^{2}/N_{data}\approx 1$, and so are the high-quality NN potentials.

On the other hand, there are satisfactory results from some chiral EFT potentials. Indeed, by using the same \emph{np} data as used in the CD-Bonn potential \cite{Bonn4}, the potentials in next-to leading (NL) and next-to-next-to leading (NNL) orders of the chiral expansion, made by Bochum-Julich-group, give a large $\chi^{2}/N_{data}$ for the data below 350 MeV. However, in another development, they set up an NNNLO potential \cite{BochumJulich2} whose parameters were fitted to the \emph{pp} and \emph{np} Nijmegen phase shifts \cite{NijmPWA} and the \emph{nn} scattering-length. This new potential gives a rather good description of the \emph{np} data with $\chi^{2}/N_{data}\geq 1.7$ and a rather bad description of the \emph{pp} data with $\chi^{2}/N_{data} \geq 2.9$ in the energy range of 0-290 MeV. 
It is notable that as the energy decreases, the dada description by the potential becomes better and better. Still, the Idaho-group CHPT potentials appear to have higher quality. Indeed, the Idaho NNNLO potential (Idaho03) \cite{Idaho2}, reproduces the \emph{np} and \emph{pp} scattering data with almost $\chi^{2}/N_{data} \approx 1.1$ and $\chi^{2}/N_{data} \approx 1.5$ for the energy range of 0-290 MeV, respectively. So, the Idaho03 potential is another high-quality potential such as NijmI, NijmII, CD-Bonn, Reid93 and Arg94 at least in describing the NN scattering data and deuteron properties.

We should stress that most potentials use one-pion-exchange-potentials (OPEP's) for the long-range part while correlated tow-pion-exchanges (TPE's) and other meson-exchanges (always OBE's) are employed for the intermediate-range part. For the short-range part, the heavy vector-boson exchanges and QCD effects or the phenomenological procedures are often used. Among the high-quality potentials, NijmII, Reid93, and Arg94 potentials are non-relativistic with local functions that couple to the (nonrelativistic) operators composed of the various spin, isospin and angular-momentum of the two-nucleon pairs. This approach is simplest for calculations in the coordinate-space. The NijmI potential also includes the $p^2$ terms attributable to the nonlocal contributions to the central force. The CD-Bonn potential is based on the relativistic meson-exchange theory and is nonlocal with including more momentum-dependent terms. In the Idaho03 potential, based on chiral perturbation theory (CHPT), mesons and quarks degrees of freedom are included and its quality is high as the NijmII, Reid93, Arg94 and CD-Bonn potentials---For some tests of above high-quality potentials in nuclear-structure calculations look, for instance, at \cite{nucl-th/9707002}, \cite{nucl-th/0407003}.

Now, according to above discussions, we try to compare some potentials in a different and fairly substantial way, which is by considering their structures directly. Before doing so, we mention a plain comparison of some potentials in \cite{nucl-th/0305035} slightly similar to the procedure we use here. Indeed, they have arrived in some effective low-momentum potentials by applying the renormalization-group (RG) methods to the potentials of Paris80, Bonn-A and CD-Bonn, Nijm I and Nijm II, Arg94, and Idaho03. Then, the resultant potentials have compared with the model-independent RG potentials that reproduce the experimental phase shifts up to $T_{lab}$=350 MeV. The last comparison confirms the results outlined above from confronting the potentials with data, nearly.

\section{Reducing Some Potentials into Reid Potential}
\subsection{The Basic Sketch}
Among various NN potentials mentioned in the last sections, we here consider the forms of Ried68 potential \cite{Reid68} and its extended version to higher orders by Day that is Day81 potential (or Full-Reid potential) \cite{Reid68-Day}, Reid93 potential \cite{Nijm93}, UrbanaV14 (Urb84) potential \cite{UV14}, ArgonneV18 (Arg94) potential \cite{AV18} and Nijm93, NijmI, NijmII potentials \cite{Nijm93}. These potentials are almost the phenomenological and boson-exchange ones, where the latter is the most important candidate for giving a true picture of NN interaction nowadays with more confirmations form chiral EFT as well. In another moment, we try to extend above list to include more potentials.

Reid in 1968 parameterized potentials in each partial-wave up to $J\leq2$ separately \cite{Reid68}. He used a central potential for the singlet- and triplet-uncoupled states while for triplet-coupled states, a potential with central, tensor and first order spin-orbit forces was used as
\begin{equation} \label{GrindEQ__1_}
V=V_{C}(r)+ V_{T}(r) S_{12}+ V_{SO}(r)\vec{L}.\vec{S}
\end{equation}
where $S_{12}= 3(\vec{\sigma}_{1}.\hat{r}) (\vec{\sigma}_{2}.\hat{r})-(\vec{\sigma}_{1}.\vec{\sigma}_{2} )$, $\vec{S}=(\vec{\sigma}_{1}+\vec{\sigma}_{2})/2$ and $\vec{L}.\vec{S}$ are the usual tensor, spin and spin-orbit operators, respectively. In Reid68, for the long-range OPEP
\begin{equation} \label{eq01B}
V_{OPEP} = \left(\frac{g_{pi}^{2}}{12} \right) m_{pi}c^2 \left(\frac{m_{pi}}{M} \right)^{2} (\vec{\tau}_{1}.\vec{\tau}_{2}) \left[(\vec{\sigma}_{1}.\vec{\sigma}_{2}) + S_{12} \left(1+\frac{3}{x} + \frac{3}{x^{2}} \right)\right] \frac{e^{-x}}{x}.
\end{equation}
where $m_{pi}=$138.13 MeV and $M=$ 938.903 MeV are used for the pion and nucleon mass respectively, $g_{pi}^{2}=14$ with $g_{pi}$ for the pion-nucleon coupling constant, and $x=\mu r$ with $\mu=m_{pi} c/{\hbar}=r_0^{-1}$ ($\mu=0.7 fm^{-1}$ here) is the inter-nucleon distance measured in the unit of the pion Compton's wavelength. There, to remove $x^{-2}$ and $x^{-3}$ behaviors at small $x$, a short-range interaction is also subtracted. The lack of the soft-core versions is that the potentials still have a $x^{-1}$ singularity at origin. The intermediate-range potentials are given by a sum of the Yukawa-type functions as $e^{-n x}/x$, where $n$ is an integer. Meanwhile, the SR repulsive part is given by an average of both very hard-core and (Yukawa) soft-core potentials. In 1981, Day extends the Reid68 potential roughly for the states with $J\geq3$ up to $J=5$ \cite{Reid68-Day}.

Now, for a structural comparison of the potentials, we reduce the mentioned potentials for all uncoupled and coupled states to the Reid potential structure. Because the prime Reid potential includes three terms as central, tensor and spin-orbit (\ref{GrindEQ__1_}); so after the reduction schema, all terms in the potentials reduce to these two central and non-central parts. The most important reason for doing so is that because not only the main terms in a potential are these three terms but also, by having a similar operator form for all potentials, one can somehow compare the potential structures.

\subsection{Reducing UrbanaV14 Potential into Reid Potential}
The UrbanaV14 potential \cite{UV14} is still using in some nuclear-structure calculations. Its two-nucleon interaction reads
\begin{equation} \label{GrindEQ__2_}
V_{Urb.}=\sum_{i=1}^{14} \left(V_{L}^{i}(r) + V_{M}^{i}(r) + V_{S}^{i}(r)\right) O_{i},
\end{equation}
where $V_{L}$, $V_{M}$ and $V_{S}$ stand for the LR, MR and SR part potentials, and $O_i$'s are 14 conveniently chosen operators that we indicate them as $c$ (for central), $\sigma$ (for spin), $\tau$ (for isospin), $\sigma \tau$, $t$ (for tensor), $t \tau$, $ls$ (for spin-orbit), $ls \tau$, $ll$ (for quadratic-orbit), $ll \sigma$ , $ll \tau$, $ll \sigma \tau$, $ls2$ (for quadratic spin-orbit) and $ls2 \tau$, respectively. \\
The LR part potential reads
\begin{equation} \label{GrindEQ__3_}
V_{L} =V_{\pi}^{\sigma \tau}(r)(\vec{\sigma}_{1}.\vec{\sigma}_{2})(\vec{\tau}_{1}.\vec{\tau}_{2}) + V_{\pi}^{t\tau}(r)S_{12} (\vec{\tau}_{1}.\vec{\tau}_{2}),
\end{equation}
and for MR part, the contribution is
\begin{equation} \label{GrindEQ__4_}
\begin{split}
 V_{M} & =T_{\pi}^{2}(r)\big(I^{c}+I^{\sigma} (\vec{\sigma}_{1}.\vec{\sigma}_{2})+I^{\tau} (\vec{\tau}_{1}.\vec{\tau}_{2})+I^{\sigma \tau} (\vec{\sigma}_{1} .\vec{\sigma}_{2})(\vec{\tau}_{1}.\vec{\tau}_{2}) \\
 &+I^{t} S_{12}+I^{t\tau} S_{12} (\vec{\tau}_{1}.\vec{\tau}_{2})+I^{ll} L^{2}+I^{ll\sigma} L^{2} (\vec{\sigma}_{1}.\vec{\sigma}_{2})+I^{ll\tau} L^{2} (\vec{\tau}_{1}.\vec{\tau}_{2}) \\
 &+I^{ll\sigma \tau} L^{2} (\vec{\sigma}_{1}.\vec{\sigma}_{2})(\vec{\tau}_{1}.\vec{\tau}_{2})+I^{ls2} (\vec{L}.\vec{S})^{2}+I^{ls2\tau} (\vec{L}.\vec{S})^{2} (\vec{\tau}_{1}.\vec{\tau}_{2})\big),
\end{split}
\end{equation}
and also for SR part, it becomes
\begin{equation} \label{GrindEQ__5_}
\begin{split}
V_{S} &=W(r) \big(S^{c}+S^{\sigma} (\vec{\sigma}_{1}.\vec{\sigma}_{2})+S^{\tau} (\vec{\tau}_{1}.\vec{\tau}_{2})+S^{\sigma \tau} (\vec{\sigma}_{1}.\vec{\sigma}_{2})(\vec{\tau}_{1}.\vec{\tau}_{2}) \\
& +S^{ls} (\vec{L}.\vec{S})+S^{ls\tau} (\vec{L}.\vec{S})(\vec{\tau}_{1}.\vec{\tau}_{2})+S^{ll} L^{2}+S^{ll\sigma} L^{2} (\vec{\sigma}_{1}.\vec{\sigma}_{2}) \\
& +S^{ll\tau} L^{2} (\vec{\tau}_{1}.\vec{\tau}_{2})+I^{ll\sigma \tau} L^{2} (\vec{\sigma}_{1}.\vec{\sigma}_{2})(\vec{\tau}_{1}.\vec{\tau}_{2})+I^{ls2} (\vec{L}.\vec{S})^{2} +I^{ls2\tau} (\vec{L}.\vec{S})^{2} (\vec{\tau}_{1}.\vec{\tau}_{2})\big) \\
& +\grave{W}(r)\big(\grave{S}^{ls} (\vec{L}.\vec{S})+\grave{S}^{ls\tau} (\vec{L}.\vec{S})(\vec{\tau}_{1}.\vec{\tau}_{2})\big).
\end{split}
\end{equation}
One should note that in above relations,
\begin{equation} \label{GrindEQ__5a_}
V_{\pi}^{\sigma \tau}(r)= 3.488 \frac{e^{-0.7r} }{0.7r} \left(1-e^{-cr^{2}} \right),
\end{equation}
\begin{equation} \label{GrindEQ__5b_}
V_{\pi}^{t\tau}(r)= 3.488 \left(1+\frac{3}{0.7r}+\frac{3}{(0.7r)^{2}} \right) \frac{e^{-0.7r}}{0.7r} \left(1-e^{-cr^{2}} \right)^{2}= 3.488 T_{\pi}(r),
\end{equation}
\begin{equation} \label{GrindEQ__5c_}
W(r)=\left(1+\exp \left(\frac{r-R}{a} \right)\right)^{-1}, \quad \grave{W}(r)=\left(1+\exp \left(\frac{r-\grave{R}}{\grave{a}} \right)\right)^{-1},
\end{equation}
where the cutoff parameter $c$ and the strengths $I^{i}$, $S^{i}$, $\grave{S}^{i}$ are determined by fitting to the scattering data (phase shifts). Indeed, the parameter values of $c=0.2 fm^{-2}$, $R=0.5 fm$, $a=0.2 fm$, $\grave{R}=0.36 fm$, $\grave{a}=0.17 fm$ are used in the Urb84 potential. \\
In our reduction schema, we now estimate the expectation values for all operators in a particular state and so, as in the Reid potential, finally have just a radial function of $r$ for uncoupled states. Therefore, for an uncoupled state, e.g. $^{3}P_{0}$, after a little calculation, we get
\begin{equation} \label{GrindEQ__6_}
V({}^{3}P_{0})=2694.69 W(r)+4400 \grave{W}(r)-3.6 T_{\pi}^{2}(r)+V_{\pi}^{\sigma \tau}(r)-4V_{\pi}^{t\tau}(r),
\end{equation}
and for a coupled state, e.g. $^{3}S_{1}-^{3}D_{1}$, at the end, we get
\begin{equation} \label{GrindEQ__7_}
\begin{split}
V({}^{3}S_{1}-{}^{3}D_{1}) & =\left(2399.99 W(r)-6.8008 T_{\pi}^{2}(r)-2 V_{\pi}^{\sigma \tau}(r)\right) \\
& +\left(0.75 T_{\pi}^{2}(r)-3 V_{\pi}^{t\tau}(r)\right) S_{12}+(80 W(r)) \vec{L}.\vec{S}.
\end{split}
\end{equation}
It is noticeable that for the coupled states we consider $\ell=j-1$, and that the coefficients in the resultant relations are coming from the expectation values of the operators during the reduction into the desired form.

\subsection{Reducing ArgonneV18 Potential into Reid Potential}
ArgonneV14 potential \cite{AV14} has some improvements with respect to the Urb81 potential \cite{UV14} in describing data as we hinted in the subsection \ref{2.2.1}, and that Arg94 is a real high-quality potential. Among 18 operators of Arg94, 14 operators are those in Urb81 and other four operators are three charge-asymmetry operators of $T$ (for the isospin-operator of $T_{12}=3\tau_{z1}\tau_{z2}-\vec{\tau}_{1}.\vec{\tau}_{2}$), $\sigma T$, $t T$ and one charge-asymmetry operator of $\tau z$ as $\tau_{z1}+\tau_{z2}$. In addition, a more complete electromagnetic interaction than that used in the Nijmegen PWA93 \cite{NijmPWA} is included.

The potential is written as a sum of an electromagnetic (EM) part, an OPE part and the remaining (R) intermediate- and short-range phenomenological parts. The EM part, in turn, for \emph{pp}, \emph{np} and \emph{nn} interactions, depended on the case, includes one- and two-photon Coulomb terms, the Darwin-Foldy term, vacuum polarization and magnetic-moment interactions, each with a proper form factor. For charge-dependent OPE part, the neutron-proton and neutral- and charged-pion mass differences, the same as in the Nijmegen PWA93, are included as well. For the SR and MR parts, the potential is similar to Urb81 but with the Yukawa and tensor functions and a Woods-Saxon function more improved than (\ref{GrindEQ__5a_}), (\ref{GrindEQ__5b_}), (\ref{GrindEQ__5c_}) next to four sets of the strengths to fit the scattering data and a regularization condition at the origin. In general, Arg94 potential includes more intricacies and improvements than the Urb81 potential. For details see \cite{AV18}.

Therefore, the Arg84 reduction is similar to the Urb81 reduction. However, because of the four extra operators and a full electromagnetic interaction as well as further subtleties, a little more lengthy calculation is required. It is also necessary to mention that, as the previous case, all terms including the operators, functions and constants, regardless the meanings and implications of the individual terms, reduce or absorb to the chosen Reid form.

\subsection{Reducing Nijmegen Potentials into Reid Potential}
\subsubsection{Nijm93, NijmI and NijmII Potentials}
All these potentials \cite{Nijm93} are based on the Nijm78 potential \cite{Nijm78} with some differences as we mentioned in the subsection \ref{2.2.2} concisely. The basic potentials are OBE's with momentum-dependent central terms and the exponential form factors. In general, for the LR part, OPE's with including the pion mass differences are considered. Indeed, the \emph{pp} and \emph{np} OPE potentials read
\begin{equation} \label{GrindEQ__8a_}
V_{OPE}(pp)=f_{\pi}^{2} V(m_{\pi_{0}}),
\end{equation}
\begin{equation} \label{GrindEQ__8b_}
V_{OPE}(np)=-f_{\pi}^{2} V(m_{\pi_{0}}) \pm 2f_{\pi}^{2} V(m_{\pi_{\pm}}),
\end{equation}
in which
\begin{equation} \label{GrindEQ__8c_}
V(m_{pi})=\left(\frac{m_{pi}}{m_{\pi_{\pm}}}\right)^{2} \frac{1}{3} m_{pi} c^{2} \left[\phi_C^1(m_{pi}, r) (\vec{\sigma}_{1}.\vec{\sigma}_{2}) + 3 \phi_T^0(m_{pi}, r) S_{12} \right],
\end{equation}
and
\begin{equation} \label{GrindEQ__8d_}
\phi_{C}^{1}(r)=\phi_{C}^{0}(r)-4\pi \delta^{3}(m_{pi} \vec{r}),
\end{equation}
where without the form factors the latter is used instead of $\phi_{C}^{0}(r)$, and that the tensor (spin-orbit) functions $\phi_{T}^{0}$ ($\phi_{SO}^{0}$) are written in derivatives of the central function $\phi_{C}^{0}$. On should note that $f_{\pi}^{2}$ is for the pion-nucleon coupling constant and that the plus (minus) sign in (\ref{GrindEQ__8b_}) is for the total isospin of $T=1(0)$. \\
For the remaining MR and SR parts, the potential's structure, in coordinate-space, reads
\begin{equation} \label{GrindEQ__9_}
\begin{split}
V_{Nijm.} &= V_{C}(r)+V_{SS}(r) (\vec{\sigma}_{1}.\vec{\sigma}_{2})+V_{T}(r) S_{12}+V_{SO}(r) \vec{L}.\vec{S} \\
& +V_{SOA}(r) \vec{L}.\vec{A}+V_{Q12}(r) Q_{12},
\end{split}
\end{equation}
where the potential forms are assumed to be same in all partial waves; $\vec{L}.\vec{A}$ with $\vec{A}=(\vec{\sigma}_{1}-\vec{\sigma}_{2})/2$ is called the charge-symmetry operator and
\begin{equation} \label{GrindEQ__9b_}
Q_{12} =\big[(\vec{\sigma}_{1}.\vec{L})(\vec{\sigma}_{2}.\vec{L})+(\vec{\sigma}_{2}.\vec{L})(\vec{\sigma}_{1}.\vec{L})\big]/2= \left[(\vec{L}.\vec{S})^{2}-\delta_{LJ} L^{2} \right],
\end{equation}
is the quadratic spin-orbit operator whose presence can be simulated by introducing nonlocal potentials. In fact, the Nijm93 and NijmI potentials have a little non-locality in their central parts, which is
\begin{equation} \label{GrindEQ__10_}
V_{C}(r)=V_{C}(r)-\frac{1}{2M_{red}} \big[\nabla^{2} V_P(r)+V_P(r) \nabla^{2}\big],
\end{equation}
with $M_{red}=(m_{\pi_0}+2m_{\pi_\pm})/3\equiv \bar{m}$ as the average pion-mass, while in the NijmII potential, $V_P(r)\equiv 0$. It is notable that the antisymmetric spin-orbit part (SOA), in principle, does not use in these potentials.

Thus, in the reduction schema, for the Nijm93 and NijmI nonlocal potentials, we must add for the uncoupled states the expectation value of the second term in (\ref{GrindEQ__10_}) as well. On the other hand, for the singlet-coupled states, the tensor and spin-orbit terms become zero, and in the uncoupled states, except for $^{3} P_{0}$, $\left\langle \delta_{LJ} L^{2} \right\rangle$ is not zero.\\
Now, one can easily estimate the expectation values especially for the second term of (\ref{GrindEQ__10_}) by having $V_P(r)$ and using the direct Laplacian in the spherical coordinate for a state with definite angular-momentum.\\
For reducing the potentials to the three terms of (\ref{GrindEQ__1_}), we note that since for all coupled states, $L\ne J$; therefore $\left\langle L^{2} \delta_{LJ} \right\rangle$ is zero. So, in general, for reduction we write
\begin{equation} \label{GrindEQ__11a_}
V_{Central}=V_{C}(r)+V_{SS}(r)\left\langle (\vec{\sigma}_{1}.\vec{\sigma}_{2})\right\rangle -\frac{1}{2M_{red}} \left\langle[\nabla^{2} V_P(r)+V_P \nabla^{2} ]\right\rangle,
\end{equation}
\begin{equation} \label{GrindEQ__11b_}
V_{Tensor} = V_{T}(r),
\end{equation}
\begin{equation} \label{GrindEQ__11c_}
V_{Spin-Orbit} =V_{SO}(r)+V_{Q12}(r)\langle \vec{L}.\vec{S}\rangle.
\end{equation}

\subsubsection{Reid93 Potential}
For the Reid93 potential \cite{Nijm93}, the OPE tail is the same as (\ref{GrindEQ__8b_}), while in (\ref{GrindEQ__8a_}) $\phi_{C}^{0}$ is used instead of $\phi_{C}^{0}$ except for S-waves. The potential, for each partial-wave, is parameterized with the same central, tensor and spin-orbit operators as Reid68 \cite{Reid68} besides some combinations of the following functions with arbitrary masses and cutoff parameters,
\begin{equation} \label{GrindEQ__12a_}
  \bar{Y}(p)= p\bar{m} \phi_{c}^{0} (p\bar{m}, r), \quad \bar{Z}(p)= p\bar{m} \phi_{T}^{0} (p\bar{m}, r), \quad \bar{W}(p)= p\bar{m} \phi_{SO}^{0} (p\bar{m}, r),
\end{equation}
where $p$ is an integer. There are also some coefficients multiplying the linear combinations of above functions in each partial-wave to fit the scattering data. The potential is now regularized at the origin, has a momentum-space version and is extended for the high partial waves.

\section{Discussions and Results}
In the table \ref{table1}, some two-nucleon states considered here with their quantum numbers are given. In our reduction schema, there are three potential types, i.e. the central (for all states), tensor and spin-orbit, where the last two are only present in the coupled states. In the CI Reid68 potential \cite{Reid68}, just the states up to $J\le2$ are included and for the $J>2$ states, only in the tensor part OPEP is used. B. D. Day extended the Reid68 potential up to the $J\le 5$ states \cite{Reid68-Day}; and for the $J>5$ states, he took the tensor part of OPEP; and for the spin-orbit part from $J\ge 5$ on, he set zero. The CD Reid93 potential has the states up to $J=9$ in the central and tensor parts and for the spin-orbit potentials in the states from $J\ge 5$ on, he set zero as was done also by Day when he extended the Reid68 potential to the higher partial waves. The CD Nijm93, NijmI and NijmII potentials \cite{Nijm93} have the same states as the Reid93 potential. The CI UrbanaV14 potential \cite{UV14} has the states up to $F (J=3)$; and the CD ArgonneV18 potential \cite{AV18} has all three potential types up to the higher states.

The Arg94, Reid93 and Urb84 potentials do not use the direct meson exchanges for MR's and SR's but the phenomenological parametrization are chosen. Arg94 uses the local functions of the Woods-Saxon type and special Yukawa's with the exponential cutoffs; whereas Reid93 employees local Yukawas with multiples of the pion masses similar to the original Reid68 potential. The new features of the Reid93 potential with respect to the Reid68 potential is that in Reid93, the fitting is to the new data of Nijmegen-group \cite{NijmPWA} and $1/r$ singularity in all partial waves is removed by introducing a dipole form factor. In the Urb81 potential, for the MR part, the local functions are the usual Yukawas with exponential cutoffs, where the cutoff parameters are determined by fitting to scattering data, and in the short-range part, the special Woods-Saxon potentials are used. Still, at the very short distances, the potentials are regularized by the exponential (Arg94, Nijm93, NijmI, NijmII) or by the dipole (Reid93) form factors that are all local functions. The three Nijmegen potentials are based on the Nijm78 potential, which is framed from the estimated OBE amplitudes next to the contributions of the Pomeron and some tensor Regge trajectories. In fact, the NijmII potential uses the totally local approximations for all OBE contributions, and the Nim93 and NijmI potentials keep some nonlocal terms in the central force components while their tensor forces are local totally. The Non-localities in the central force have only a very moderate impact in the nuclear-structure calculations compared with the non-localities in the tensor force.\\
According to the discussions so far, clearly the Reid68 and Reid93 potential forms are similar and also the Urb81 and Arg94 potential forms are alike as well as three Nijm93, NijmI, NijmII potential forms together. In the second step, the Reid potentials (especially Reid93) have more likenesses with the Urb81 and Arg94 potentials. It is also notable that the LR OPEP's are almost same for all potentials, except for few subtleties as taking the pion-mass differences.

In the Figures. 1, 2 and 3, the central, tensor and spin-orbit parts of the considered potentials, reduced into the Reid potential, are respectively given for some \emph{np} states (without any preference) between $J=0$ and $J=8$, which the latter is almost the highest fitted wave for the potentials here. The figures are plotted in the range that the potentials have definite values and so, the ranges in which the differences are not clear are neglected. In the CI potentials, we only set the present potential in that special case. Although reproducing the phase shifts and some other results from the calculations with these potentials are almost similar, the potentials are largely different. \\
With a glance to the figures, a close likeness of the Reid68 potential to the Reid93 potential, the Urb81 potential to the Arg94 potential, and the Nijm93, NijmI, NijmII potentials together is obvious. That is of course reasonable and predictable mainly from their structural similarities. It is obvious from the figures that for the LR part almost all potential shapes converge as it is of course predictable from their almost similar OPEP's. One can simply see which potentials are "softer" in the MR and SR parts. So, we note that the Nijm93 and NijmI potentials, which have some non-localities, are softer.

The looseness of the expansion by Day from the Reid68 potential is clear from the figures in that the Day expansion was to hold only satisfactory results in the nuclear calculations and was not based on any tight physical ground. The degree of the potential softness is obvious from the plots as well. The dependence to the even or odd value of the NN relative angular-momentum, which is a space-exchange marker, is also clear from the figures. For instance, in the $^{1}D_{2}$ channel with an even $L$ and in the $^{1}F_{3}$ channel with an odd $L$, one can easily see from the Figure 1 that the Reid68 and Reid93 potentials have tendency to oppose each other. The same is valid for the three Nijmegen potentials, which in turn that mean that the spatial exchanges are strong. For the tensor and spin-orbit potentials in the Figures 2 and 3, one can also see that for each state with either an even or an odd $J$, a special procedure is dominant and the differences are discussable from various point of views.

In the Figures 4, 5 and 6, three groups of the similar potentials, for the \emph{np} states, from $J=0$ up to $J=2$, are compared. In the Figure 4, the Reid potentials (Day81, Reid93) are pictured for some channels. In general, the differences between these two potentials return to two important adjustments mentioned above. The presence of a softer core in the Reid93 potential is obvious compared with the singular SR part of the Reid68 potential. The small differences in the Figure 5 are also expectable because of the small differences in the structure of the Urb81 and Arg94 potentials. The same is true for the three Nijmegen potentials in the Figure 6. The charge-dependence of the CD potentials is also showed in the Figure 7 for the $^{1}S_{0}$ central potential and the tensor potential in the $^{3}P_{2}-^{3}F_{3}$ state as well as the spin-orbit potential in the $^{3}P_{2}$ channel. It is obvious from the figures that the charge-dependence is for \emph{pp} and \emph{np} systems, and for the \emph{nn} system is almost same with \emph{pp}. As a final illustration, in the Figure 8, the dependence on the orbital angular-momentum for $^{3}S_{1}$, $^{3}D_{1}$, $^{3}P_{2}$ and $^{3}F_{2}$ states, for the \emph{np} system, are pictured. The plots demonstrate an explicit dependence on $L$ or, in other words, the spatial exchanges in the potentials, and so on.

In summary, the likenesses and differences in the figures are related to the structural and theoretical bases of the potentials as well as the external conditions such as fitting to the special scattering database. The almost different figures, although slightly, show that at least some basic physical assumptions of these models and potentials, which give almost similar results, should be wrong. That is because the similar forms reproducing the similar results for the nuclear force are somewhat ambiguous.

\section{Concluding Comments}
In the recent decades, many NN potentials have been presented. Most potentials use OPE's for the long-range part, while the TPE's and special OBE's are used in the intermediate-range part. For the short-range part, the heavy-meson exchanges, the QCD effects and the phenomenological procedures are often used. The potential's precision and quality are explored through various methods. The most important method for the potential-quality measurement (PQM) is giving satisfactory results in nuclear-structure calculations. Finding out $\chi^{2}/N_{data}$ is another usual method that, as already discussed, has its own problems and difficulties. Based on these standards, several high-precision charge-dependent NN potentials are built as we mentioned some of them.

A main conclusion that one can deduce from the comparison done here is that a definite and fixed form for the NN interaction is still a critical challenge. In fact, as we have different models and forms for the strong nuclear force, which almost all give similar results while have different structures, then the nuclear force will obviously become meaningless. A certain statement is that although some quantitative correspondences are present among the potentials, there are some other quantitative differences. Generally speaking, one can assign the quantitative differences to the theoretical and structural differences of the potentials. Various interaction ingredients such as the meson and/or quark and gluon exchanges, various phenomenological parts, and mainly the base employed models, result in partially different results. For example, using different Yukawa and Woods-Saxon functions, the form factors to regularize the potentials at the origin and in general the functions used in the various parts of the potentials, make many differences explainable. Meanwhile, the likening features are fitting the scattering data and deuteron properties that in turn make the closeness more reasonable. Although the difficulties are important in their place, however, they are not so big to stop applying some potentials in the nuclear-structure calculations. The people, who use the potentials so, by noting the comparison sketched here, may find more satisfactory reasons to the present discrepancies in the potential shapes and forms.

In summary, the differences could be arisen from the involved approximations and the failures of our knowledge on the nuclear forces. Therefore, it seems that the models and potentials in which many guesses (such as selecting the special potentials, merely fitting to the data and so on) are used, are only a temporary way for solving the NN interaction problem. Efforts for discovering more a fundamental theory and a probable definite and fixed form for the potential, as an important question in nuclear physics, are in progress yet.\\
By the way, although nowadays the chiral effective field theory potentials describe almost well two- and few-nucleon systems both quantitatively and qualitatively and stand as the best so far candidate to describe this long-standing issue in nuclear physics, there are some unsolved problems even in these conventional frameworks. The perturbative character and proper renormalization of the chiral potentials as well as the three- and few-nucleon forces wait to be addressed carefully. Fortunately, nowadays and for future, Holographic QCD as born from the string/gauge, AdS/CFT, correspondence, seems to be a new promising viewpoint to the nuclear physics problems and especially the nuclear force issue \cite{0901.0012}, \cite{0901.4449}.
\begin{table}[p]
\centering
\begin{tabular}{||l||l||c||}
\hline
\ \ \ \textit{\textbf{Potential type (state)}} & \ \ \ \textit{\textbf{Central}} & \textit{\textbf{Tensor and Spin-Orbit}} \\
\hline
\hline
\hline
{$J=0,\, S=0,\, T=1,\, L=0$}  & ${{}^{1} S_{0}\ (pp,np,nn)}$ & - \\
{$J=0,\, S=1,\, T=1,\, L=1$}  & ${{}^{3} P_{0}\ (pp,np,nn)}$ & - \\
\hline
\hline
${J=1,\, S=0,\, T=0,\, L=1}$      & ${{}^{1} P_{1}\ (np)}$               & - \\
${J=1,\, S=1,\, T=1,\, L=1}$      & ${{}^{3} P_{1}\ (pp,np,nn)}$         & - \\
${J=1,\, S=1,\, T=0,\, L=0,L=2}$  & ${{}^{3} S_{1}-{}^{3} D_{1}\ (np)}$  & ${{}^{3} S_{1} -{}^{3} D_{1}\ (np)}$ \\
\hline
\hline
${J=2,\, S=0,\, T=1,\, L=2}$      & ${{}^{1} D_{2}\ (pp,np,nn)}$               & - \\
${J=2,\, S=1,\, T=0,\, L=2}$      & ${{}^{3} D_{2}\ (np)}$                     & - \\
${J=2,\, S=1,\, T=1,\, L=1,L=3}$  & ${{}^{3} P_{2}-{}^{3} F_{2}\ (pp,np,nn)}$  & ${{}^{3} P_{2}-{}^{3} F_{2}\ (pp,np,nn)}$ \\
\hline
\hline
${J=3,\, S=0,\, T=0,\, L=3} $     & ${{}^{1} F_{3}\ (np)}$               & - \\
${J=3,\, S=1,\, T=1,\, L=3}$      & ${{}^{3} F_{3}\ (pp,np,nn)}$         & - \\
${J=3,\, S=1,\, T=0,\, L=2,L=4}$  & ${{}^{3} D_{3}-{}^{3} G_{3}\ (np)}$  & ${{}^{3} D_{3}-{}^{3} G_{3}\ (np)}$ \\
\hline
\hline
${J=4,\, S=0,\, T=1,\, L=4}$     & ${{}^{1} G_{4}\ (pp,np,nn)}$               & - \\
${J=4,\, S=1,\, T=0,\, L=4}$     & ${{}^{3} G_{4}\ (np)}$                     & - \\
${J=4,\, S=1,\, T=1,\, L=3,L=5}$ & ${{}^{3} F_{4}-{}^{3} H_{4}\ (pp,np,nn)}$  & ${{}^{3} F_{4}-{}^{3} H_{4}\ (pp,np,nn)}$ \\
\hline
\hline
\end{tabular}
\caption{\small Two nucleon states from $J=0$ up to $J=9$ and potential types in our reduction plan. For other higher states, the process is similar with $J=5$ states on shown by the Latin letters \emph{H}, \emph{I}, \emph{K}, \emph{L}, \emph{M}, \emph{N} and so on.} \label{table1}
\end{table}

\begin{figure}[p]
    \centering
    \begin{subfigure}[b]{\textwidth}
          \centering
          \begin{subfigure}[b]{0.47\textwidth}
                  \centering
                  \includegraphics[width=\textwidth,height=0.24\textheight]{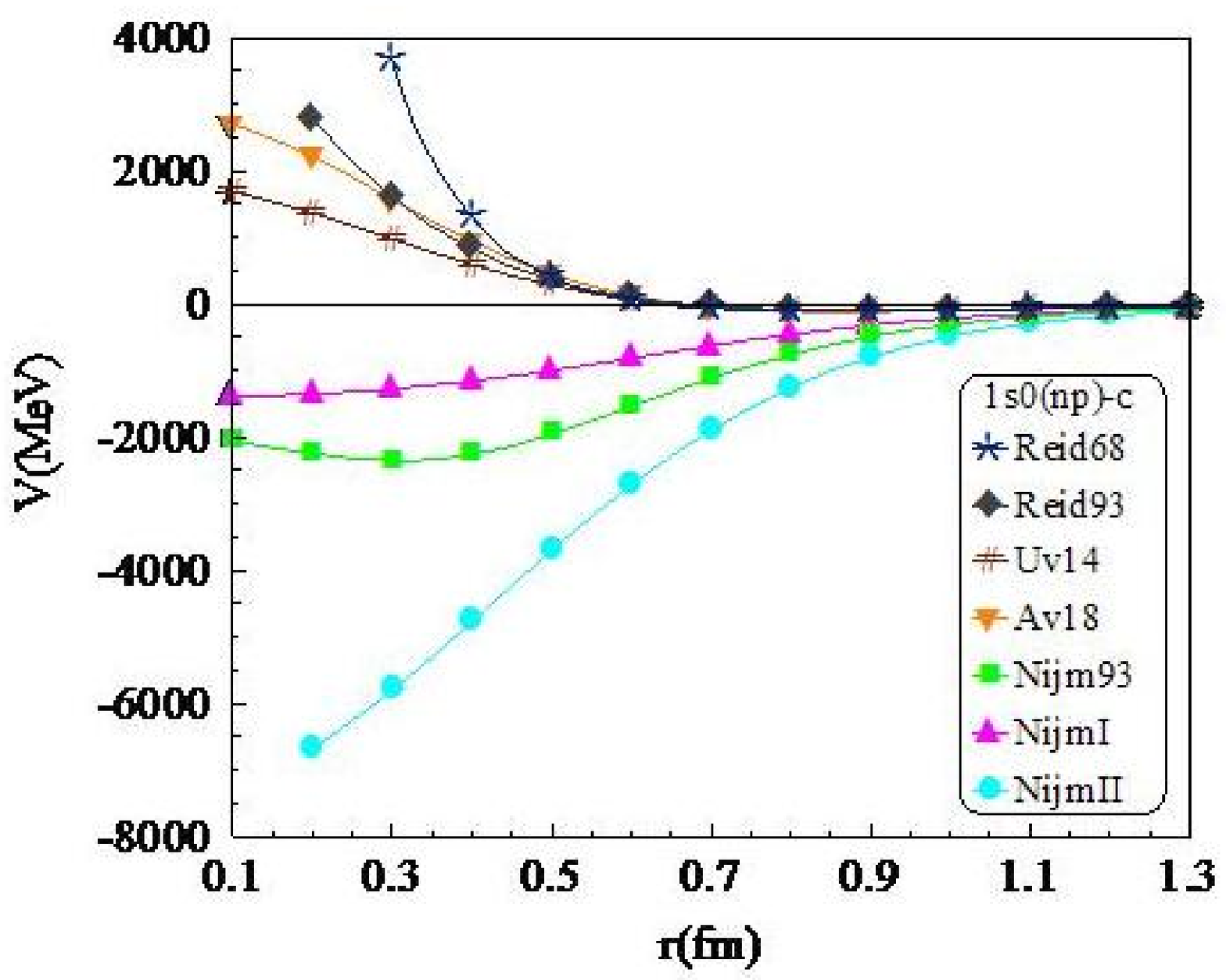}
          \end{subfigure}
          ~
          \begin{subfigure}[b]{0.47\textwidth}
                  \centering
                  \includegraphics[width=\textwidth,height=0.24\textheight]{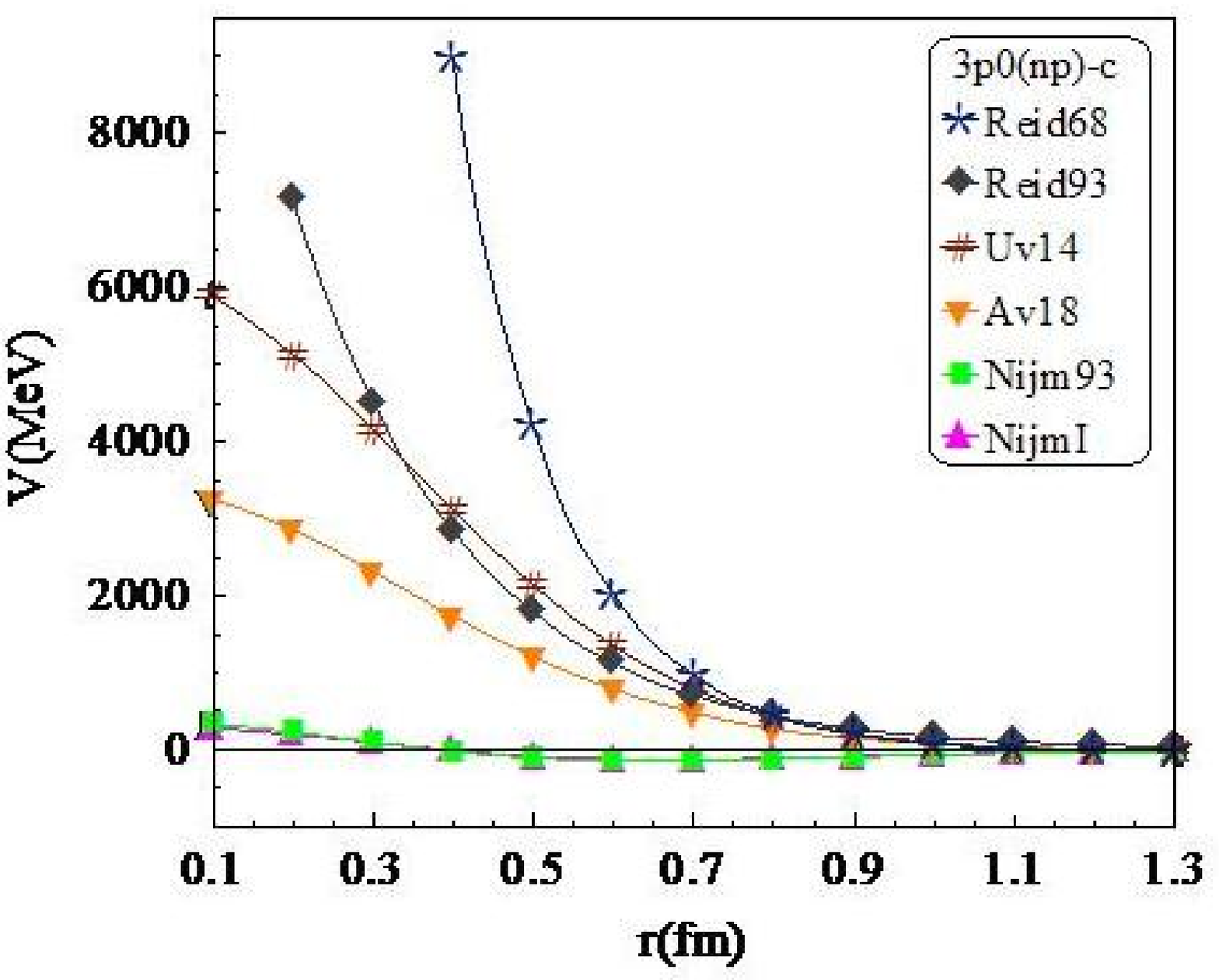}
          \end{subfigure}
    \end{subfigure}
    \begin{subfigure}[b]{\textwidth}
          \centering
          \begin{subfigure}[b]{0.47\textwidth}
                  \centering
                  \includegraphics[width=\textwidth,height=0.24\textheight]{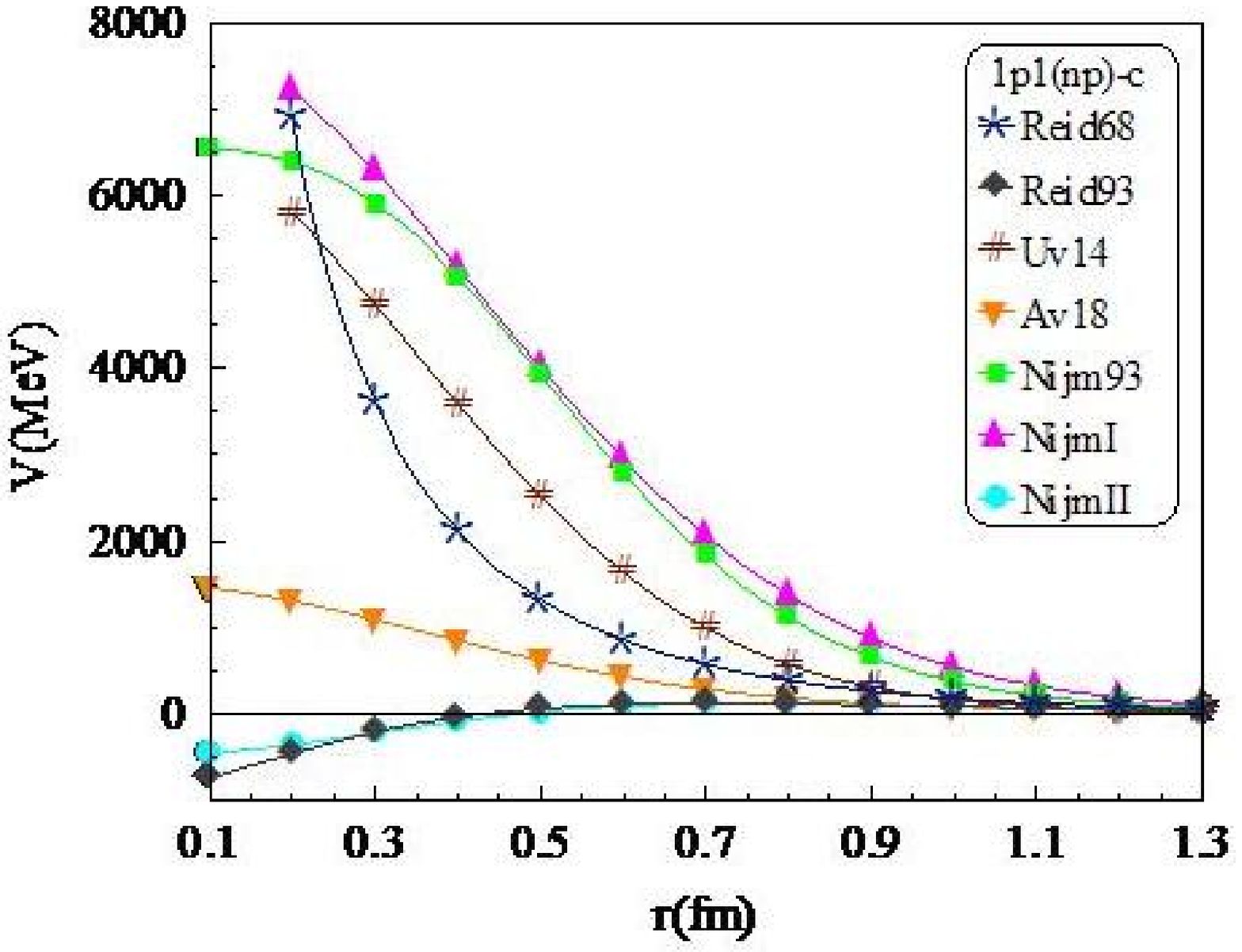}
          \end{subfigure}
          ~
          \begin{subfigure}[b]{0.47\textwidth}
                  \centering
                  \includegraphics[width=\textwidth,height=0.24\textheight]{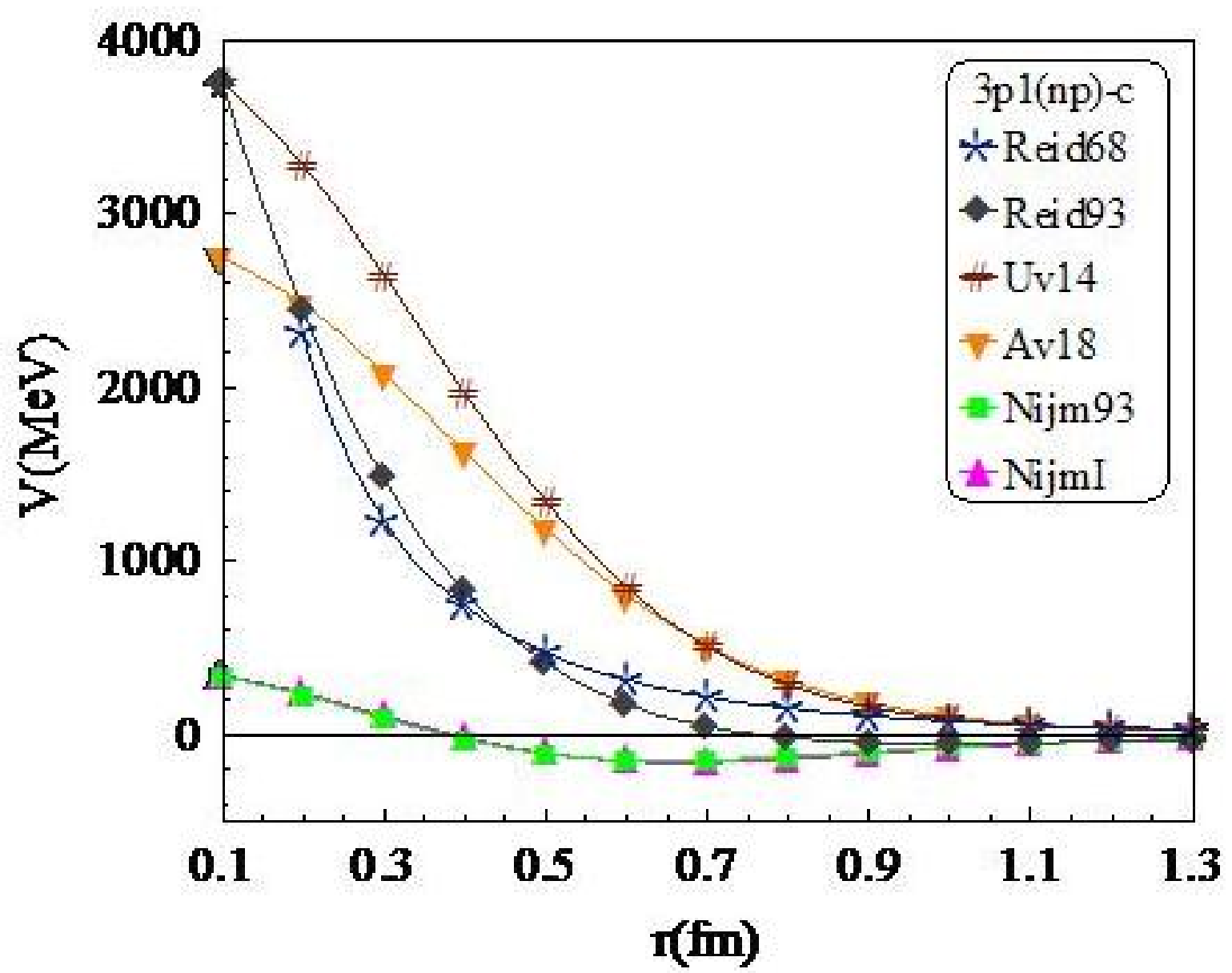}
          \end{subfigure}
    \end{subfigure}
    \caption{\textit{The central potentials of various potential forms in the states from $J=0$ up to $J=8$, for np system}.} \label{Fig1.}
\end{figure}
\setcounter{figure}{0}
\begin{figure}[p]
 \renewcommand{\figurename}{Continuation of Figure}
    \centering
    \begin{subfigure}[b]{\textwidth}
          \centering
          \begin{subfigure}[b]{0.47\textwidth}
                  \centering
                  \includegraphics[width=\textwidth,height=0.24\textheight]{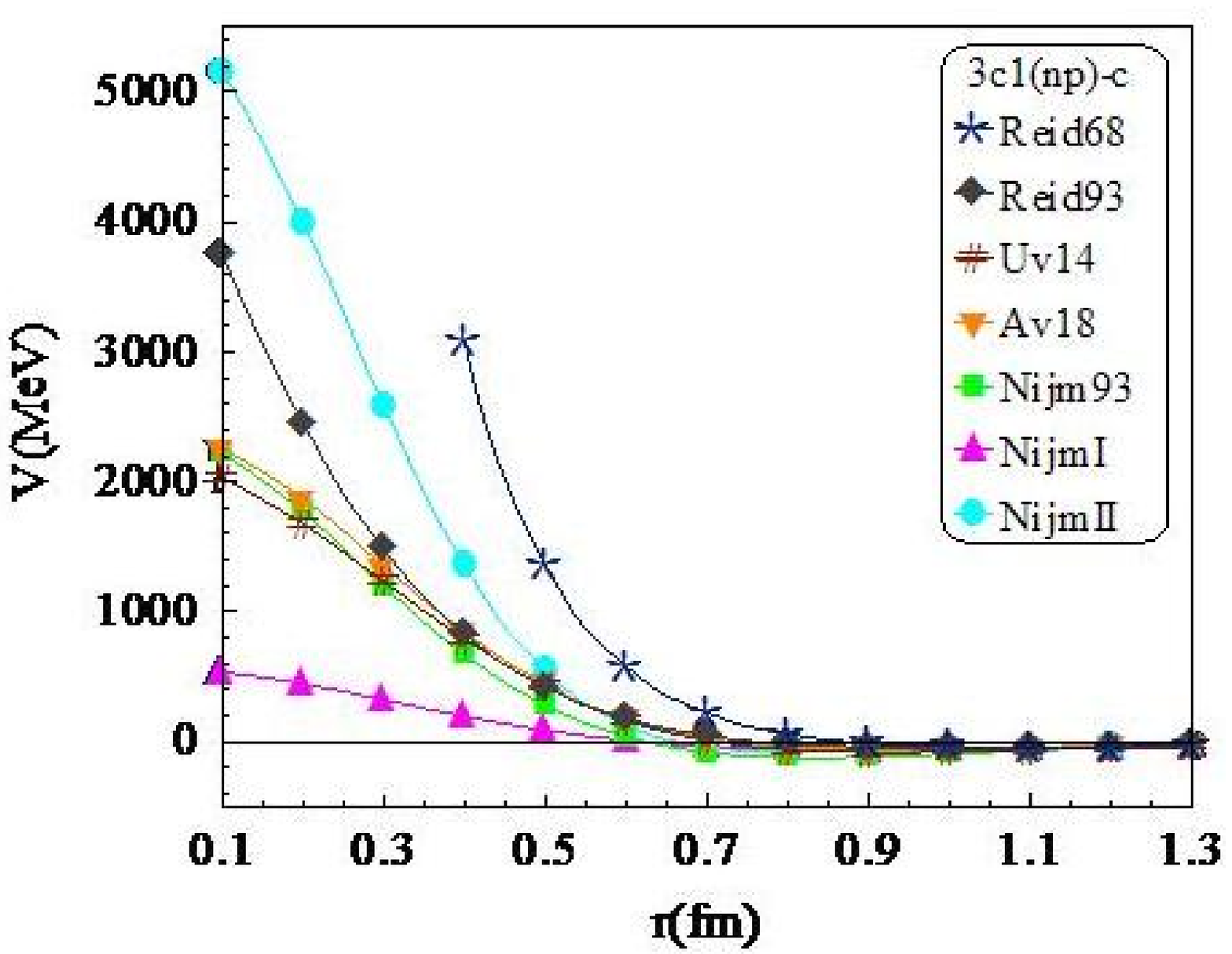}
          \end{subfigure}%
          ~
          \begin{subfigure}[b]{0.47\textwidth}
                  \centering
                  \includegraphics[width=\textwidth,height=0.24\textheight]{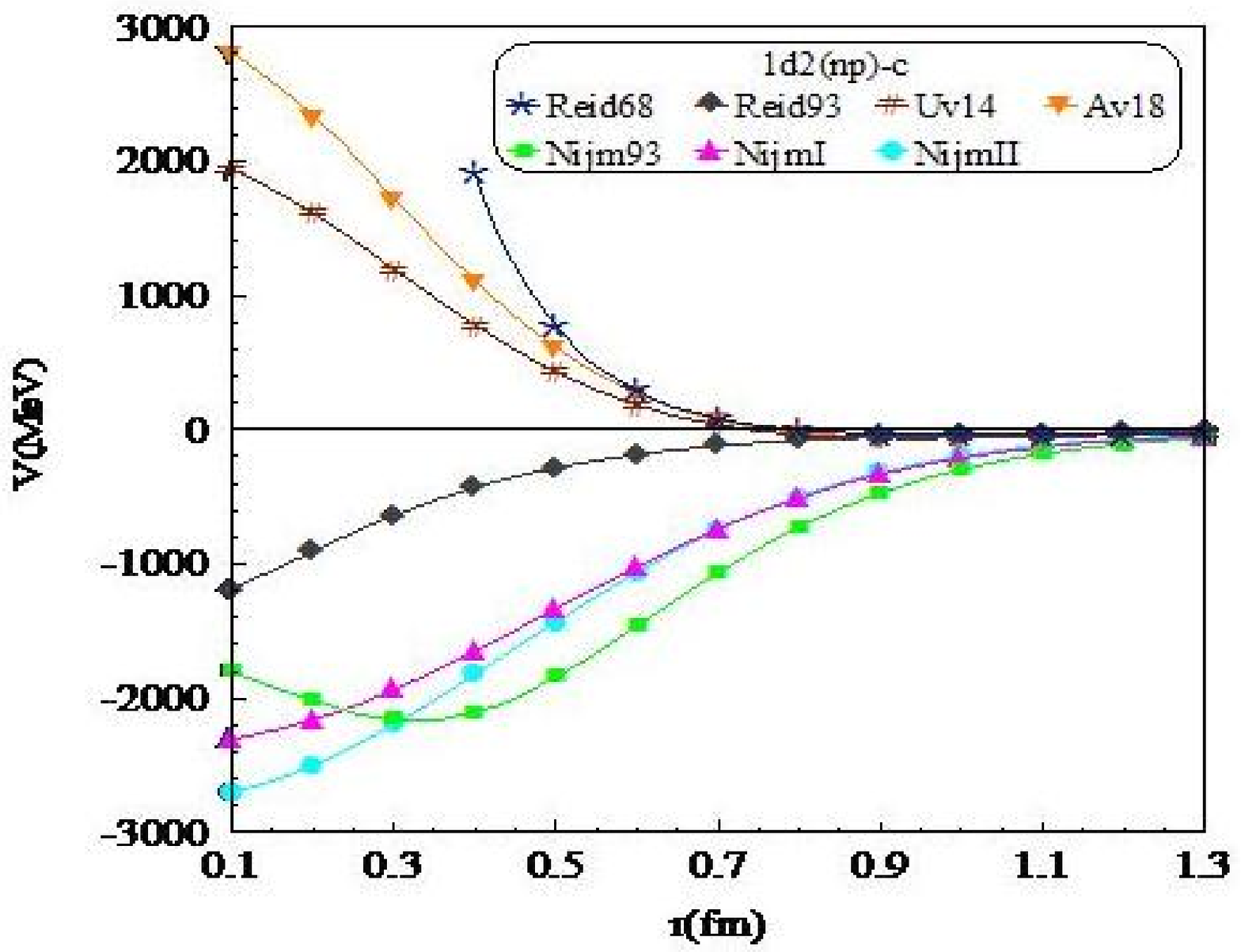}
          \end{subfigure}
    \end{subfigure}
     \begin{subfigure}[b]{\textwidth}
          \centering
          \begin{subfigure}[b]{0.47\textwidth}
                  \centering
                  \includegraphics[width=\textwidth,height=0.24\textheight]{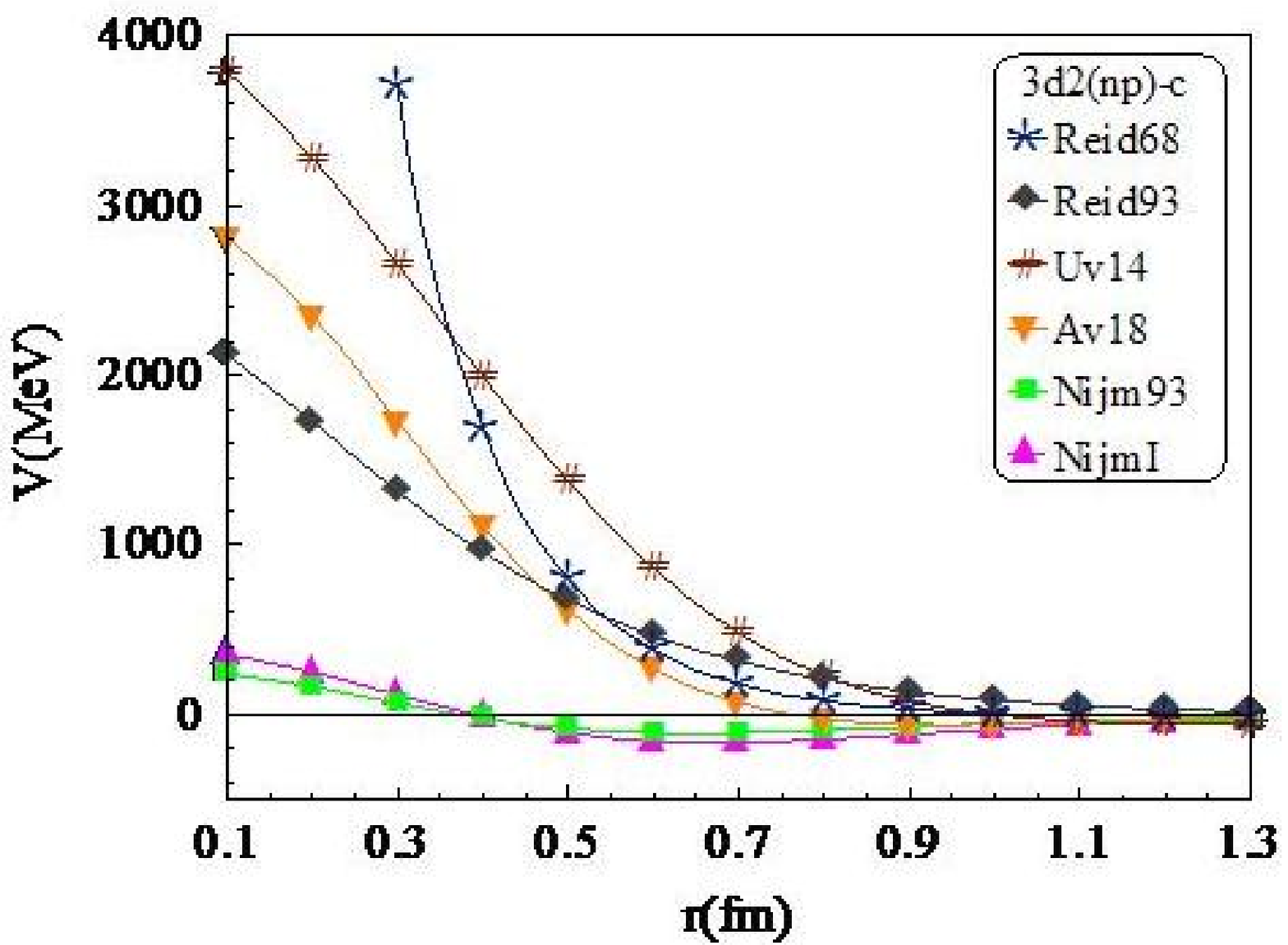}
          \end{subfigure}%
          ~
          \begin{subfigure}[b]{0.47\textwidth}
                  \centering
                  \includegraphics[width=\textwidth,height=0.24\textheight]{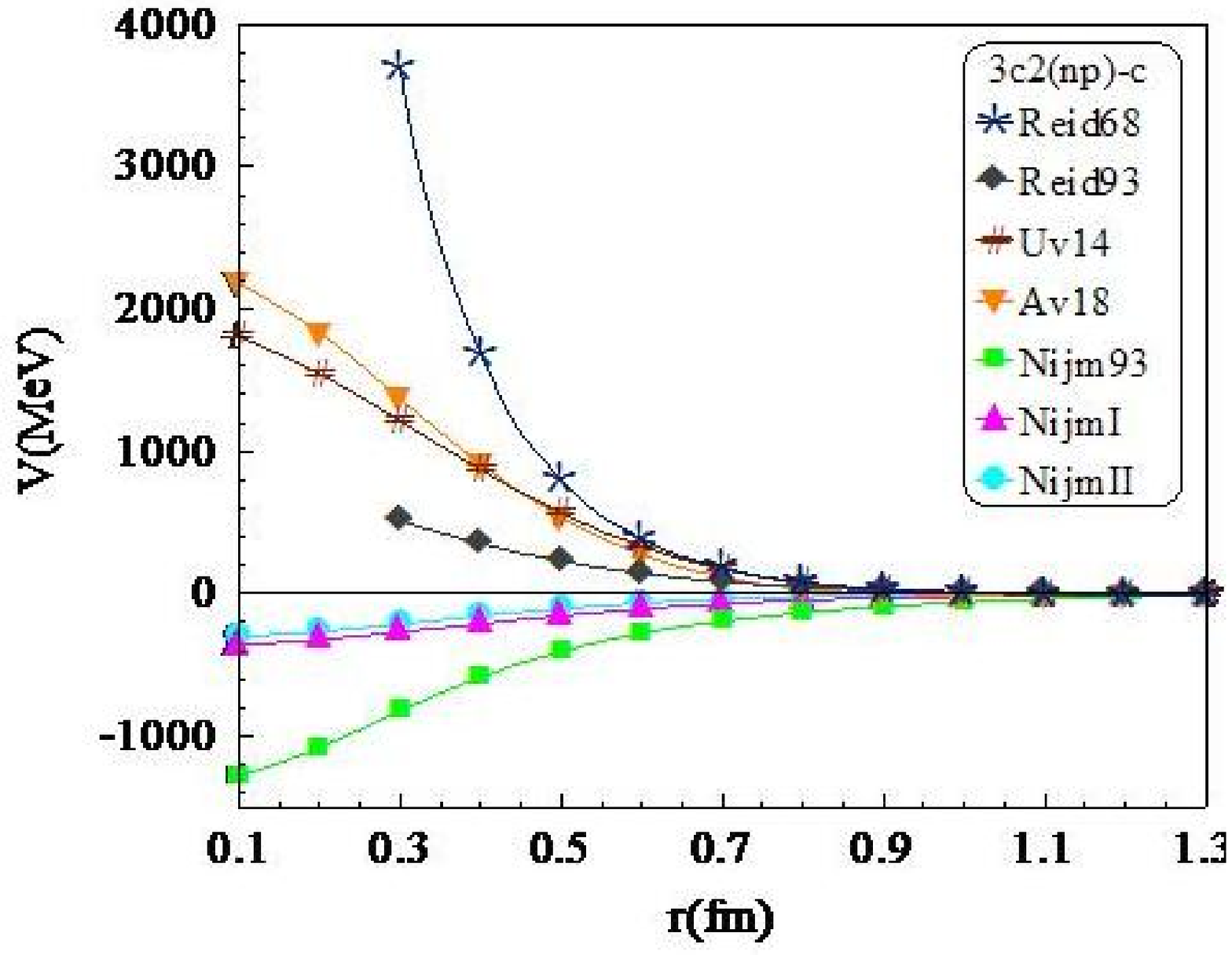}
          \end{subfigure}
    \end{subfigure}
\begin{subfigure}[b]{\textwidth}
          \centering
          \begin{subfigure}[b]{0.47\textwidth}
                  \centering
                  \includegraphics[width=\textwidth,height=0.24\textheight]{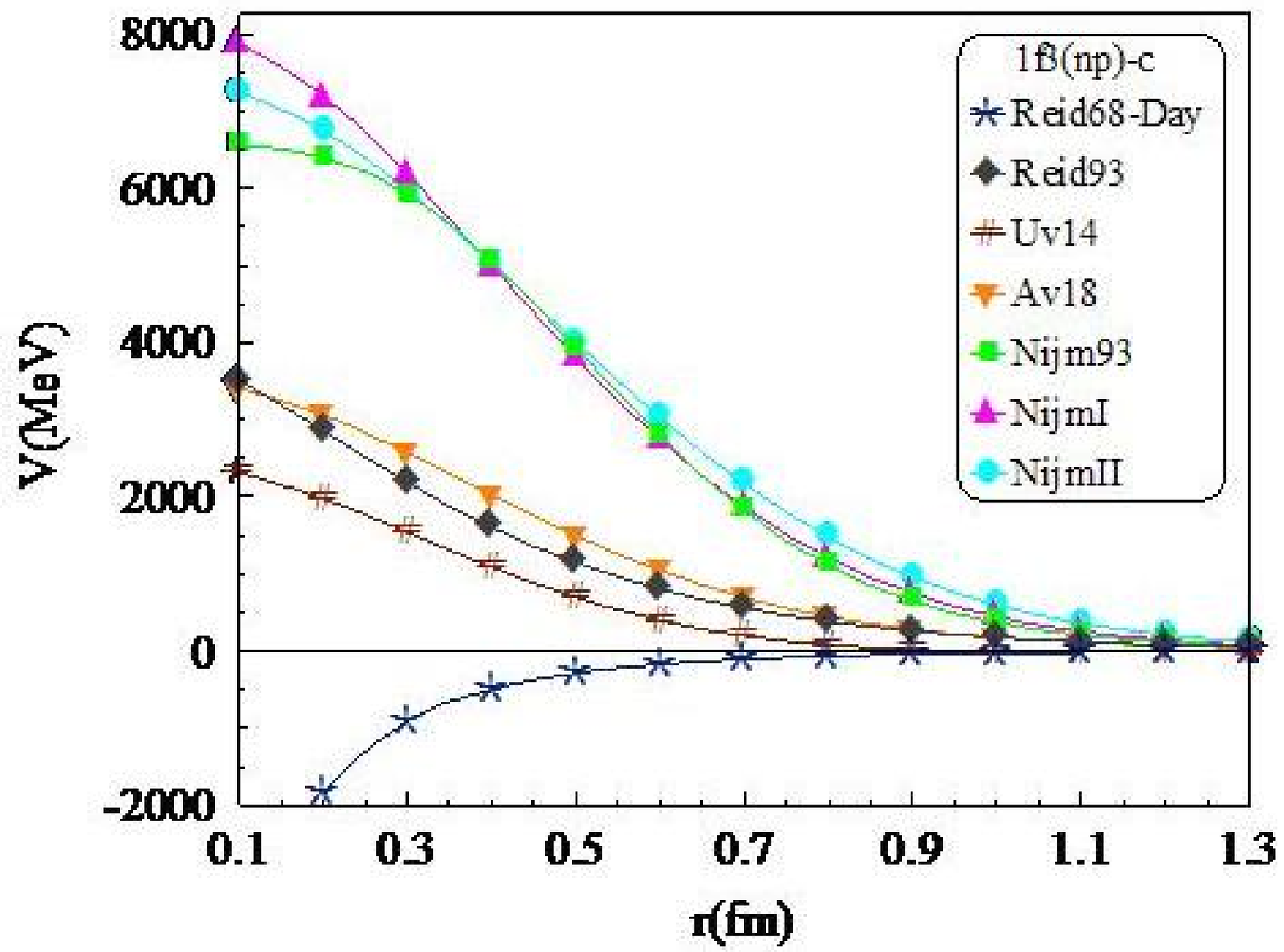}
          \end{subfigure}%
          ~
          \begin{subfigure}[b]{0.47\textwidth}
                  \centering
                  \includegraphics[width=\textwidth,height=0.24\textheight]{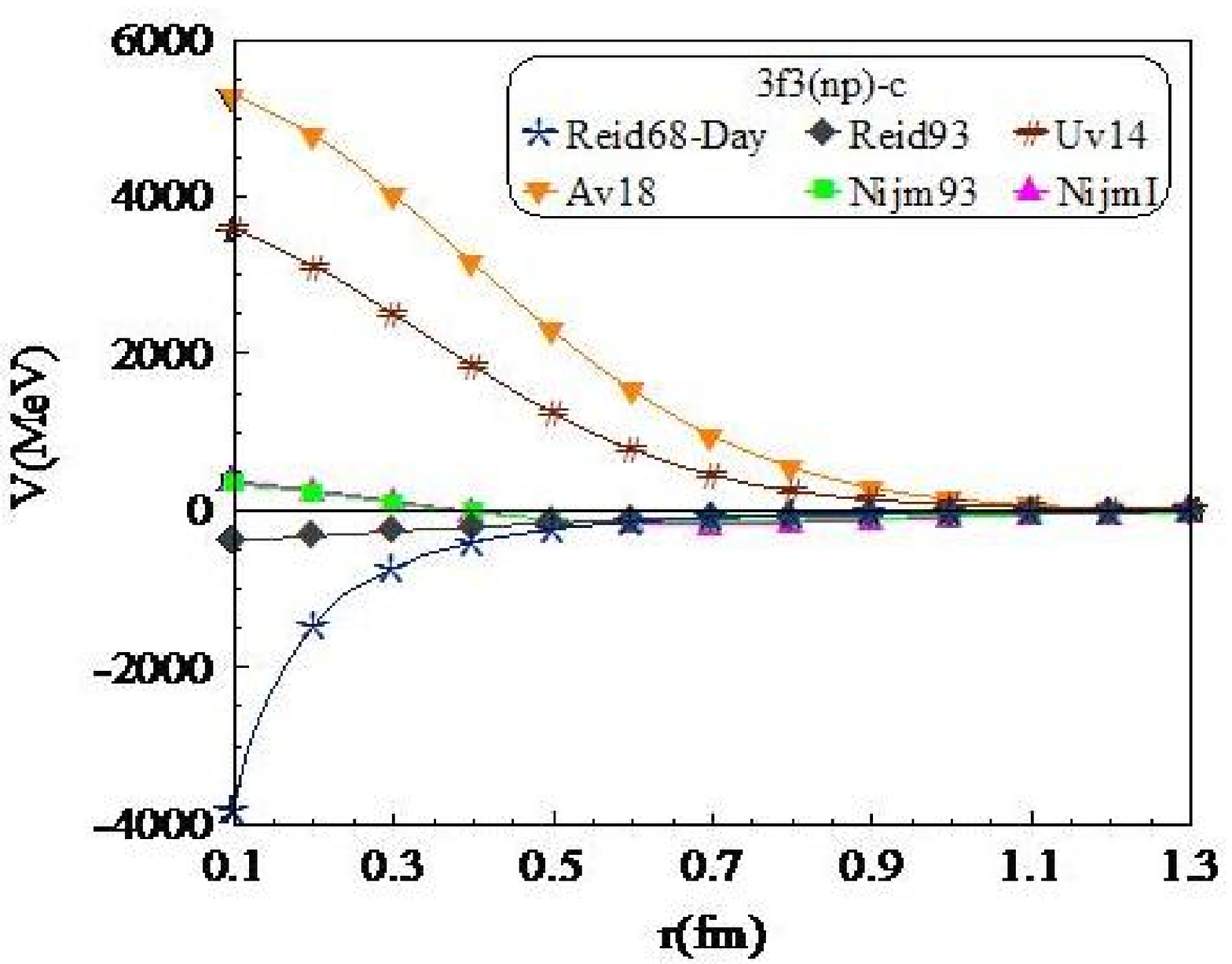}
          \end{subfigure}
    \end{subfigure}
        \begin{subfigure}[b]{\textwidth}
          \centering
          \begin{subfigure}[b]{0.47\textwidth}
                  \centering
                  \includegraphics[width=\textwidth,height=0.24\textheight]{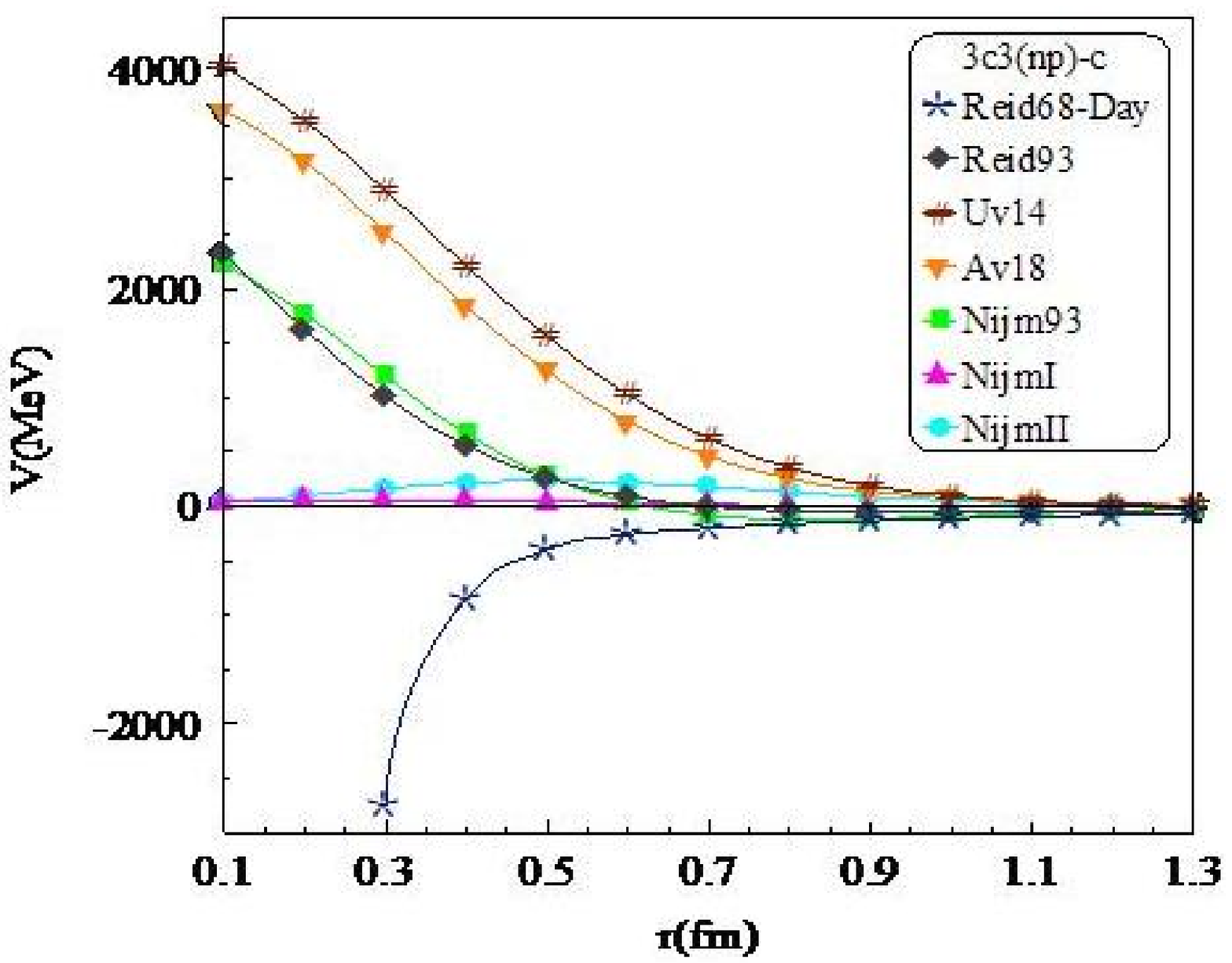}
          \end{subfigure}%
          ~
          \begin{subfigure}[b]{0.47\textwidth}
                  \centering
                  \includegraphics[width=\textwidth,height=0.24\textheight]{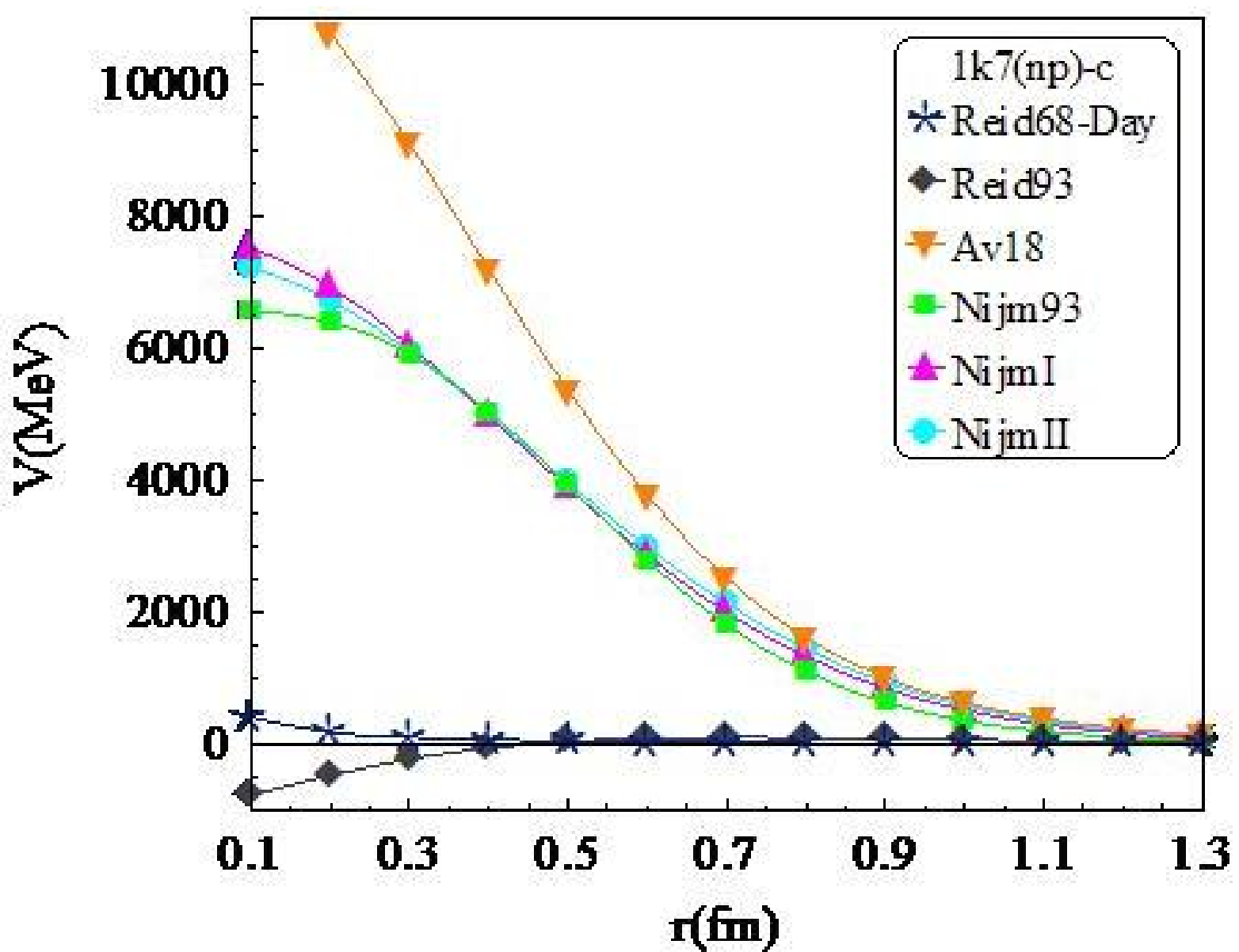}
          \end{subfigure}
    \end{subfigure}
\caption{} \label{Fig1.}
\end{figure}
    \setcounter{figure}{0}
\begin{figure}[p]
\renewcommand{\figurename}{Continuation of Figure}
    \centering
      \begin{subfigure}[b]{\textwidth}
          \centering
          \begin{subfigure}[b]{0.47\textwidth}
                  \centering
                  \includegraphics[width=\textwidth,height=0.24\textheight]{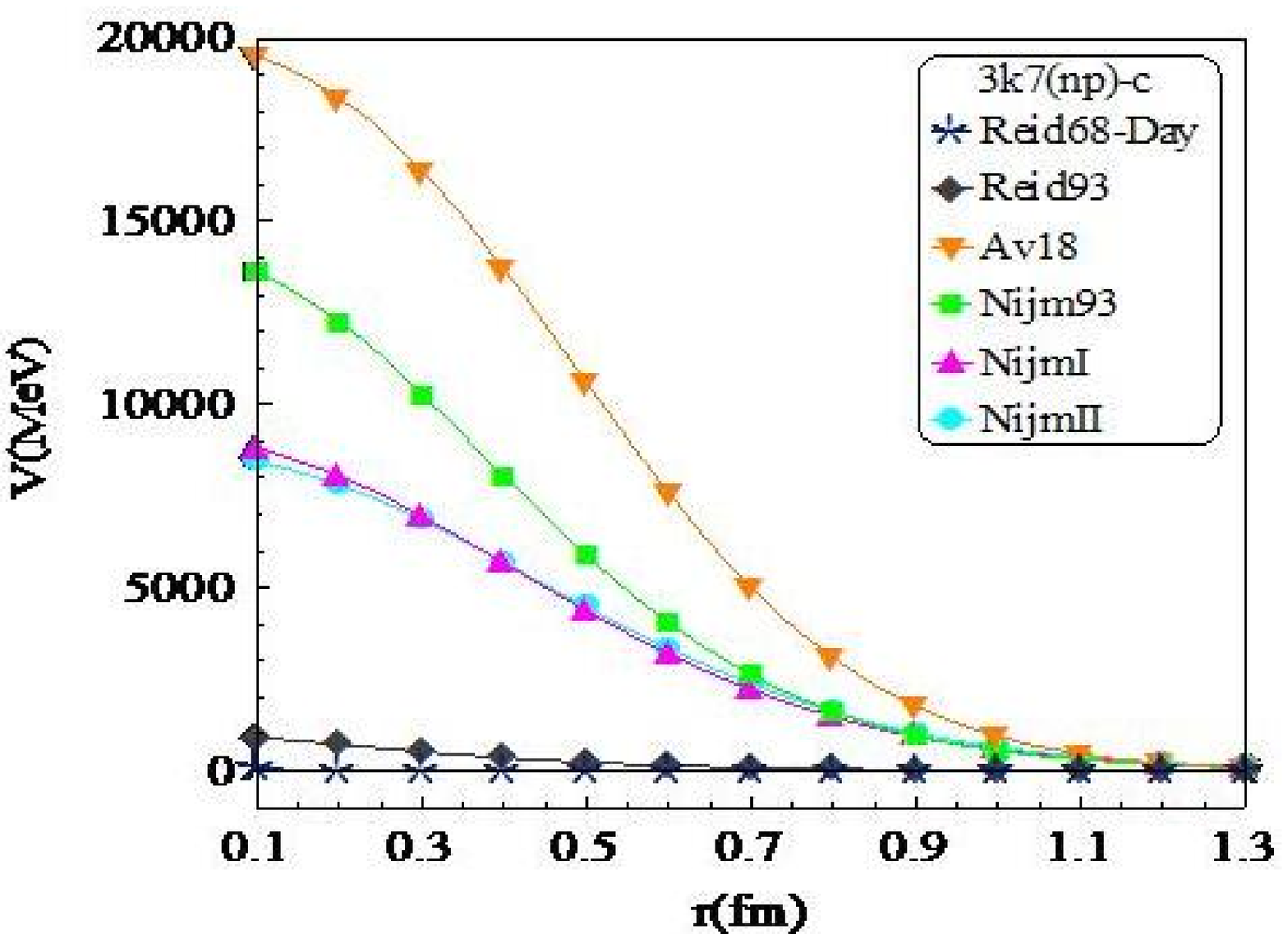}
          \end{subfigure}%
          ~
          \begin{subfigure}[b]{0.47\textwidth}
                  \centering
                  \includegraphics[width=\textwidth,height=0.24\textheight]{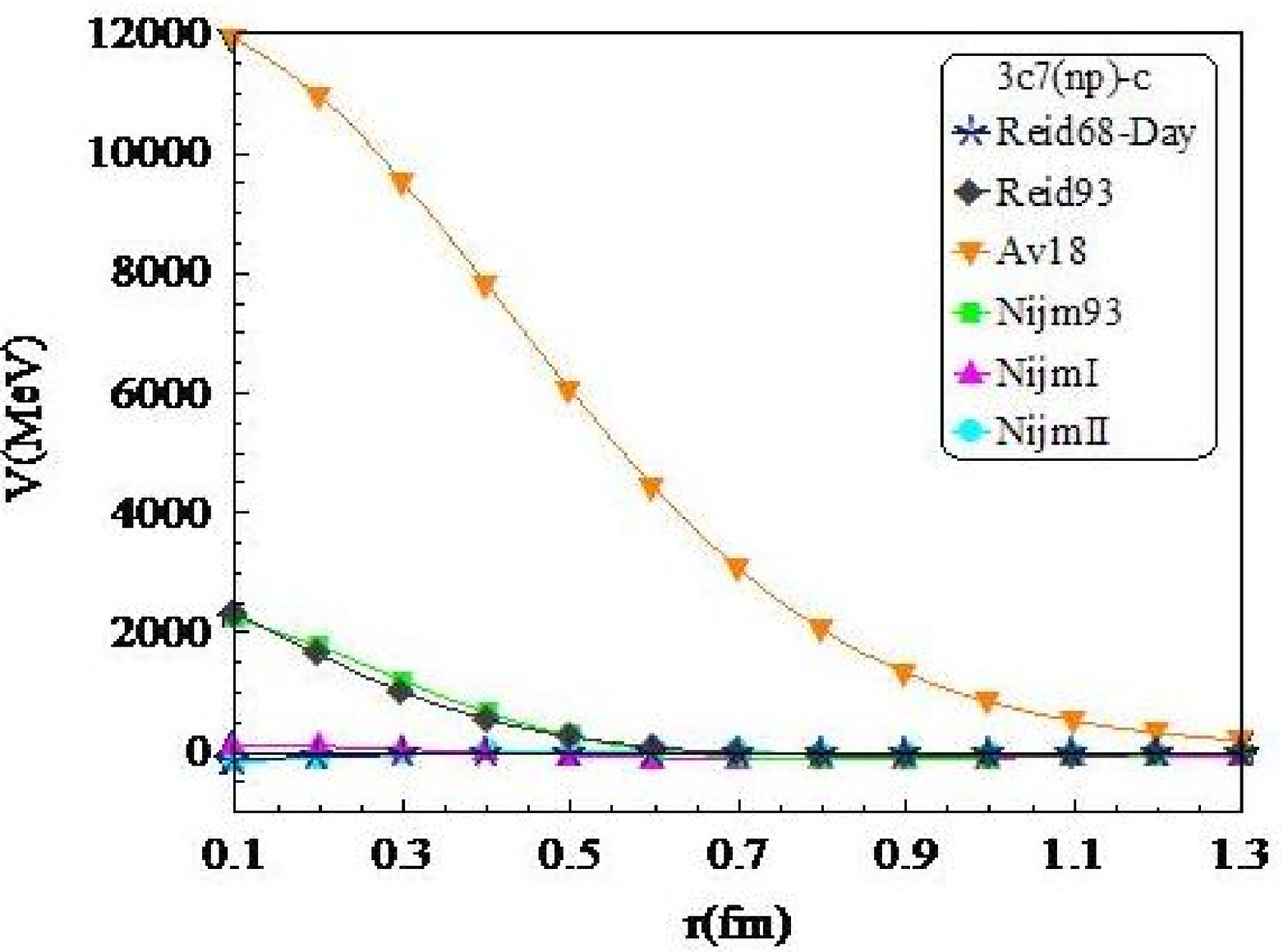}
          \end{subfigure}
    \end{subfigure}
         \begin{subfigure}[b]{\textwidth}
          \centering
          \begin{subfigure}[b]{0.47\textwidth}
                  \centering
                  \includegraphics[width=\textwidth,height=0.24\textheight]{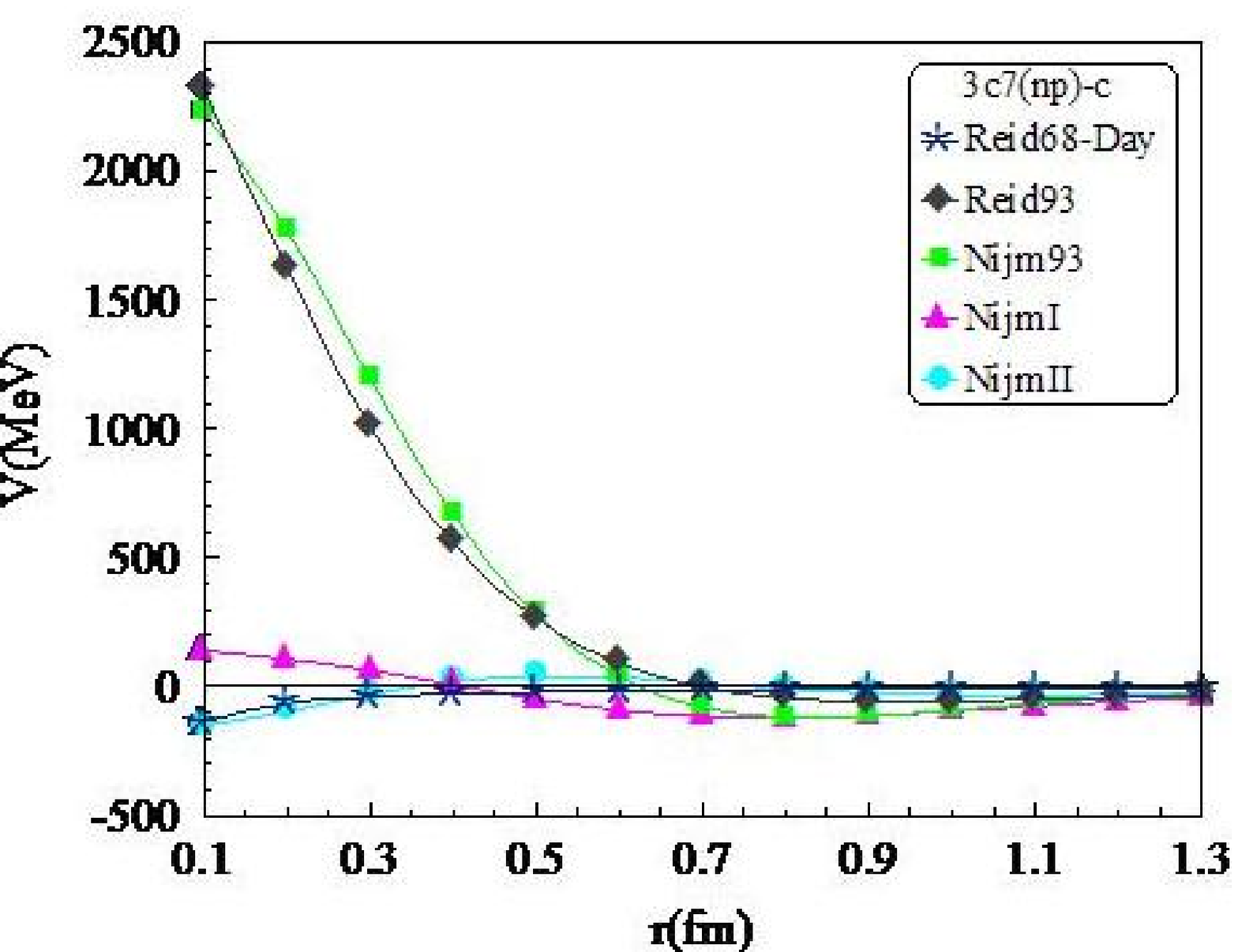}
          \end{subfigure}%
          ~
          \begin{subfigure}[b]{0.47\textwidth}
                  \centering
                  \includegraphics[width=\textwidth,height=0.24\textheight]{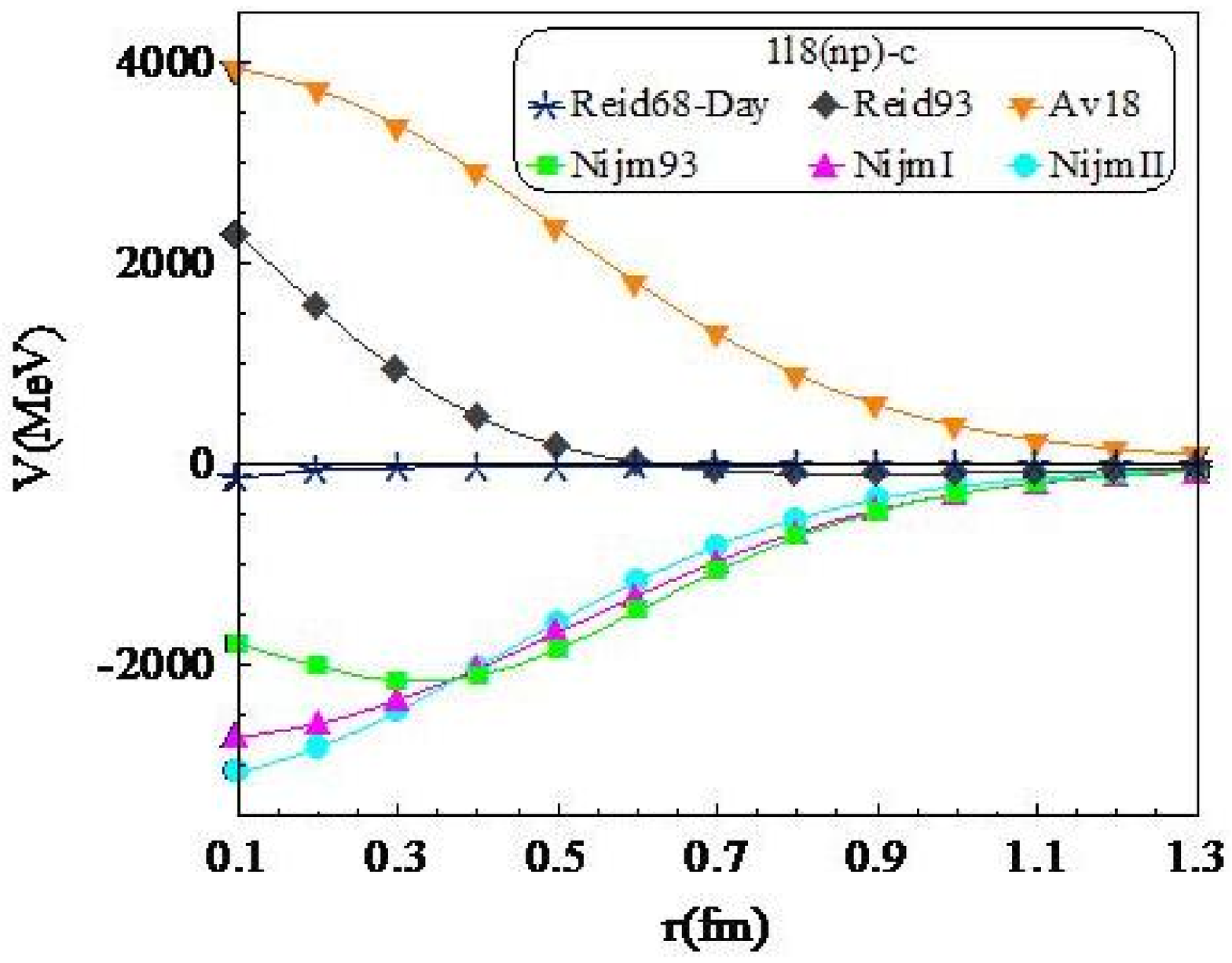}
          \end{subfigure}
    \end{subfigure}
        \begin{subfigure}[b]{\textwidth}
          \centering
          \begin{subfigure}[b]{0.47\textwidth}
                  \centering
                  \includegraphics[width=\textwidth,height=0.24\textheight]{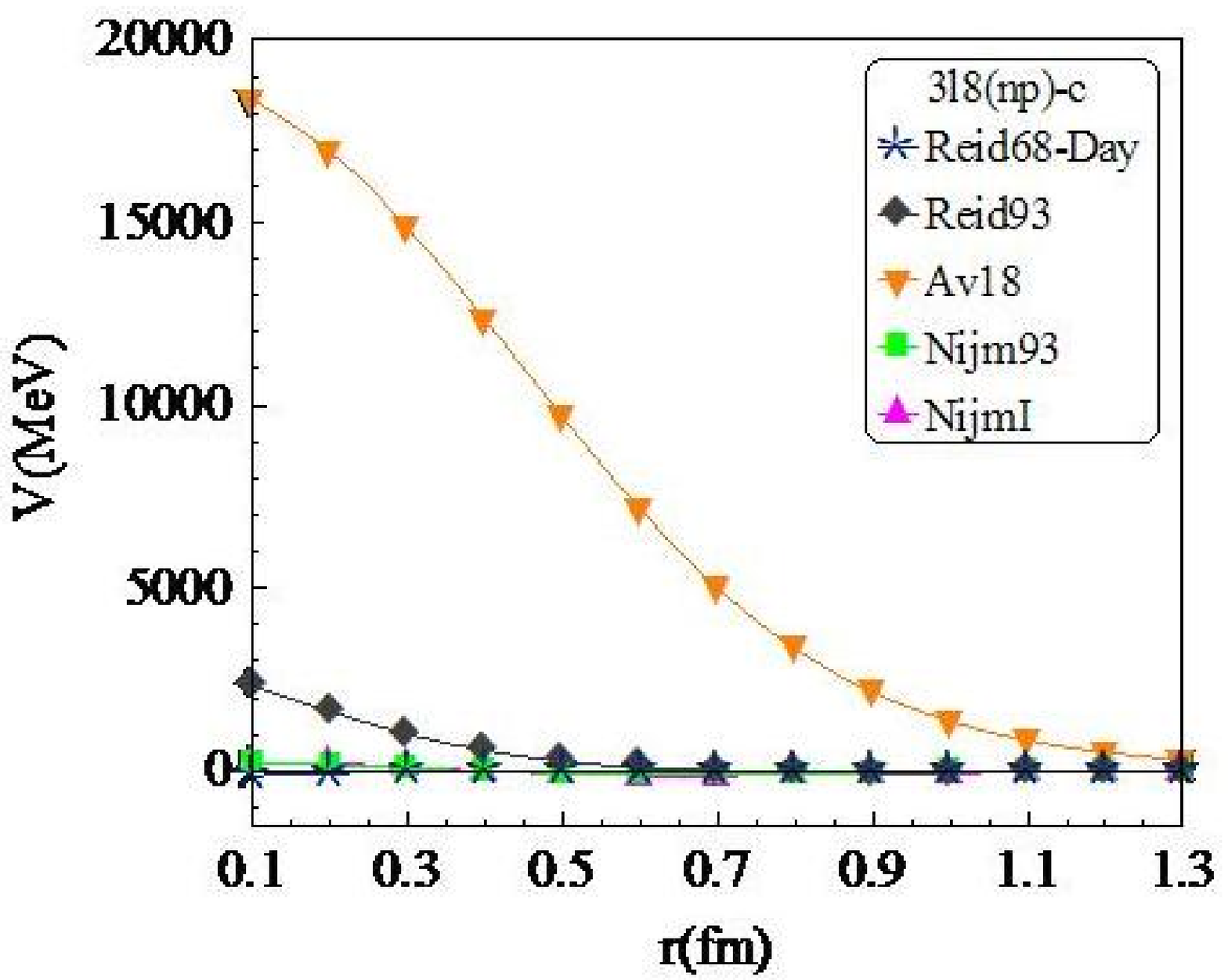}
          \end{subfigure}%
          ~
          \begin{subfigure}[b]{0.47\textwidth}
                  \centering
                  \includegraphics[width=\textwidth,height=0.24\textheight]{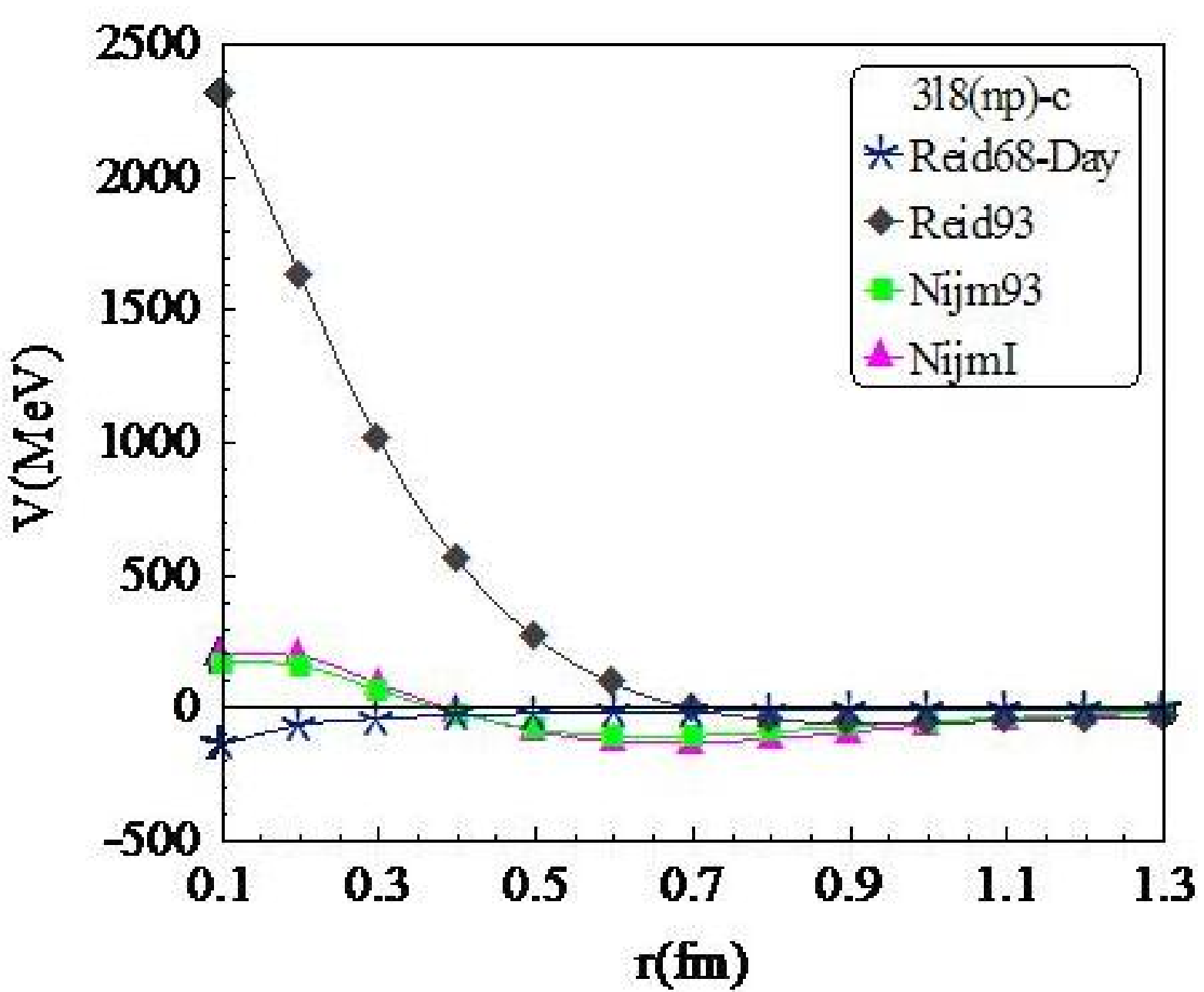}
          \end{subfigure}
    \end{subfigure}
        \begin{subfigure}[b]{\textwidth}
          \centering
          \begin{subfigure}[b]{0.47\textwidth}
                  \centering
                  \includegraphics[width=\textwidth,height=0.24\textheight]{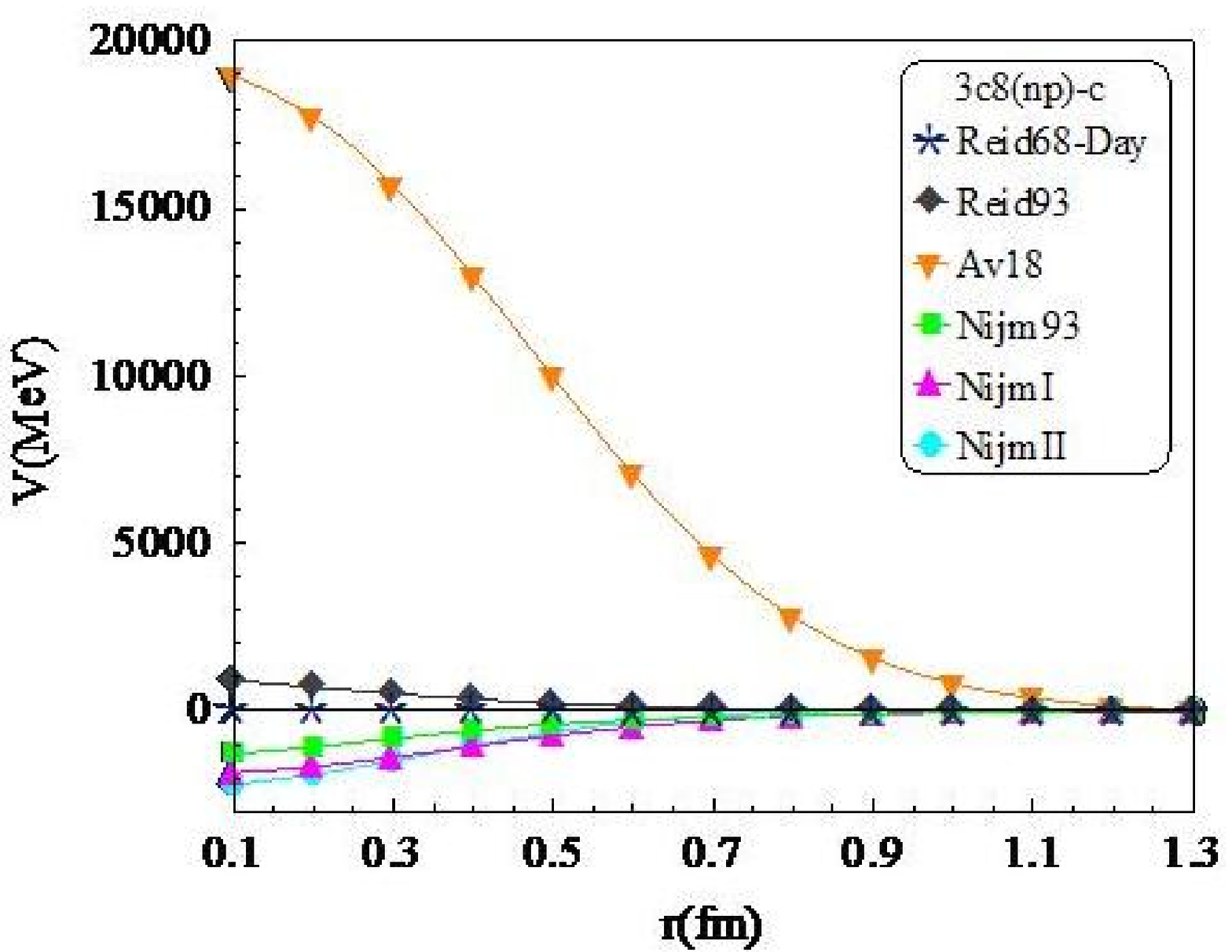}
          \end{subfigure}%
          ~
          \begin{subfigure}[b]{0.47\textwidth}
                  \centering
                  \includegraphics[width=\textwidth,height=0.24\textheight]{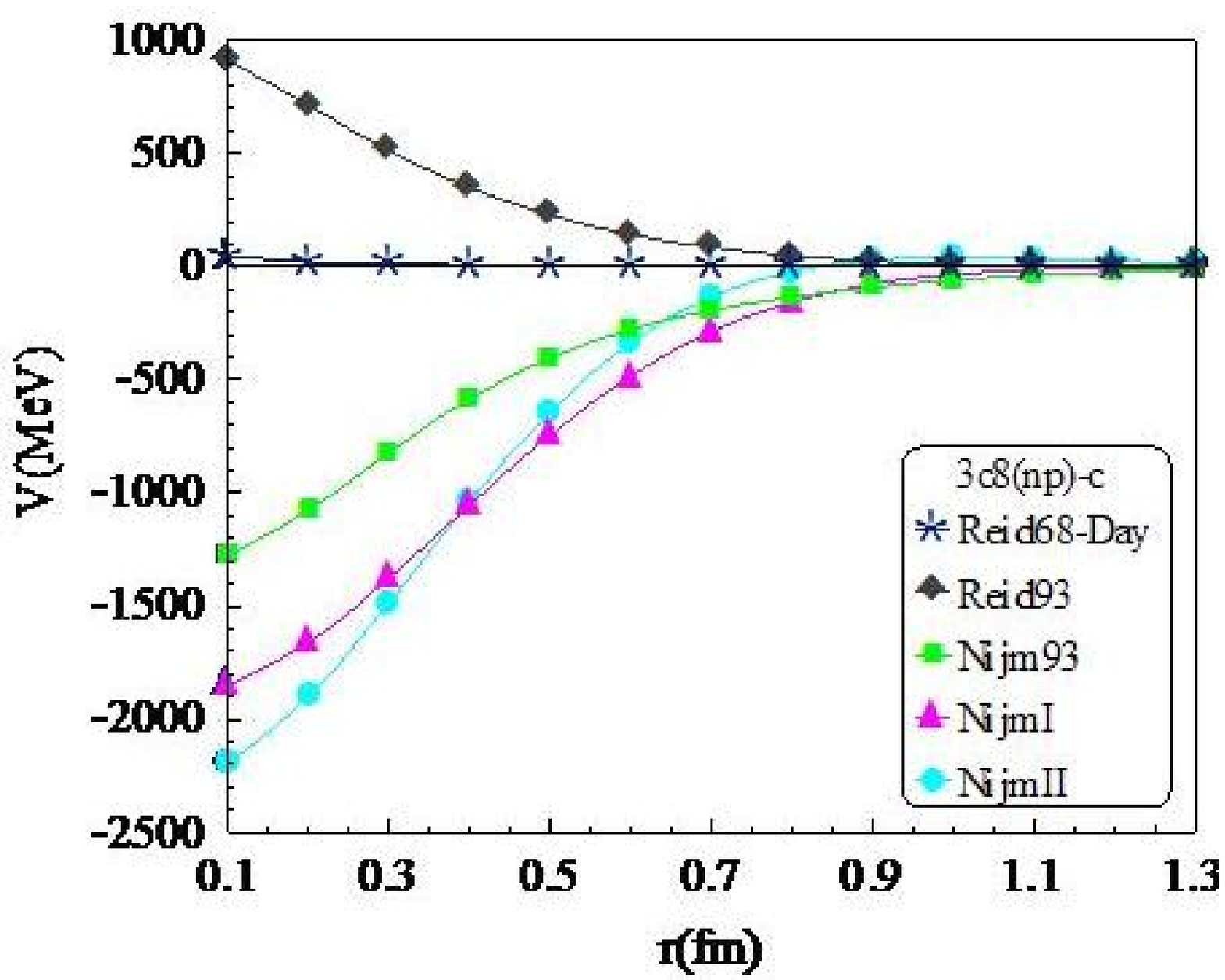}
          \end{subfigure}
    \end{subfigure}
\caption{} \label{Fig1.}
\end{figure}

\begin{figure}[p]
    \centering
      \begin{subfigure}[b]{\textwidth}
          \centering
          \begin{subfigure}[b]{0.47\textwidth}
                  \centering
                  \includegraphics[width=\textwidth,height=0.24\textheight]{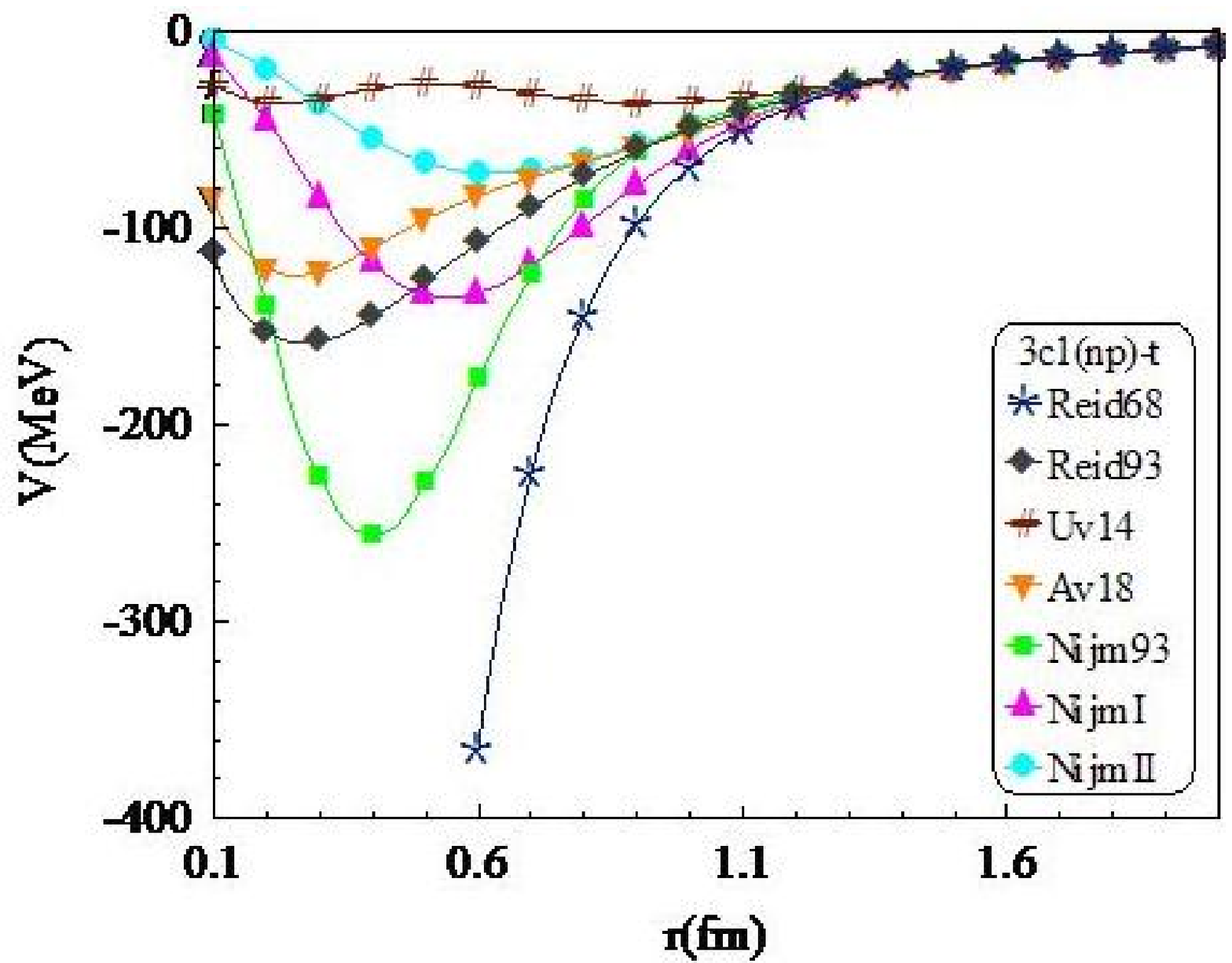}
          \end{subfigure}%
          ~
          \begin{subfigure}[b]{0.47\textwidth}
                  \centering
                  \includegraphics[width=\textwidth,height=0.24\textheight]{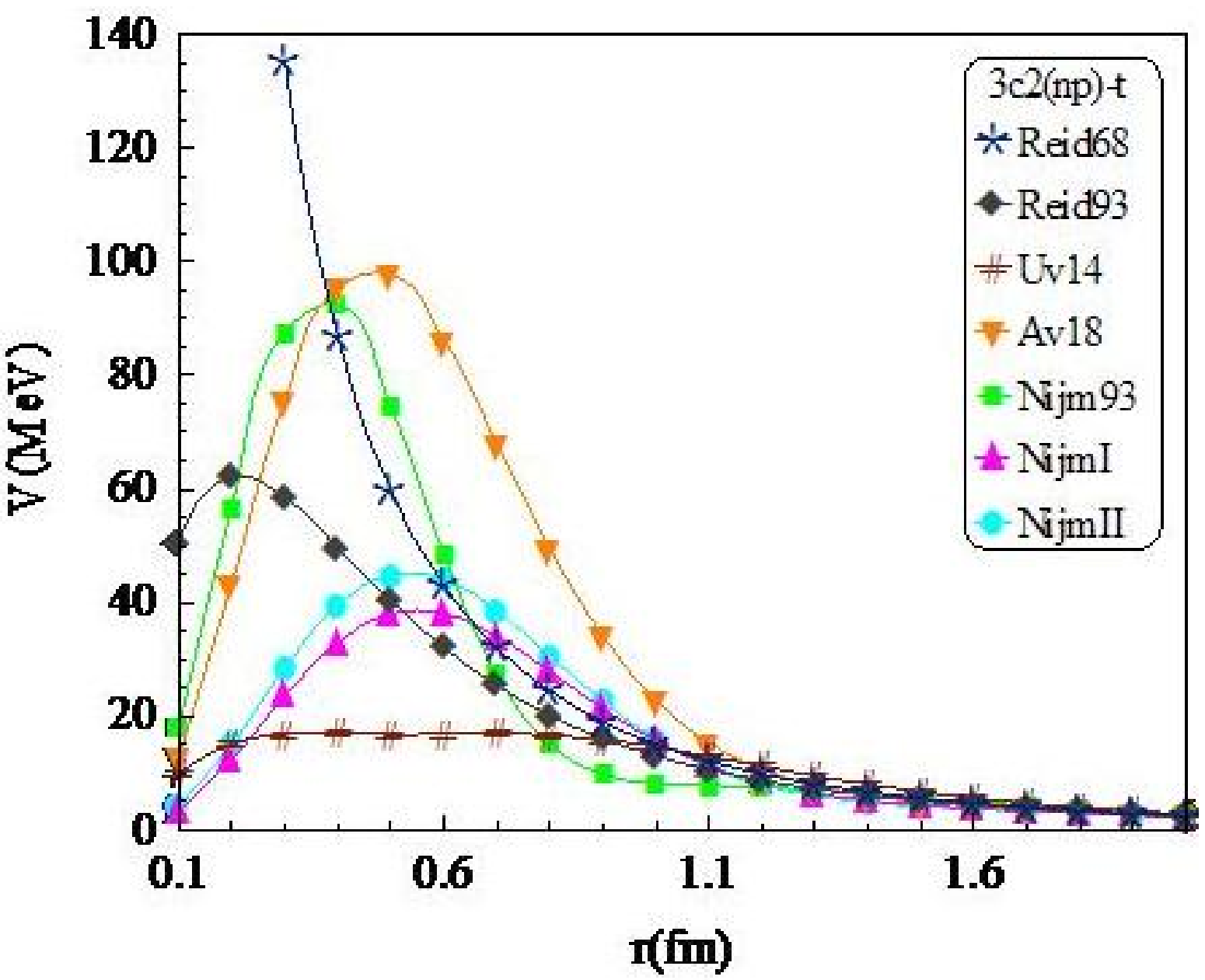}
          \end{subfigure}
    \end{subfigure}
         \begin{subfigure}[b]{\textwidth}
          \centering
          \begin{subfigure}[b]{0.47\textwidth}
                  \centering
                  \includegraphics[width=\textwidth,height=0.24\textheight]{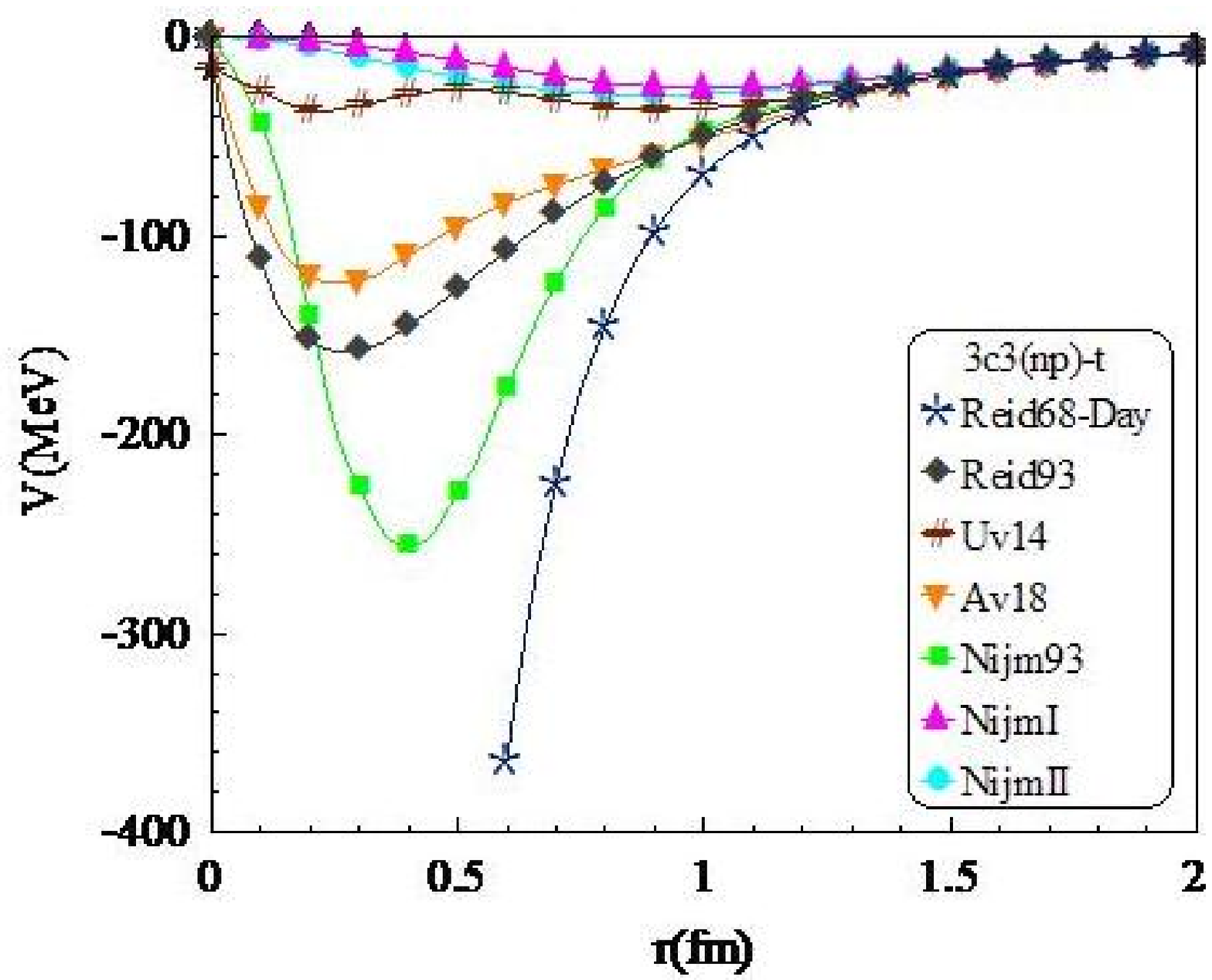}
          \end{subfigure}%
          ~
          \begin{subfigure}[b]{0.47\textwidth}
                  \centering
                  \includegraphics[width=\textwidth,height=0.24\textheight]{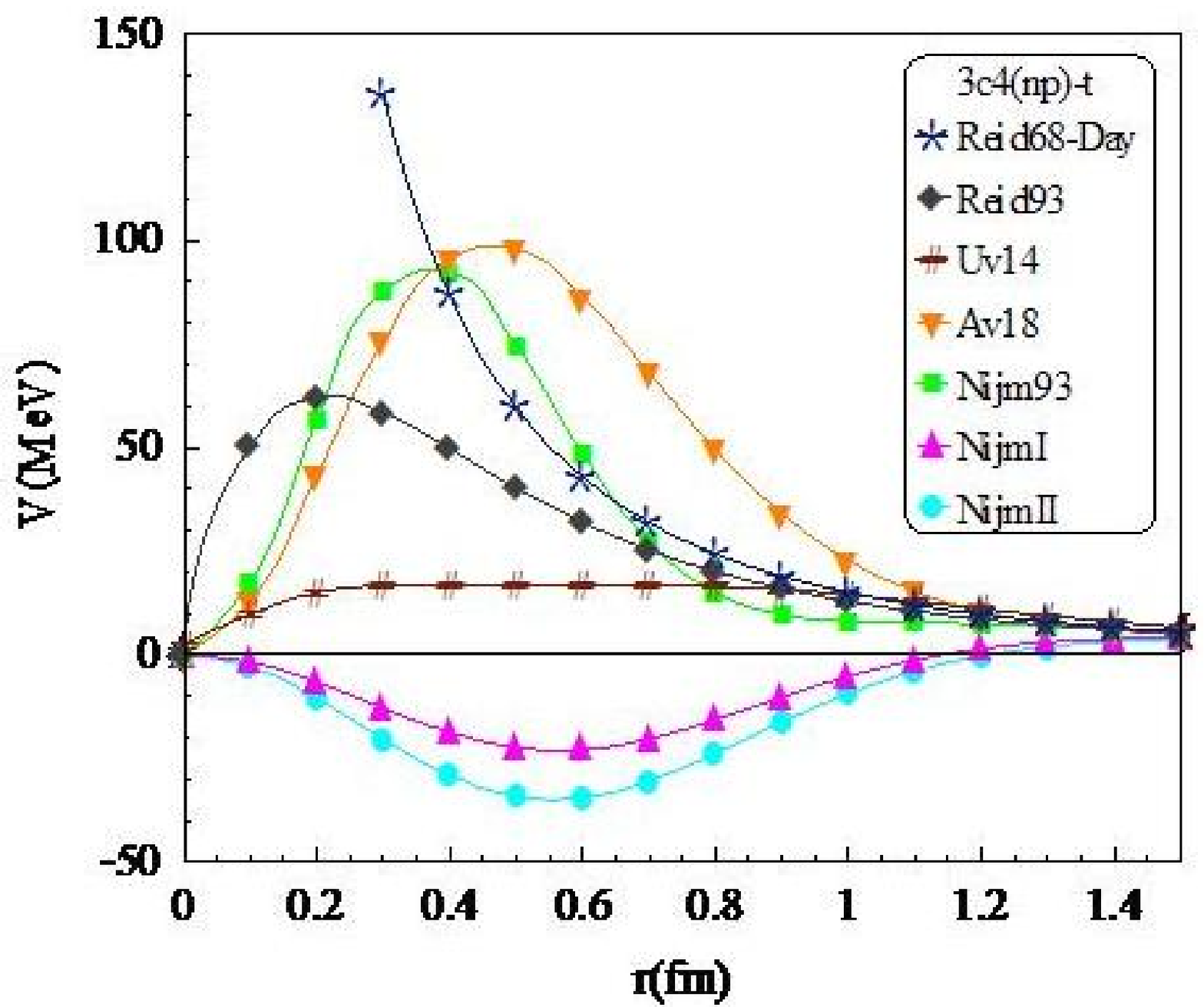}
          \end{subfigure}
    \end{subfigure}
        \begin{subfigure}[b]{\textwidth}
          \centering
          \begin{subfigure}[b]{0.47\textwidth}
                  \centering
                  \includegraphics[width=\textwidth,height=0.24\textheight]{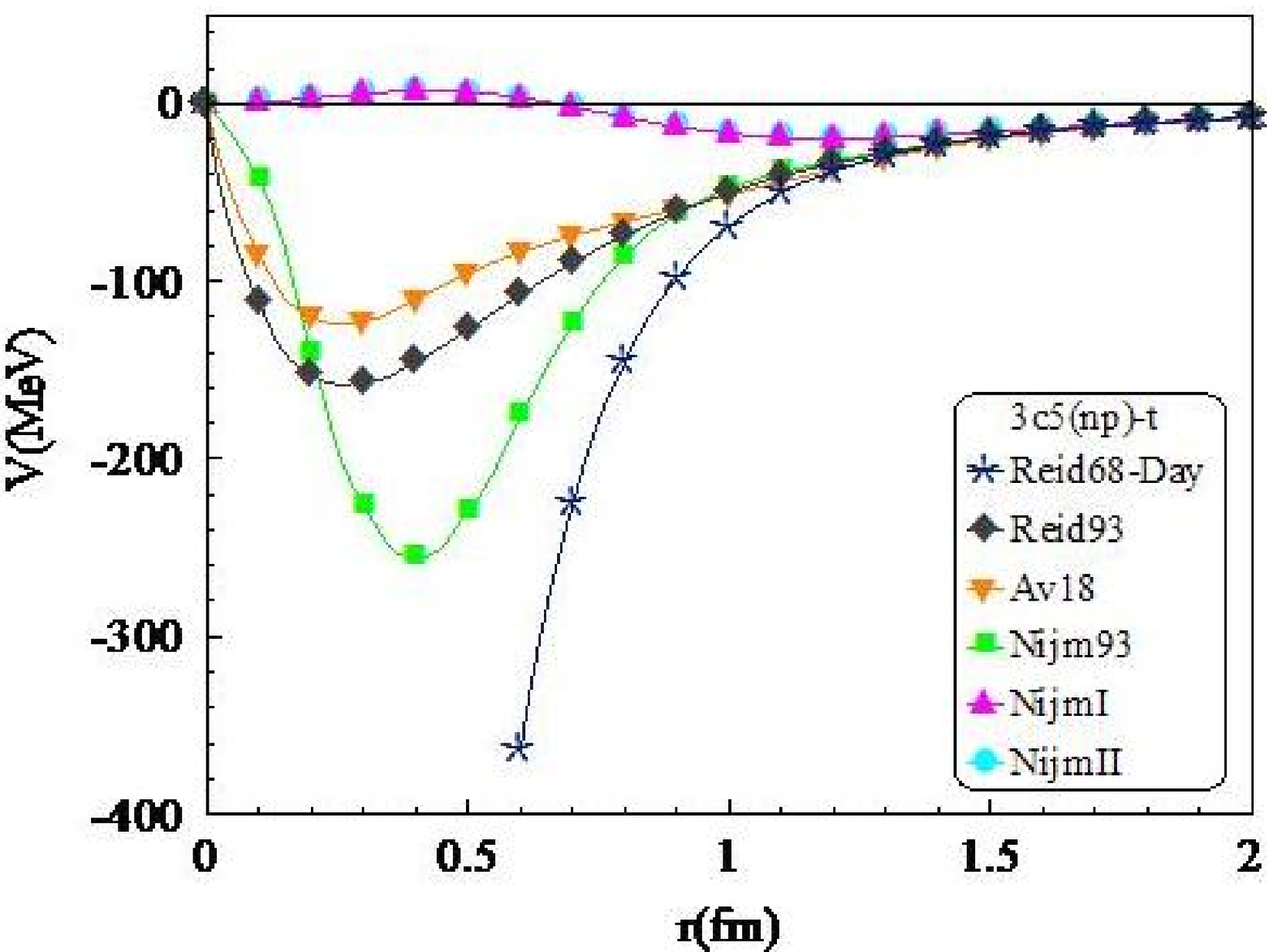}
          \end{subfigure}%
          ~
          \begin{subfigure}[b]{0.47\textwidth}
                  \centering
                  \includegraphics[width=\textwidth,height=0.24\textheight]{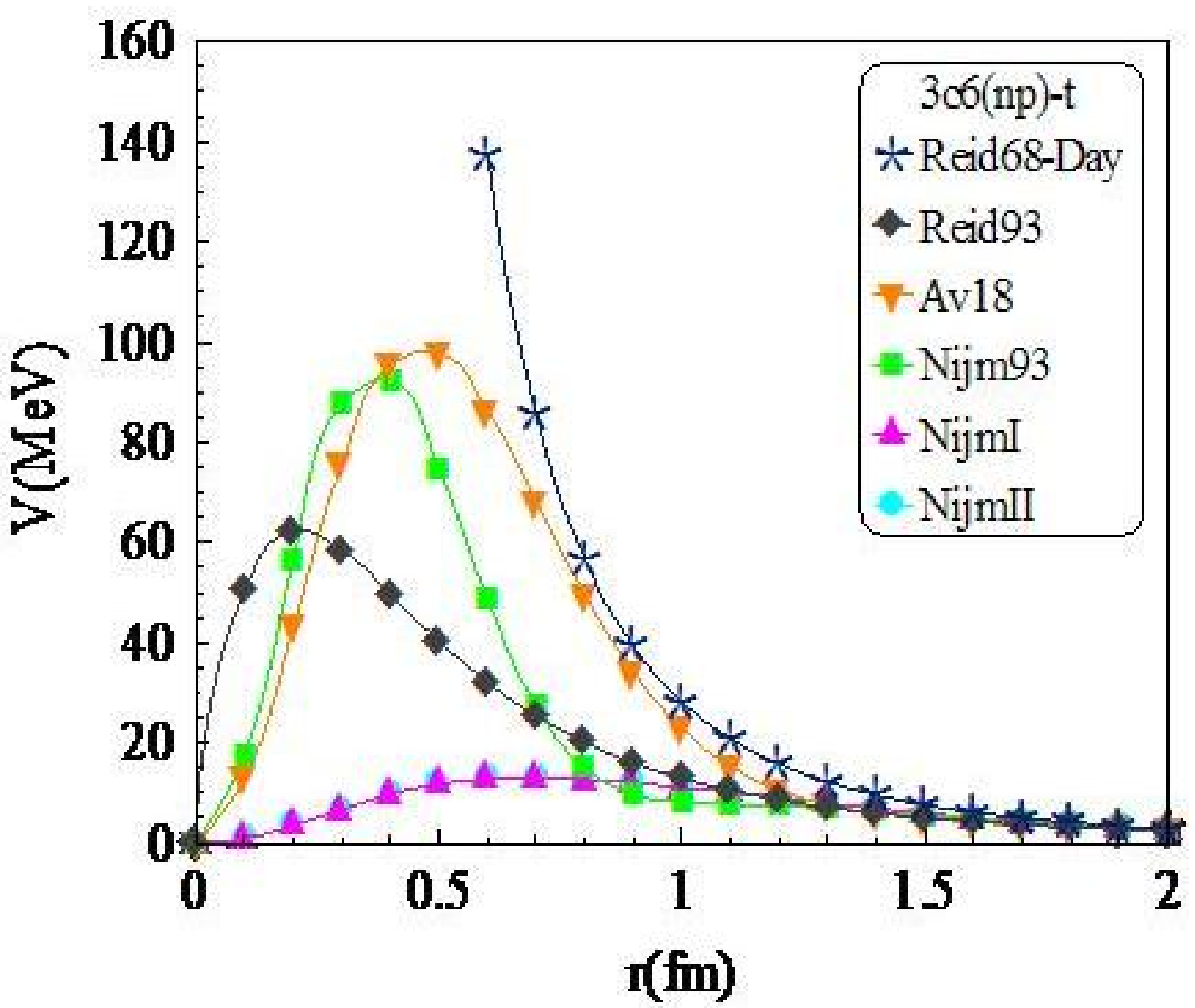}
          \end{subfigure}
    \end{subfigure}
        \begin{subfigure}[b]{\textwidth}
          \centering
          \begin{subfigure}[b]{0.47\textwidth}
                  \centering
                  \includegraphics[width=\textwidth,height=0.24\textheight]{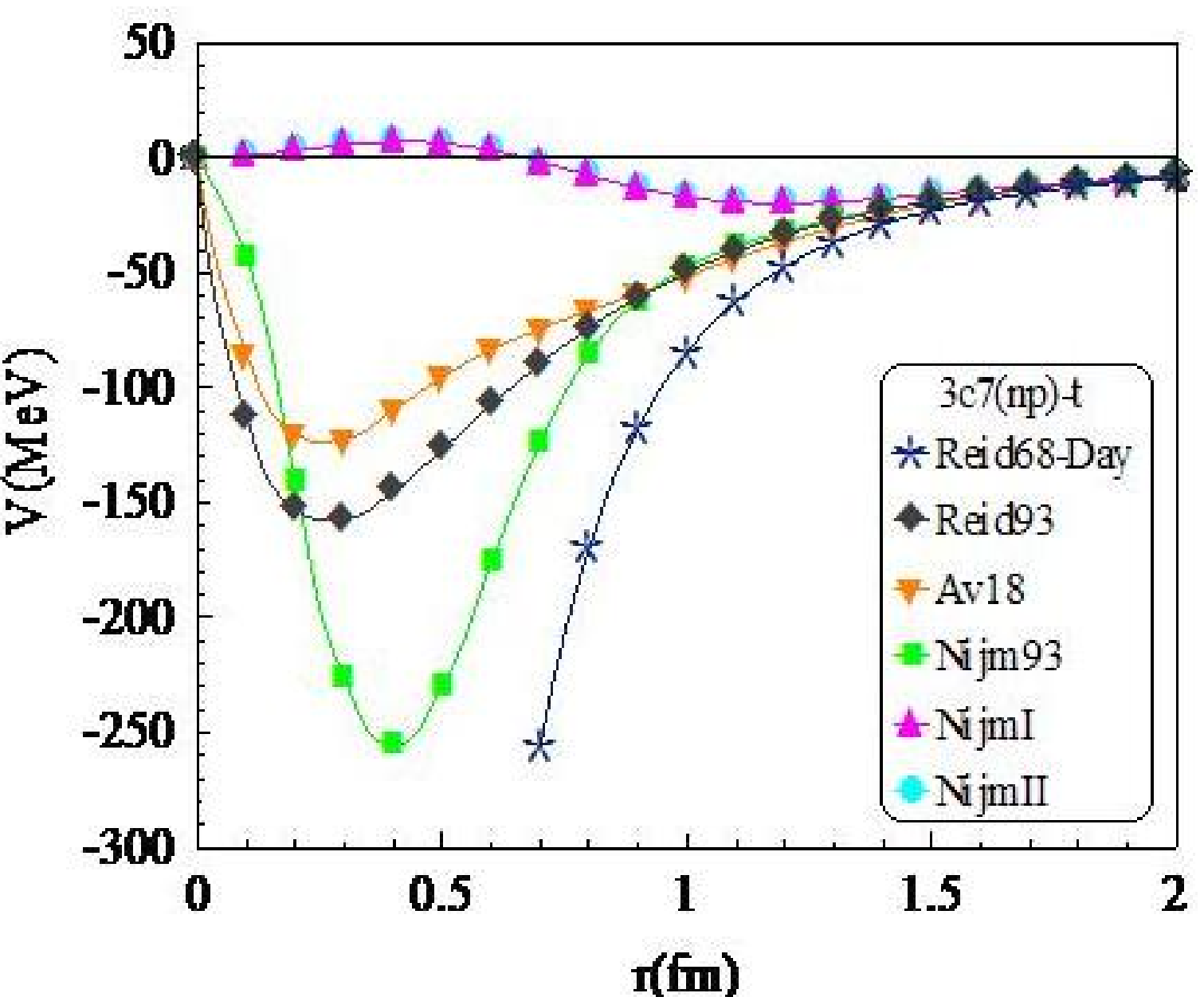}
          \end{subfigure}%
          ~
          \begin{subfigure}[b]{0.47\textwidth}
                  \centering
                  \includegraphics[width=\textwidth,height=0.24\textheight]{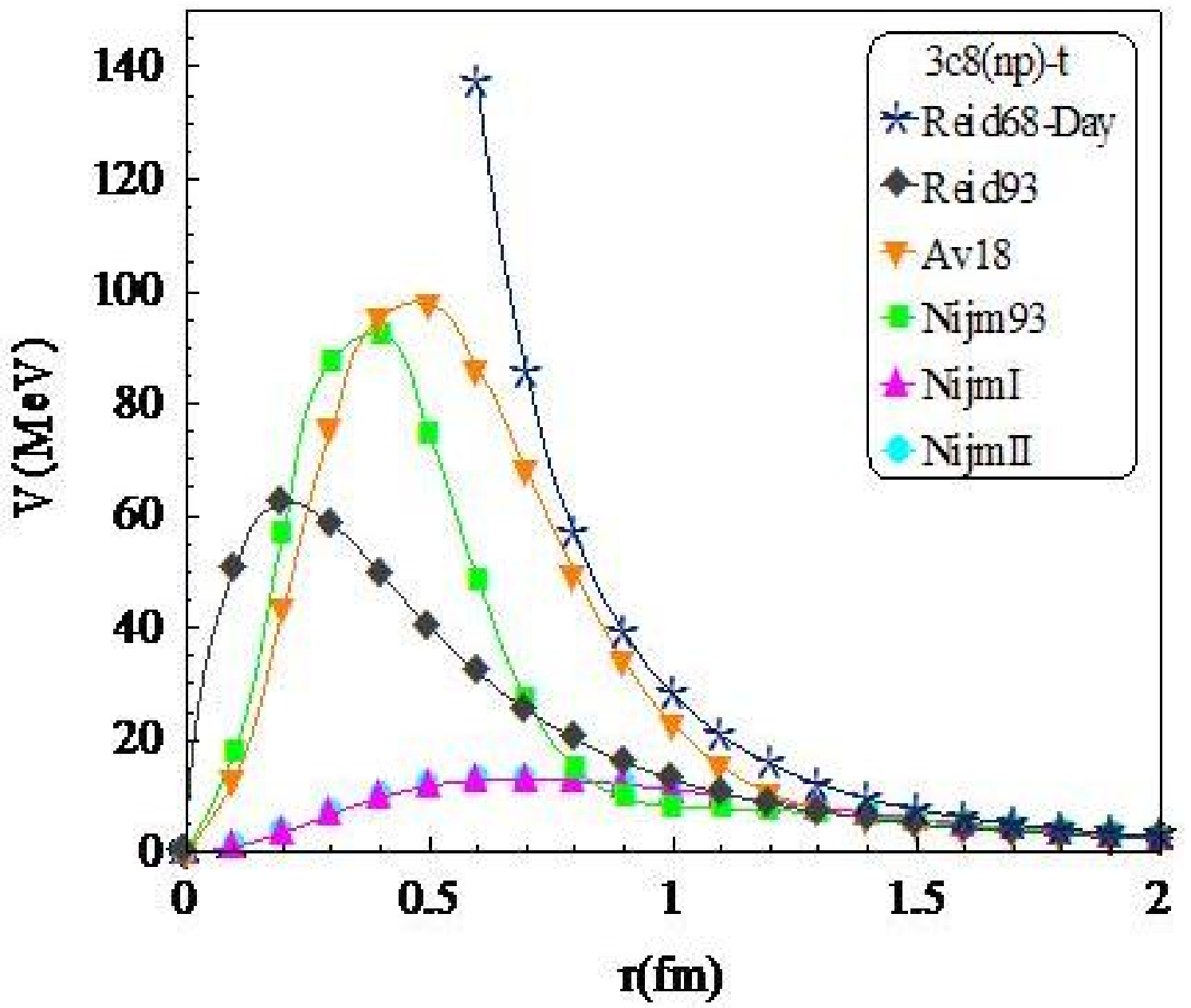}
          \end{subfigure}
    \end{subfigure}
\caption{\textit{The tensor potentials of various potential forms in the states from $J=1$ up to $J=8$, for np system}.} \label{Fig2.}
\end{figure}

\begin{figure}[p]
    \centering
      \begin{subfigure}[b]{\textwidth}
          \centering
          \begin{subfigure}[b]{0.47\textwidth}
                  \centering
                  \includegraphics[width=\textwidth,height=0.24\textheight]{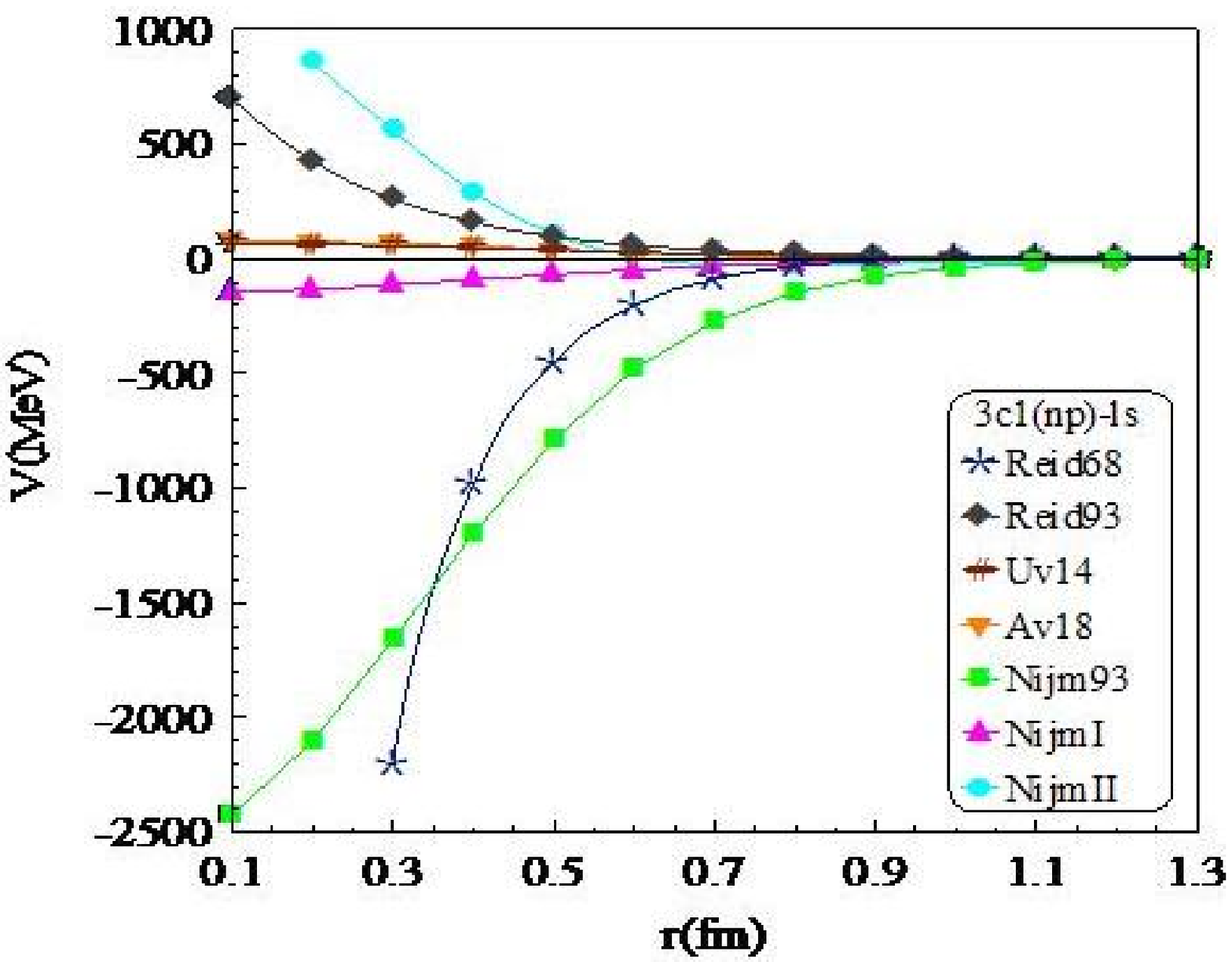}
          \end{subfigure}%
          ~
          \begin{subfigure}[b]{0.47\textwidth}
                  \centering
                  \includegraphics[width=\textwidth,height=0.24\textheight]{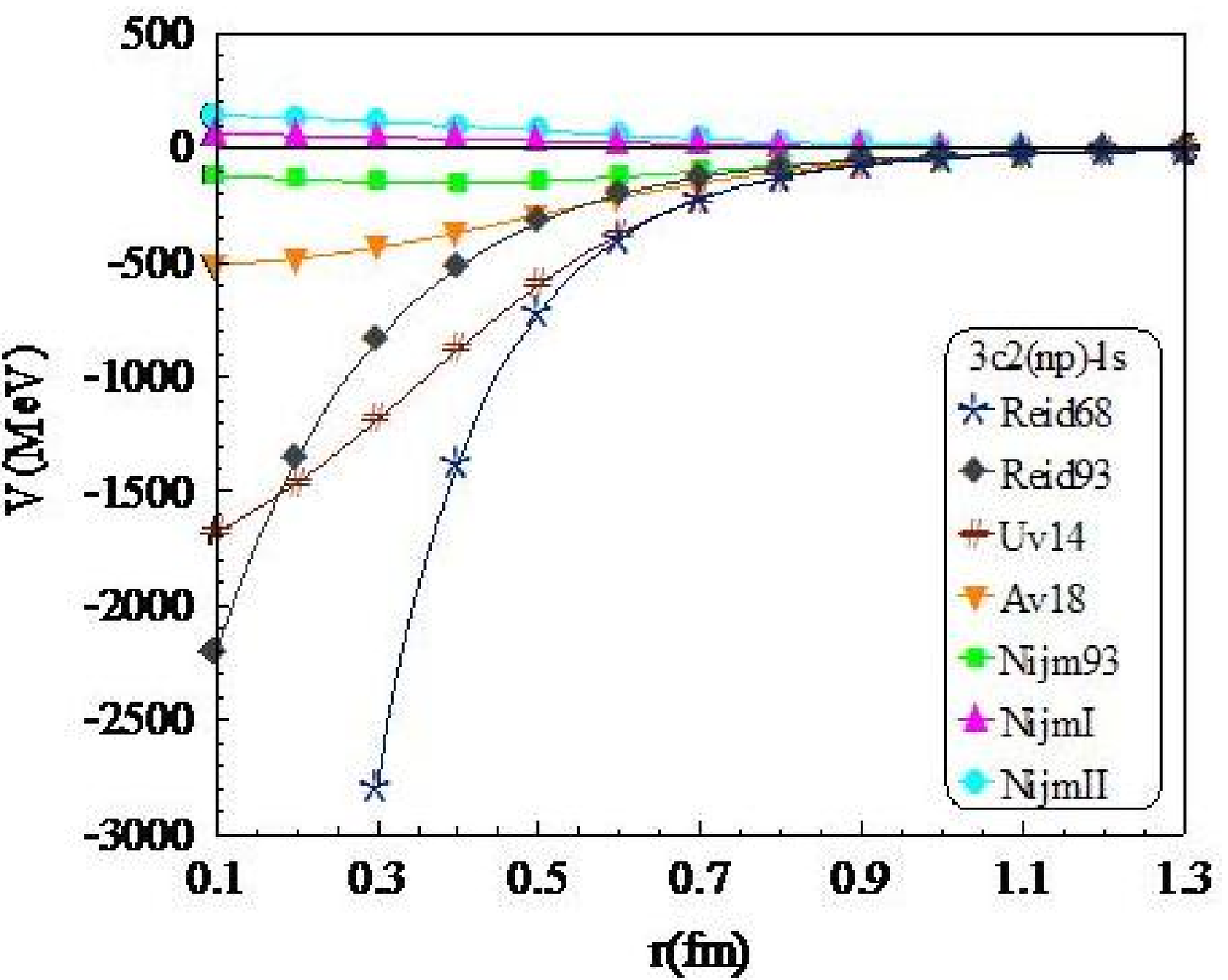}
          \end{subfigure}
    \end{subfigure}
         \begin{subfigure}[b]{\textwidth}
          \centering
          \begin{subfigure}[b]{0.47\textwidth}
                  \centering
                  \includegraphics[width=\textwidth,height=0.24\textheight]{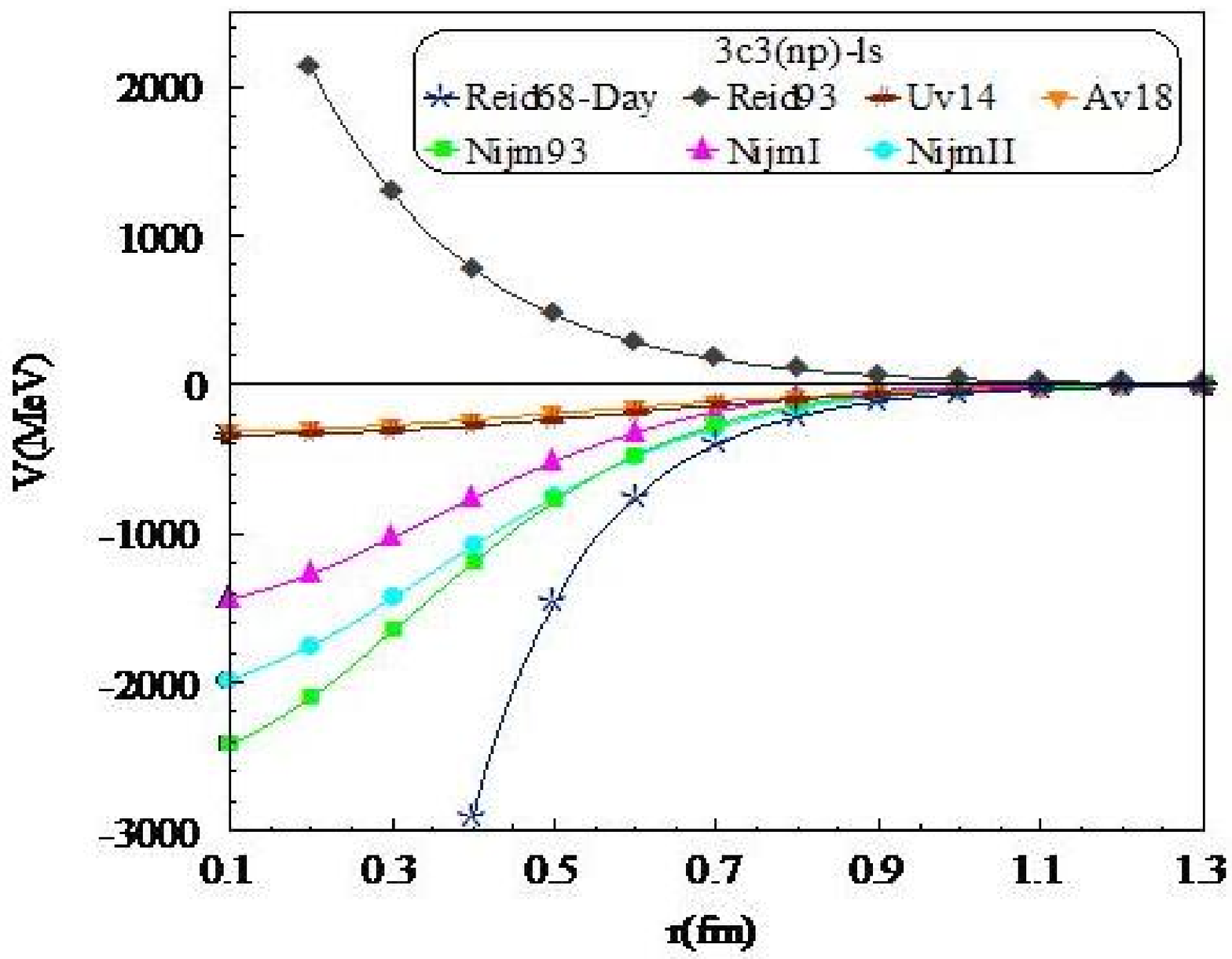}
          \end{subfigure}%
          ~
          \begin{subfigure}[b]{0.47\textwidth}
                  \centering
                  \includegraphics[width=\textwidth,height=0.24\textheight]{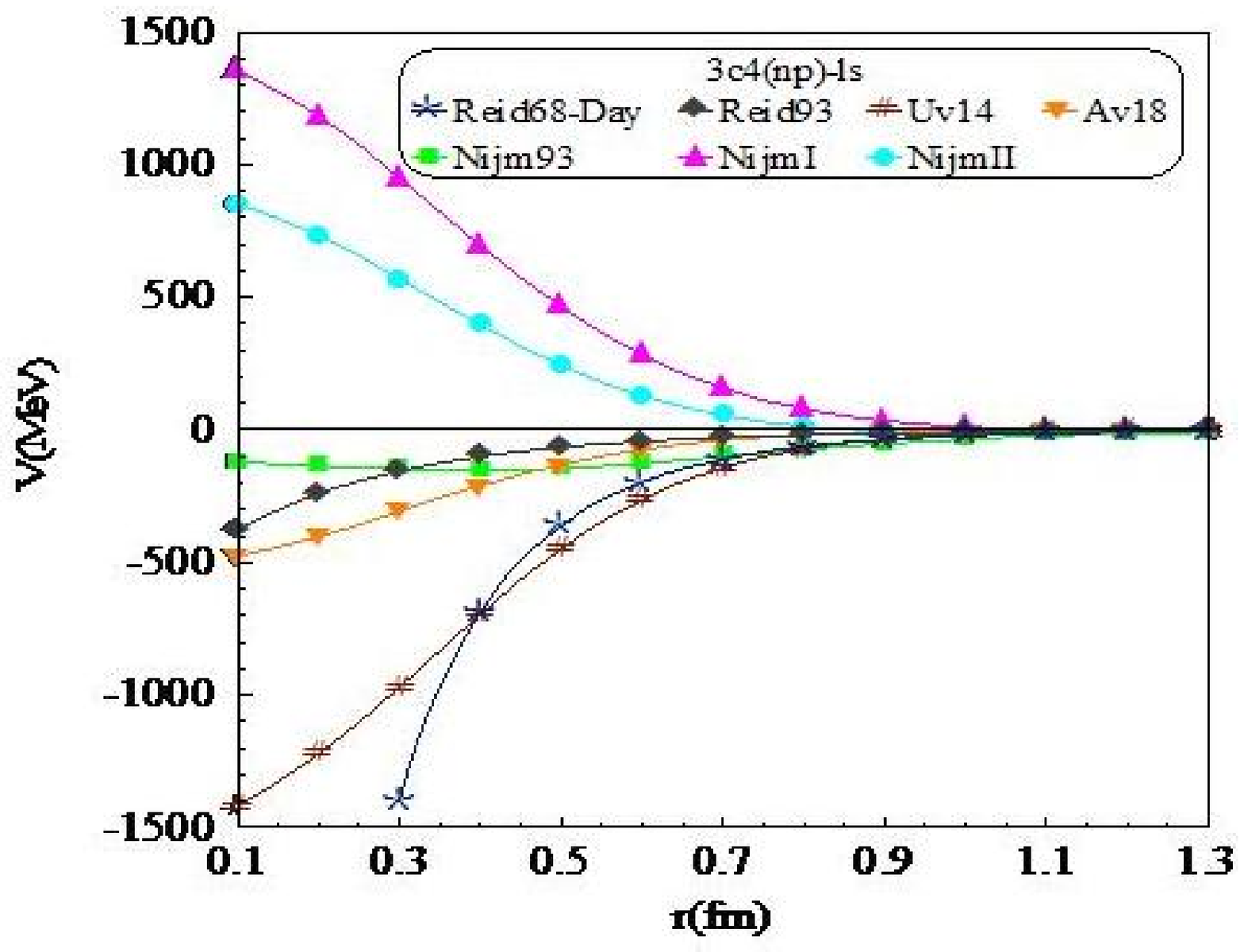}
          \end{subfigure}
    \end{subfigure}
        \begin{subfigure}[b]{\textwidth}
          \centering
          \begin{subfigure}[b]{0.47\textwidth}
                  \centering
                  \includegraphics[width=\textwidth,height=0.24\textheight]{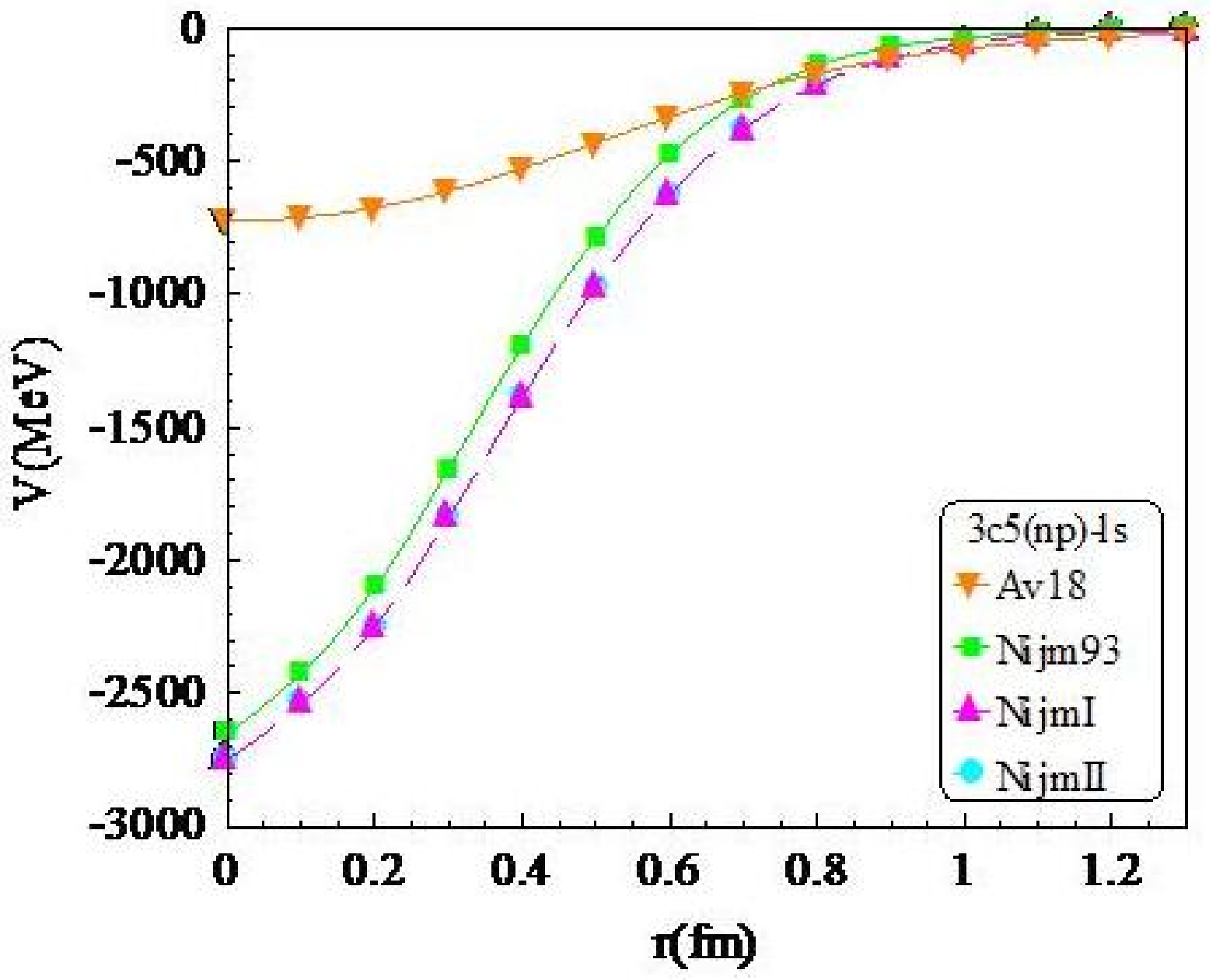}
          \end{subfigure}%
          ~
          \begin{subfigure}[b]{0.47\textwidth}
                  \centering
                  \includegraphics[width=\textwidth,height=0.24\textheight]{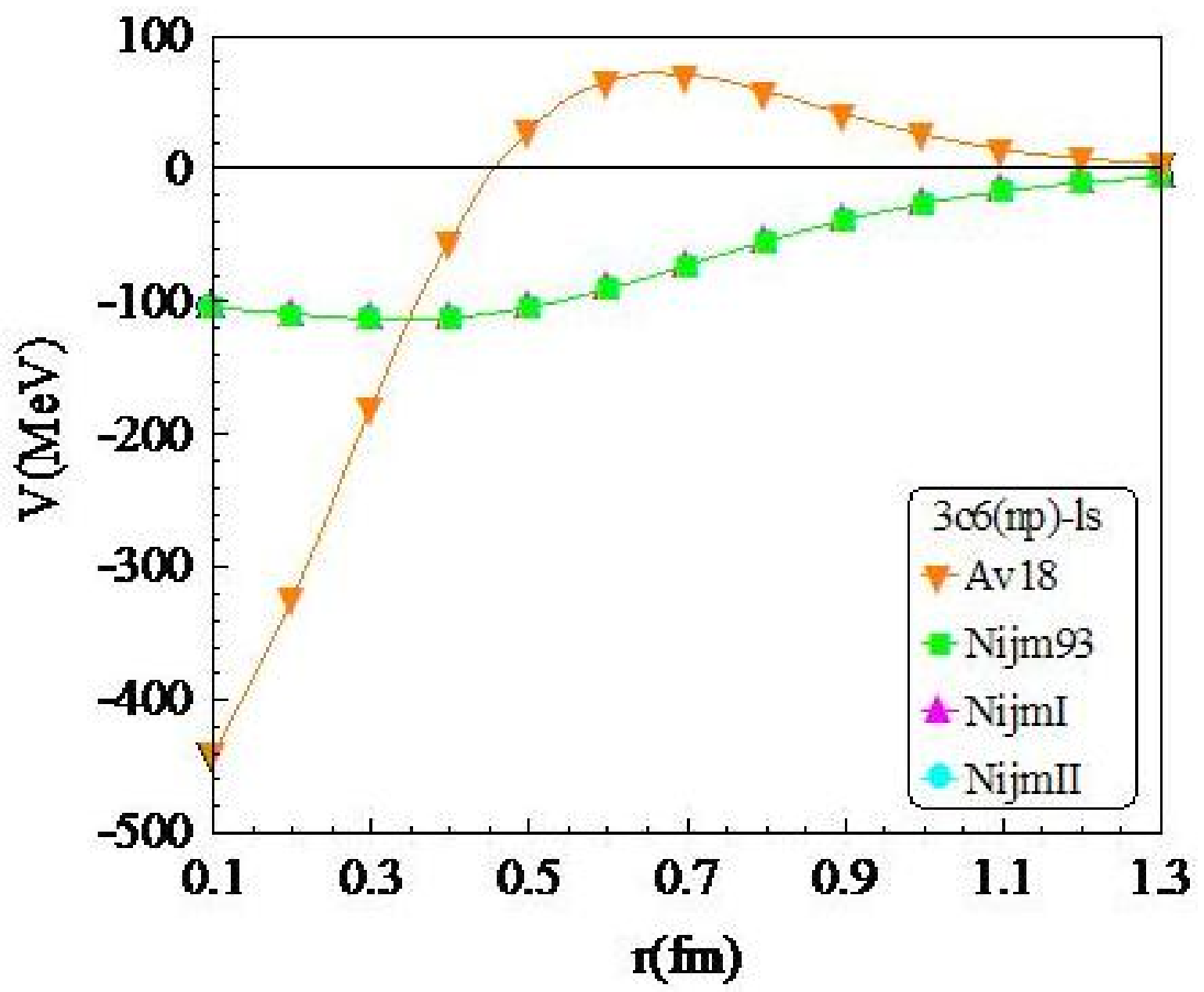}
          \end{subfigure}
    \end{subfigure}
        \begin{subfigure}[b]{\textwidth}
          \centering
          \begin{subfigure}[b]{0.47\textwidth}
                  \centering
                  \includegraphics[width=\textwidth,height=0.24\textheight]{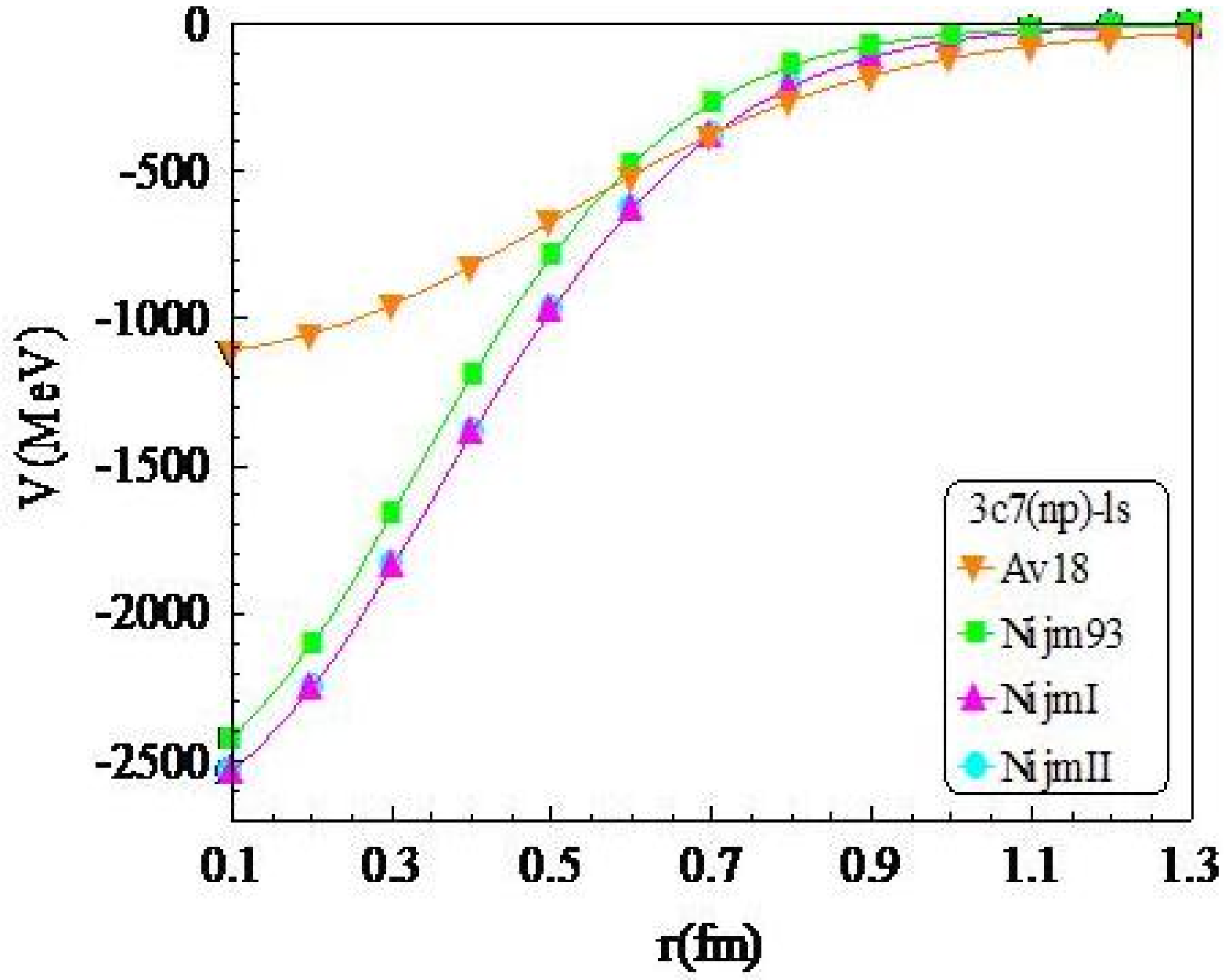}
          \end{subfigure}%
          ~
          \begin{subfigure}[b]{0.47\textwidth}
                  \centering
                  \includegraphics[width=\textwidth,height=0.24\textheight]{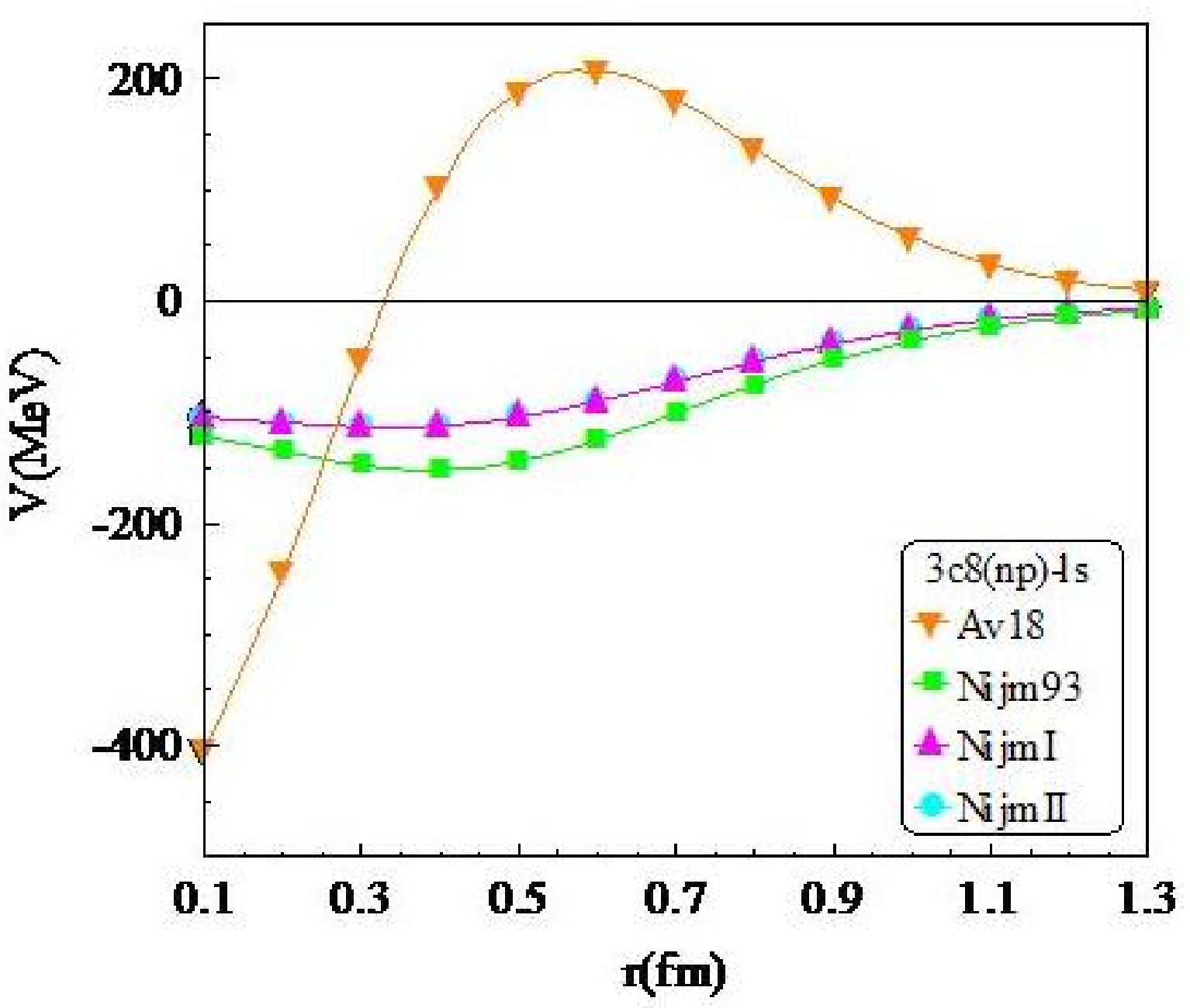}
          \end{subfigure}
    \end{subfigure}
\caption{\textit{The spin-orbit potentials of various potential forms in the states from $J=1$ up to $J=8$, for np system}.} \label{Fig3.}
\end{figure}

\begin{figure}[p]
    \centering
      \begin{subfigure}[b]{\textwidth}
          \centering
          \begin{subfigure}[b]{0.31\textwidth}
                  \centering
                  \includegraphics[width=\textwidth,height=0.24\textheight]{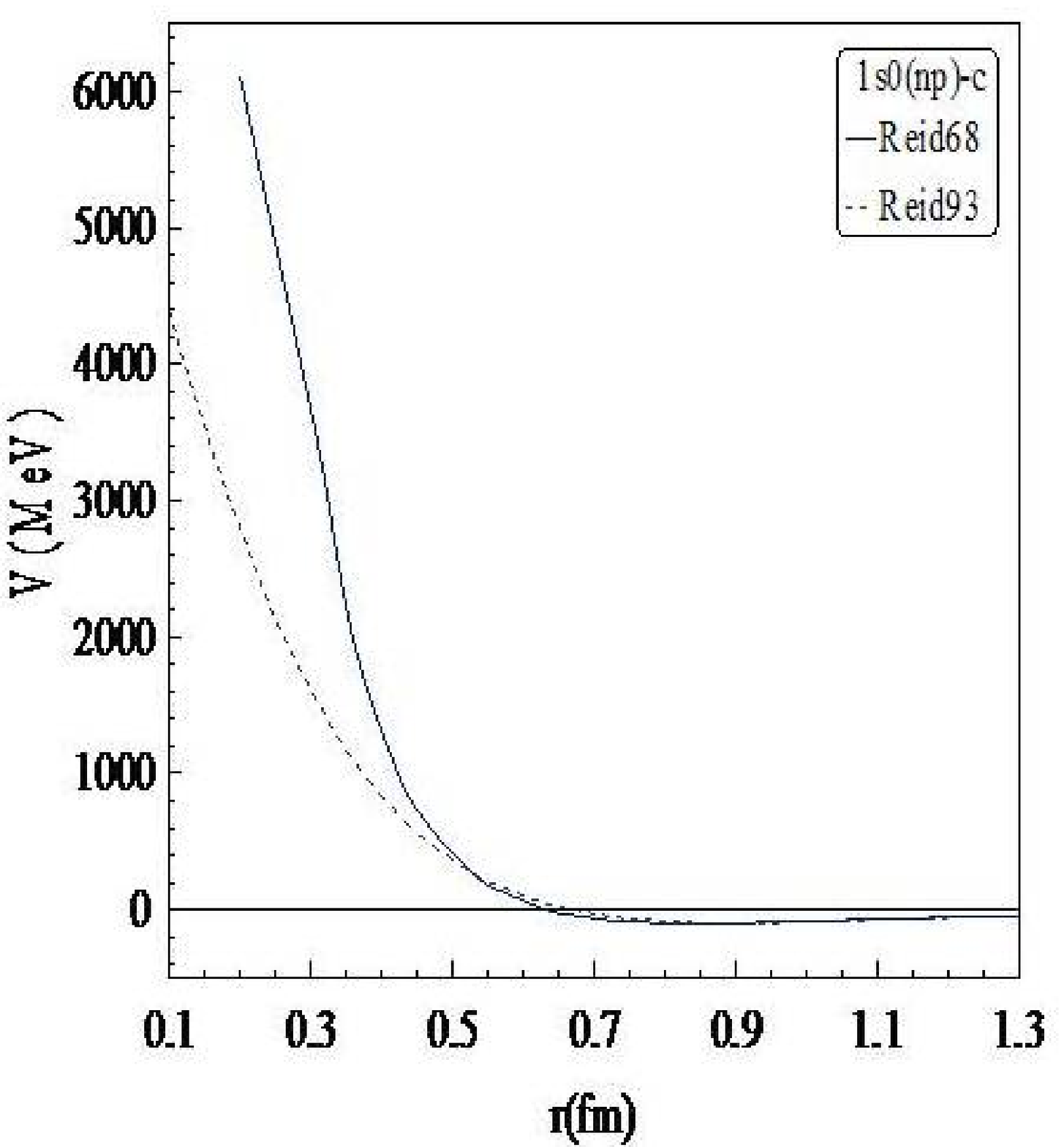}
          \end{subfigure}%
          ~
          \begin{subfigure}[b]{0.31\textwidth}
                  \centering
                  \includegraphics[width=\textwidth,height=0.24\textheight]{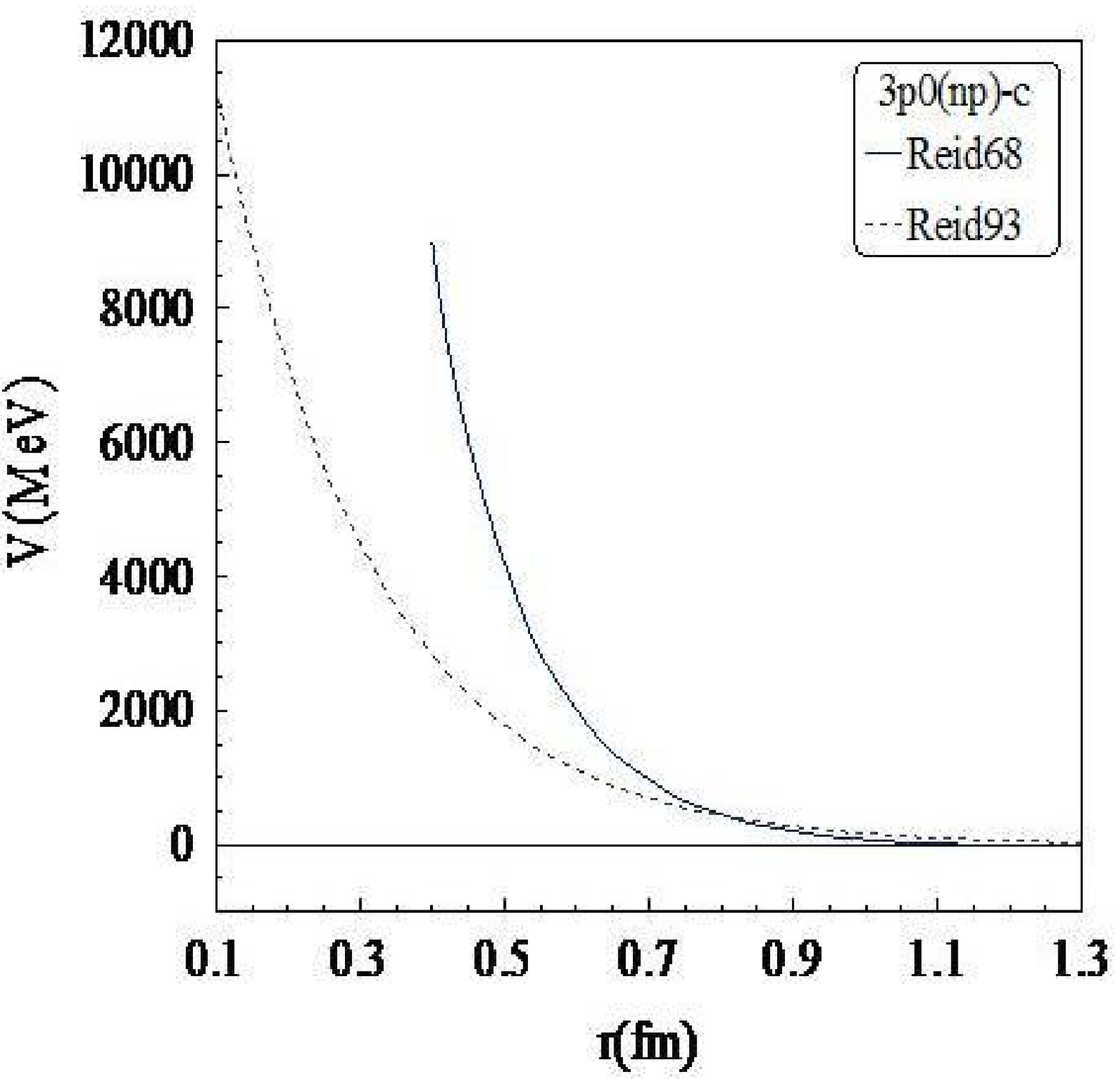}
          \end{subfigure}
           ~
          \begin{subfigure}[b]{0.31\textwidth}
                  \centering
                  \includegraphics[width=\textwidth,height=0.24\textheight]{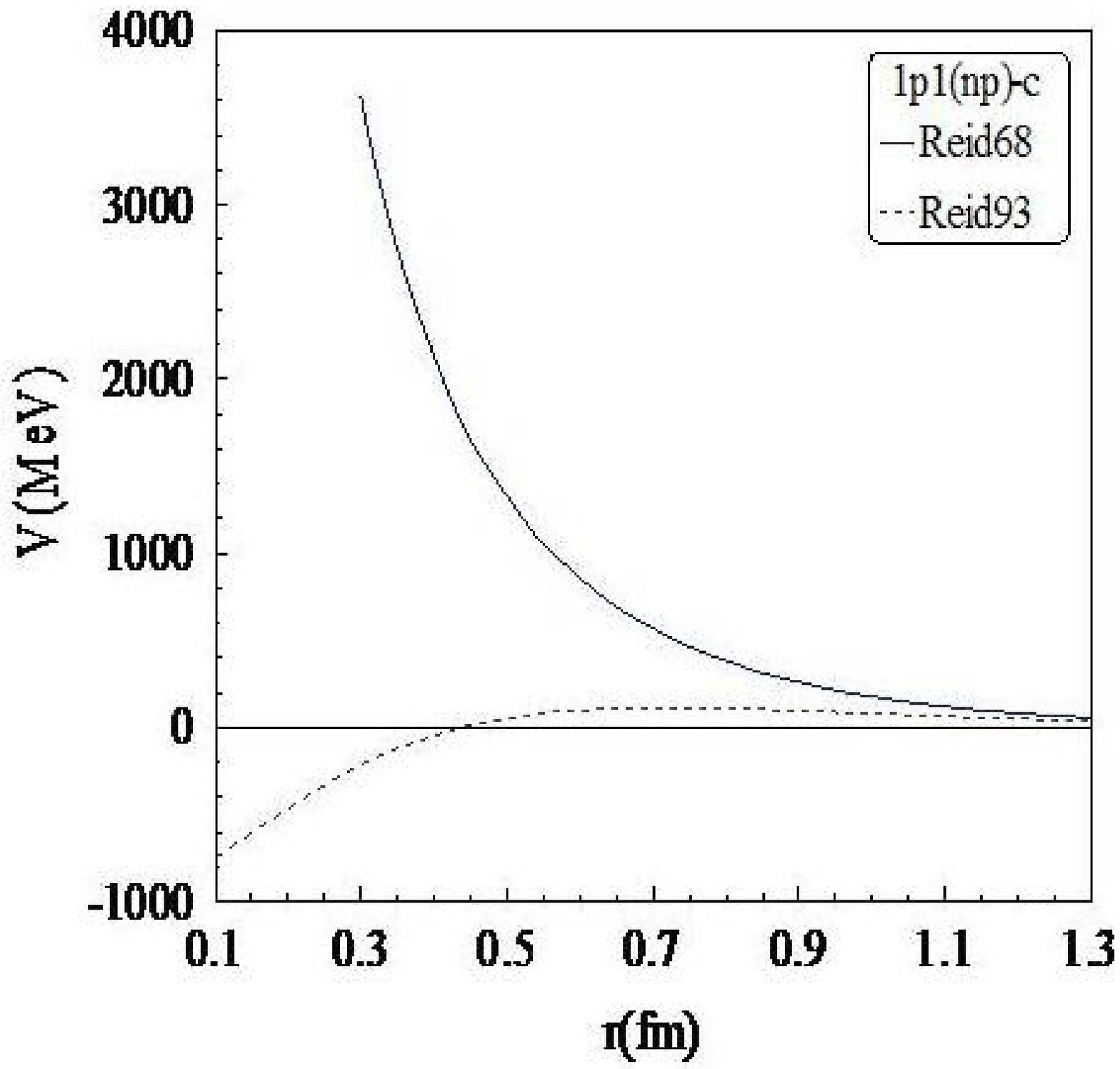}
          \end{subfigure}
    \end{subfigure}
    \begin{subfigure}[b]{\textwidth}
          \centering
           \begin{subfigure}[b]{0.31\textwidth}
                  \centering
                  \includegraphics[width=\textwidth,height=0.24\textheight]{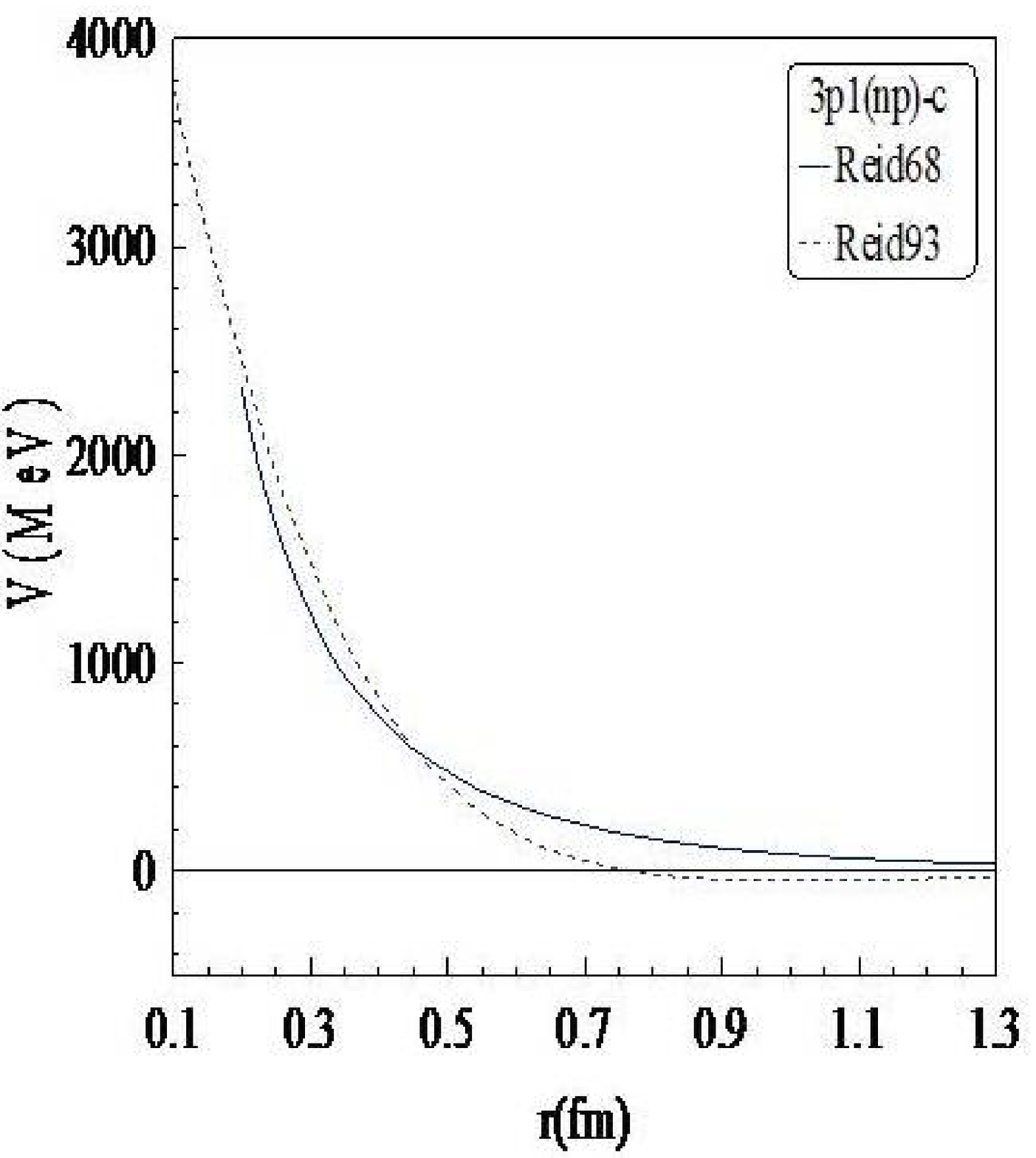}
          \end{subfigure}%
          ~
          \begin{subfigure}[b]{0.31\textwidth}
                  \centering
                  \includegraphics[width=\textwidth,height=0.24\textheight]{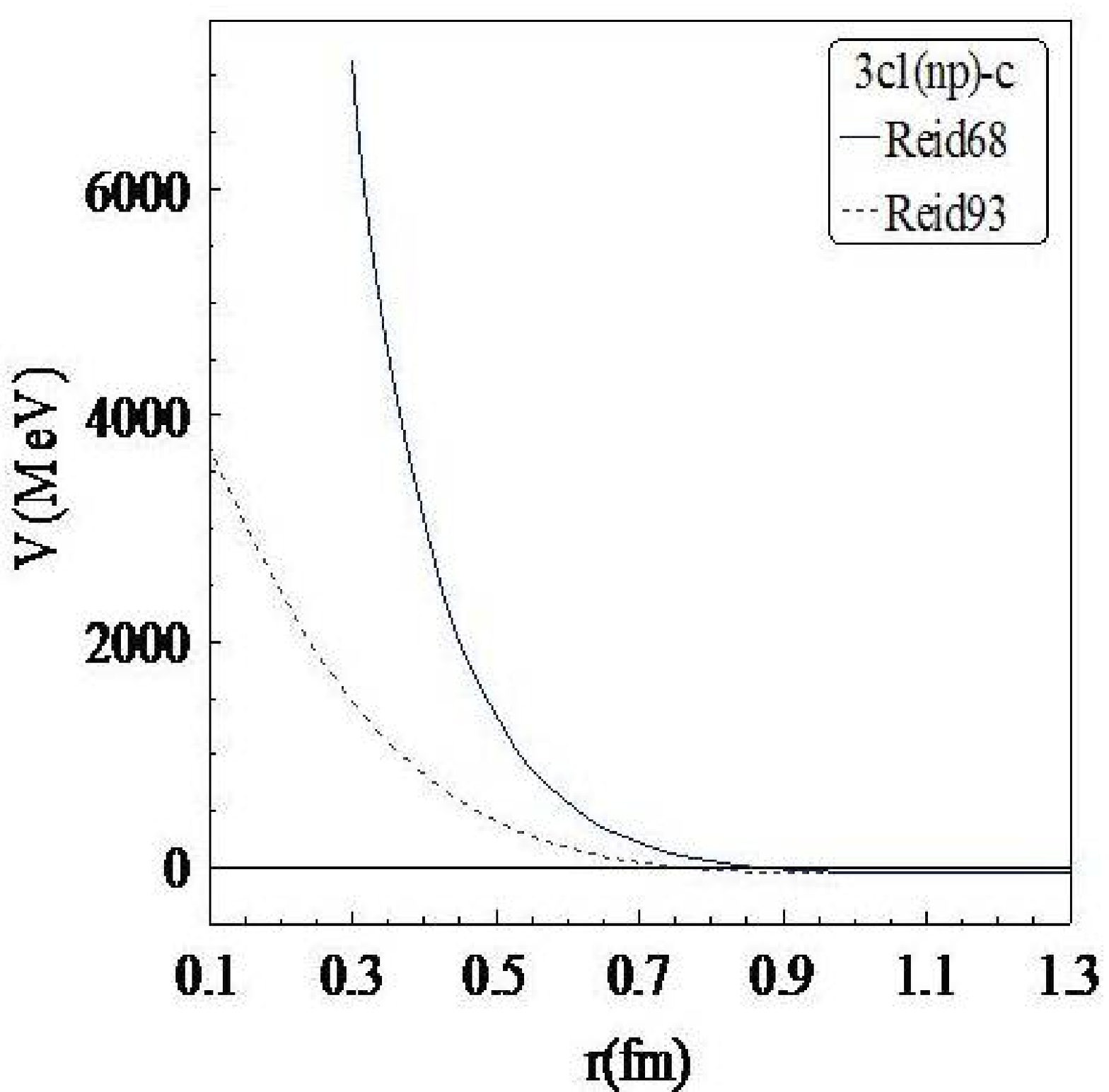}
          \end{subfigure}
           ~
          \begin{subfigure}[b]{0.31\textwidth}
                  \centering
                  \includegraphics[width=\textwidth,height=0.24\textheight]{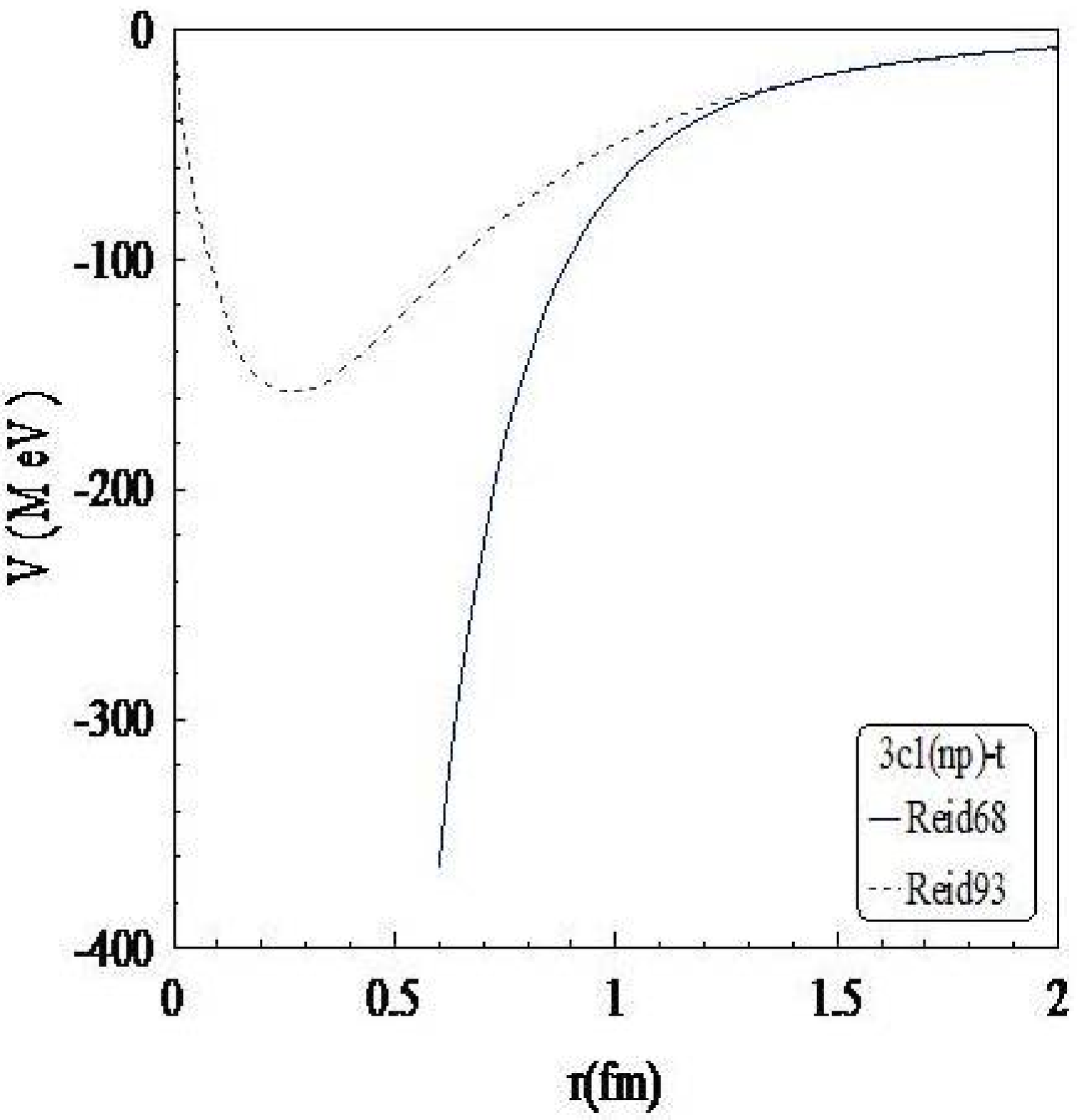}
          \end{subfigure}
   \end{subfigure}
   \begin{subfigure}[b]{\textwidth}
          \centering
        \begin{subfigure}[b]{0.31\textwidth}
                  \centering
                  \includegraphics[width=\textwidth,height=0.24\textheight]{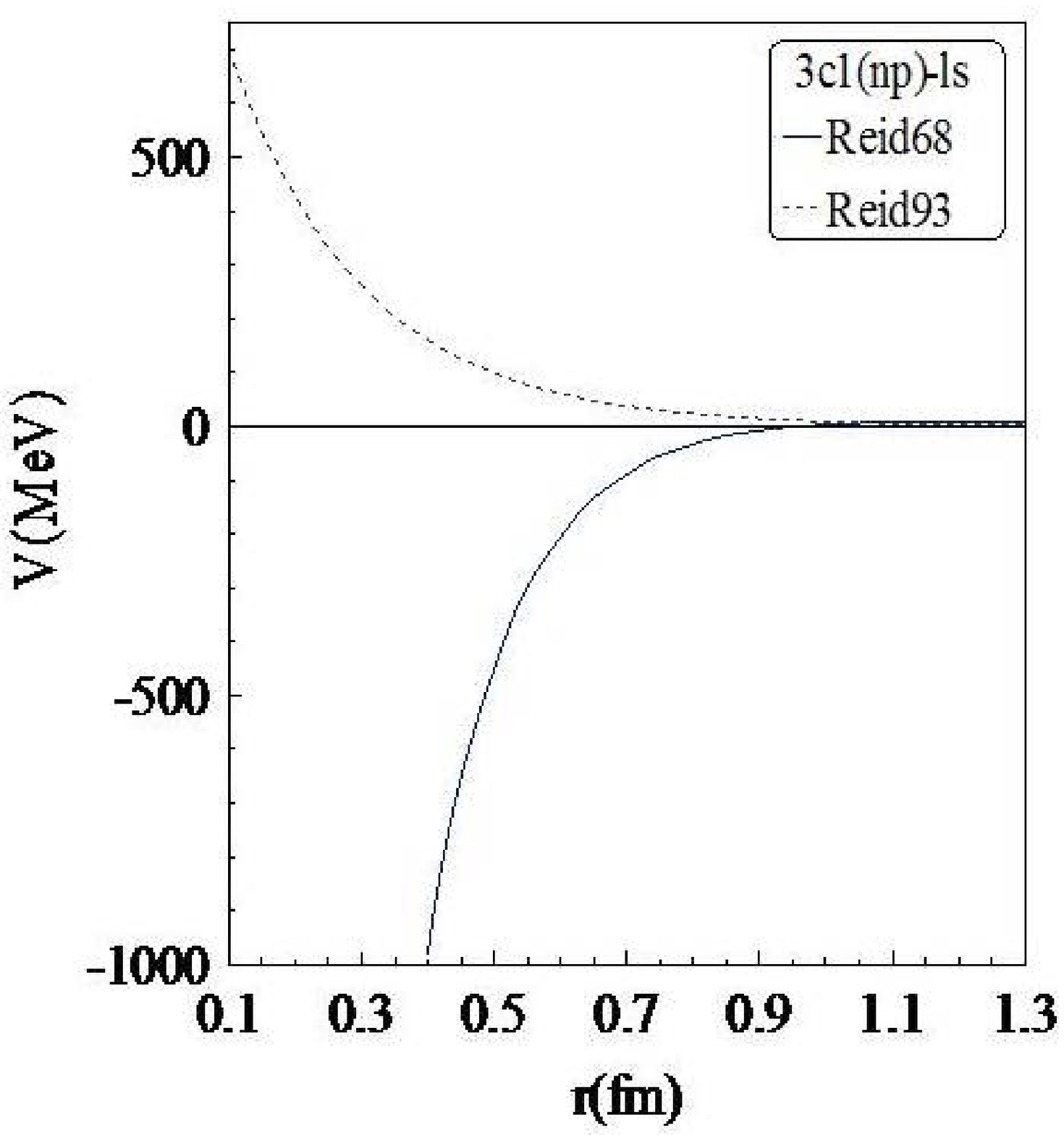}
          \end{subfigure}%
          ~
          \begin{subfigure}[b]{0.31\textwidth}
                  \centering
                  \includegraphics[width=\textwidth,height=0.24\textheight]{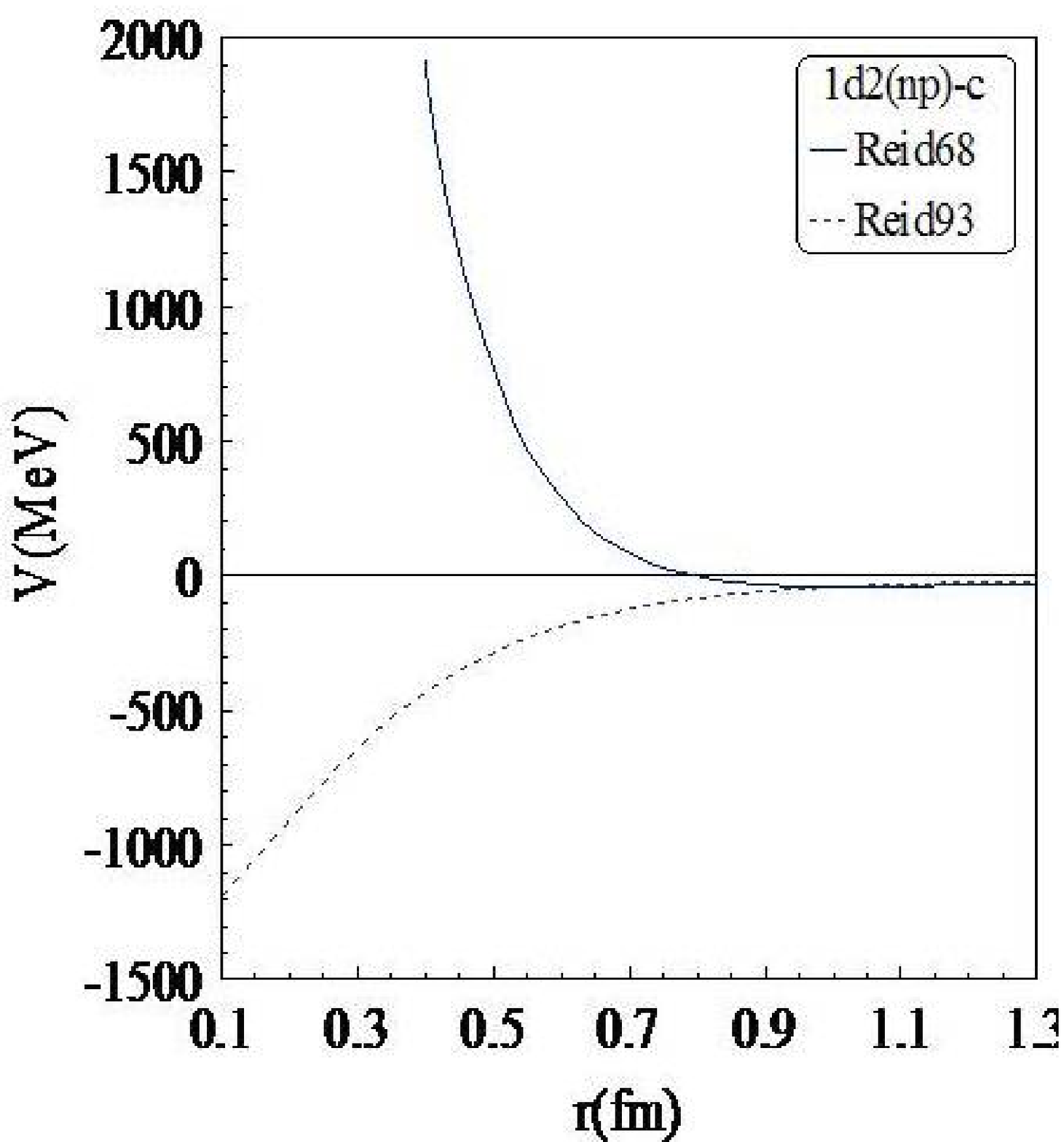}
          \end{subfigure}
           ~
          \begin{subfigure}[b]{0.31\textwidth}
                  \centering
                  \includegraphics[width=\textwidth,height=0.24\textheight]{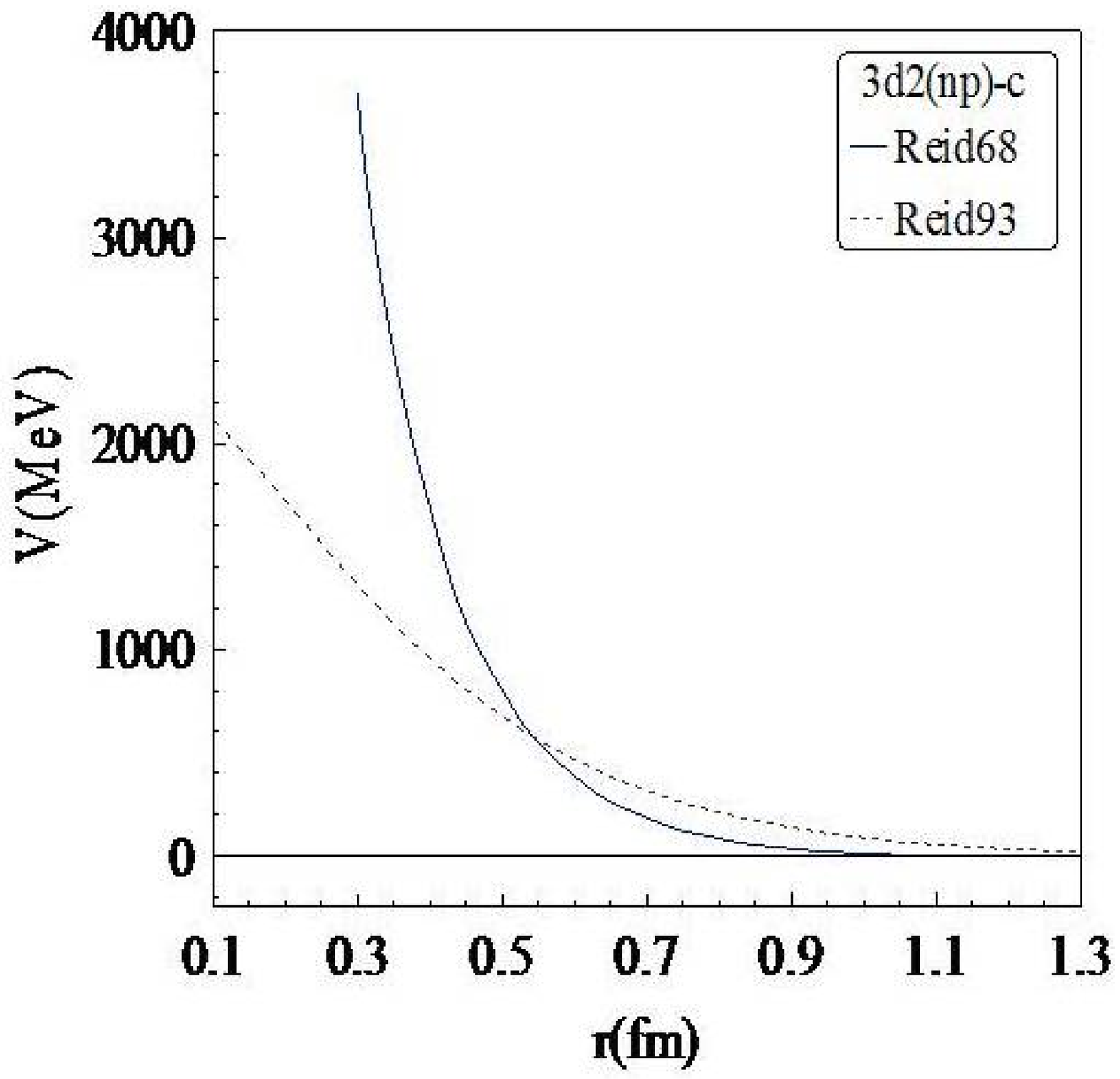}
          \end{subfigure}
   \end{subfigure}
   \begin{subfigure}[b]{\textwidth}
          \centering
        \begin{subfigure}[b]{0.31\textwidth}
                  \centering
                  \includegraphics[width=\textwidth,height=0.24\textheight]{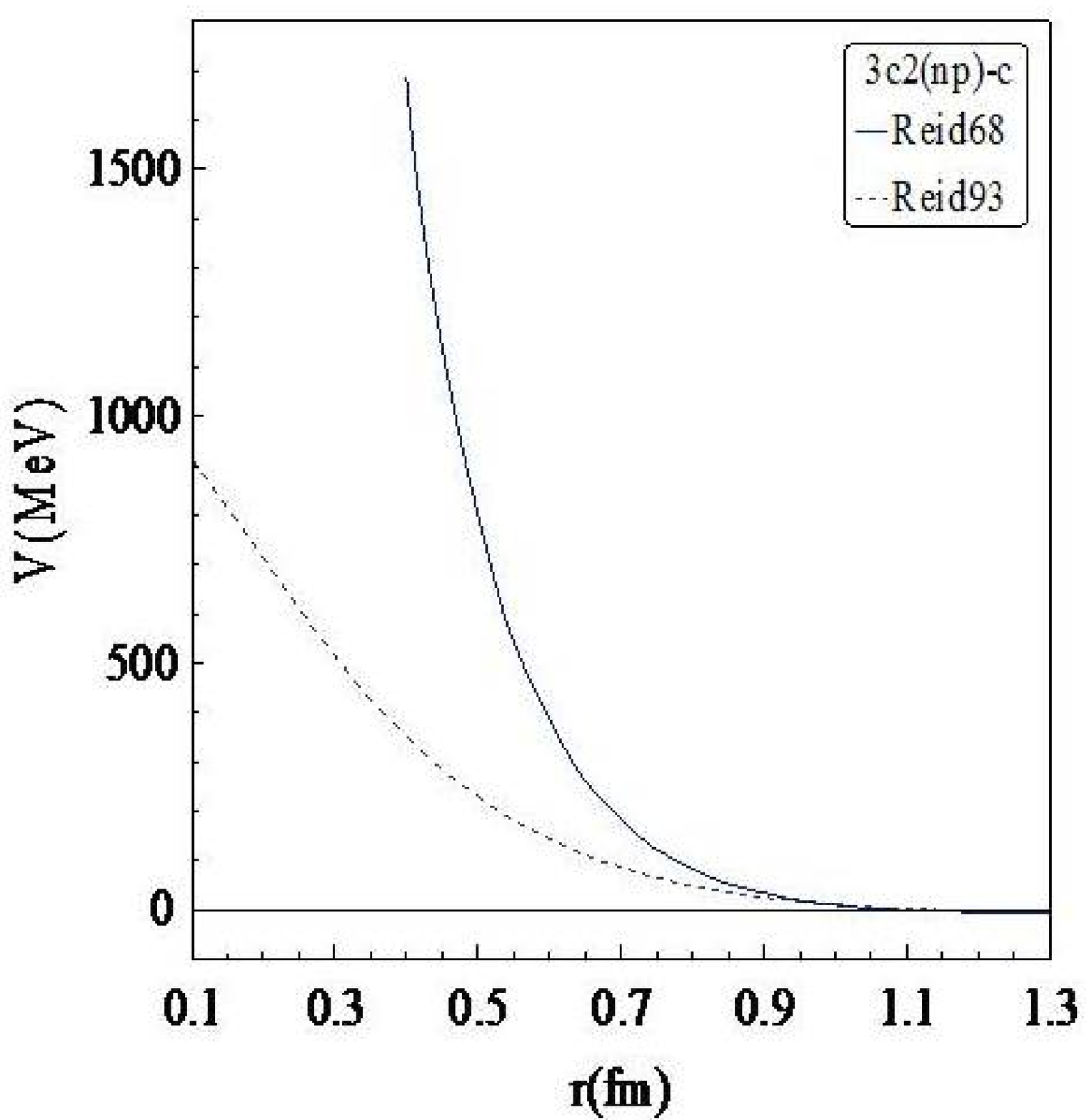}
          \end{subfigure}%
          ~
          \begin{subfigure}[b]{0.31\textwidth}
                  \centering
                  \includegraphics[width=\textwidth,height=0.24\textheight]{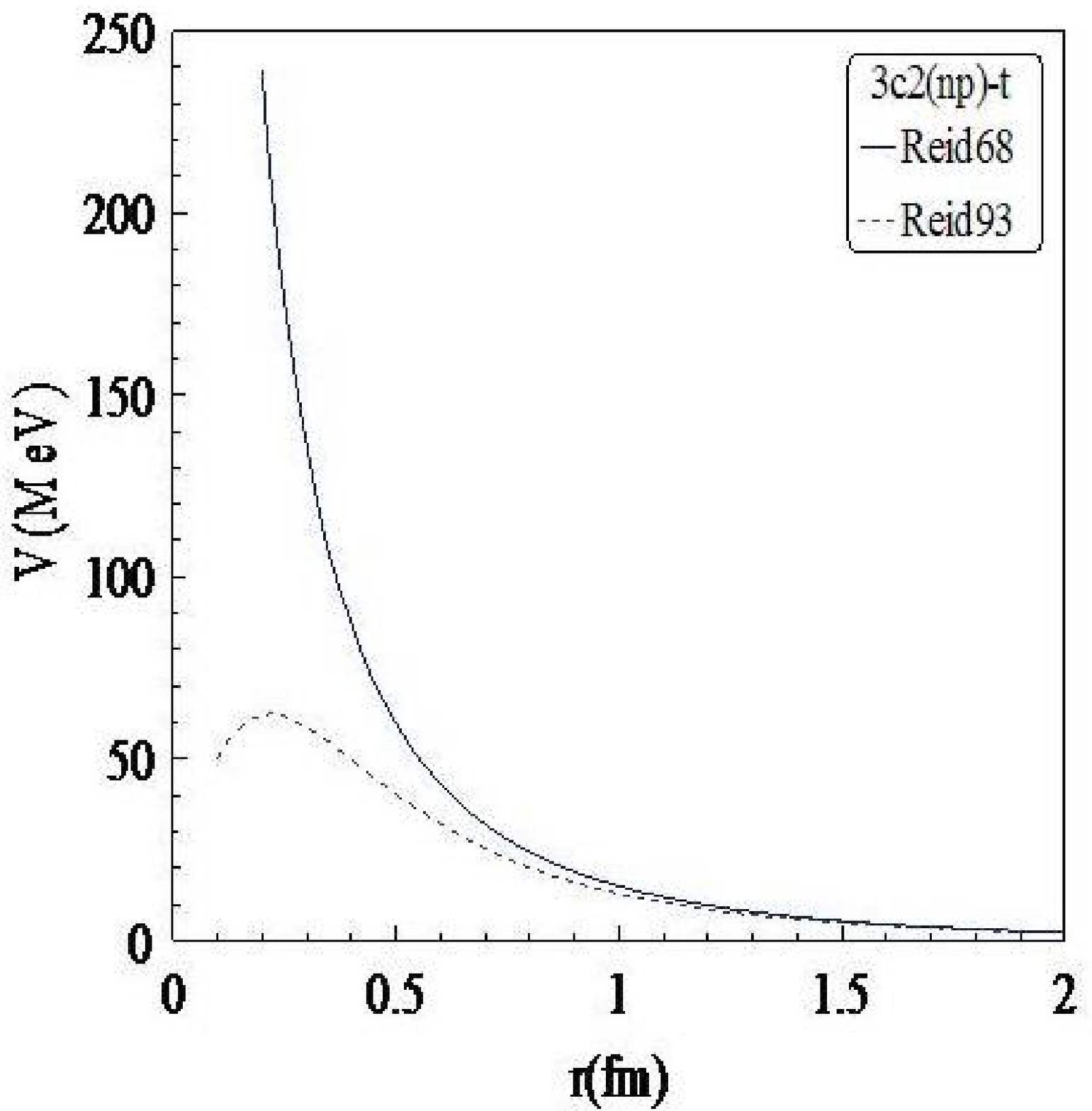}
          \end{subfigure}
           ~
          \begin{subfigure}[b]{0.31\textwidth}
                  \centering
                  \includegraphics[width=\textwidth,height=0.24\textheight]{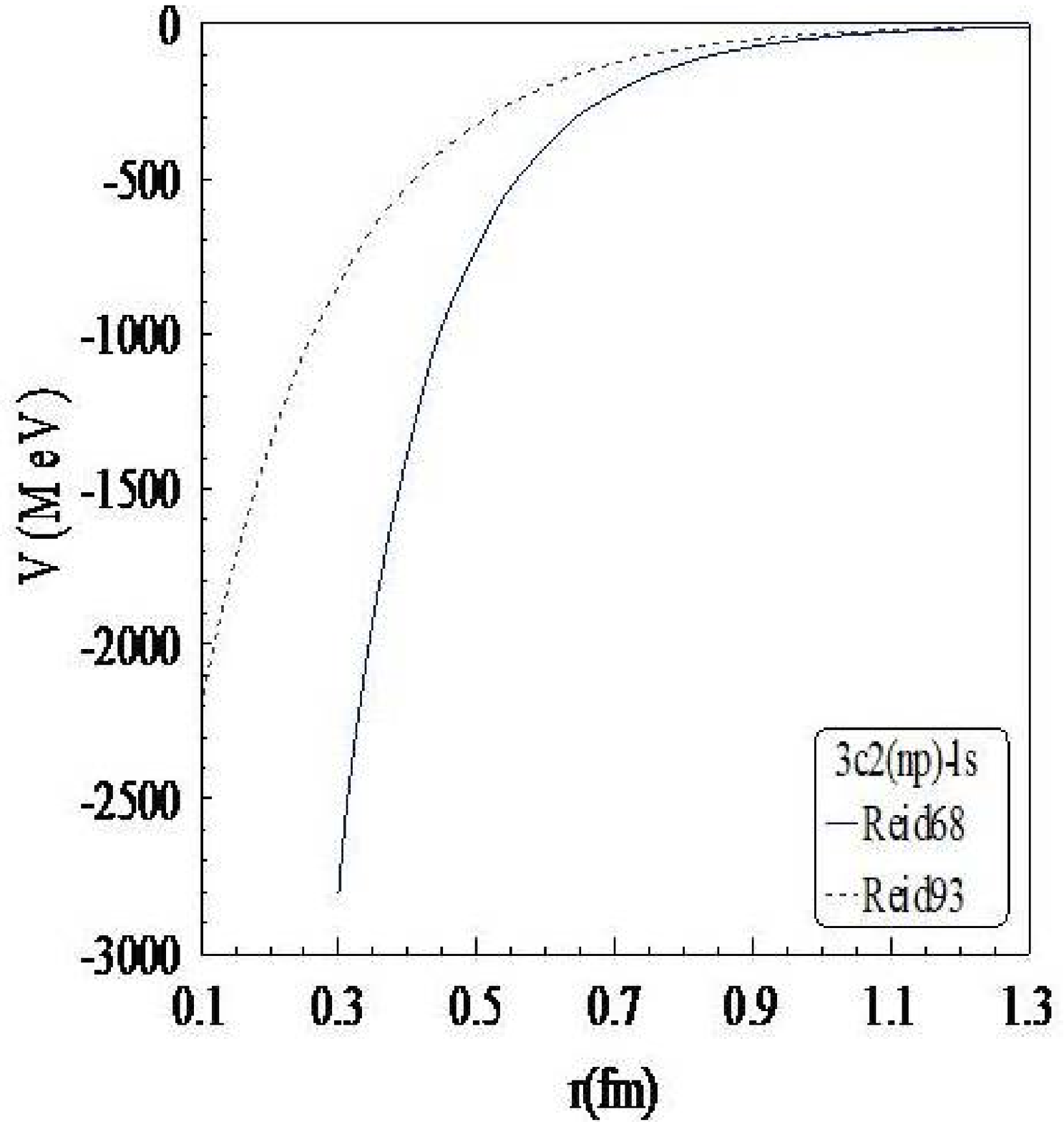}
          \end{subfigure}
    \end{subfigure}
\caption{\textit{The comparison of the central, tensor and spin-orbit potentials of Reid68 and Reid93, for the states from $J=0$ up to $J=2$, for np system}.} \label{Fig4.}
\end{figure}

\begin{figure}[p]
    \centering
      \begin{subfigure}[b]{\textwidth}
          \centering
          \begin{subfigure}[b]{0.31\textwidth}
                  \centering
                  \includegraphics[width=\textwidth,height=0.24\textheight]{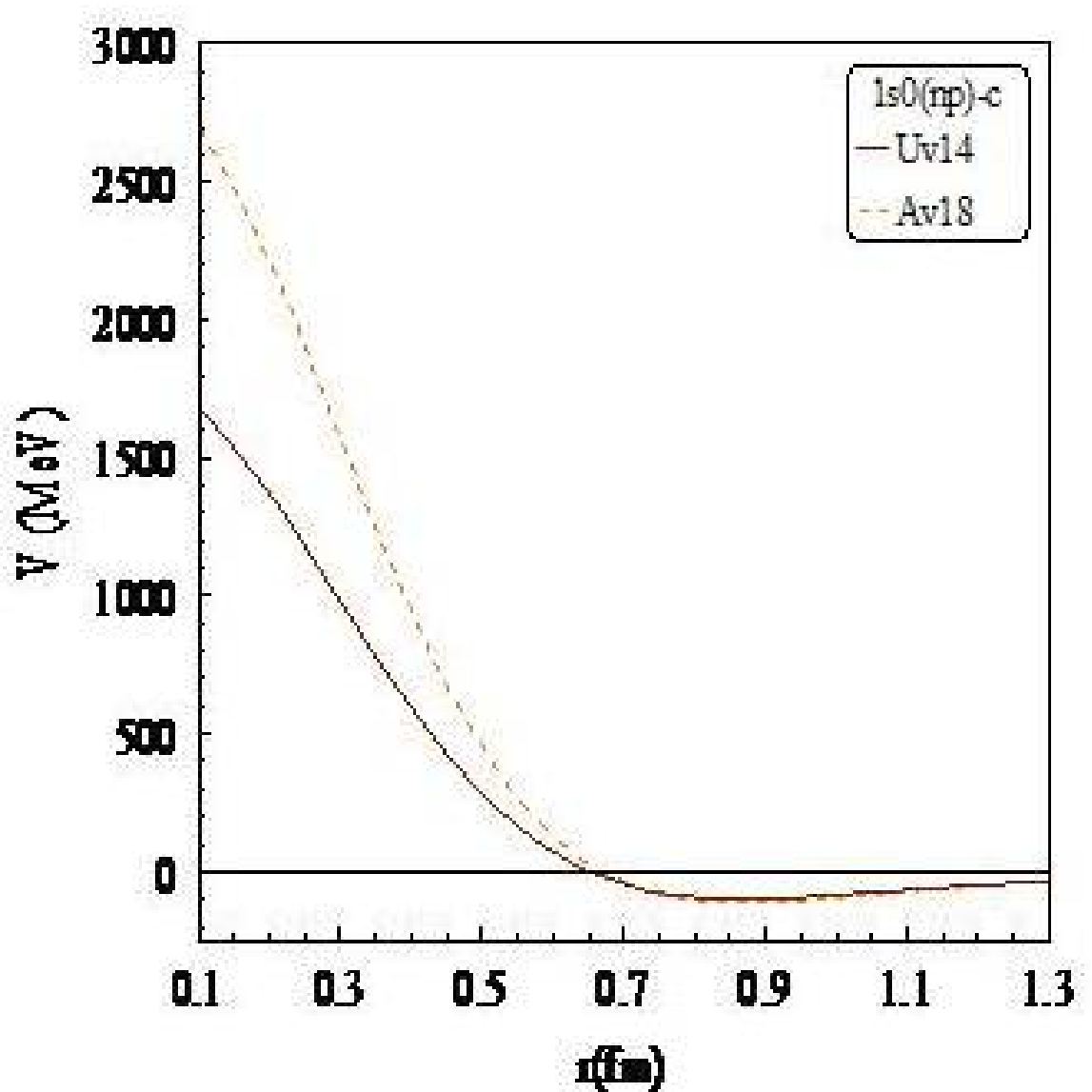}
          \end{subfigure}%
          ~
          \begin{subfigure}[b]{0.31\textwidth}
                  \centering
                  \includegraphics[width=\textwidth,height=0.24\textheight]{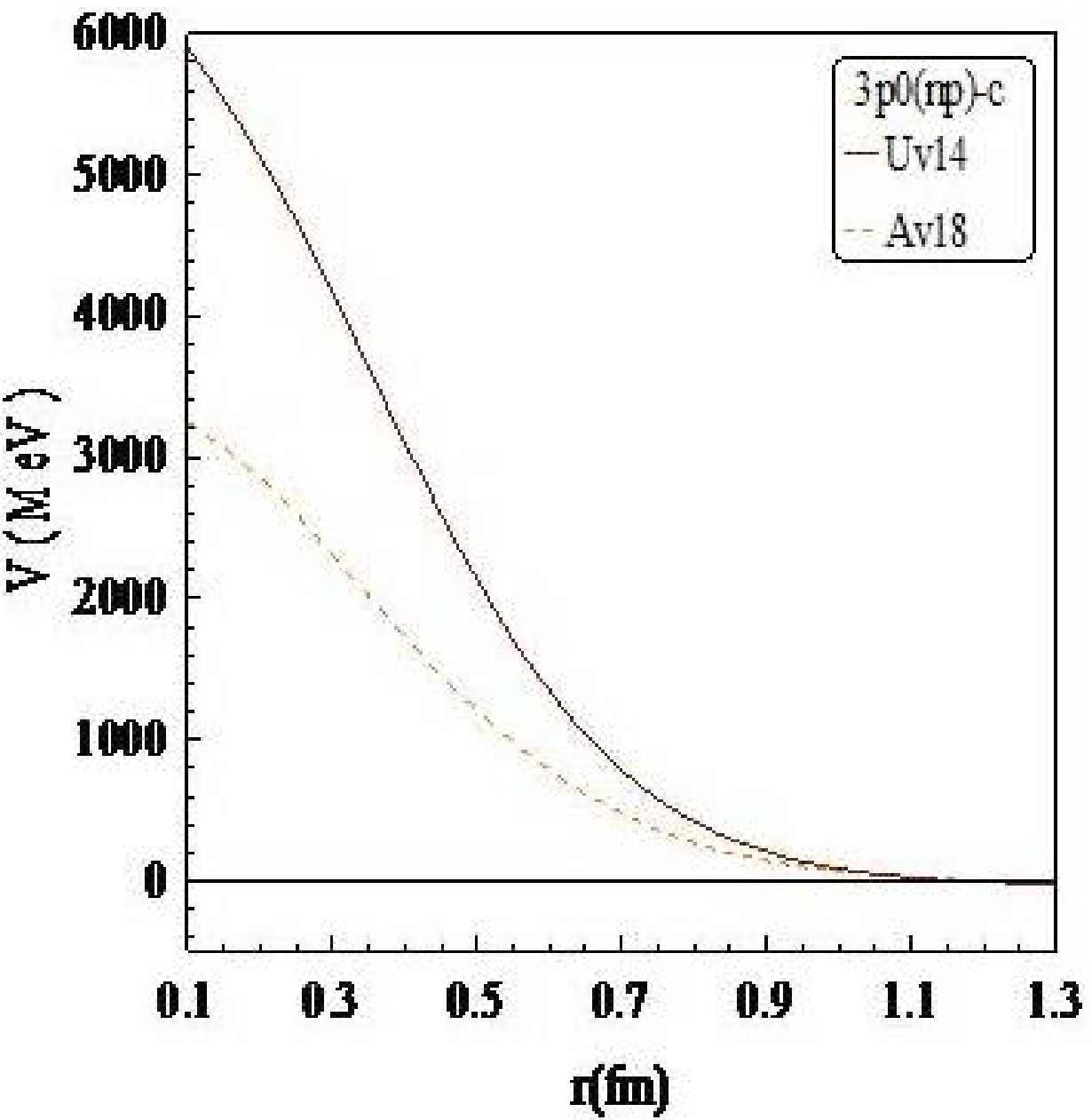}
          \end{subfigure}
           ~
          \begin{subfigure}[b]{0.31\textwidth}
                  \centering
                  \includegraphics[width=\textwidth,height=0.24\textheight]{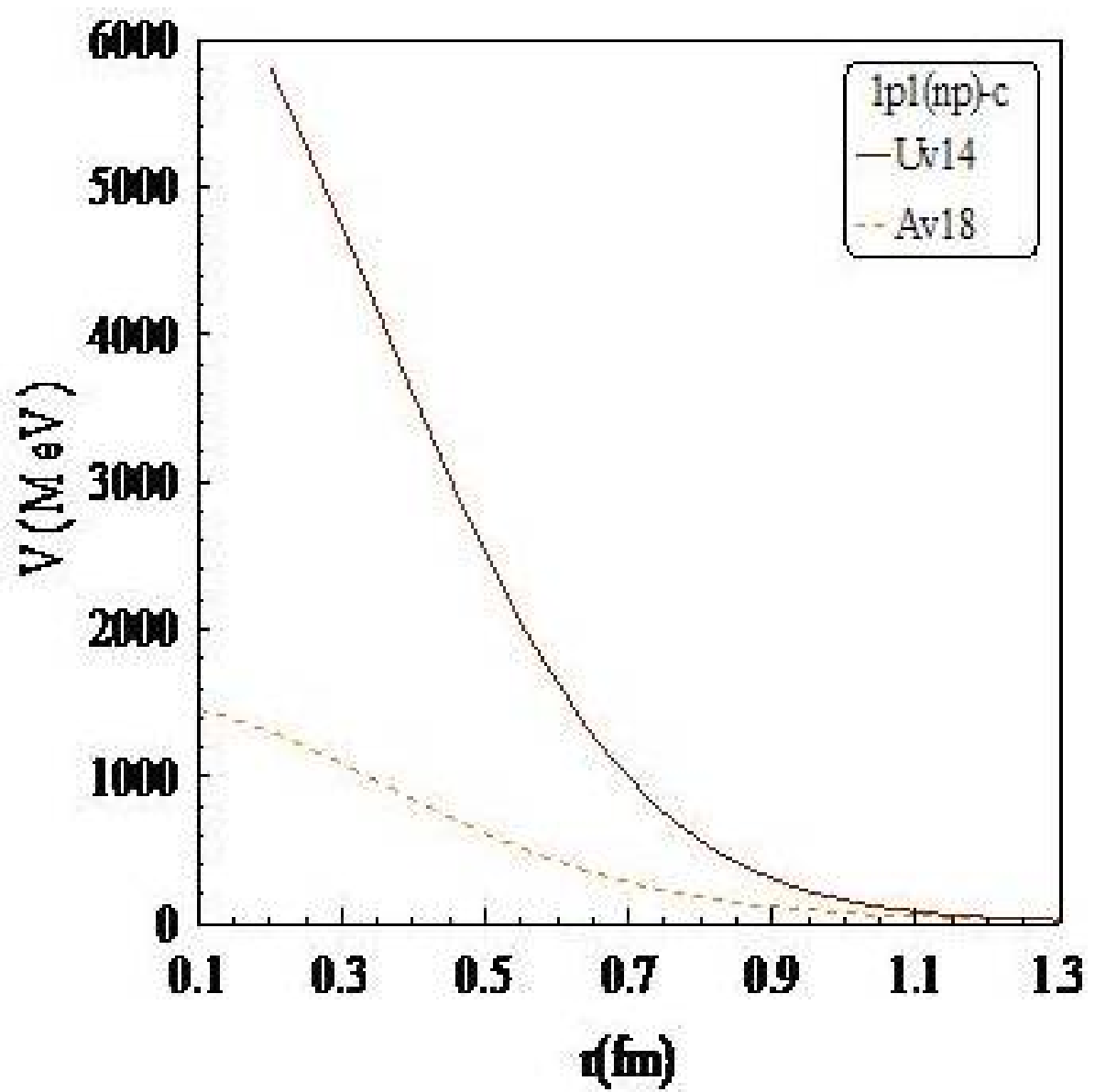}
          \end{subfigure}
    \end{subfigure}
    \begin{subfigure}[b]{\textwidth}
          \centering
           \begin{subfigure}[b]{0.31\textwidth}
                  \centering
                  \includegraphics[width=\textwidth,height=0.24\textheight]{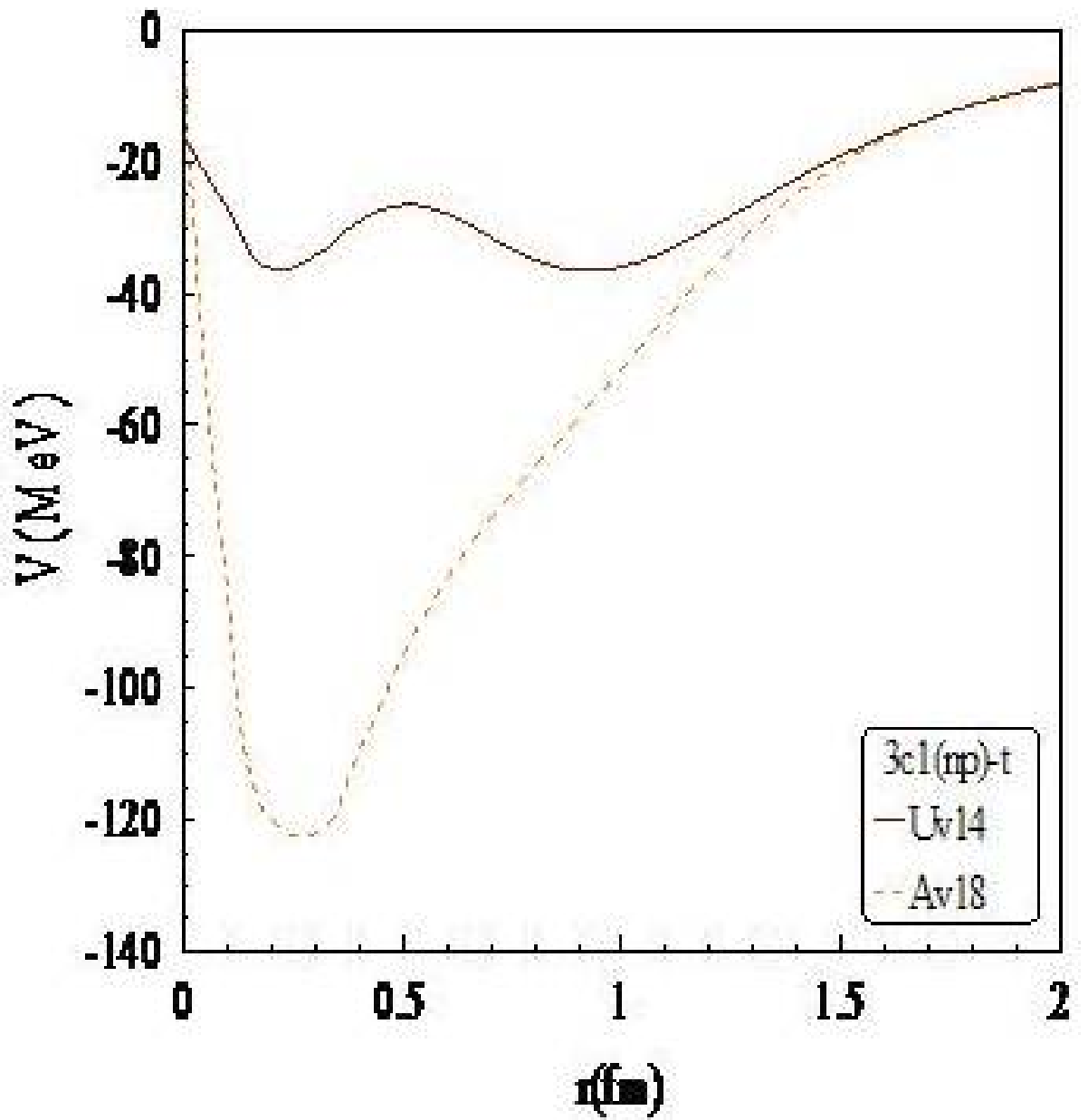}
          \end{subfigure}%
          ~
          \begin{subfigure}[b]{0.31\textwidth}
                  \centering
                  \includegraphics[width=\textwidth,height=0.24\textheight]{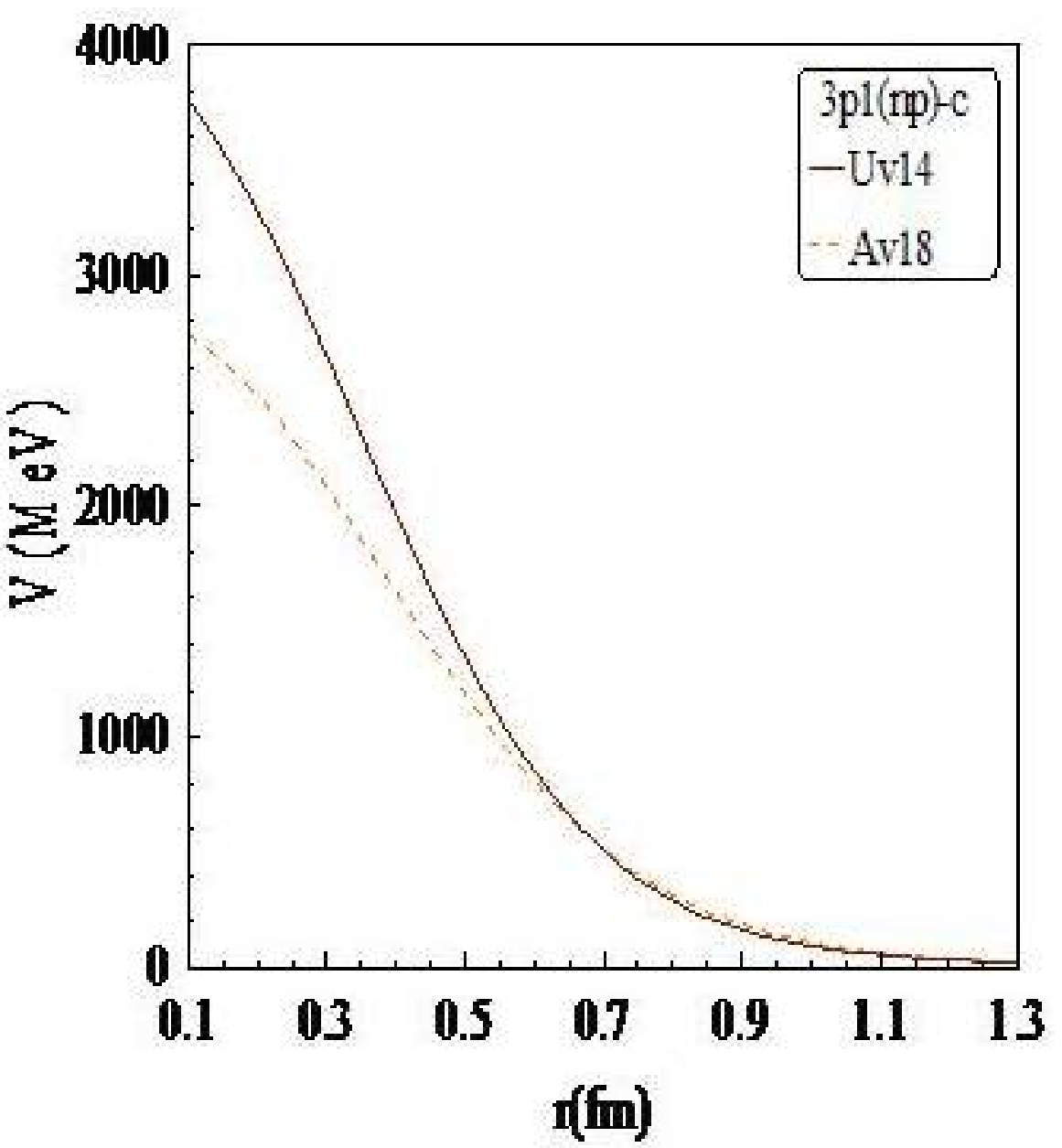}
          \end{subfigure}
           ~
          \begin{subfigure}[b]{0.31\textwidth}
                  \centering
                  \includegraphics[width=\textwidth,height=0.24\textheight]{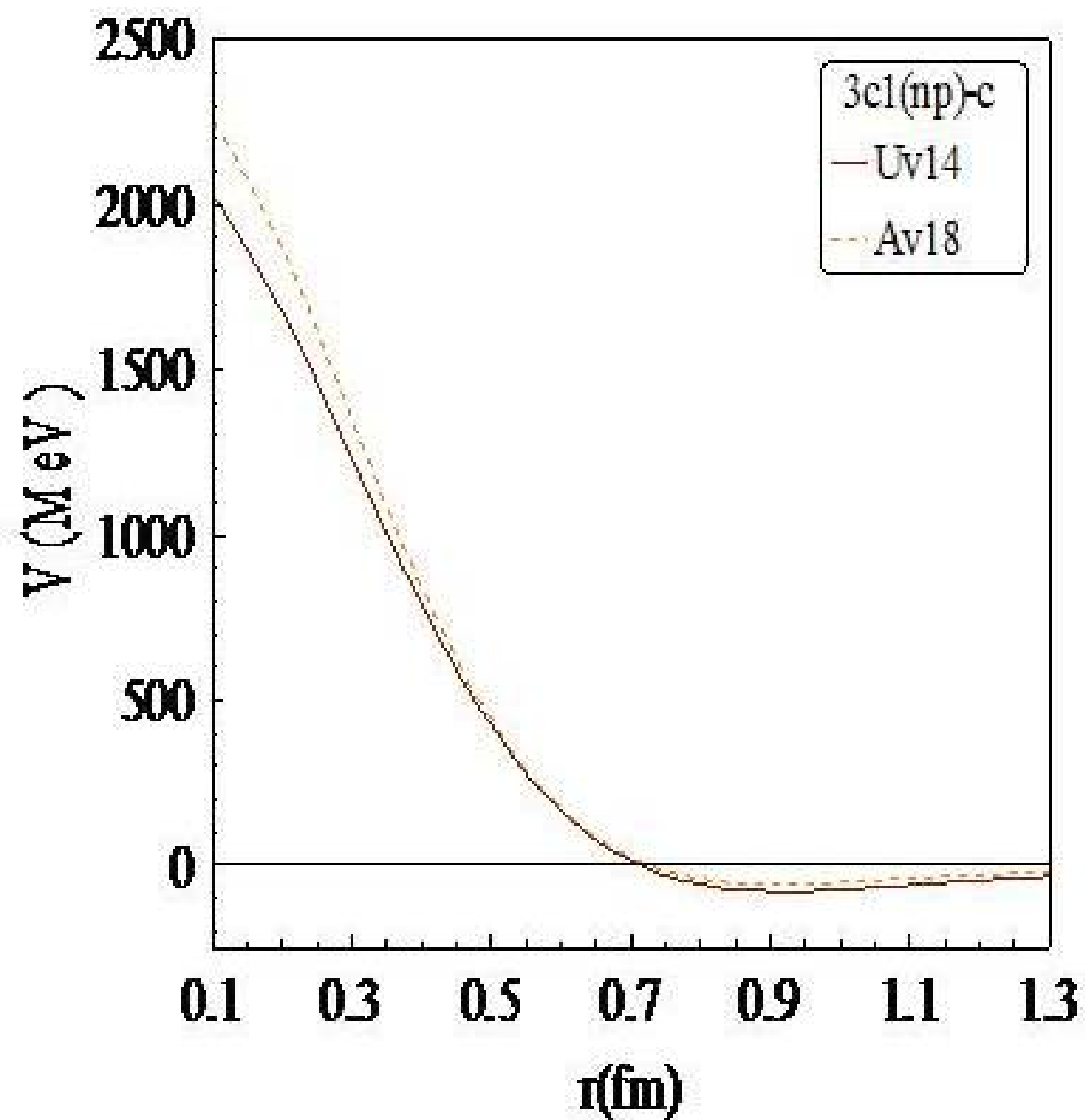}
          \end{subfigure}
   \end{subfigure}
   \begin{subfigure}[b]{\textwidth}
          \centering
        \begin{subfigure}[b]{0.31\textwidth}
                  \centering
                  \includegraphics[width=\textwidth,height=0.24\textheight]{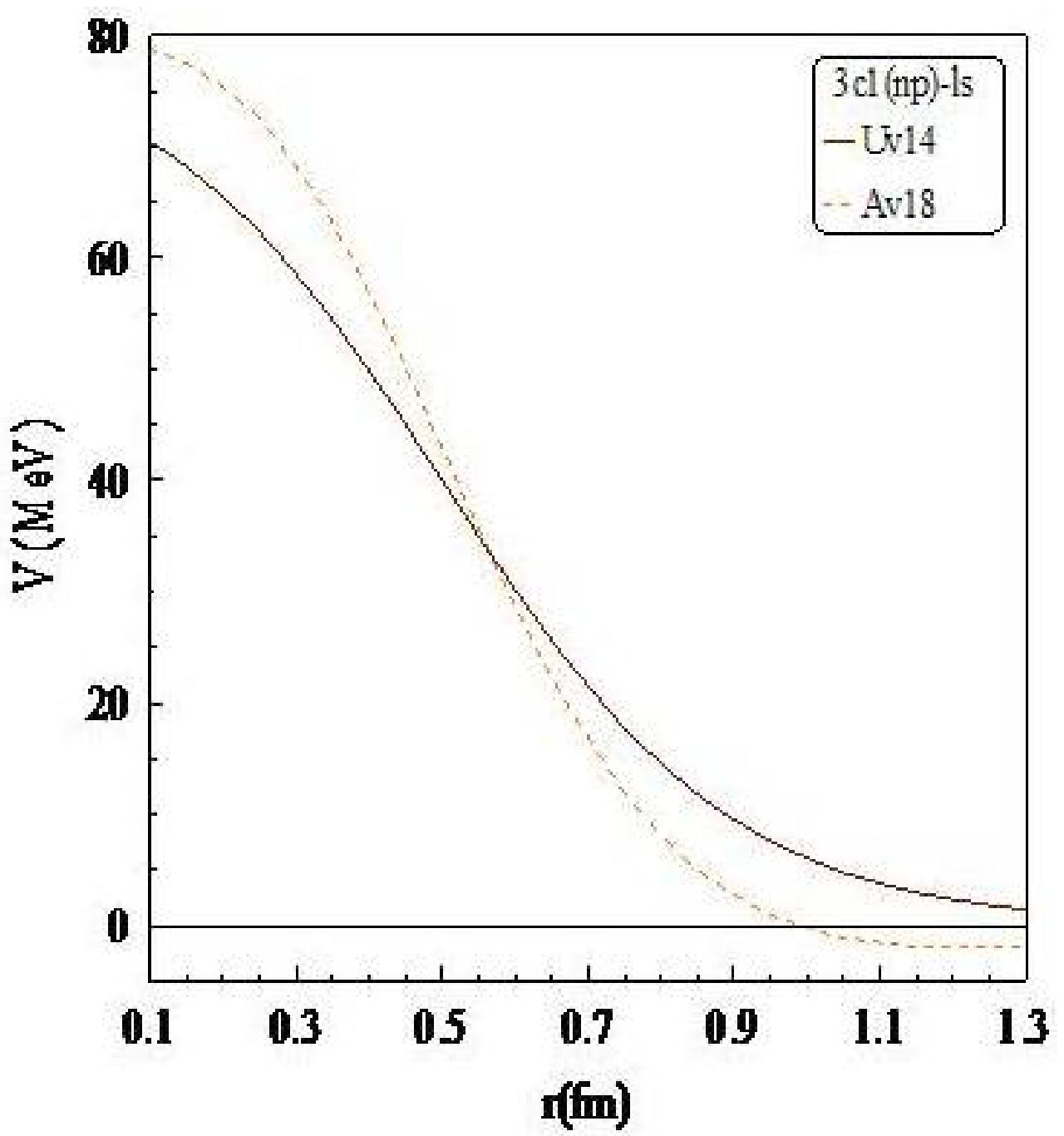}
          \end{subfigure}%
          ~
          \begin{subfigure}[b]{0.31\textwidth}
                  \centering
                  \includegraphics[width=\textwidth,height=0.24\textheight]{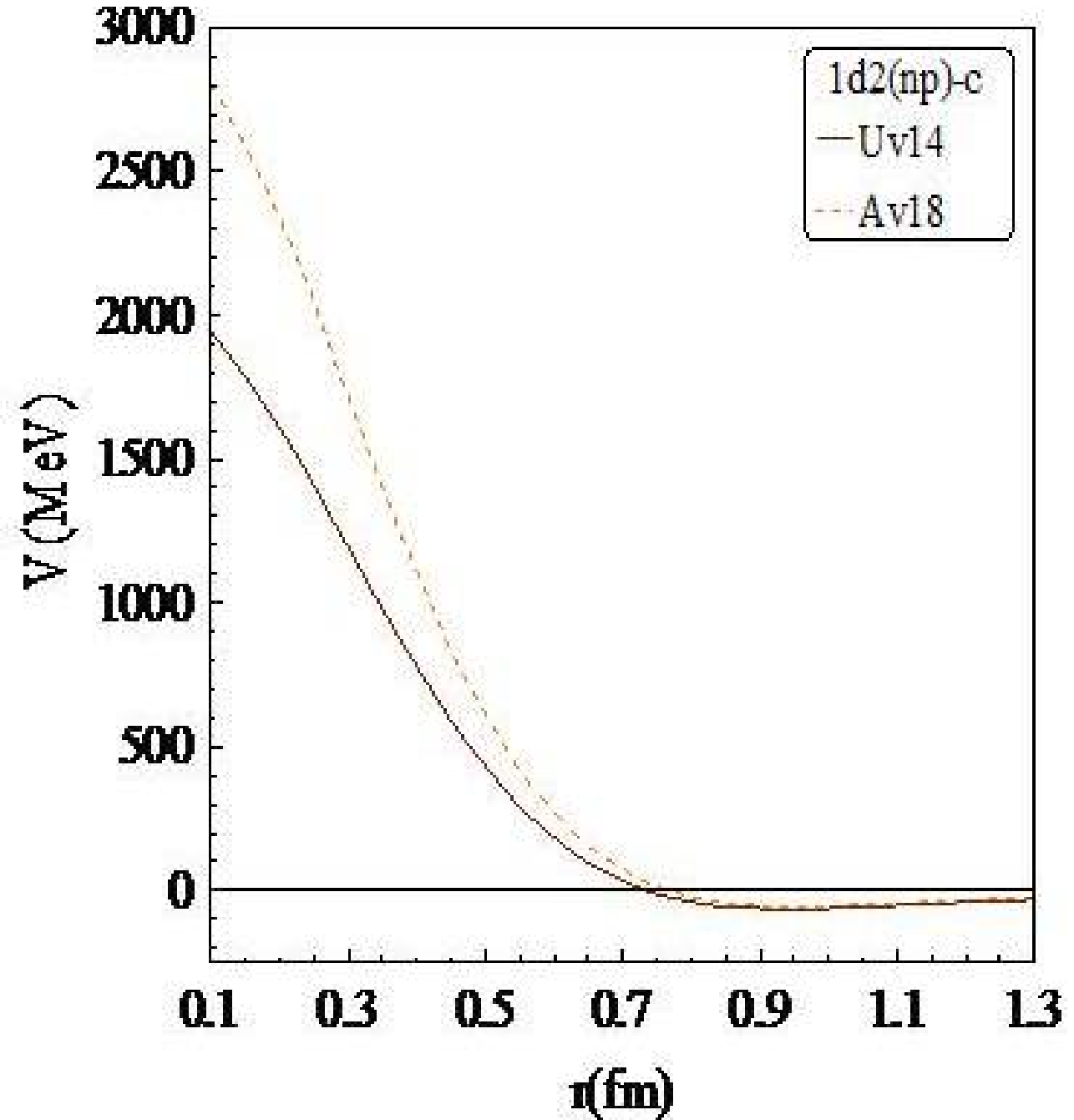}
          \end{subfigure}
           ~
          \begin{subfigure}[b]{0.31\textwidth}
                  \centering
                  \includegraphics[width=\textwidth,height=0.24\textheight]{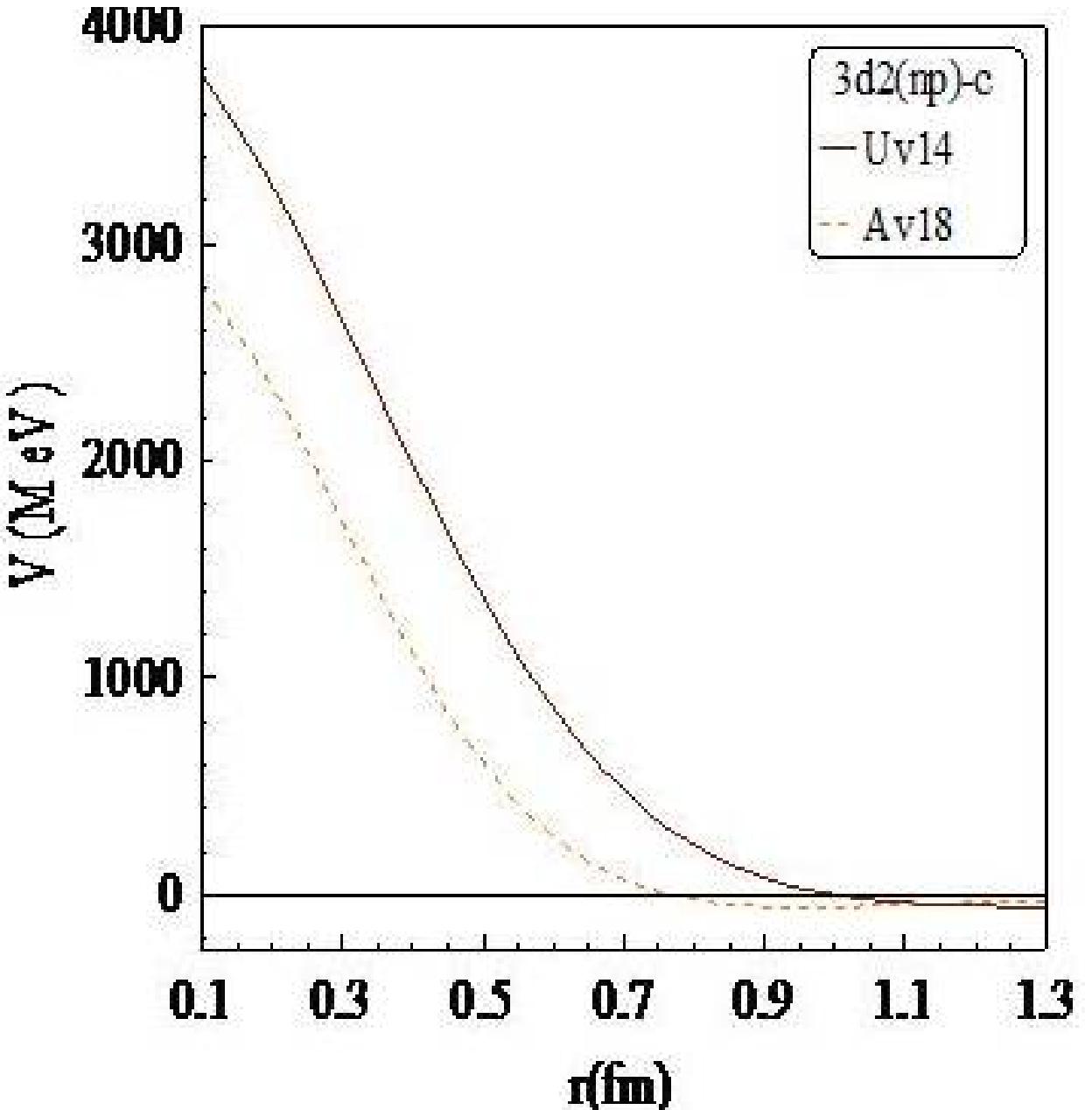}
          \end{subfigure}
   \end{subfigure}
   \begin{subfigure}[b]{\textwidth}
          \centering
        \begin{subfigure}[b]{0.31\textwidth}
                  \centering
                  \includegraphics[width=\textwidth,height=0.24\textheight]{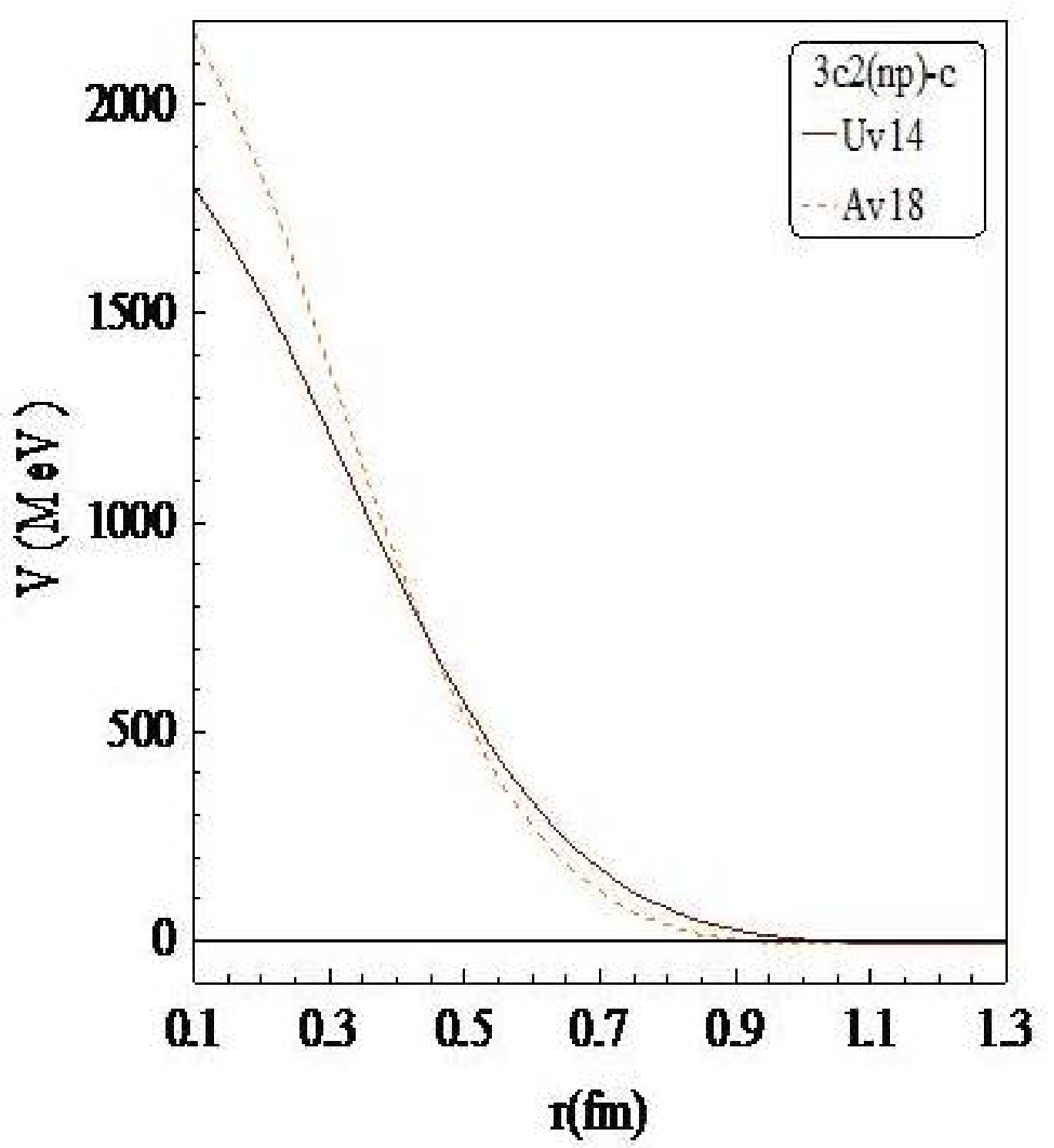}
          \end{subfigure}%
          ~
          \begin{subfigure}[b]{0.31\textwidth}
                  \centering
                  \includegraphics[width=\textwidth,height=0.24\textheight]{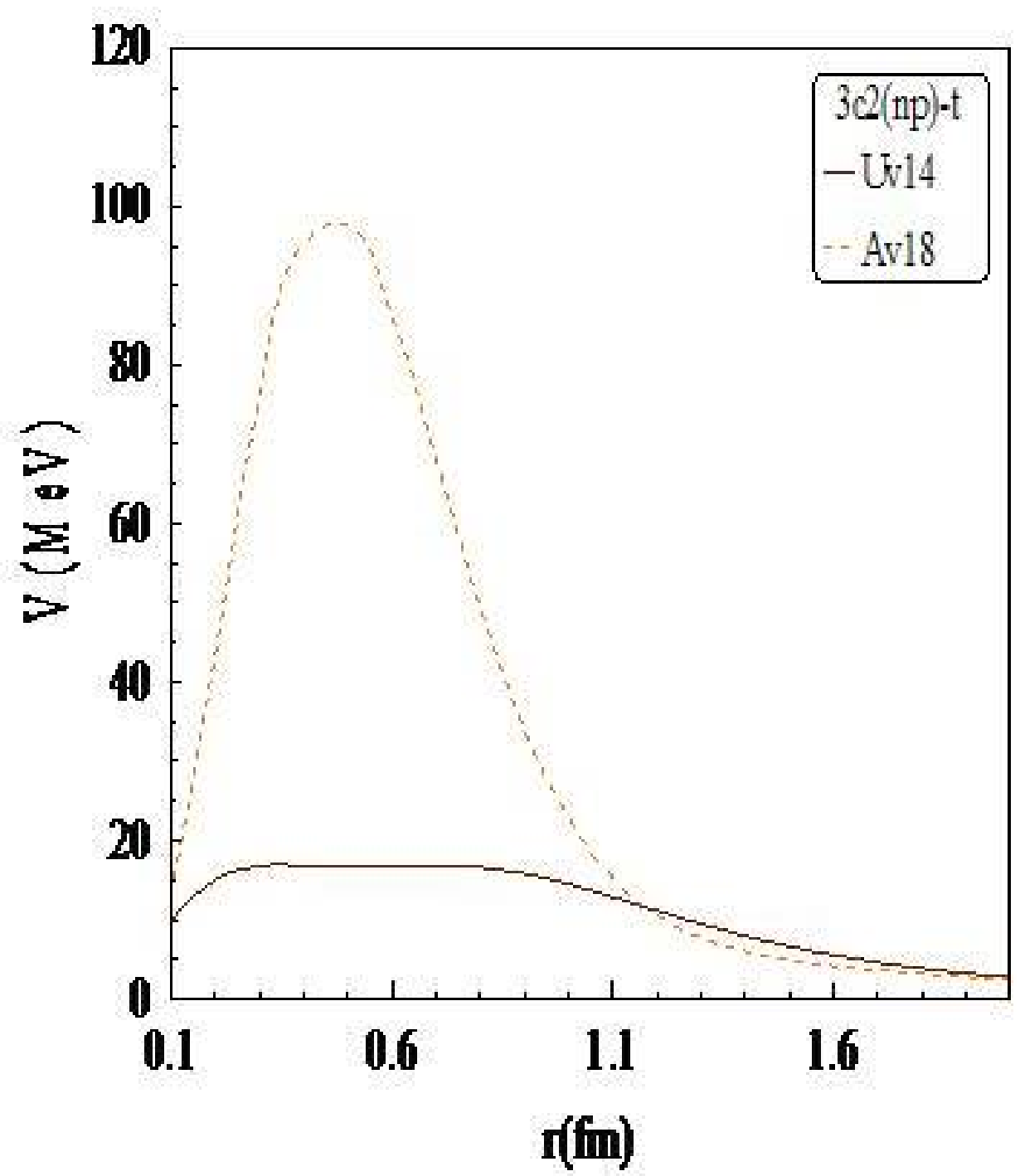}
          \end{subfigure}
           ~
          \begin{subfigure}[b]{0.31\textwidth}
                  \centering
                  \includegraphics[width=\textwidth,height=0.24\textheight]{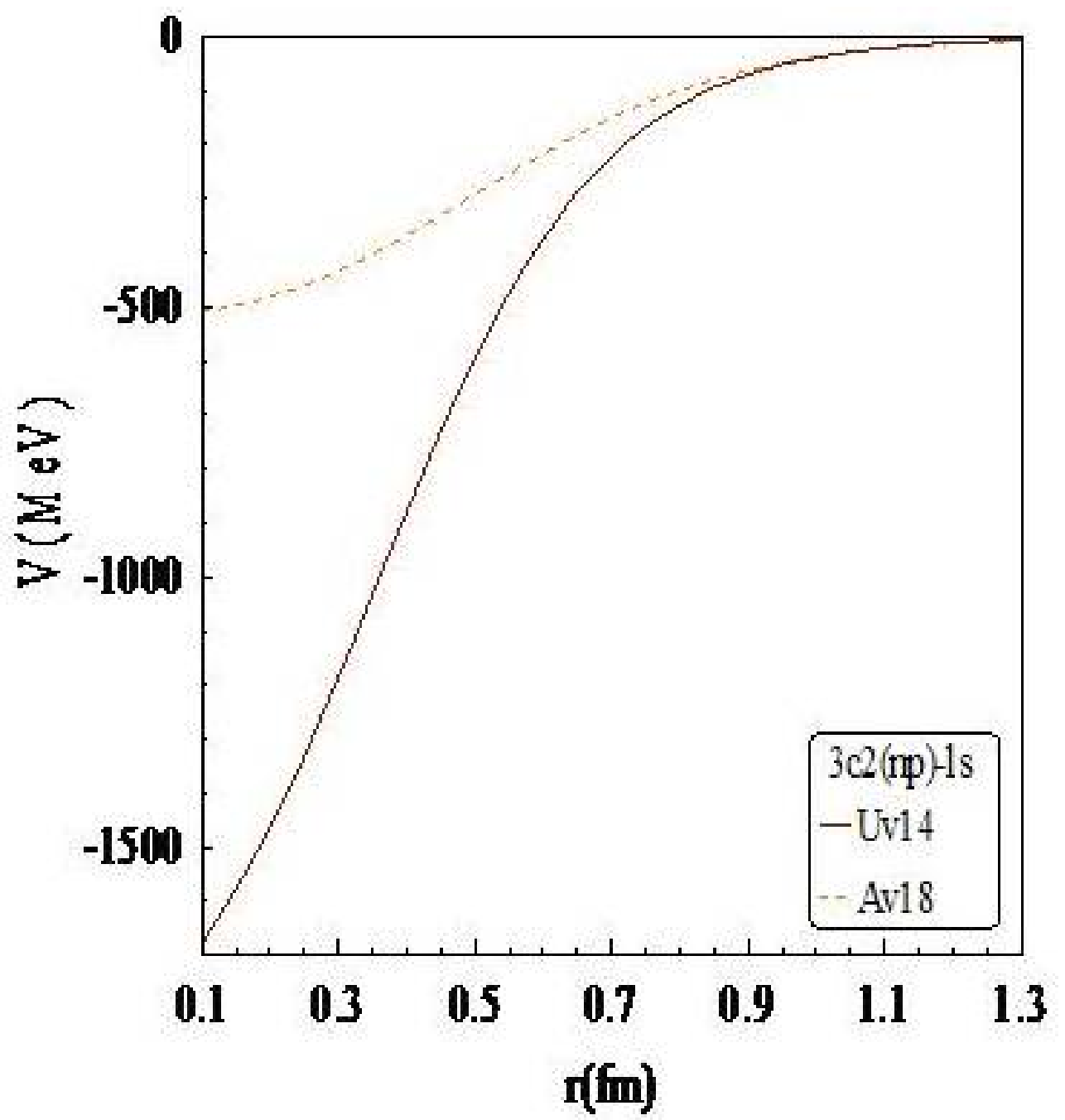}
          \end{subfigure}
    \end{subfigure}
\caption{\textit{The comparison of the central, tensor and spin-orbit potentials of Urb81 ($UV_{14}$) and Arg94 ($AV_{18}$), for the states from $J=0$ up to $J=2$, for np system}.} \label{Fig5.}
\end{figure}

\begin{figure}[p]
    \centering
      \begin{subfigure}[b]{\textwidth}
          \centering
          \begin{subfigure}[b]{0.31\textwidth}
                  \centering
                  \includegraphics[width=\textwidth,height=0.24\textheight]{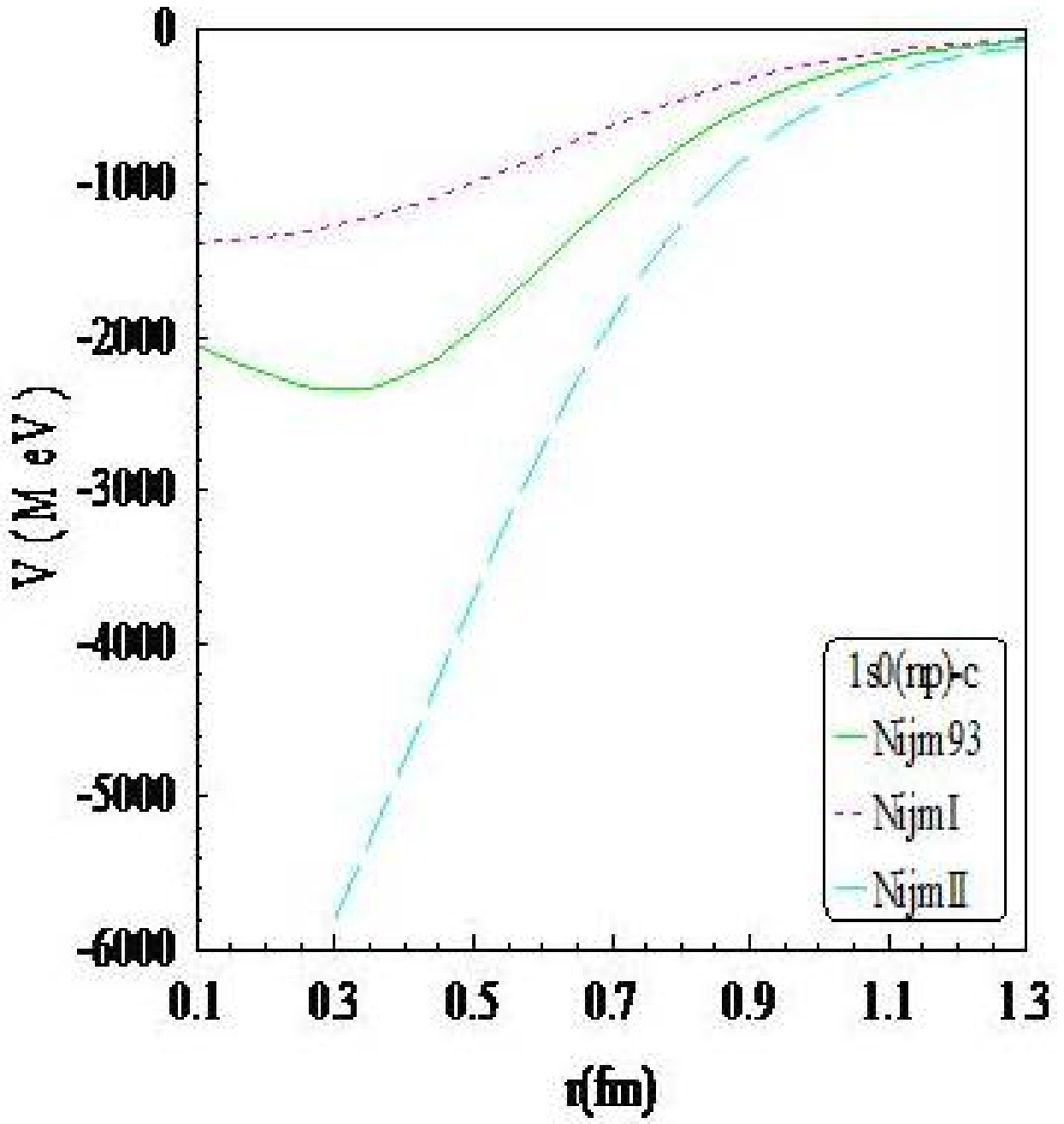}
          \end{subfigure}%
          ~
          \begin{subfigure}[b]{0.31\textwidth}
                  \centering
                  \includegraphics[width=\textwidth,height=0.24\textheight]{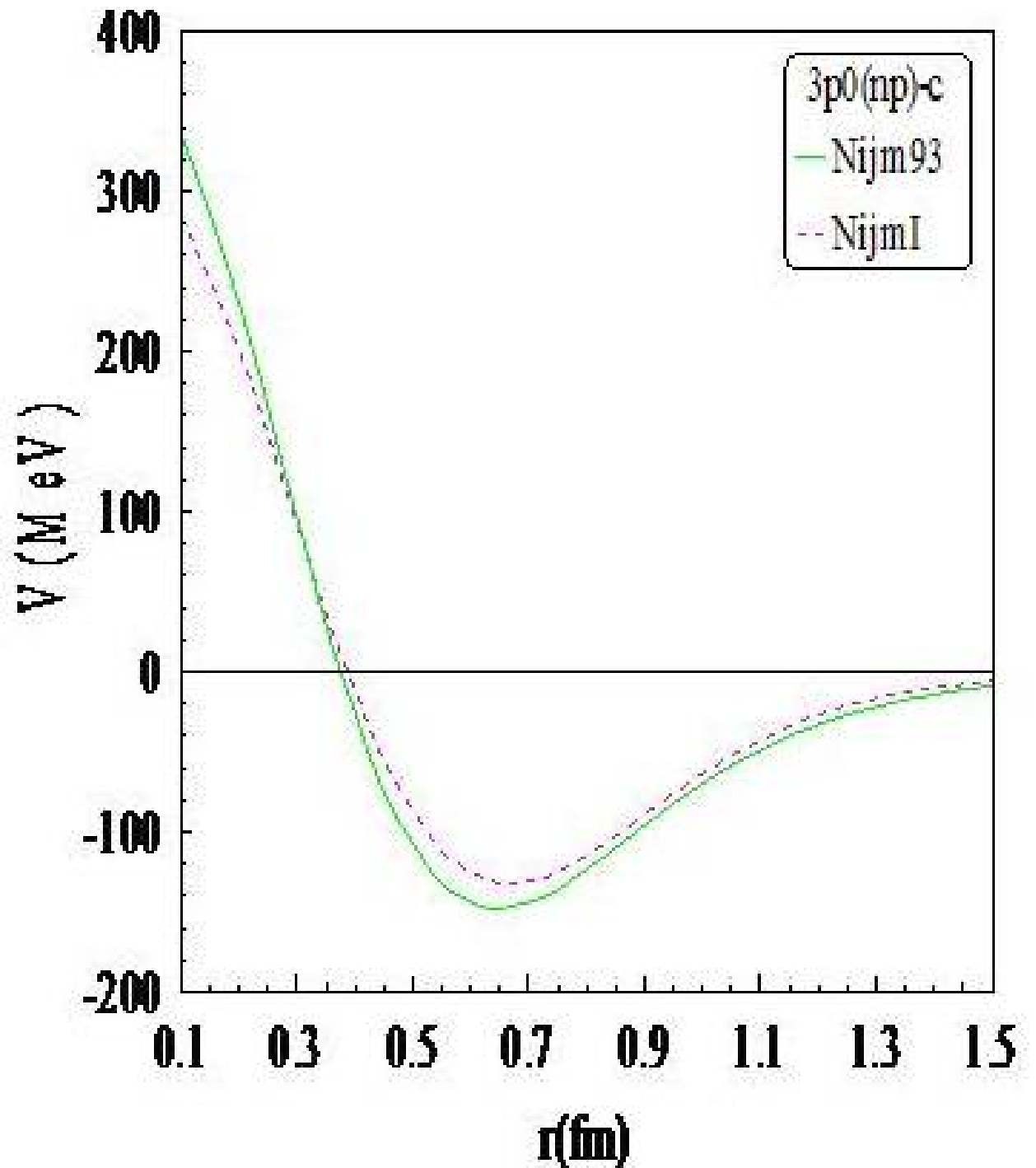}
          \end{subfigure}
           ~
          \begin{subfigure}[b]{0.31\textwidth}
                  \centering
                  \includegraphics[width=\textwidth,height=0.24\textheight]{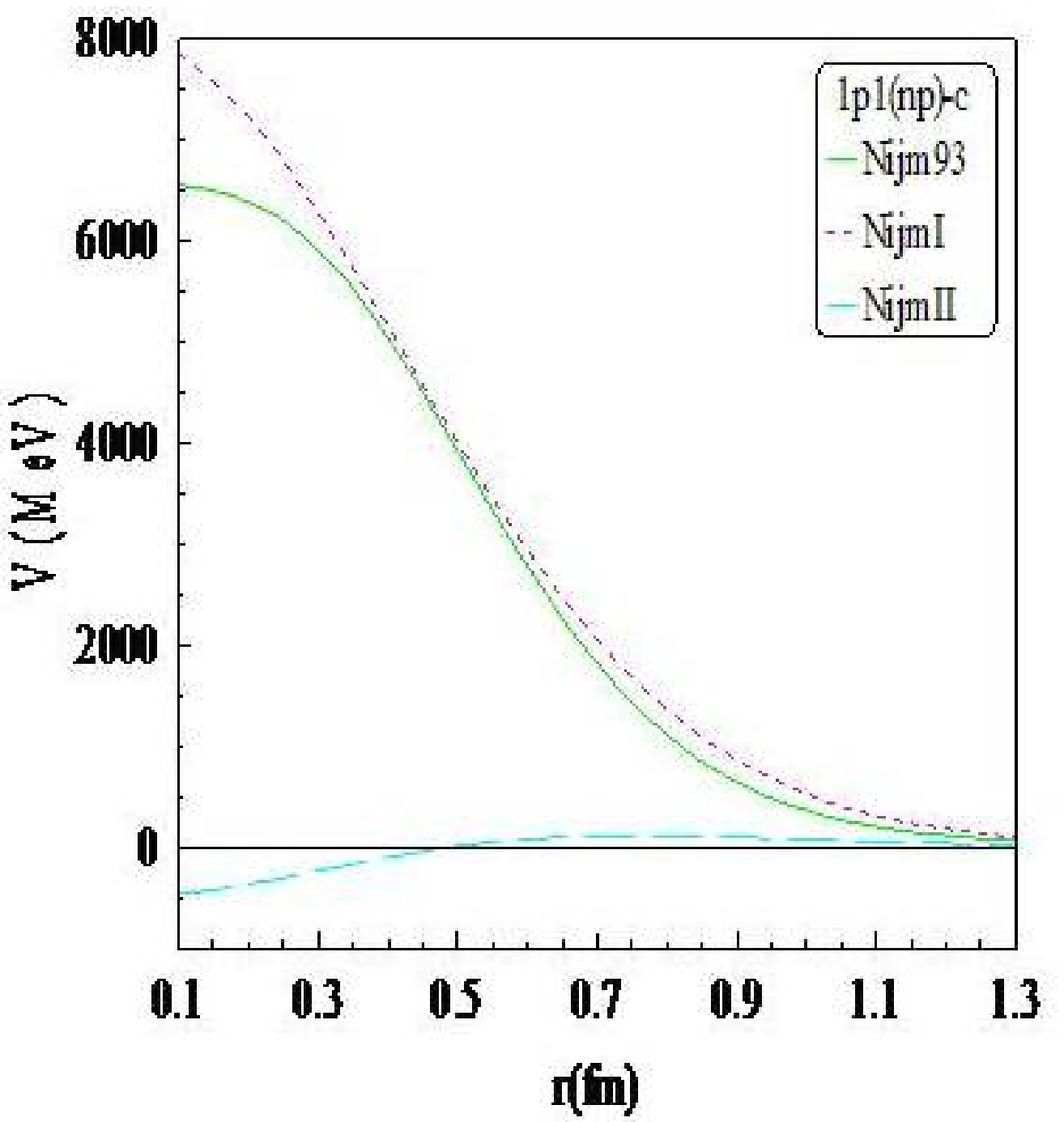}
          \end{subfigure}
    \end{subfigure}
    \begin{subfigure}[b]{\textwidth}
          \centering
           \begin{subfigure}[b]{0.31\textwidth}
                  \centering
                  \includegraphics[width=\textwidth,height=0.24\textheight]{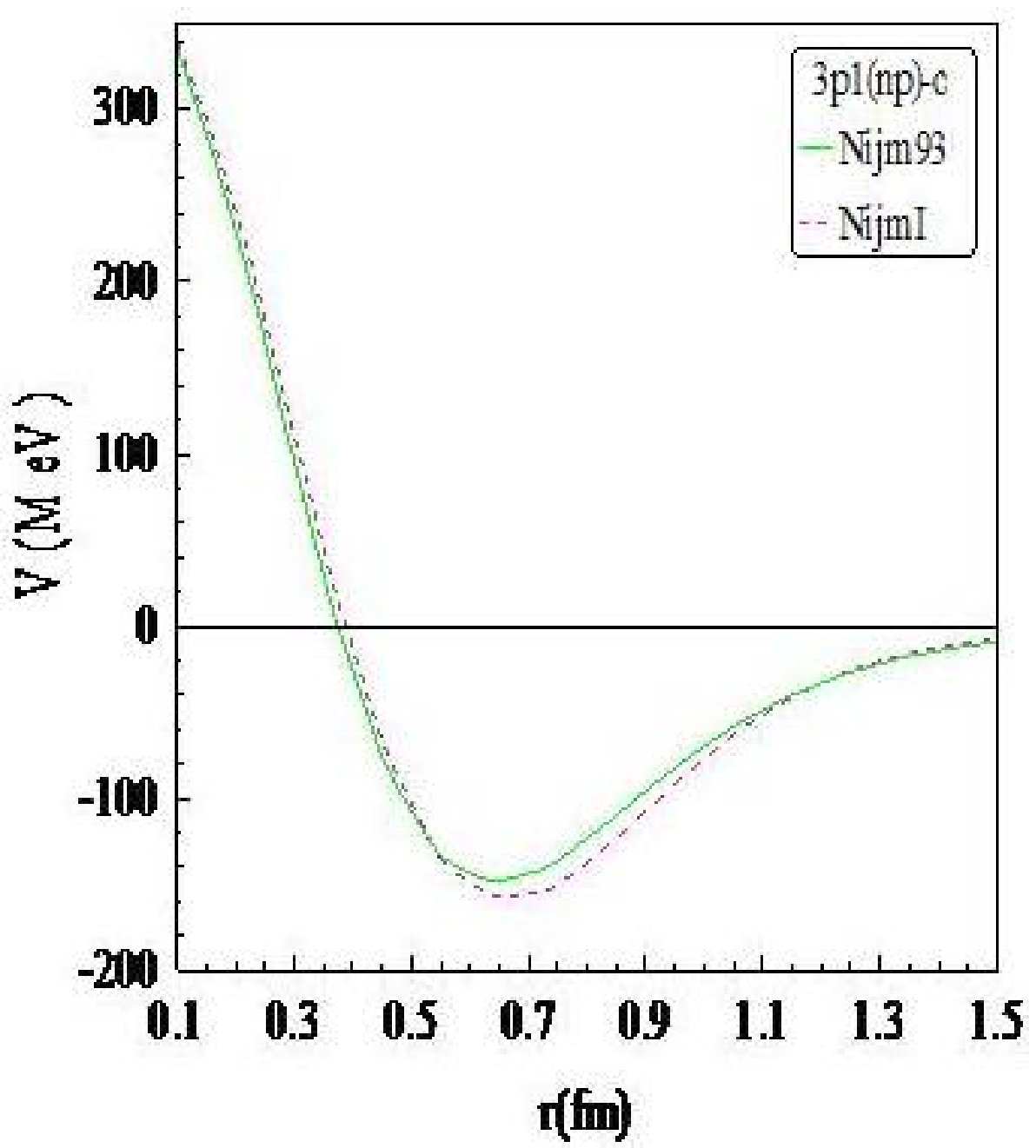}
          \end{subfigure}%
          ~
          \begin{subfigure}[b]{0.31\textwidth}
                  \centering
                  \includegraphics[width=\textwidth,height=0.24\textheight]{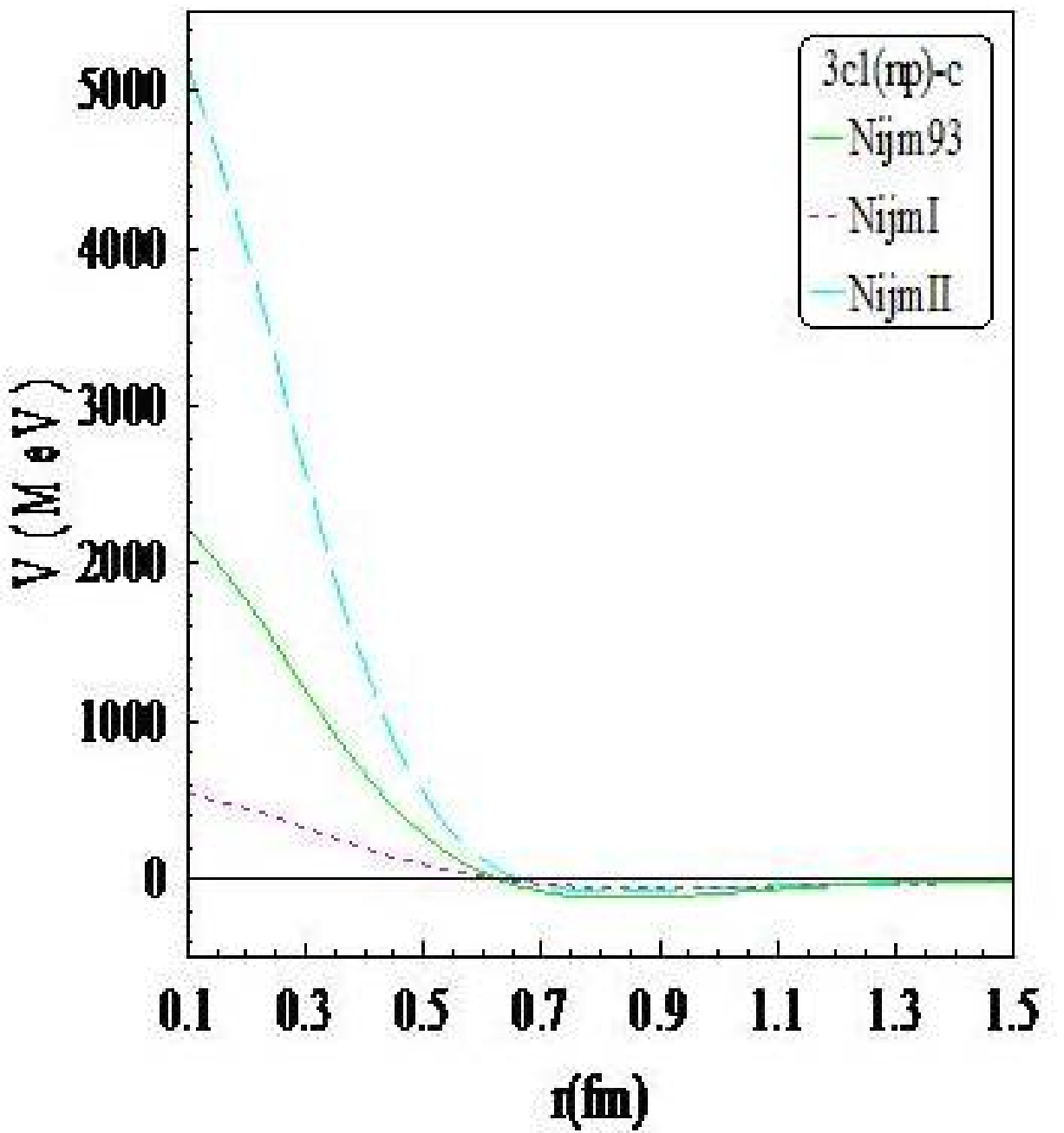}
          \end{subfigure}
           ~
          \begin{subfigure}[b]{0.31\textwidth}
                  \centering
                  \includegraphics[width=\textwidth,height=0.24\textheight]{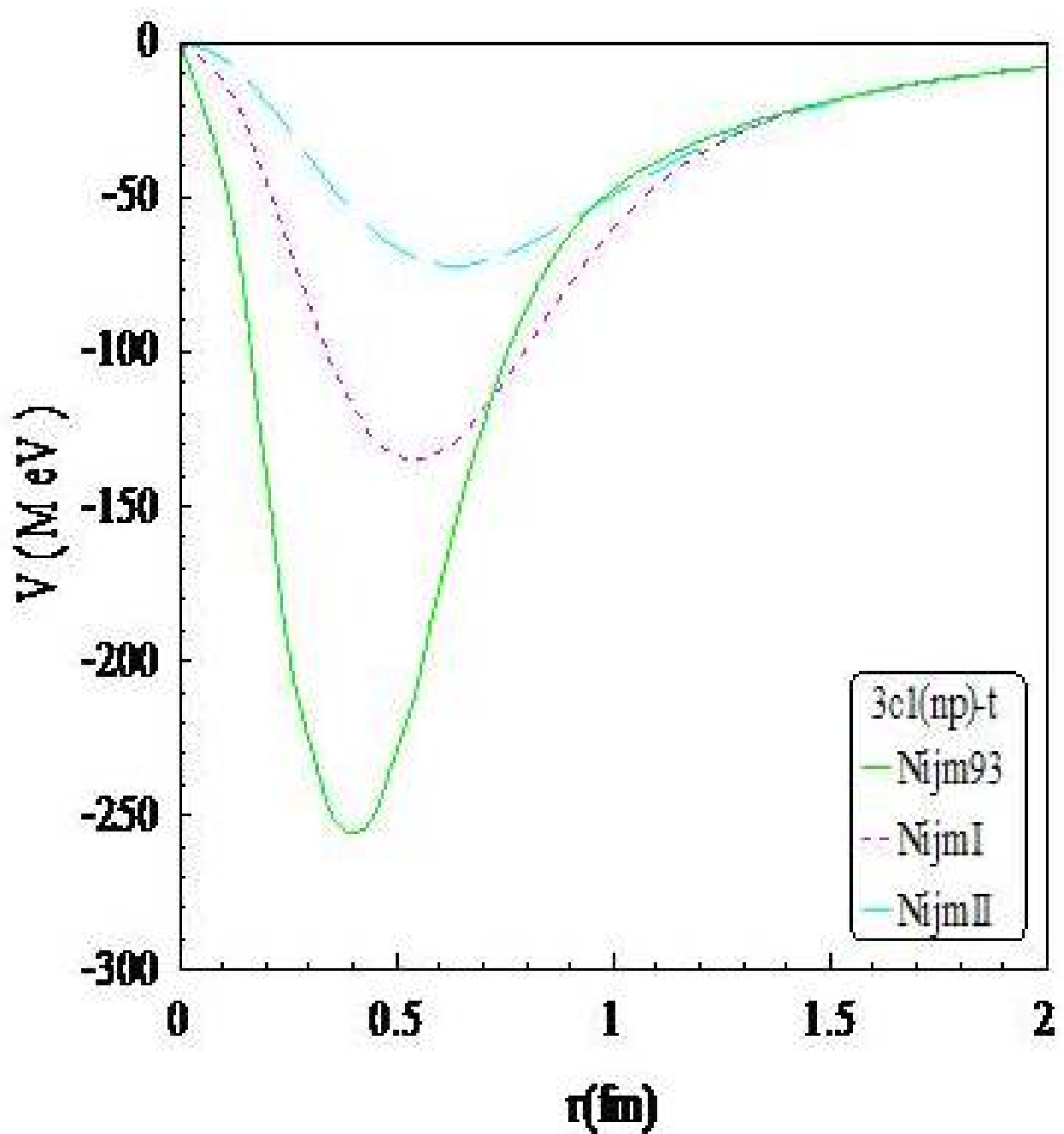}
          \end{subfigure}
   \end{subfigure}
   \begin{subfigure}[b]{\textwidth}
          \centering
        \begin{subfigure}[b]{0.31\textwidth}
                  \centering
                  \includegraphics[width=\textwidth,height=0.24\textheight]{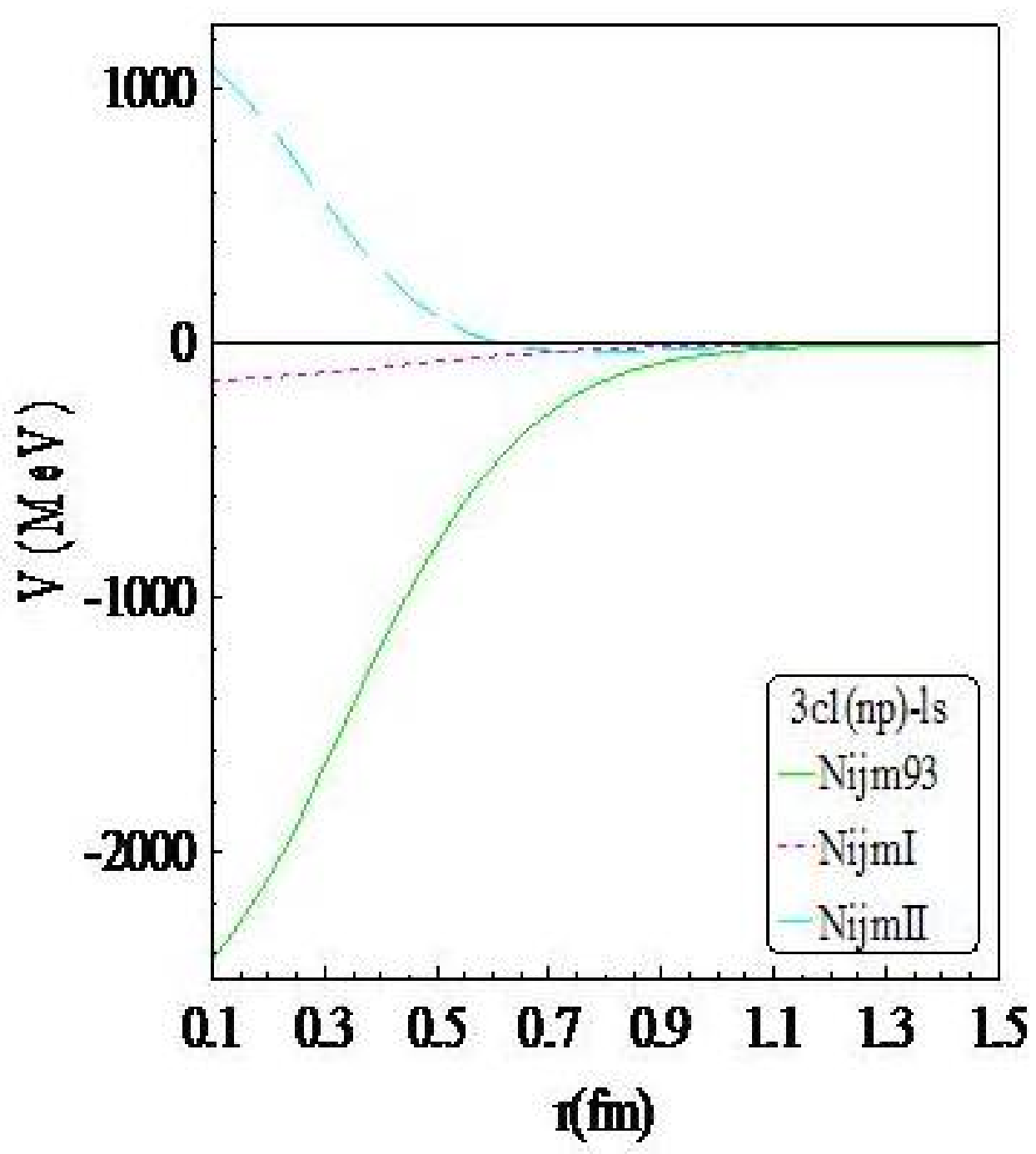}
          \end{subfigure}%
          ~
          \begin{subfigure}[b]{0.31\textwidth}
                  \centering
                  \includegraphics[width=\textwidth,height=0.24\textheight]{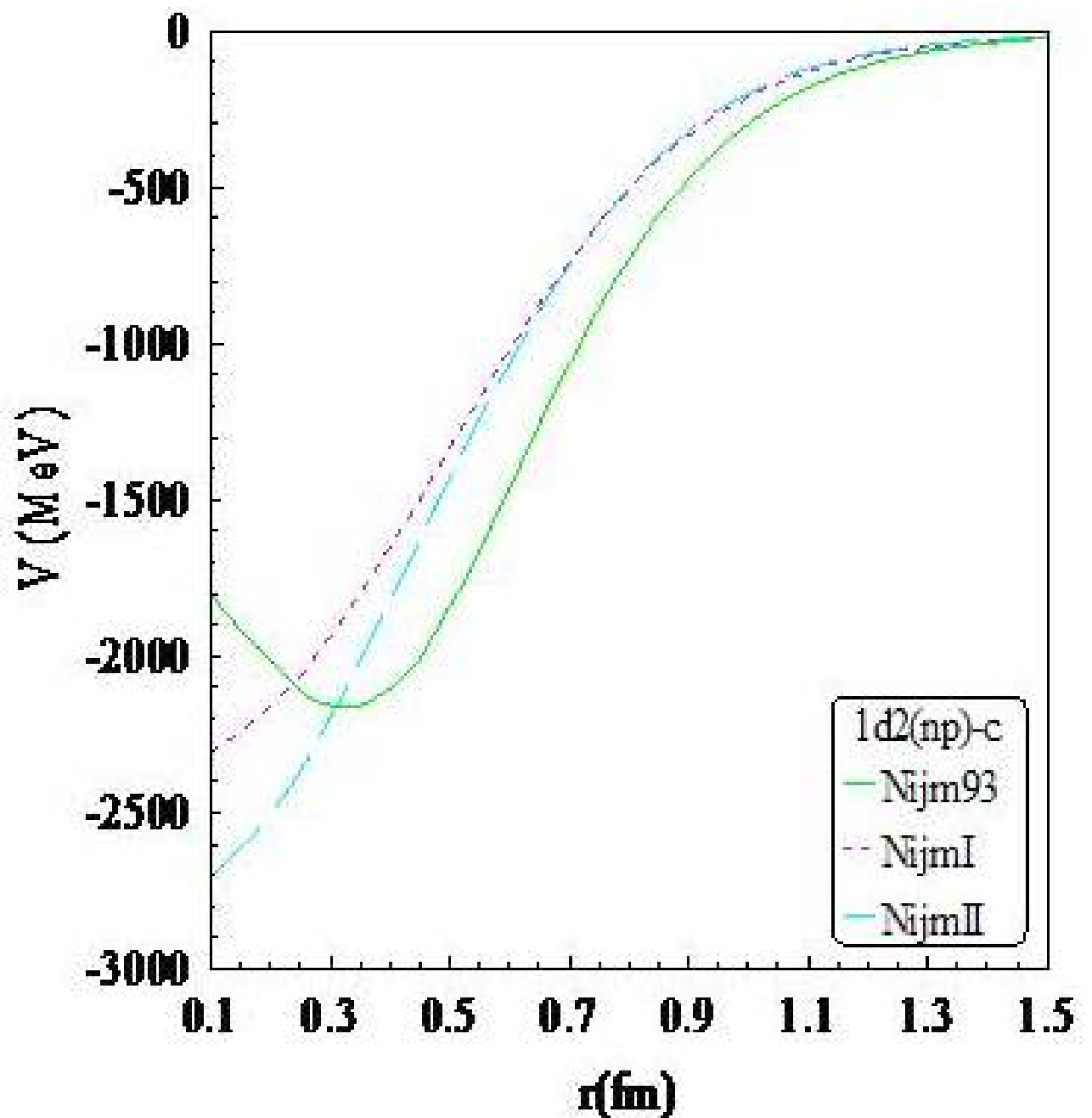}
          \end{subfigure}
           ~
          \begin{subfigure}[b]{0.31\textwidth}
                  \centering
                  \includegraphics[width=\textwidth,height=0.24\textheight]{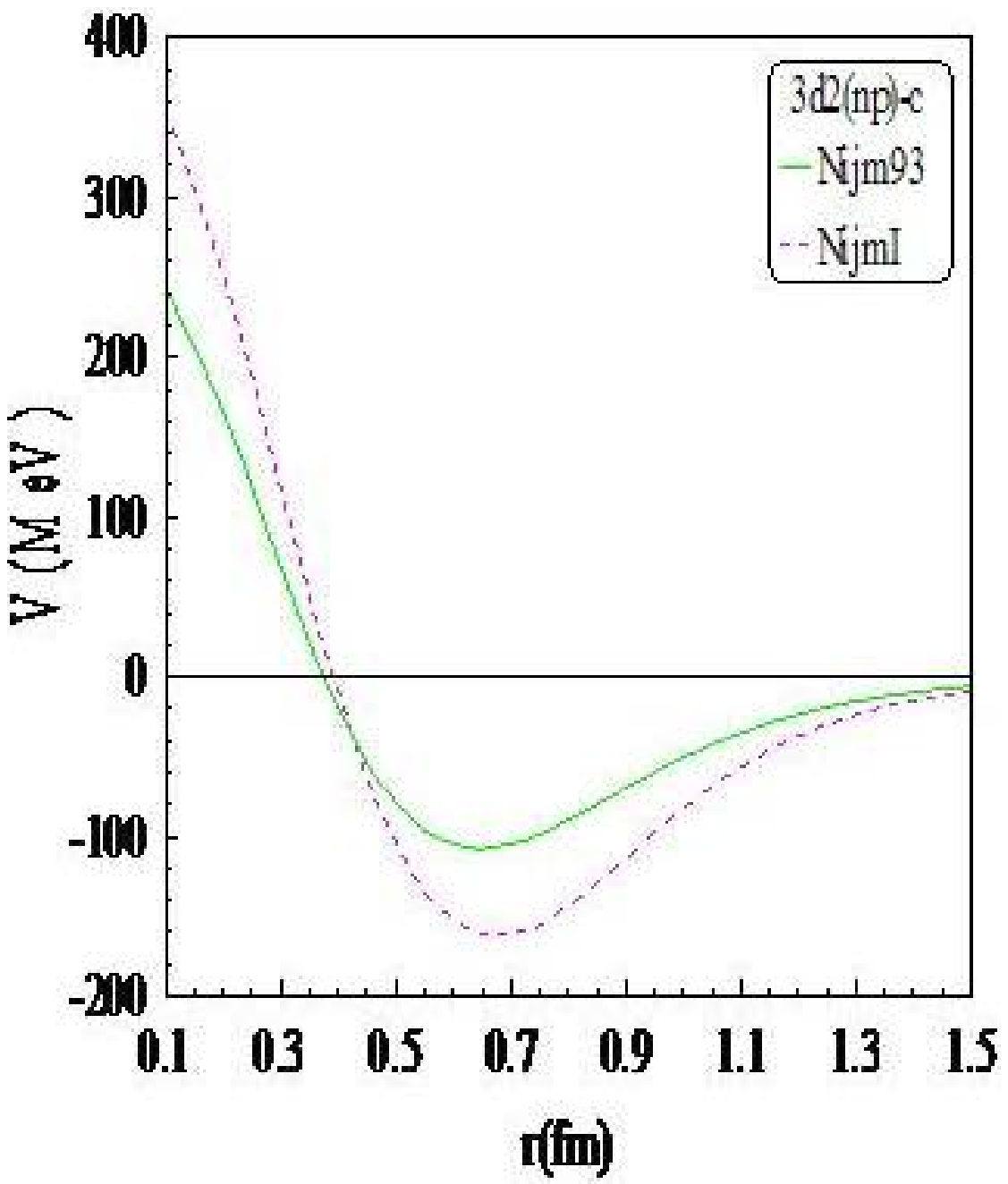}
          \end{subfigure}
   \end{subfigure}
   \begin{subfigure}[b]{\textwidth}
          \centering
        \begin{subfigure}[b]{0.31\textwidth}
                  \centering
                  \includegraphics[width=\textwidth,height=0.24\textheight]{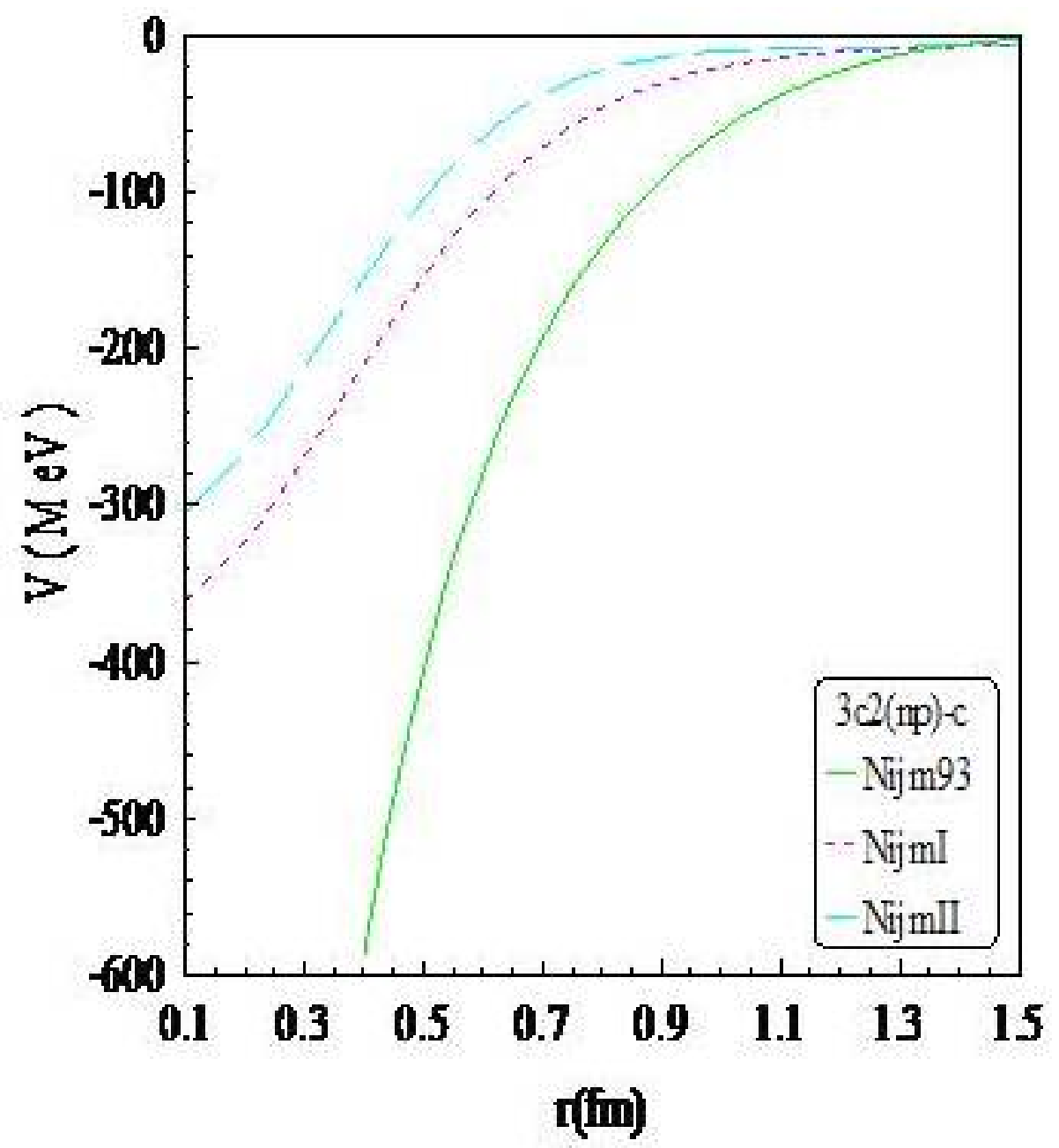}
          \end{subfigure}%
          ~
          \begin{subfigure}[b]{0.31\textwidth}
                  \centering
                  \includegraphics[width=\textwidth,height=0.24\textheight]{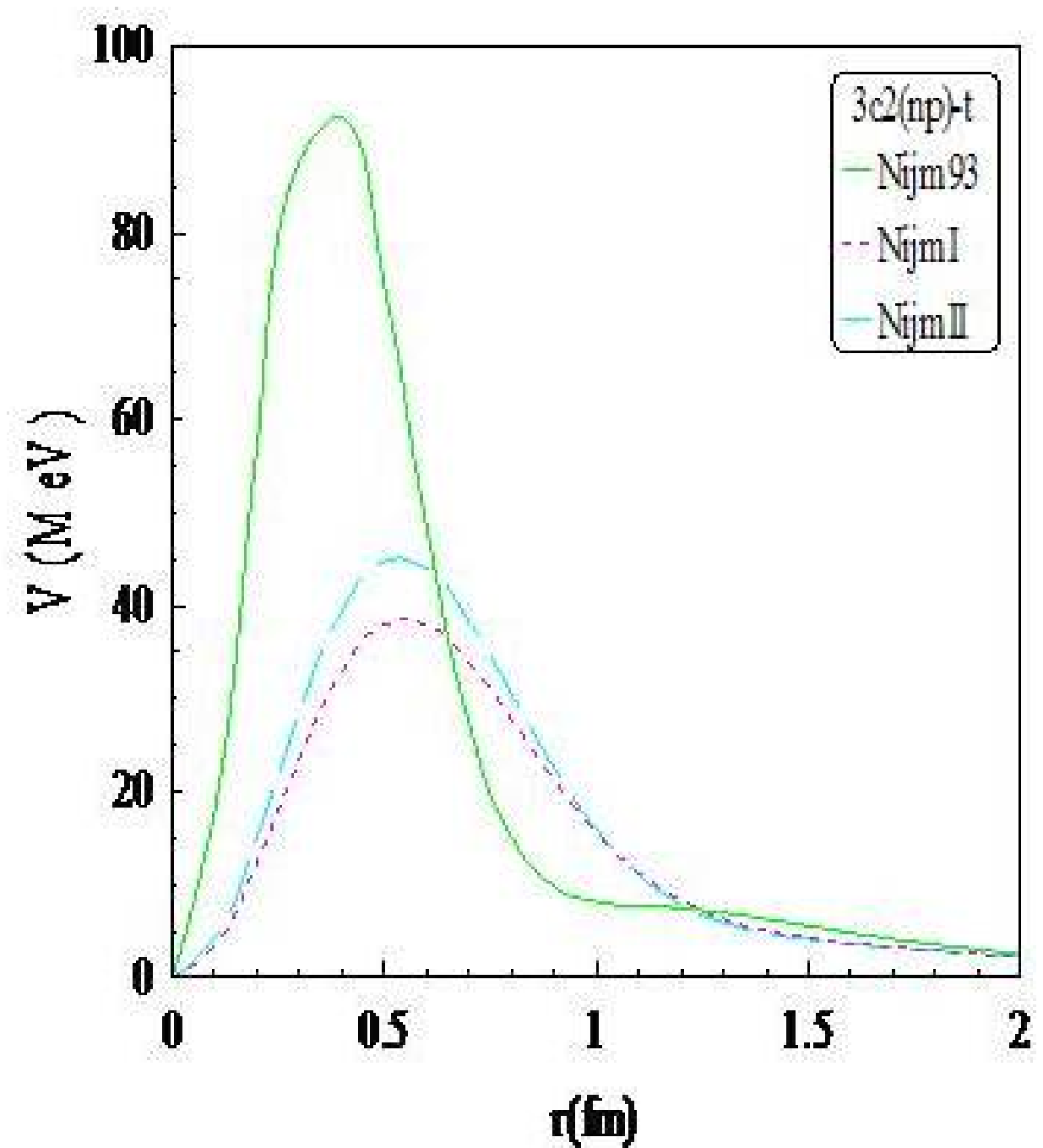}
          \end{subfigure}
           ~
          \begin{subfigure}[b]{0.31\textwidth}
                  \centering
                  \includegraphics[width=\textwidth,height=0.24\textheight]{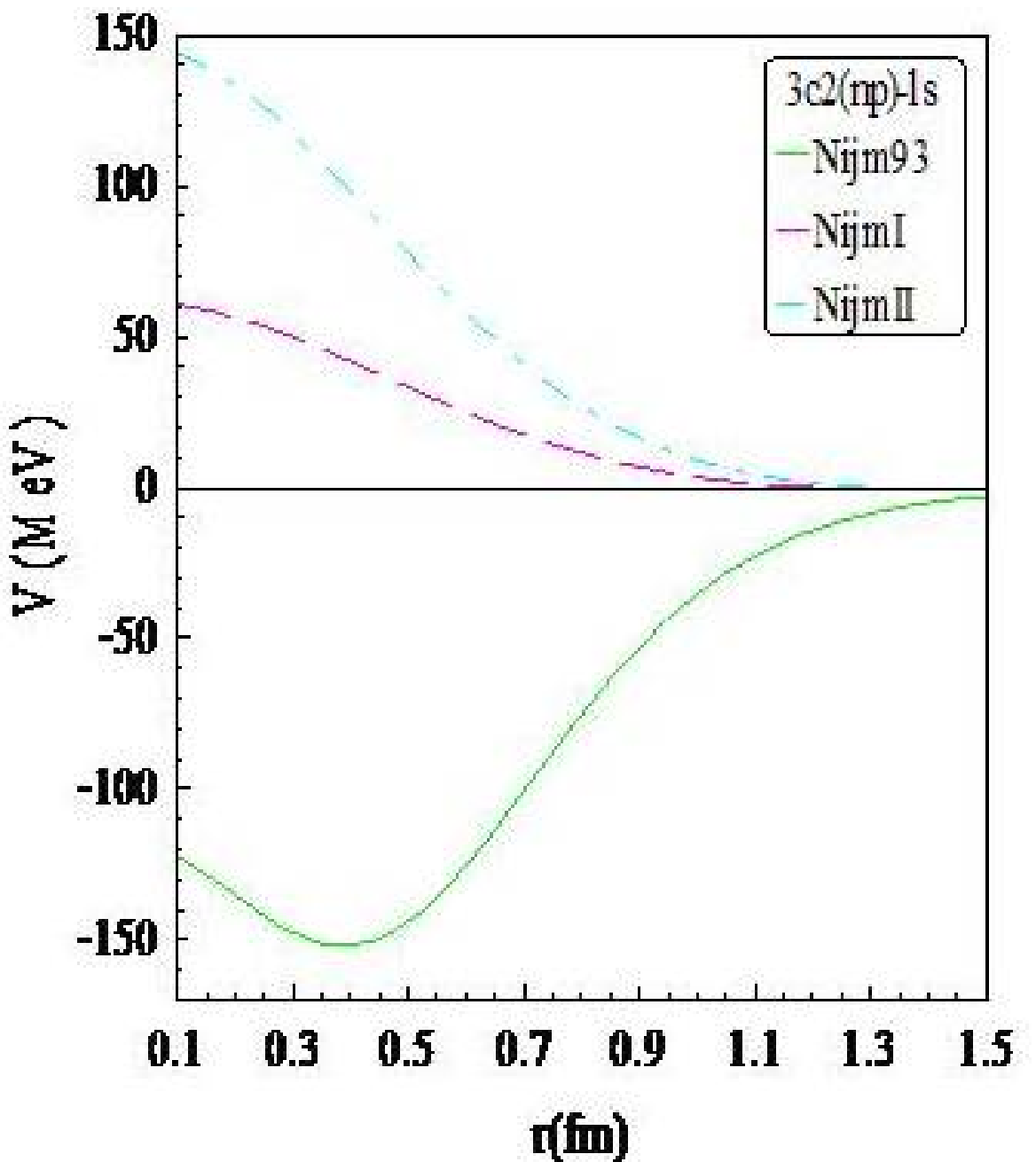}
          \end{subfigure}
    \end{subfigure}
\caption{\textit{The comparison of the central, tensor and spin-orbit potentials of Nijm93, NijmI, and NijmII reduced into the Reid potential, for the states from $J=0$ up to $J=2$, for np system}.} \label{Fig6.}
\end{figure}

\begin{figure}[p]
    \centering
      \begin{subfigure}[b]{\textwidth}
          \centering
          \begin{subfigure}[b]{0.31\textwidth}
                  \centering
                  \includegraphics[width=\textwidth,height=0.24\textheight]{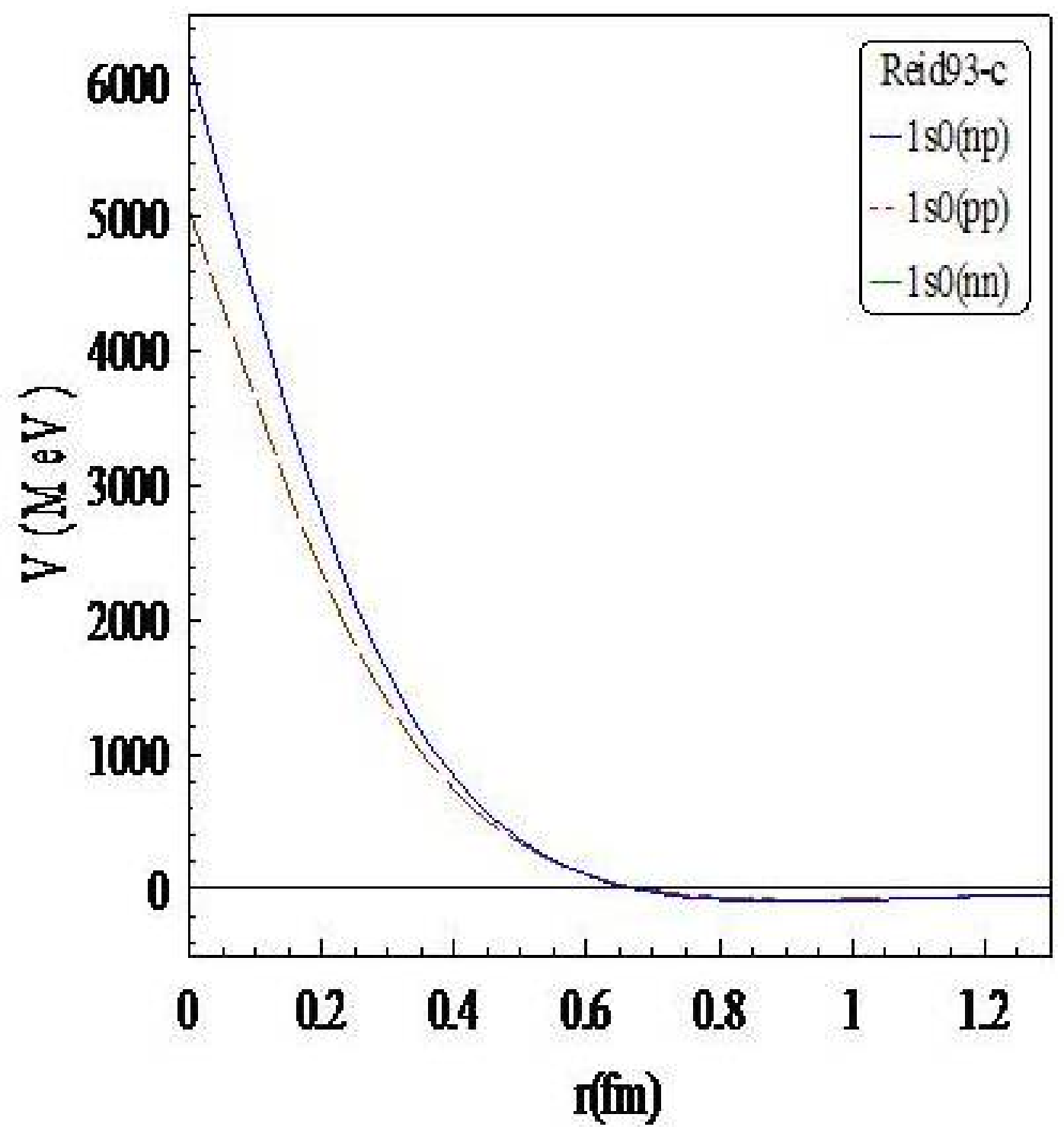}
          \end{subfigure}%
          ~
          \begin{subfigure}[b]{0.31\textwidth}
                  \centering
                  \includegraphics[width=\textwidth,height=0.24\textheight]{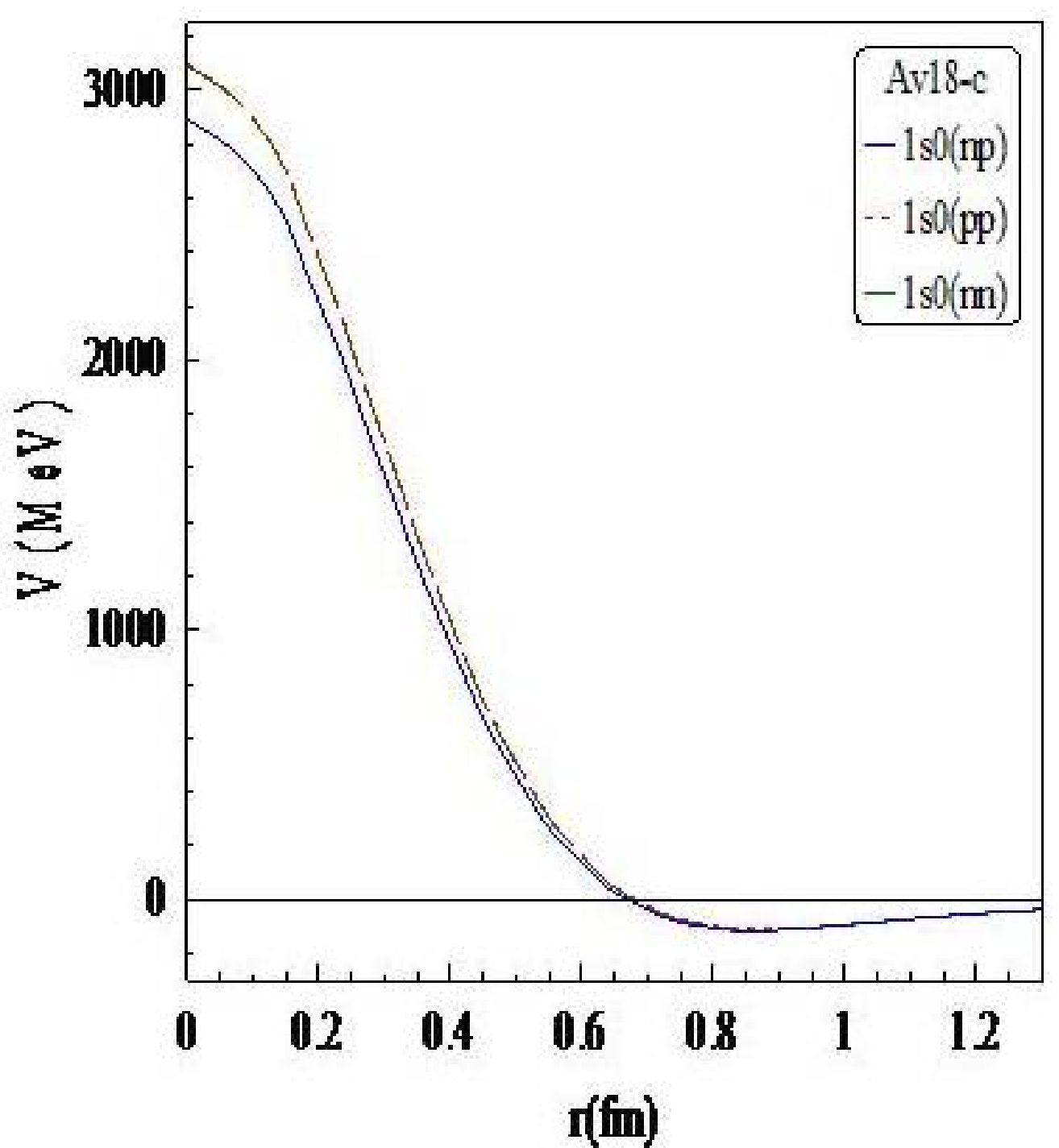}
          \end{subfigure}
           ~
          \begin{subfigure}[b]{0.31\textwidth}
                  \centering
                  \includegraphics[width=\textwidth,height=0.24\textheight]{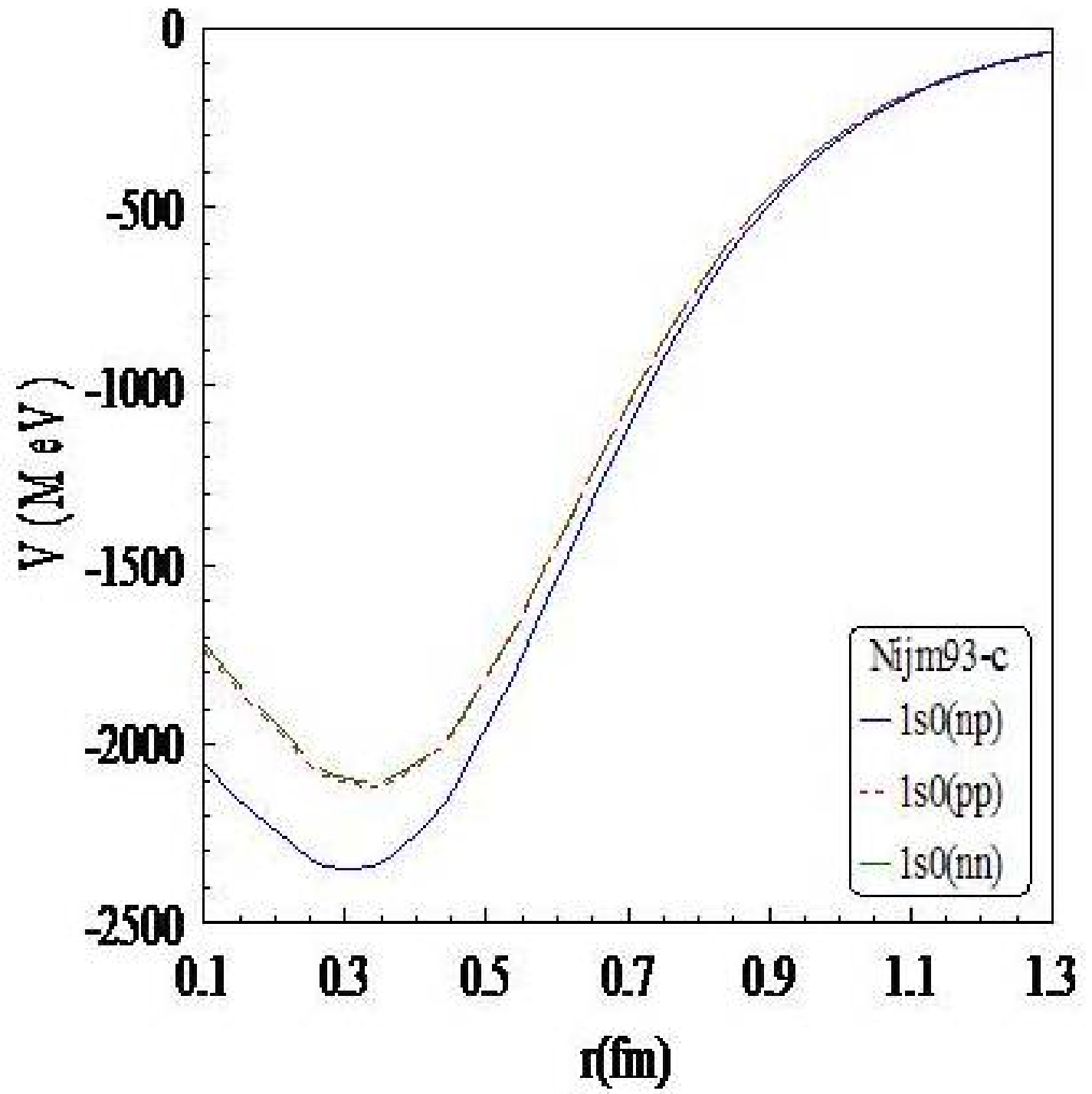}
          \end{subfigure}
    \end{subfigure}
    \begin{subfigure}[b]{\textwidth}
          \centering
           \begin{subfigure}[b]{0.31\textwidth}
                  \centering
                  \includegraphics[width=\textwidth,height=0.24\textheight]{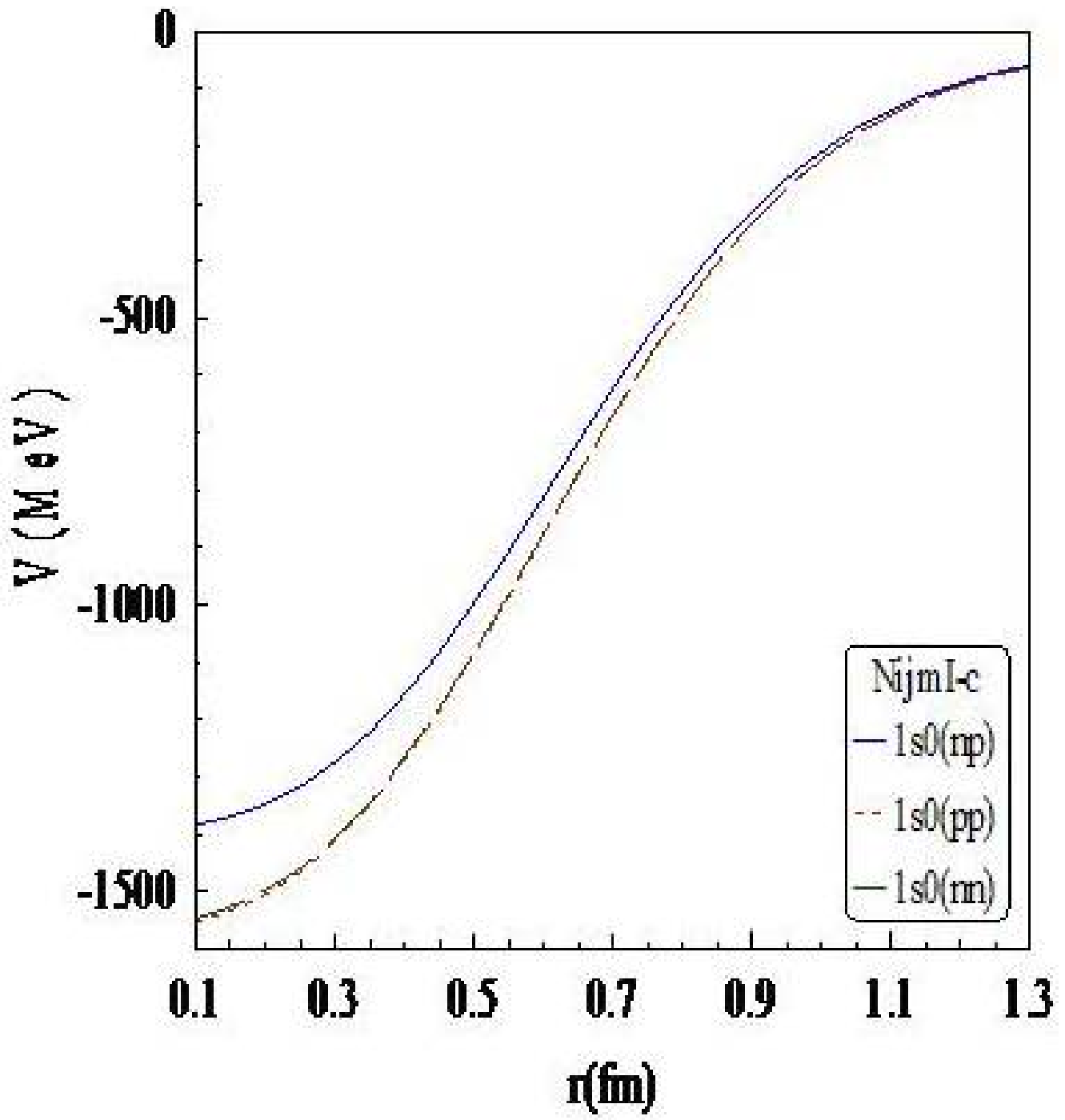}
          \end{subfigure}%
          ~
          \begin{subfigure}[b]{0.31\textwidth}
                  \centering
                  \includegraphics[width=\textwidth,height=0.24\textheight]{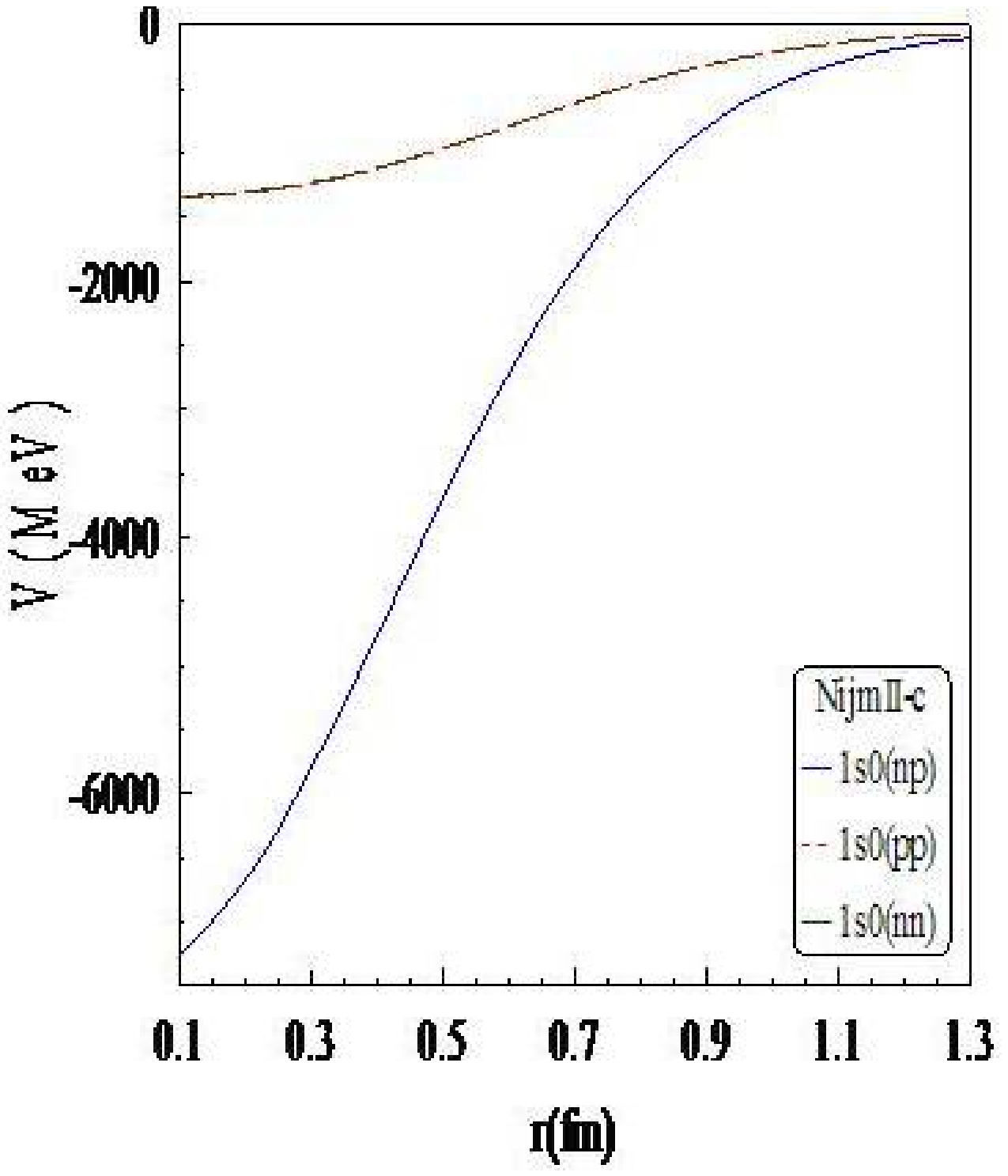}
          \end{subfigure}
           ~
          \begin{subfigure}[b]{0.31\textwidth}
                  \centering
                  \includegraphics[width=\textwidth,height=0.24\textheight]{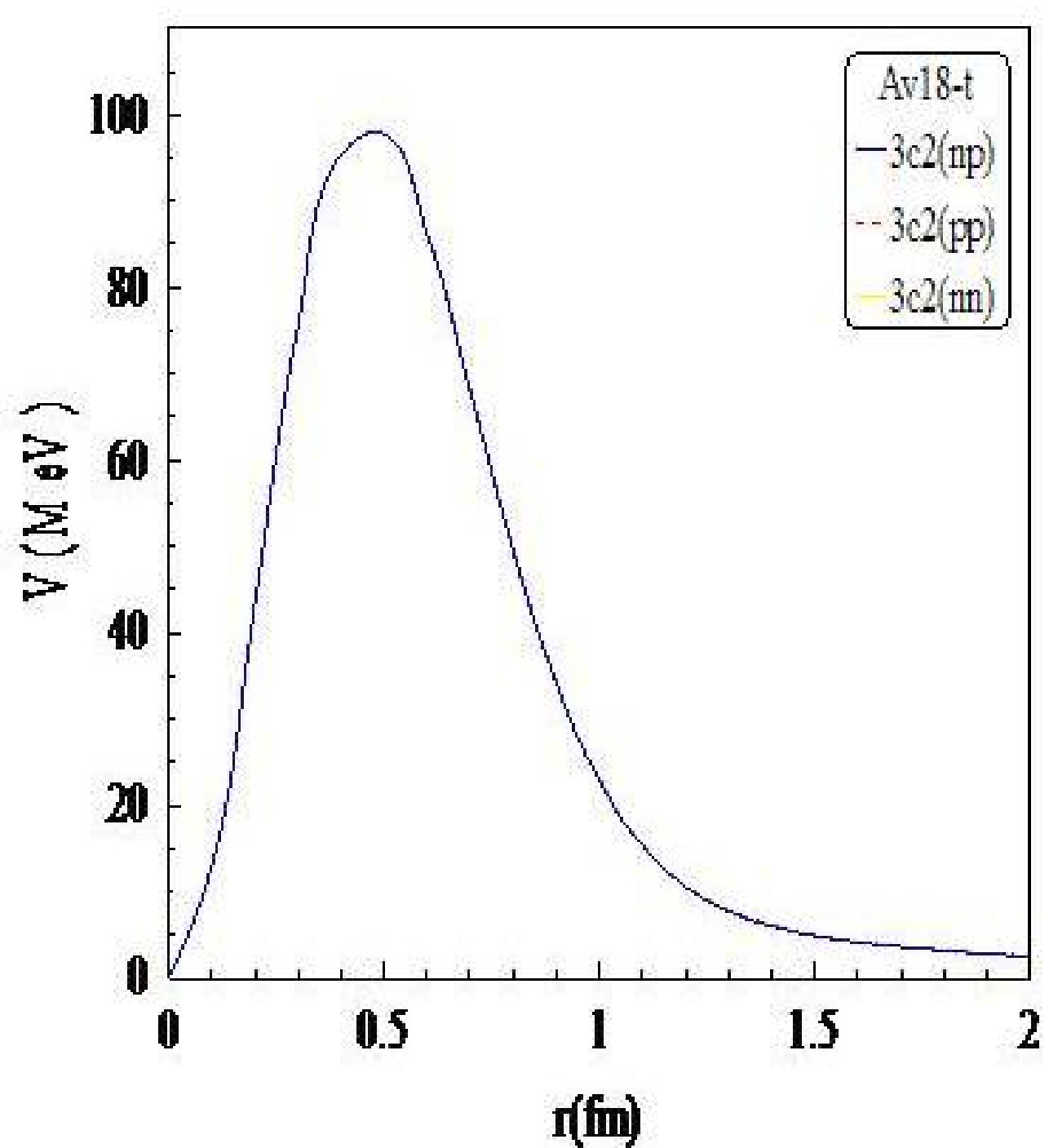}
          \end{subfigure}
   \end{subfigure}
   \begin{subfigure}[b]{\textwidth}
          \centering
        \begin{subfigure}[b]{0.31\textwidth}
                  \centering
                  \includegraphics[width=\textwidth,height=0.24\textheight]{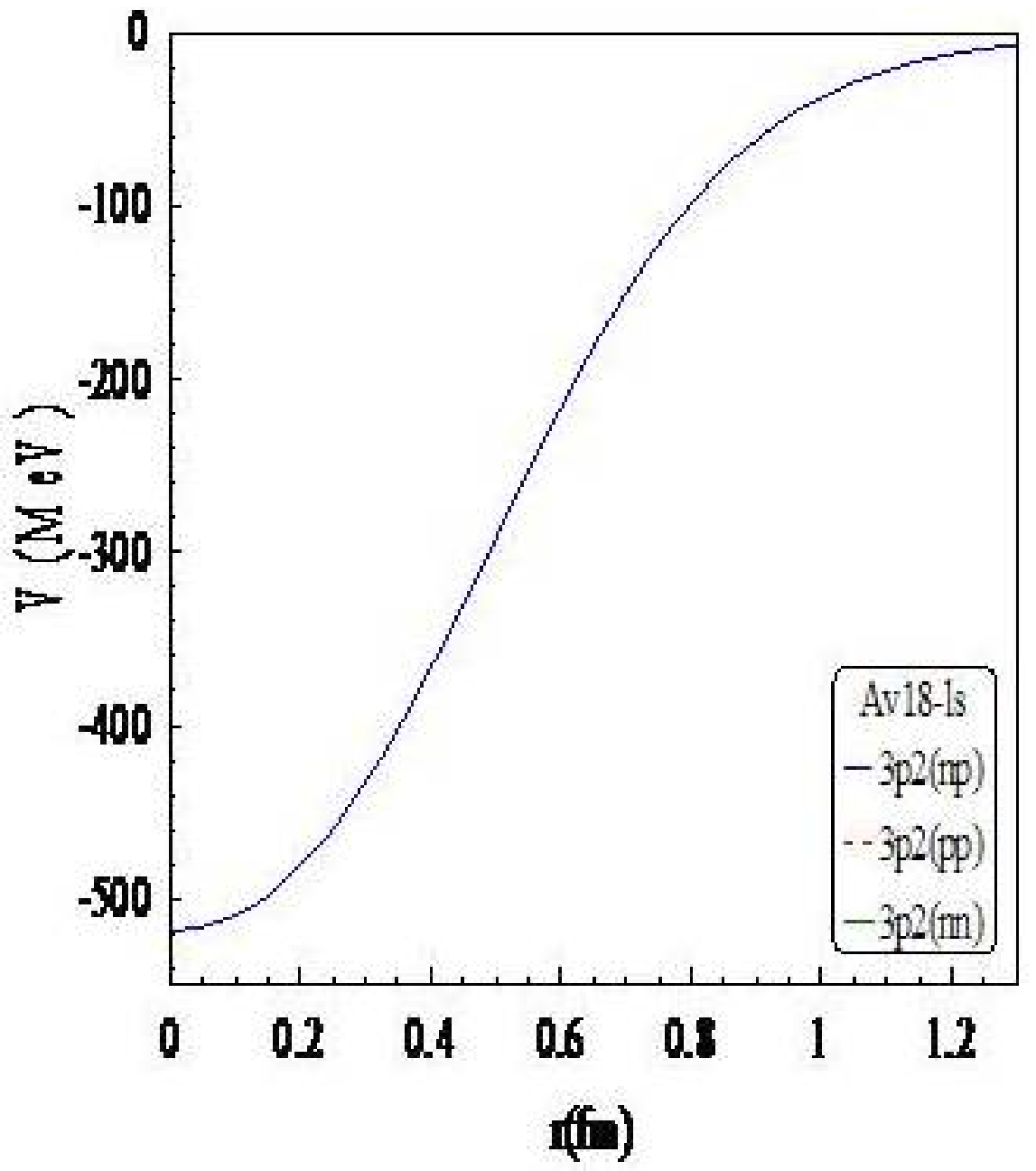}
          \end{subfigure}%
          ~
          \begin{subfigure}[b]{0.31\textwidth}
                  \centering
                  \includegraphics[width=\textwidth,height=0.24\textheight]{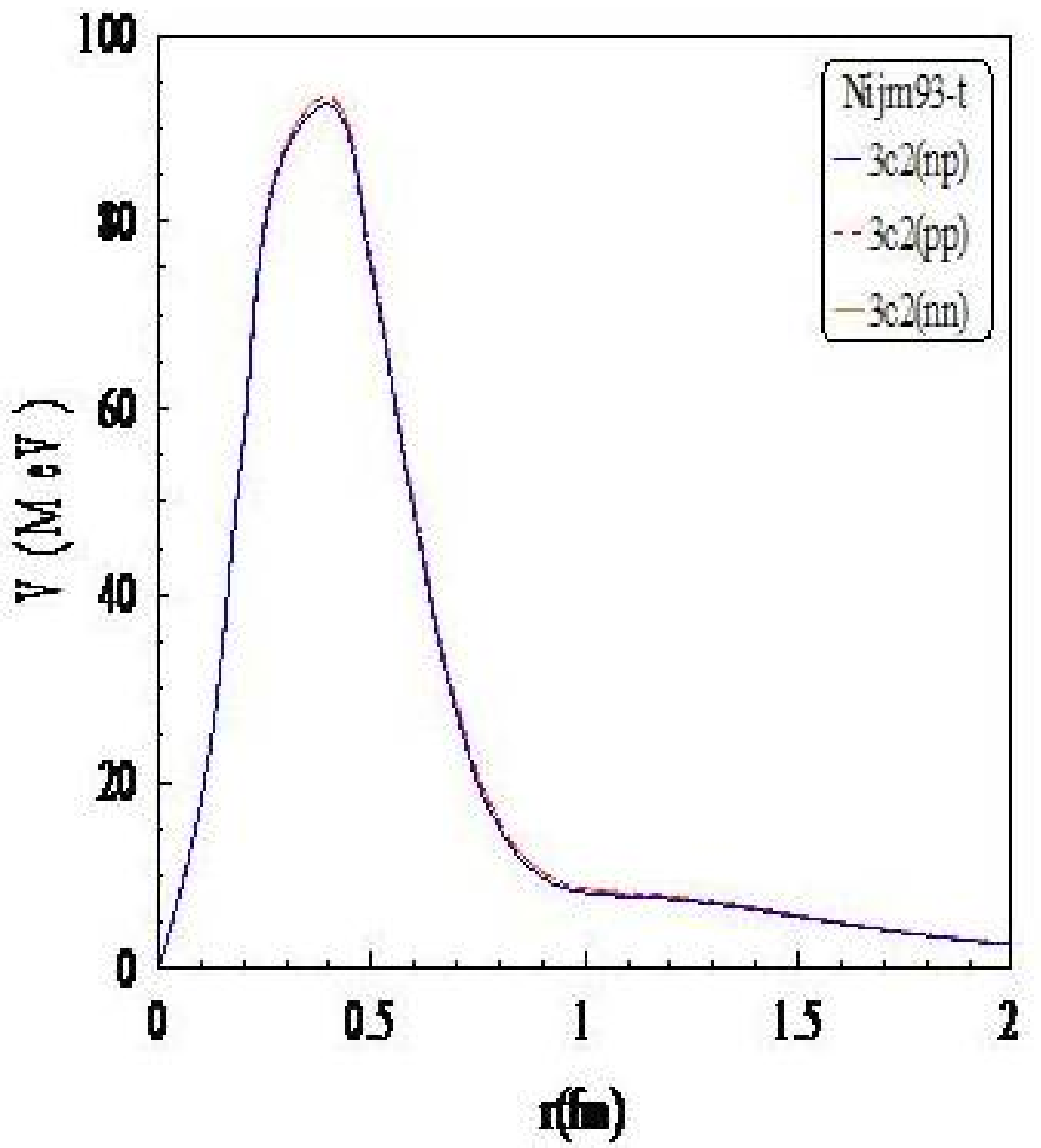}
          \end{subfigure}
           ~
          \begin{subfigure}[b]{0.31\textwidth}
                  \centering
                  \includegraphics[width=\textwidth,height=0.24\textheight]{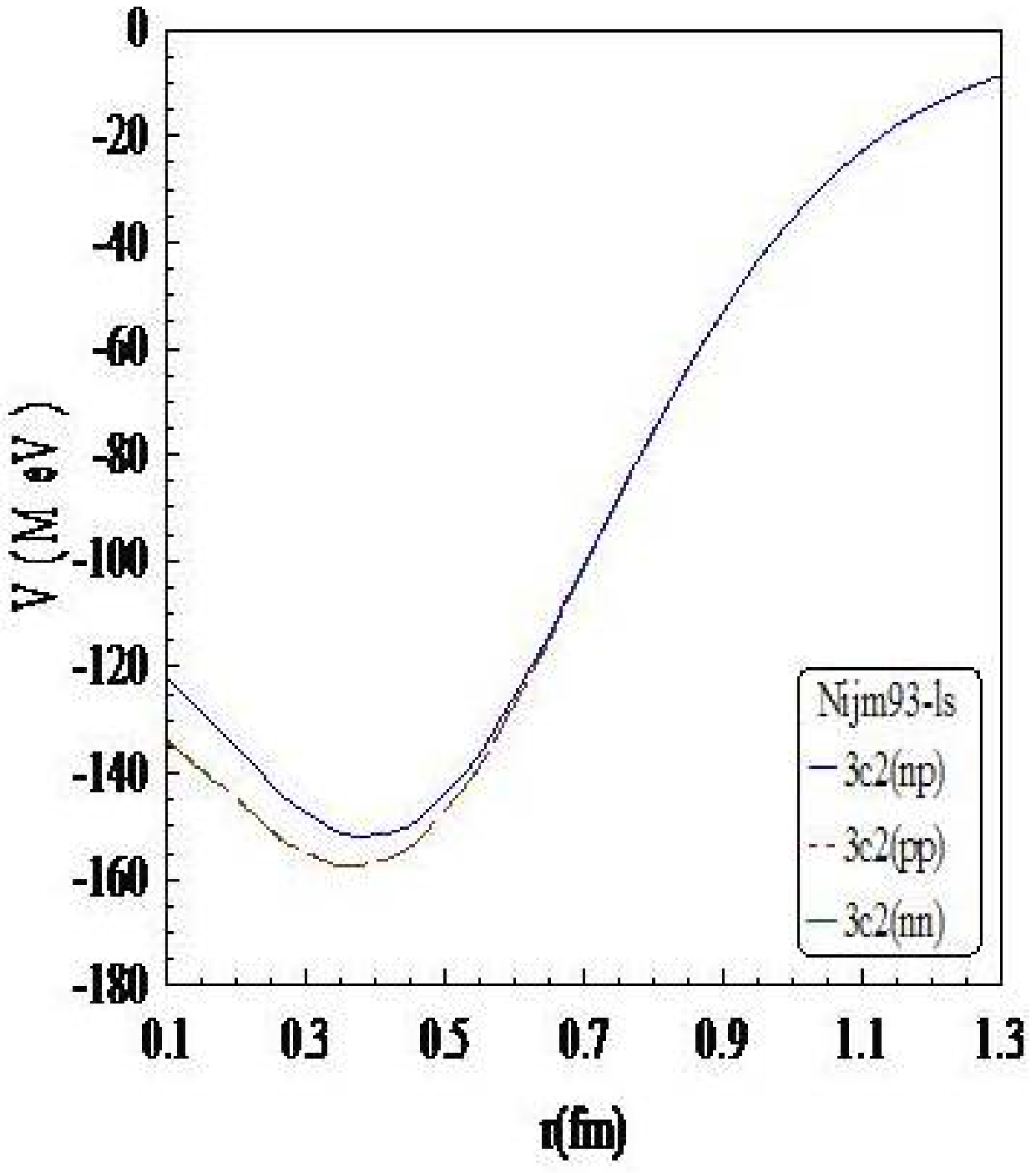}
          \end{subfigure}
   \end{subfigure}
   \begin{subfigure}[b]{\textwidth}
          \centering
        \begin{subfigure}[b]{0.31\textwidth}
                  \centering
                  \includegraphics[width=\textwidth,height=0.24\textheight]{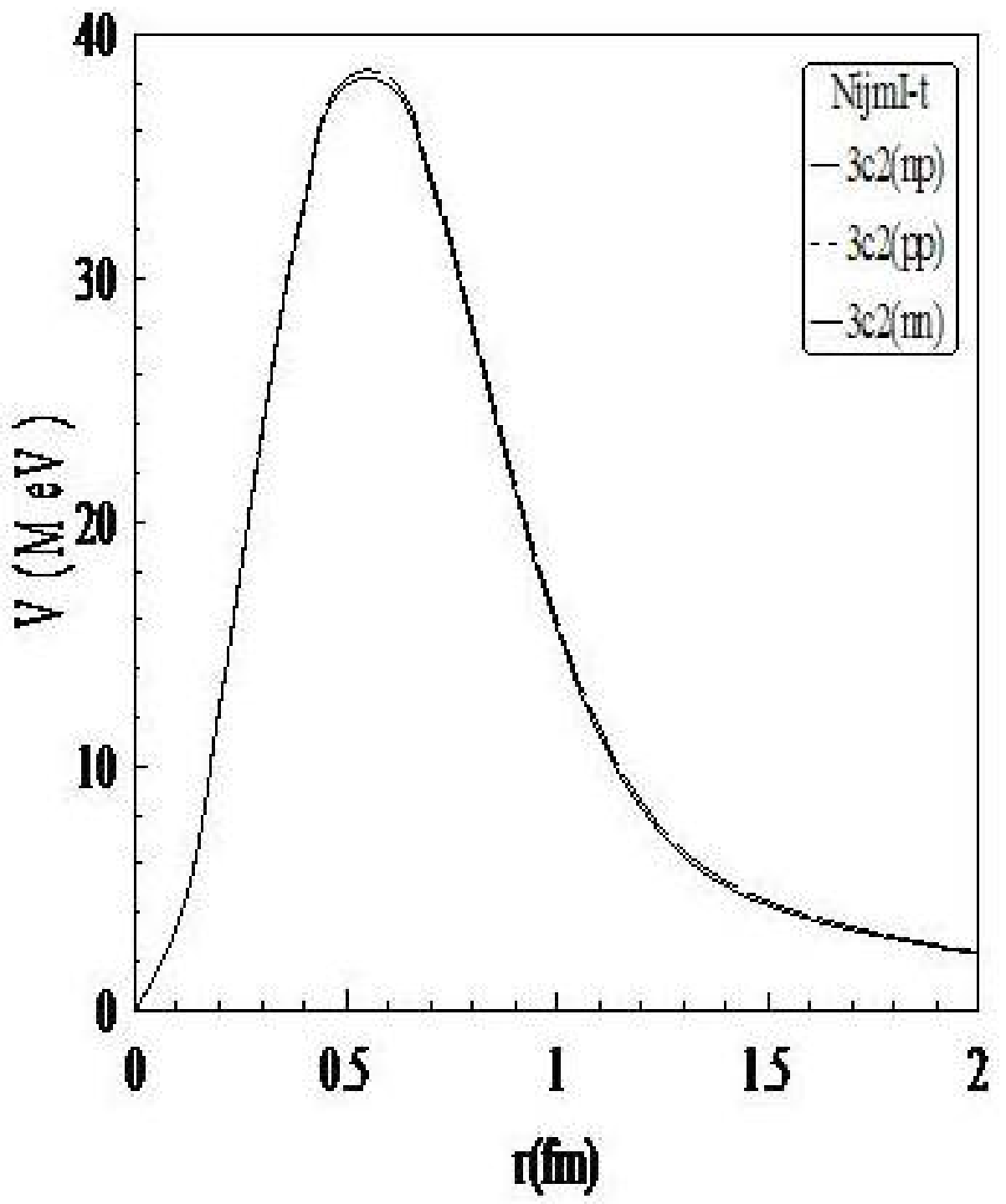}
          \end{subfigure}%
          ~
          \begin{subfigure}[b]{0.31\textwidth}
                  \centering
                  \includegraphics[width=\textwidth,height=0.24\textheight]{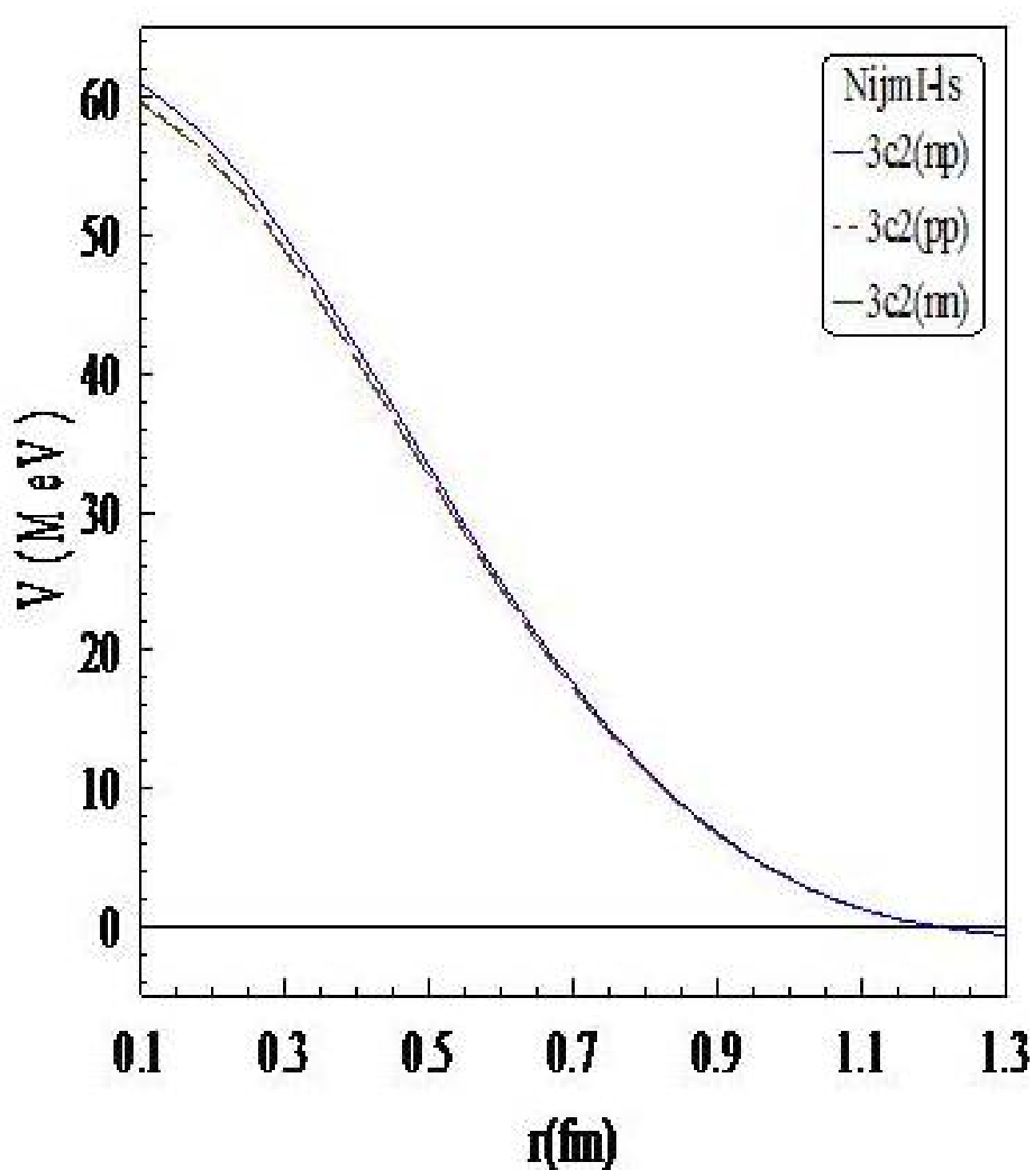}
          \end{subfigure}
           ~
          \begin{subfigure}[b]{0.31\textwidth}
                  \centering
                  \includegraphics[width=\textwidth,height=0.24\textheight]{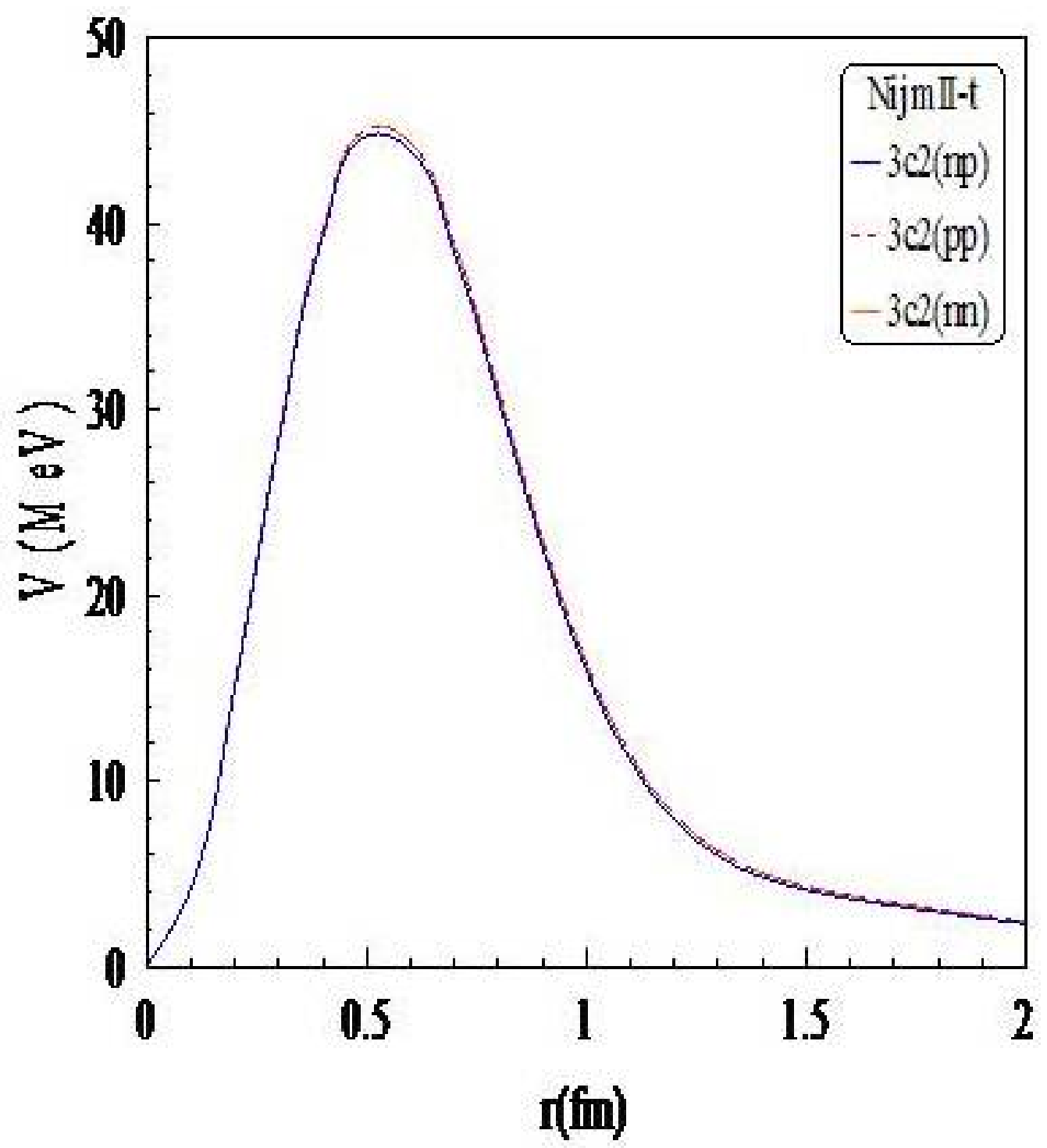}
          \end{subfigure}
    \end{subfigure}
\caption{\textit{The charge-dependence of the charge-dependent potentials reduced to the Reid potential, for the states ${}^1S_0$ (central) and ${}^3P_2-{}^3F_2 (3C2)$ (tensor and spin-orbit)}.} \label{Fig7.}
\end{figure}

\begin{figure}[p]
    \centering
      \begin{subfigure}[b]{\textwidth}
          \centering
          \begin{subfigure}[b]{0.31\textwidth}
                  \centering
                  \includegraphics[width=\textwidth,height=0.27\textheight]{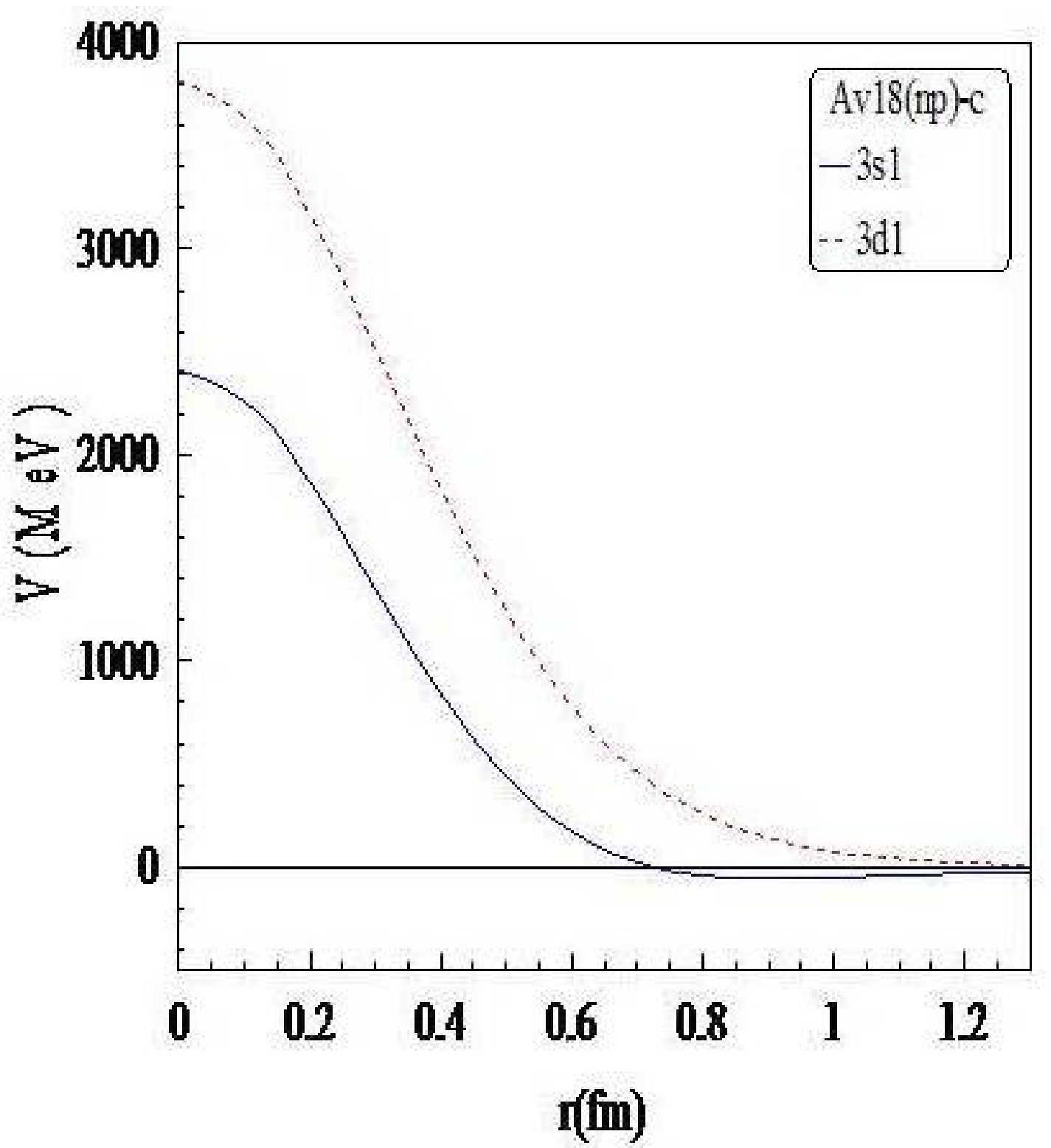}
          \end{subfigure}%
          ~
          \begin{subfigure}[b]{0.31\textwidth}
                  \centering
                  \includegraphics[width=\textwidth,height=0.27\textheight]{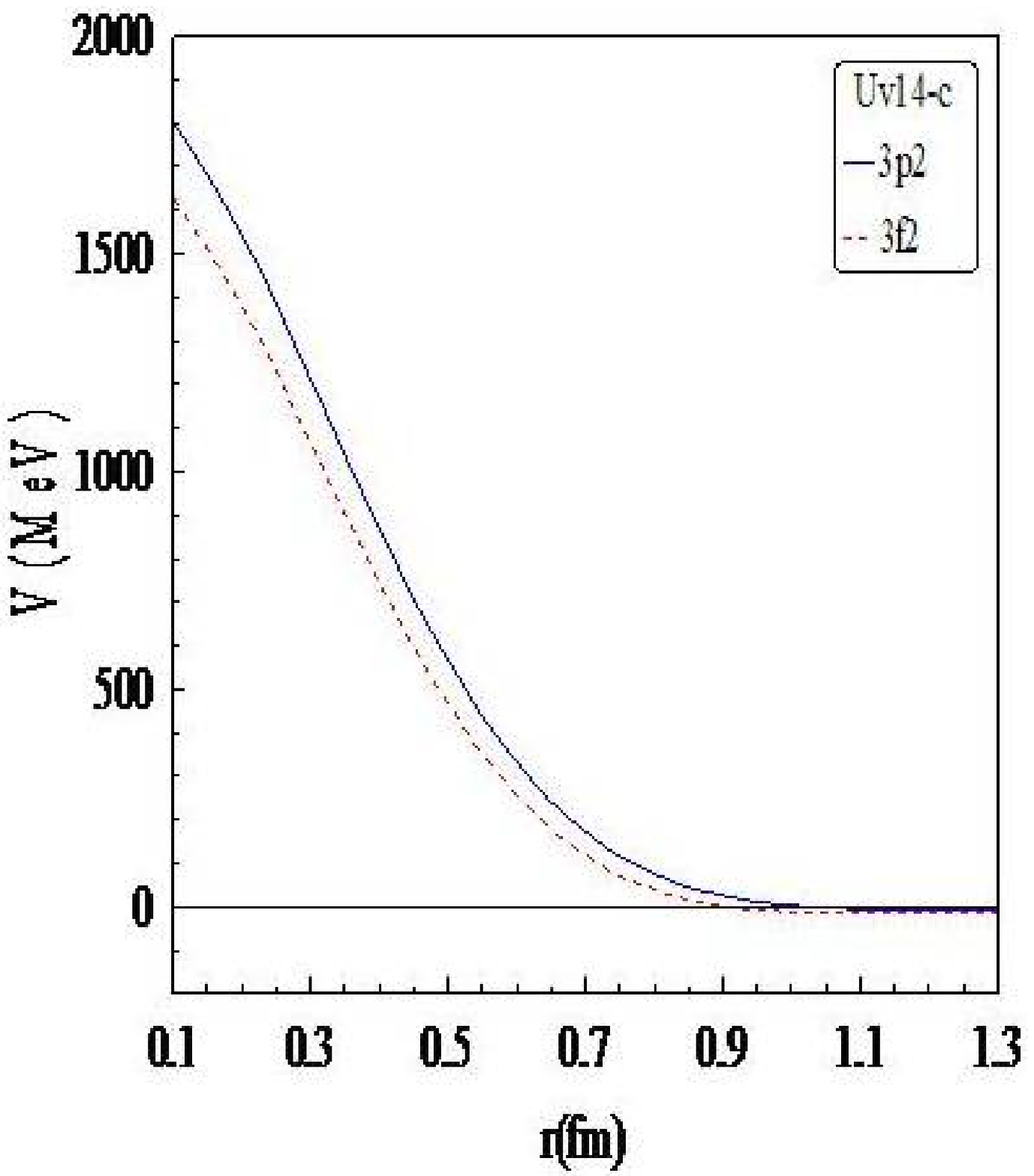}
          \end{subfigure}
           ~
          \begin{subfigure}[b]{0.31\textwidth}
                  \centering
                  \includegraphics[width=\textwidth,height=0.27\textheight]{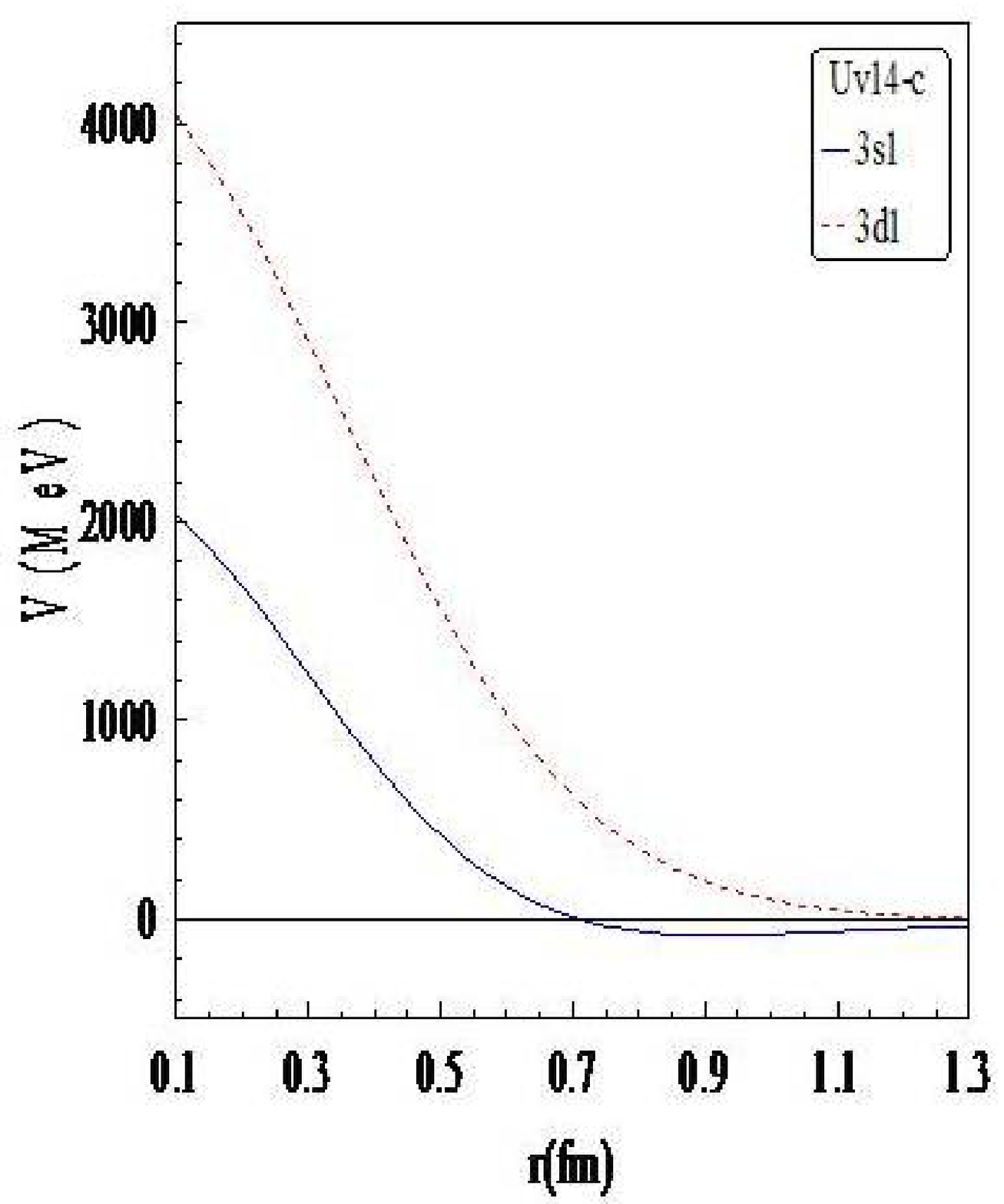}
          \end{subfigure}
    \end{subfigure}
    \begin{subfigure}[b]{\textwidth}
          \centering
           \begin{subfigure}[b]{0.31\textwidth}
                  \centering
                  \includegraphics[width=\textwidth,height=0.27\textheight]{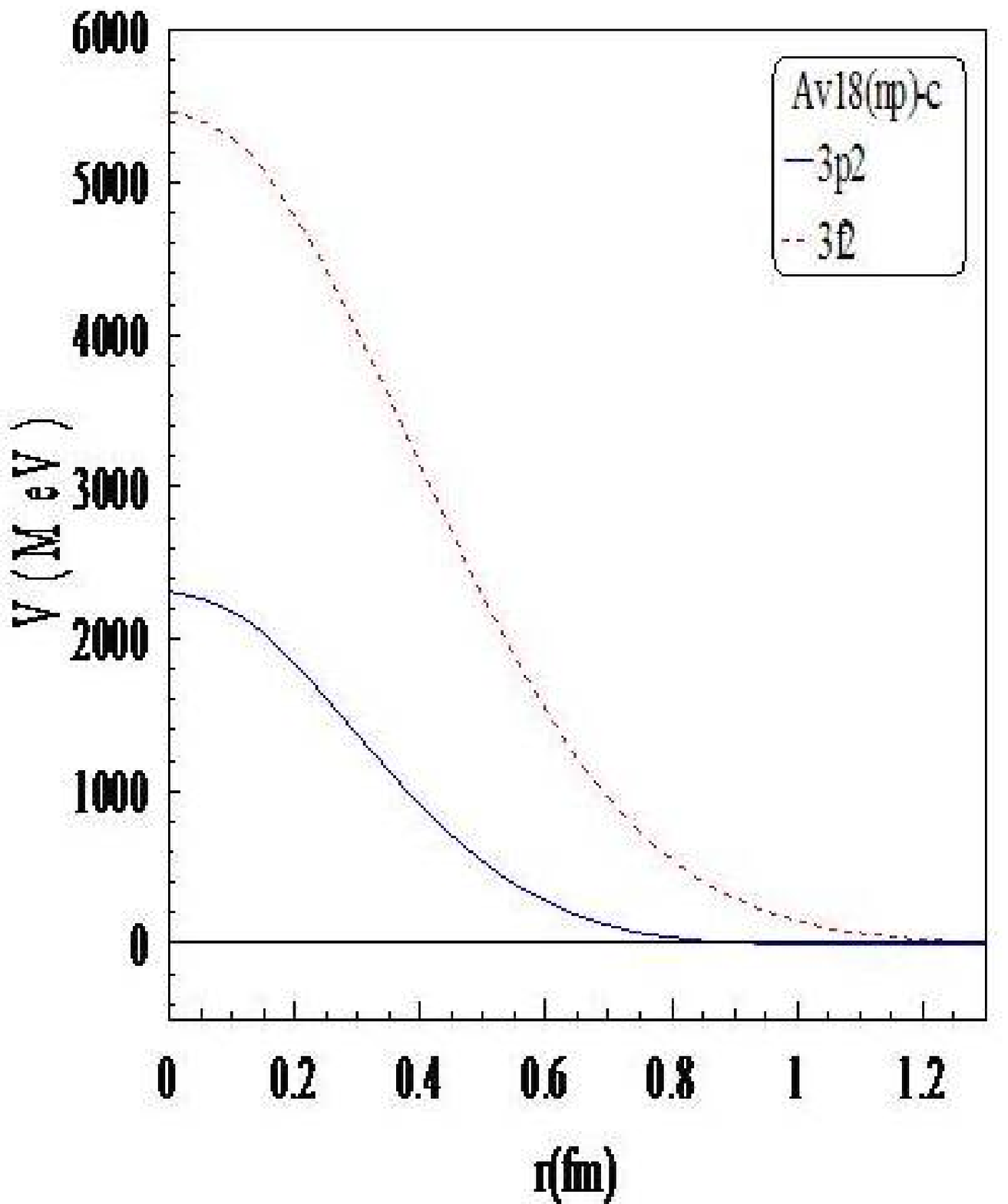}
          \end{subfigure}%
          ~
          \begin{subfigure}[b]{0.31\textwidth}
                  \centering
                  \includegraphics[width=\textwidth,height=0.27\textheight]{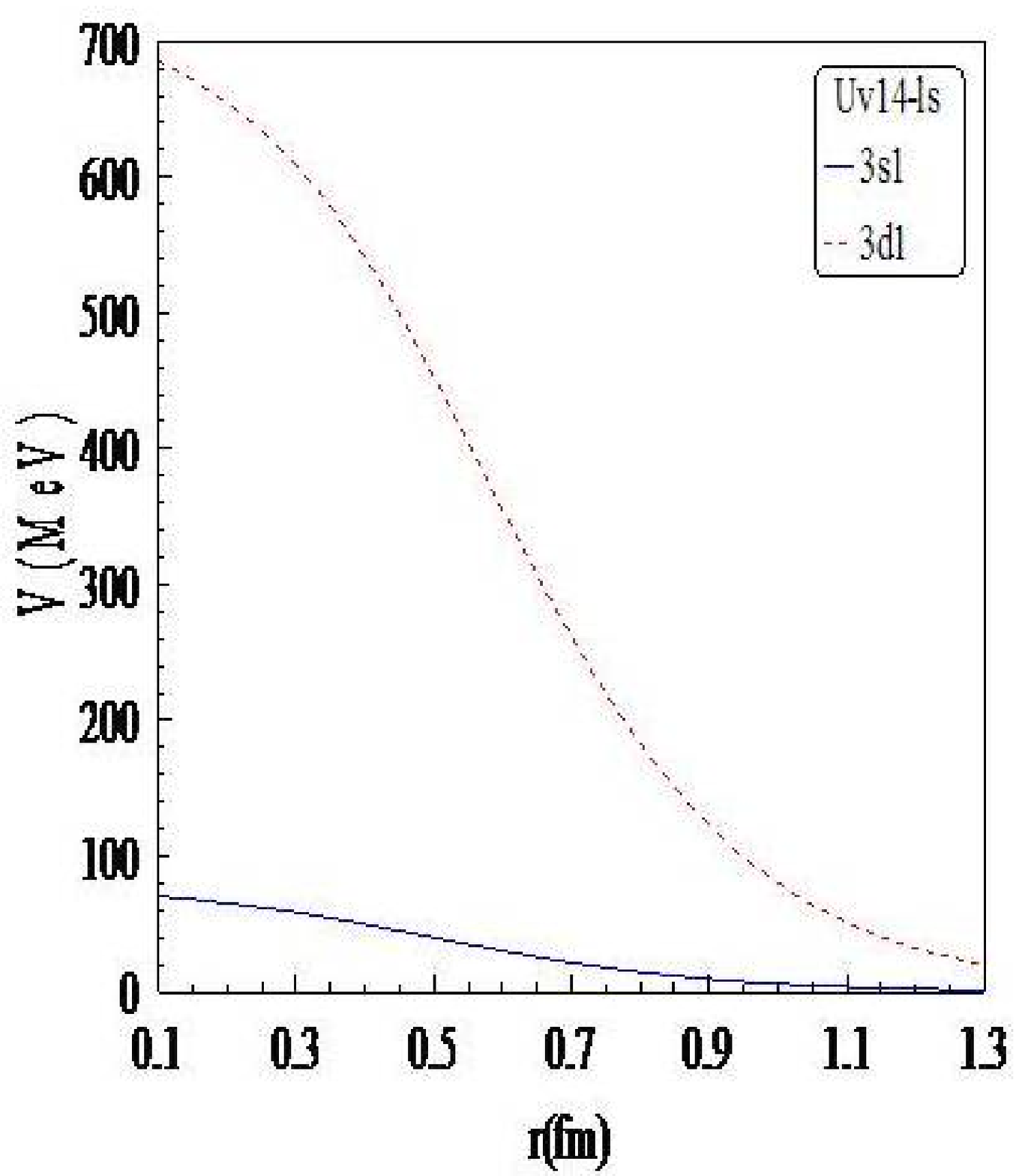}
          \end{subfigure}
           ~
          \begin{subfigure}[b]{0.31\textwidth}
                  \centering
                  \includegraphics[width=\textwidth,height=0.27\textheight]{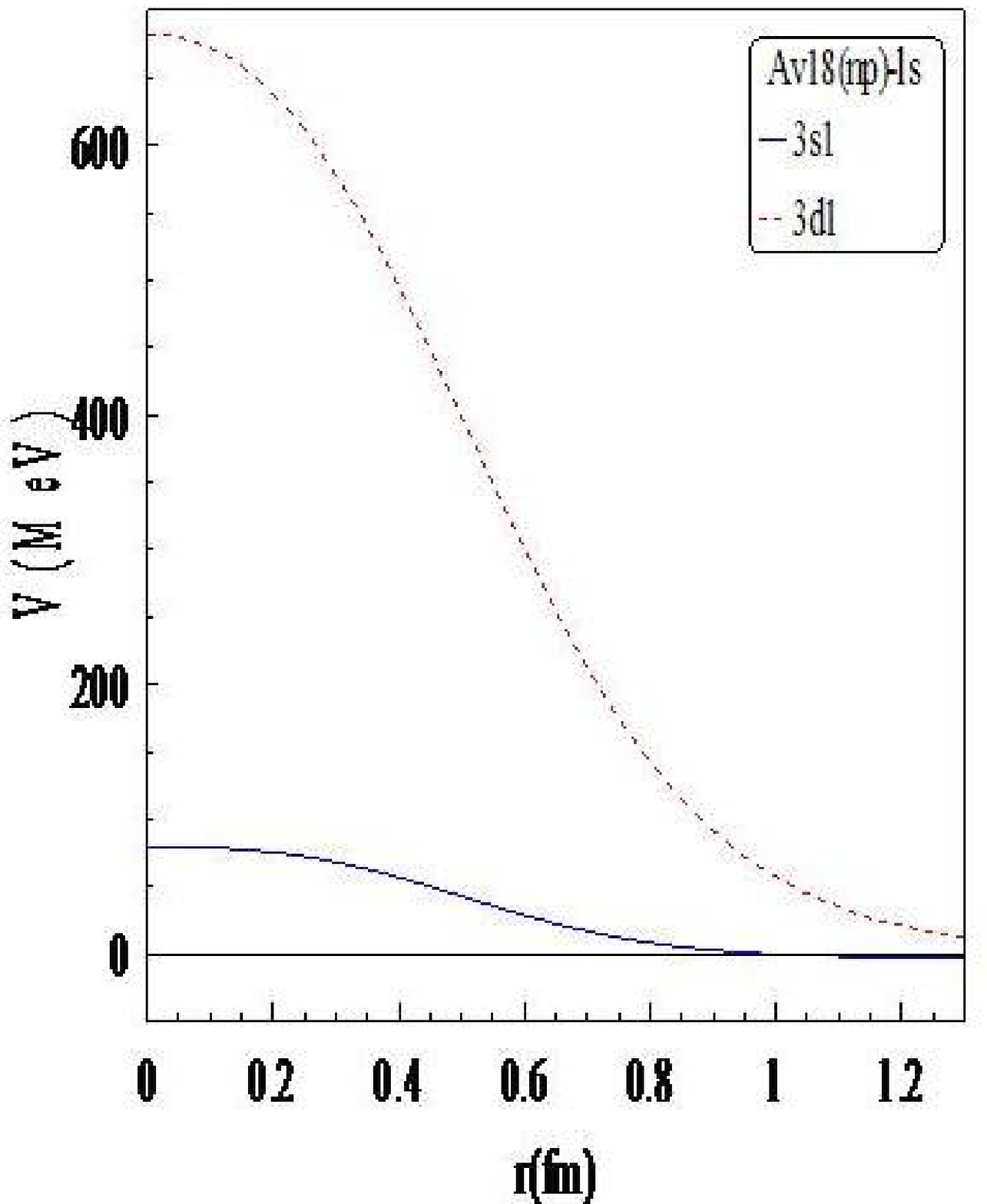}
          \end{subfigure}
   \end{subfigure}
   \begin{subfigure}[b]{\textwidth}
          \centering
        \begin{subfigure}[b]{0.31\textwidth}
                  \centering
                  \includegraphics[width=\textwidth,height=0.27\textheight]{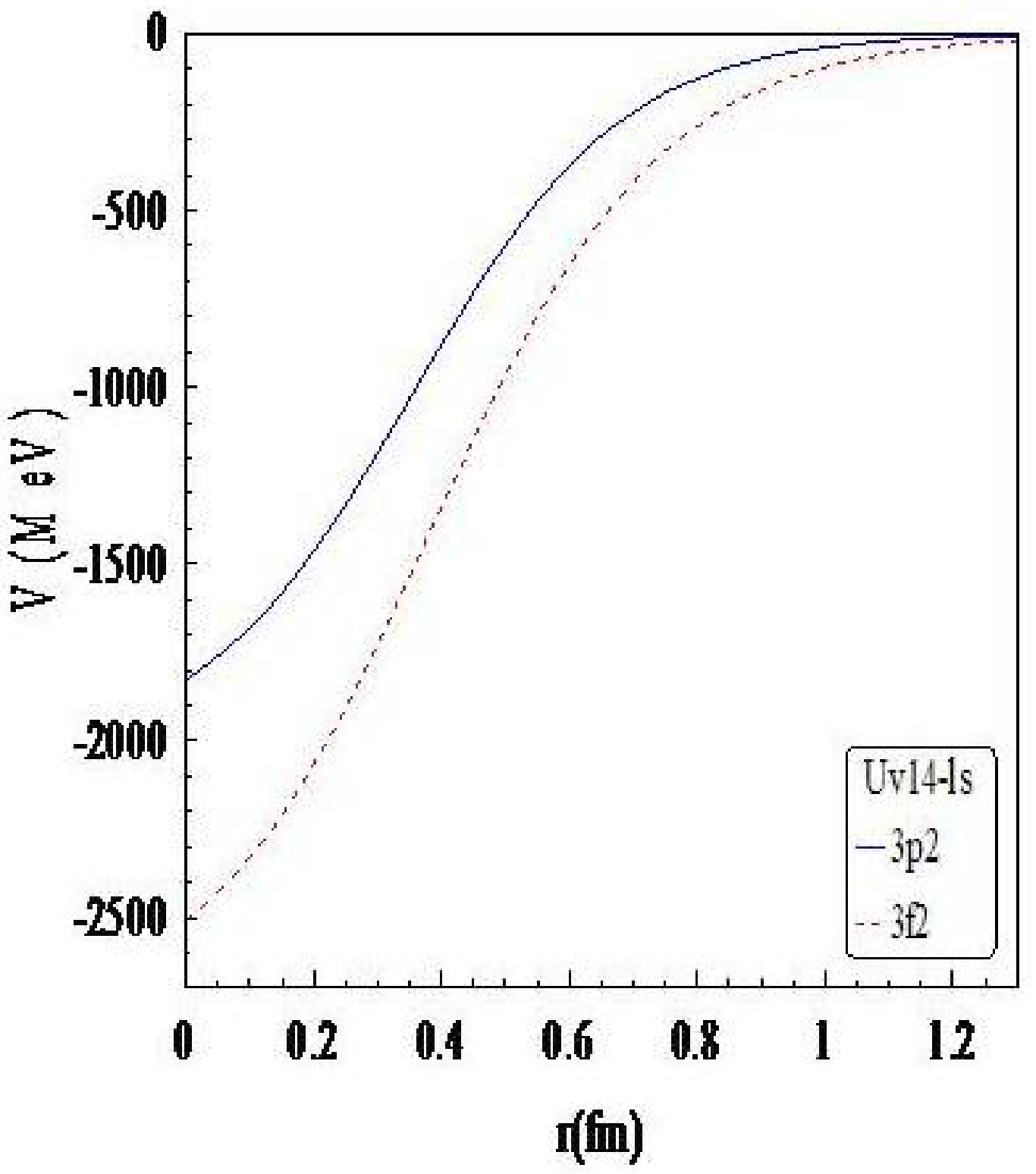}
          \end{subfigure}%
          ~
          \begin{subfigure}[b]{0.31\textwidth}
                  \centering
                  \includegraphics[width=\textwidth,height=0.27\textheight]{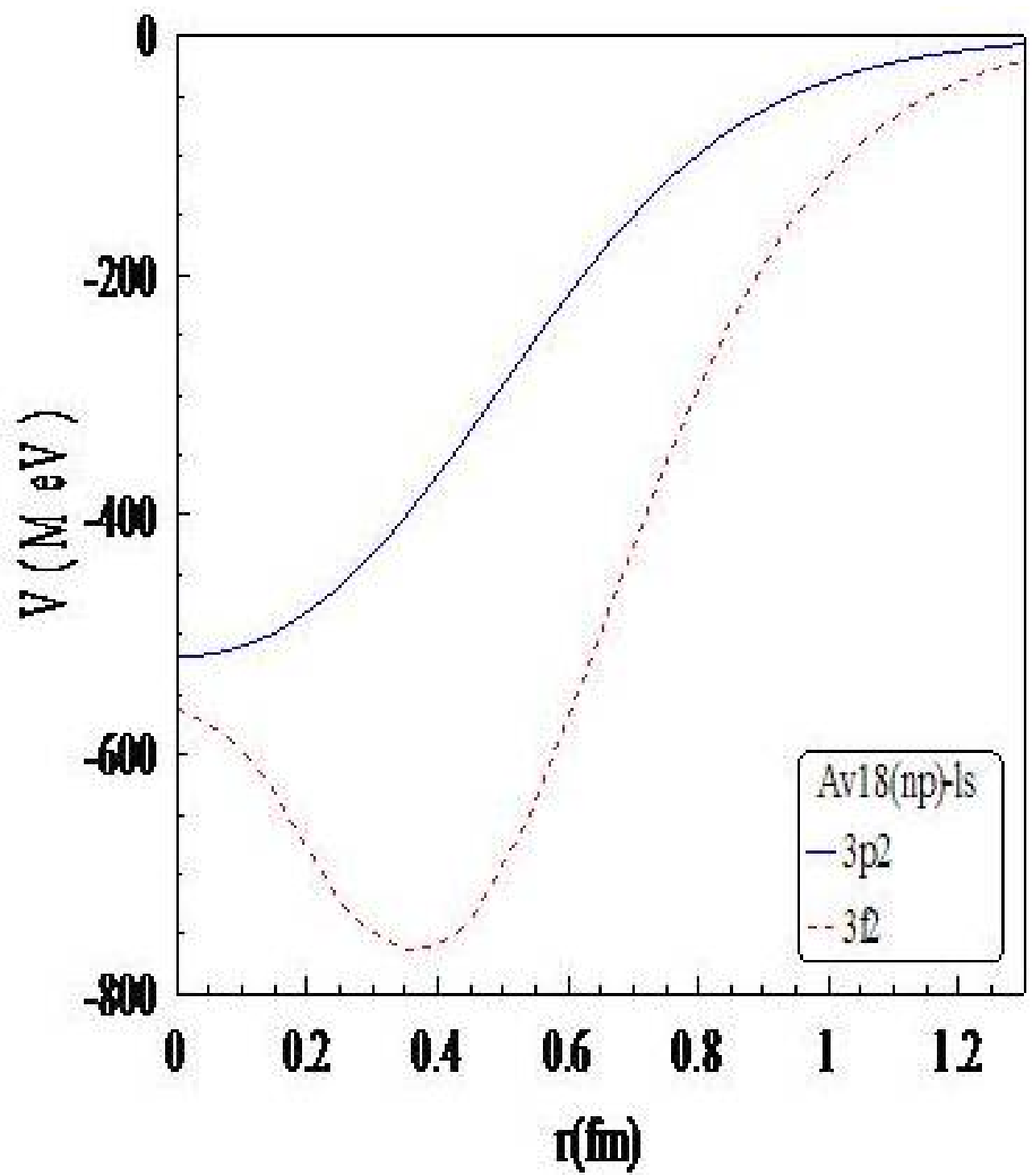}
          \end{subfigure}
   \end{subfigure}
\caption{\textit{The comparison of ${}^{3} S_{1} (\ell =0)$, ${}^{3}D_{1} (\ell =2)$ and also ${}^{3}P_{2} (\ell =1)$, ${}^{3}F_{2} (\ell =3)$ central and spin-orbit potentials of the Urb81 ($UV_{14}$) and Arg94 ($AV_{18}$) potentials reduced into the Reid potential, for np system}.} \label{Fig8.}
\end{figure}


\begin{thebibliography}{99}
\bibitem{M1N} M. Naghdi, \textit{"Nucleon-Nucleon interaction: A typical/concise review"}, \href{http://arxiv.org/abs/nucl-th/0702078}{[arXiv:nucl-th/0702078]}.
\bibitem{FredMyhrer} F. Myhrer and J. Wroldsen, \textit{"The nucleon-nucleon force and the quark degrees of freedom"}, Rev. Mod. Phys. 60, 629 (1988).
\bibitem{nucl-th/0007051} I. P. Cavalcante, M. R. Robilotta, \textit{"Nucleon-nucleon interaction in the Skyrme model"}, Phys. Rev. C 63, 044008 (2001), \href{http://arxiv.org/abs/nucl-th/0007051}{[arXiv:nucl-th/0007051]}.
\bibitem{Rashdan} M. Rashdan, \textit{"NN interaction derived from the Nambu-Jona-Lasinio model"}, Chaos, Solitons \& Fractals 18, 107 (2003).
\bibitem{Beane1} S. R. Beane and M. J. Savage, \textit{"Nucleon-Nucleon interactions on the lattice"}, Phys. Lett. B 535, 177 (2002),  \href{http://arxiv.org/abs/hep-lat/0202013}{[arXiv:hep-lat/0202013]}.
\bibitem{Kukulin} V. I. Kukulin, V. N. Pomerantsev, A. Faessler, A. J. Buchmann and E. M. Tursunov, \textit{"Moscow-type NN-potentials and three-nucleon bound states"}, Phys. Rev. C 57, 535 (1998), \href{http://arxiv.org/abs/nucl-th/9711043}{[arXiv:nucl-th/9711043]}.
\bibitem{Oxford1} C. Downum, J. Stone, T. Barnes, E. Swanson and I. Vidana, \textit{"Nucleon-Nucleon interactions from the quark model"}, AIP Conf. Proc. 1257, 538 (2010), \href{http://arxiv.org/abs/1001.3320}{[arXiv:1001.3320 [nucl-th]]}.
\bibitem{1210.0992} R. Machleidt, Q. MacPherson, E. Marji, R. Winzer, Ch. Zeoli and D. R. Entem, \textit{"Recent progress in the theory of nuclear forces"}, \href{http://arxiv.org/abs/1210.0992}{[arXiv:1210.0992 [nucl-th]]}.
\bibitem{1302.3241} E. Epelbaum, \textit{"Nuclear physics with chiral effective field theory: state of the art and open challenges"},  \href{http://arxiv.org/abs/1302.3241}{[arXiv:1302.3241 [nucl-th]]}.
\bibitem{Ordonez1} C. Ordonez, L. Ray and U. van Kolck, \textit{"Nucleon-nucleon potential from an effective chiral Lagrangian"}, Phys. Rev. Lett. 72, 1982 (1994).
\bibitem{Robilotta1} C. A. da Rocha and M. R. Robilotta, \textit{"Two pion exchange nucleon-nucleon potential: The minimal chiral model"}, Phys. Rev. C 49, 1818 (1994).
\bibitem{Munich1} N. Kaiser, R. Brockmann and W. Weise, \textit{"Peripheral nucleon-nucleon phase shifts and chiral symmetry"}, Nucl. Phys. A 625, 758 (1997), \href{http://arxiv.org/abs/nucl-th/9706045}{[arXiv:nucl-th/9706045]}.
\bibitem{Idaho1} D. R. Entem and R. Machleidt, \textit{"Accurate nucleon-nucleon potential based upon chiral perturbation theory"}, Phys. Lett. B 524, 93,(2002), \href{http://arxiv.org/abs/nucl-th/0108057}{[arXiv:nucl-th/0108057]}.
\bibitem{BochumJulich2} E. Epelbaum, A. Nogga, W. Gloeckle, H. Kamada, U.-G. Meissner and H. Witala, \textit{"The two-nucleon system at next-to-next-to-next-to-leading order"}, Nucl. Phys. A 747, 362 (2005), \href{http://arxiv.org/abs/nucl-th/0405048}{[arXiv:nucl-th/0405048]}.
\bibitem{Tokyo1} M. Taketani, S. Nakamura and M. Sasaki, \textit{"On the method of the theory of nuclear forces"}, Prog. Theor. Phys. 6, 581 (1951).
\bibitem{PartoviLomon} M. H. Partovi and E. L. Lomon, \textit{"Field-theoretical nucleon-nucleon potential"}, Phys. Rev. D 2, 1999 (1970).
\bibitem{Jackson} A. D. Jackson, D. O. Riska and B. Verwest, \textit{"Meson exchange model for the nucleon-nucleon interaction"}, Nucl. Phys. A 249, 397
(1975).
\bibitem{deTourrei0l} R. de Torreil and D. W. L .Sprang, \textit{"Construction of a nucleon-nucleon soft-core potential"}, Nucl. Phys. A 201, 193 (1973).
\bibitem{Funabashi01} T. Obinata and M. Wada, \textit{"Nonstatic one-boson-exchange potential with retardation"}, Prog. Theor. Phys. 53, 732 (1975).
\bibitem{Cottingham} W. N. Cottingham, M. Lacombe, B. Loiseau, J. M. Richardand R. Vinhman, \textit{"Nucleon-Nucleon interaction from pion-nucleon phase-shift analysis"}, Phys. Rev. D 8, 800 (1973).
\bibitem{Machleidt} R. Machleidt, K. Holinde and Ch. Elster, \textit{"The bonn meson-exchange model for the nucleon-nucleon interaction"}, Phys. Rep. 149, 1 (1987).
\bibitem{Minelli} T. A. Minelli, A. Pascolini and C. Villi, \textit{"The Padua model of the nucleon and the nucleon-nucleon potential"}, Il Nuovo Cimento A, 104, 1589 (1991).
\bibitem{Nijm78} M. M. Nagels, T. A. Rijken and J. J. de Swart, \textit{"Low-energy nucleon-nucleon potential from Regge-pole theory"}, Phys. Rev. D 17, 768 (1978).
\bibitem{Jaede} L. Jaede and H. V. von Geram, \textit{"A nonlinear approach to NN interactions using self-interacting meson fields"}, Phys. Rev. C 55, 57 (1997), \href{http://arxiv.org/abs/nucl-th/9604002}{[arXiv:nucl-th/9604002]}.
\bibitem{Virgina1} F. Gross, J. W. Van Orden and K. Holinde, \textit{"Relativistic one-boson-exchange model for the nucleon-nucleon interaction"}, Phys. Rev. C 45, 2094 (1992).
\bibitem{Bochum1} D. Pluemper, J. Flender and M. F. Gari, \textit{"Nucleon-nucleon interaction from meson exchange and nucleonic structure"}, Phys. Rev. C 49, 2370 (1994).
\bibitem{Tubingen1} O. Plohl, C. Fuchs and E. N. E. van Dalen, \textit{"Model independent study of the Dirac structure of the nucleon-nucleon interaction"}, Phys. Rev. C 73, 014003 (2006), \href{http://arxiv.org/abs/nucl-th/0509049}{[arXiv:nucl-th/0509049]}.
\bibitem{HamadaJohnston} T. Hamada and I.D. Johnston, \textit{"A potential model representation of two-nucleon data below 315 MeV"}, Nucl. Phys. 34, 382 (1962).
\bibitem{Yale} K. E. Lassila, M. H. Hull, H. M. Ruppel, F. A. McDonald and G. Breit, \textit{"Note on a nucleon-nucleon potential"}, Phys. Rev. 126, 881 (1962).
\bibitem{Reid68} R. V. Reid, \textit{"Local phenomenological nucleon-nucleon potentials"}, Ann. Phys. (NY) 50, 411 (1968).
\bibitem{Reid68-Day} B. D. Day, \textit{"Three-body correlations in nuclear matter"}, Phys. Rev. C 24, 1203 (1981).
\bibitem{Nijm93} V. G. J. Stoks, R.A. M. Klomp, C. P. F. Terheggen and J.J. de Swart, \textit{"Construction of high-quality nucleon-nucleon potential models"}, Phys. Rev. C 49, 2950 (1994), \href{http://arxiv.org/abs/nucl-th/9406039}{[arXiv:nucl-th/9406039]}.
\bibitem{UV14} I. E. Lagaris and V. R. Pandharipande, \textit{"Phenomenological two-nucleon interaction operator"}, Nucl. Phys. A 359, 331 (1981).
\bibitem{AV14} R.  B. Wiringa, R. A. Smith and T. L. Ainsworth, \textit{"Nucleon-nucleon potentials with and without $\Delta(1232)$ degrees of freedom"}, Phys. Rev. C 29, 1207 (1984).
\bibitem{AV18} R. B. Wiringa, V. G. J. Stoks, R. Schiarilla, \textit{"Accurate nucleon-nucleon potential with charge-independence breaking"}, Phys. Rev. C 51, 38 (1995), \href{http://arxiv.org/abs/nucl-th/9408016}{[arXiv:nucl-th/9408016]}.
\bibitem{nucl-th/9509024} J. J. de Swart, R. A. M. M. Klomp, M. C. M. Rentmeester and Th. A. Rijken, \textit{"The Nijmegen potentials"}, Few Body Syst. Suppl. 8, 438 (1995), , \href{http://arxiv.org/abs/nucl-th/9509024}{[arXiv:nucl-th/9509024]}.
\bibitem{Serra} M. Serra, T. Otsuka, Y. Akaishi, P. Ring and Sh. Hirose, \textit{"Relativistic mean field models and nucleon-nucleon interactions"}, Prog. Theor. Phys. 113, 1009 (2005).
\bibitem{nucl-th/0108041} S. K. Bogner, T. T. S. Kuo, A. Schwenk, D. R. Entem and R. Machleidt, \textit{"Towards a model independent low momentum nucleon nucleon interaction"}, Phys. Lett. B 576, 265 (2003), \href{http://arxiv.org/abs/nucl-th/0108041}{[arXiv:nucl-th/0108041]}.
\bibitem{VinhMau1} R. Vinh Mau, C. Semay, B. Loiseau and M. Lacombe, \textit{"Nuclear forces and quark degrees of freedom"}, Phys. Rev. Lett. 67, 1392 (1991).
\bibitem{JapanQCDNN} Y. Fujiwara, C. Nakamoto and Y. Suzuki, \textit{"Unified description of NN and YN interactions in a quark model with Effective meson-Exchange potentials"}, Phys. Rev. Lett. 76, 2242 (1996).
\bibitem{Shimizu001} K. Shimizu, S. Takeuchi and A. J. Buchmann, \textit{"Study of nucleon nucleon and hyperon nucleon interaction"}, Prog. Theor. Phys. Suppl. 137, 43 (2000).
\bibitem{nucl-th/0212044} A. Valcarce, F. Fernandez and P. Gonzalez, \textit{"NN interaction in chiral constituent quark models"}, Few Body Syst. Suppl. 15, 25 (2003), \href{http://arxiv.org/abs/nucl-th/0212044}{[arXiv:nucl-th/0212044]}.
\bibitem{1110.3761} F. Gross, T. D. Cohen, E. Epelbaum and R. Machleidt, \textit{"Conference discussion of the nuclear force"}, Few-Body Syst. 50, 31 (2011), \href{http://arxiv.org/abs/1110.3761}{[arXiv:1110.3761 [nucl-th]]}.
\bibitem{Machleidt00} R. Machleidt, \textit{"The Meson theory of nuclear forces and nuclear structure"}, Adv. Nucl. Phys. 19, 189 (1989).
\bibitem{Paris2} M. Lacombe, B. L. Seau, J. M. Richard, R. Vinhman, J. Côté, P. Pirés and R. de Tourreil, \textit{"Parametrization of the Paris N-N potential"}, Phys. Rev. C 21, 861 (1980).
\bibitem{Bonn4} R. Machleidt, \textit{"The high-precision, charge-dependent Bonn nucleon-nucleon potential (CD-Bonn)"}, Phys. Rev. C 63, 024001 (2001),  \href{http://arxiv.org/abs/nucl-th/0006014}{[arXiv:nucl-th/0006014]}.
\bibitem{nucl-th/9301019} R. Machleidt and G.Q. Li, \textit{"Nucleon-Nucleon potentials in comparison: physics or polemics?"}, Phys.Rept. 242, 5 (1994),  \href{http://arxiv.org/abs/nucl-th/9301019}{[arXiv:nucl-th/9301019]}.
\bibitem{nucl-th/9211013} V. Stoks and J. J. de Swart, \textit{"Comparison of potential models with the pp scattering data below 350 MeV"}, Phys. Rev. C 47, 761 (1993), \href{http://arxiv.org/abs/nucl-th/9211013}{[arXiv:nucl-th/9211013]}.
\bibitem{deTourreil} R. de Tourreil, B. Rouben and D. W. L. Sprung, \textit{"Super-soft-core nucleon-nucleon interaction with $\pi$-, $\rho$- and $\omega$-exchange contributions"}, Nucl. Phys. A 242, 445 (1975).
\bibitem{Funabashi02} T. Obinata and M. Wada, \textit{"Nonstatic one-boson-exchange potential with retardation and nuclear matter- Including velocity-dependent tensor potential-"}, Prog. Theor. Phys. 57, 1984 (1977). 
\bibitem{Haidenbauer} J. Haidenbauer and K. Holinde, \textit{"Application of the Bonn potential to proton-proton scattering"}, Phys. Rev. C 40, 2465 (1989).
\bibitem{nucl-th/9411002} V. Stoks and J. J. de Swart, \textit{"Comparison of potential models with the $np$ scattering data below 350 MeV"}, Phys. Rev. C 52, 1698 (1995), \href{http://arxiv.org/abs/nucl-th/9411002}{[arXiv:nucl-th/9411002]}.
\bibitem{NijmPWA} V. G. J. Stoks, R. A. M. Klomp, M. C. M. Rentmeester and J .J. Swart, \textit{"Partial-wave analysis of all nucleon-nucleon scattering data below 350 MeV"}, Phys. Rev. C 48, 792 (1993).
\bibitem{NijmPWA3} M. C. M. Rentmeester, R. G. E. Timmermans and J. J. de Swart \textit{"Partial-wave analyses of all proton-proton and neutron-proton data below 500 MeV"}, \href{http://arxiv.org/abs/nucl-th/0410042v}{[arXiv:nucl-th/0410042]}.
\bibitem{0706.2195} R. A. Arndt, W. J. Briscoe, I. I. Strakovsky and R.L. Workman, \textit{"Updated analysis of NN elastic scattering to 3 GeV"}, Phys. Rev. C. 76, 025209 (2008), \href{http://arxiv.org/abs/0706.2195}{[arXiv:0706.2195 [nucl-th]]}.
\bibitem{1304.0895} R. Navarro Perez, J. E. Amaro and E. Ruiz Arriola, \textit{"Partial wave analysis of nucleon-nucleon scattering below pion production threshold"}, \href{http://arxiv.org/abs/1304.0895}{[arXiv:1304.0895 [nucl-th]]}.
\bibitem{NijmESC2} Th. A. Rijken, \textit{"Extended-soft-core baryon-baryon model I. nucleon-nucleon scattering (ESC04)"}, Phys. Rev. C 73, 044007 (2006), \href{http://arxiv.org/abs/nucl-th/0603041}{[arXiv:nucl-th/0603041]}.
\bibitem{AV18pq} R. B. Wiringa, A. Arriaga and V. R. Pandharipande, \textit{"Quadratic momentum dependence in the nucleon-nucleon interaction"}, Phys. Rev. C. 68, 054006 (2003), \href{http://arxiv.org/abs/nucl-th/0306018}{[arXiv:nucl-th/0306018]}.
\bibitem{Hamburg2a} L. Jaede and H. V. VonGeramb, \textit{"Nucleon-nucleon scattering observables from the solitary boson exchange potential"}, Phys. Rev. C 57, 496 (1998), \href{http://arxiv.org/abs/nucl-th/9707023}{[arXiv:nucl-th/9707023]}.
\bibitem{nucl-th/9706003} R. A. Arndt, C. H. Oh, I. I. Strakovsky, R. L. Workman and F. Dohrmann, \textit{"Nucleon-nucleon elastic scattering analysis to 2.5 GeV"}, Phys. Rev. C 56, 3005 (1997), \href{http://arxiv.org/abs/nucl-th/9706003}{[arXiv:nucl-th/9706003]}.
\bibitem{Moscow3} A. Faessler, V. I. Kukulin, I. T. Obukhovsky nad V. N. Pomerantsev \textit{"Description of intermediate- and short-range NN nuclear force within a covariant effective field theory"}, Ann. Phys. 320, 71 (2005), \href{http://arxiv.org/abs/nucl-th/0505026}{[arXiv:nucl-th/0505026]}.
\bibitem{Idaho2} D. R. Entem and R. Machleidt, \textit{"Accurate charge-dependent nucleon-nucleon potential at fourth order of chiral perturbation theory"}, Phys. Rev. C 68, 041001 (2003), \href{http://arxiv.org/abs/nucl-th/0304018}{[arXiv:nucl-th/0304018]}.
\bibitem{nucl-th/9707002} L. Engvik, M. Hjorth-Jensen, R. Machleidt, H. Muether and A. Polls, \textit{"Modern nucleon-nucleon potentials and symmetry energy in infinite matter"}, Nucl. Phys. A 627, 85 (1997), \href{http://arxiv.org/abs/nucl-th/9707002}{[arXiv:nucl-th/9707002]}.
\bibitem{nucl-th/0407003} L. Coraggio, A. Covello, A. Gargano, N. Itaco, T. T. S. Kuo and R. Machleidt, \textit{"Nuclear structure calculations and modern nucleon-nucleon potentials"}, Phys. Rev. C 71, 014307 (2005), \href{http://arxiv.org/abs/nucl-th/0407003}{[arXiv:nucl-th/0407003]}.
\bibitem{nucl-th/0305035} S. K. Bogner, T. T. S. Kuo and A. Schwenk, \textit{"Model independent low momentum nucleon interaction from phase shift equivalence"}, Phys. Rept. 386, 1 (2003), \href{http://arxiv.org/abs/nucl-th/0305035}{[arXiv:nucl-th/0305035]}.
\bibitem{0901.0012} K.-Y. Kim and I. Zahed, \textit{"Nucleon-Nucleon potential from holography"}, JHEP 0903, 131 (2009), \href{http://arxiv.org/abs/0901.0012}{[arXiv:0901.0012 [nucl-th]]}.
\bibitem{0901.4449} K. Hashimoto, T. Sakai and S. Sugimoto, \textit{"Nuclear force from string theory"}, Prog. Theor. Phys. 122, 427 (2009), \href{http://arxiv.org/abs/0901.4449}{[arXiv:0901.4449 [nucl-th]]}.
\end{thebibliography}
\end{document}